\documentclass[11pt,a4paper,turkish,english,nofootinbib]{revtex4}
\usepackage{lmodern}

\usepackage[T1]{fontenc}
\usepackage[latin9]{inputenc}
\usepackage{babel}
\usepackage{amsmath}
\usepackage{amssymb}
\usepackage{graphicx}
\usepackage{esint}
\usepackage[unicode=true,pdfusetitle,
 bookmarks=true,bookmarksnumbered=false,bookmarksopen=false,
 breaklinks=false,pdfborder={0 0 1},backref=false,colorlinks=false]
 {hyperref}

\makeatletter

\pdfpageheight\paperheight
\pdfpagewidth\paperwidth

\newcommand{\noun}[1]{\textsc{#1}}

\@ifundefined{textcolor}{}
{%
 \definecolor{BLACK}{gray}{0}
 \definecolor{WHITE}{gray}{1}
 \definecolor{RED}{rgb}{1,0,0}
 \definecolor{GREEN}{rgb}{0,1,0}
 \definecolor{BLUE}{rgb}{0,0,1}
 \definecolor{CYAN}{cmyk}{1,0,0,0}
 \definecolor{MAGENTA}{cmyk}{0,1,0,0}
 \definecolor{YELLOW}{cmyk}{0,0,1,0}
 }

\usepackage{latexsym}\usepackage{bm}

\makeatother

\makeatother

\begin{document}

\title{MASSIVE HIGHER DERIVATIVE GRAVITY THEORIES}

\author{\.{I}brahim Güllü }

\email{e075555@metu.edu.tr}

\selectlanguage{english}%

\affiliation{Department of Physics,\\
 Middle East Technical University, 06531, Ankara, Turkey}
\begin{abstract}
In this thesis massive higher derivative gravity theories are analyzed
in some detail. One-particle scattering amplitude between two covariantly
conserved sources mediated by a graviton exchange is found at tree-level
in $D$ dimensional (Anti)-de Sitter and flat spacetimes for the
most general quadratic curvature theory augmented with the Pauli-Fierz
mass term. From the amplitude expression, the Newtonian potential
energies are calculated for various cases. Also, from this amplitude
and the propagator structure, a three dimensional unitary theory is
identified. In the second part of the thesis, the found three dimensional
unitary theory is studied in more detail from a canonical point of
view. The general higher order action is written in terms of gauge-invariant
functions both in flat and de Sitter backgrounds. The analysis is
extended by adding static sources, spinning masses and the gravitational
Chern-Simons term separately to the theory in the case of flat spacetime.
For all cases the microscopic spectrum and the masses are found. In
the discussion of curved spacetime, the masses are found in the relativistic
and non-relativistic limits. In the Appendix, some useful calculations
that are frequently used in the bulk of the thesis are given.\tableofcontents{}\end{abstract}
\maketitle

\newpage

\begin{acknowledgments}
I express sincere appreciation to {\it Prof. Dr. Bayram Tekin} for his guidance and insight throughout the research. 
I  thank him to teach me how to think and cope with research problems. I am also grateful to him for giving me the feeling of 
creating and proposing research areas. I am grateful to my collaborator {\it Tahsin \c{C}agr{\i} \c{S}i\c{s}man} for studying and 
doing research with me. I also thank {\it TUB\.{I}TAK} for supporting me as a research student in two projects, 110T339 
({\it The Classical and Quantum Mechanical Structures of Higher Derivative Gravity Theories}) and 104T177 
({\it Research on the Mathematical Background of Nonperturbative Superstring Theories}), during my PhD work. 
I also sincerely thank all the jury members for participating in my thesis defense and reading and doing corrections in my thesis. 
 
To my girlfriend, {\it P{\i}nar Yaran}, I offer sincere thanks for her great understanding and motivation throughout the writing process. 
I am also grateful to my friends {\it Ertan Zafero\u{g}lu}, {\it Ozan Keysan}, {\it H\"useyin Kapukaya} and {\it Kurtulu\c{s} Y{\i}lmaz} 
for their moral support and for sharing life with me during the most difficult part of my PhD work.

I am indebted to my family for their moral and logistic supports, encouragement and love throughout the 
time that has gone for getting my PhD degree.  
\end{acknowledgments}

\newpage

\part{Introduction }

\section{Introduction}

General Relativity (GR) \cite{Einstein} can be thought as a modification
(albeit a major one) of Newtonian gravity. The history of gravity
starts with Newton's $\frac{1}{r^{2}}$ force law in $1687$. The
precise form of this law is 
\begin{equation}
F=-\frac{Gm_{1}m_{2}}{r^{2}},\label{newton}
\end{equation}
where $G$ is the universal Newton's constant and $m_{i}$ are the
masses of the interacting particles. The minus sign indicates that
the force is attractive. The Celestial mechanics gave successful verifications
of this law. A well-known example of the success of Newton's theory
is the prediction of a new planet. After the discovery of the orbit
of Uranus there were attempts to calculate this orbit theoretically.
However, these attempts did not work exactly since there were deviations
between the calculated and observed orbits \cite{Grosser,Goldhaber}.
Using Newton's law (\ref{newton}) the location of Neptune was predicted
and just after this calculation it was observed exactly at that location
\cite{Grosser,Goldhaber}. Hence an {}``outer'' mass was responsible
for the perturbations of the orbit of Uranus. Newton's law, with the
help of {}``dark matter'' (Neptune in this case) worked well for
the case of Uranus. After this resolution, there were studies of all
planets. When Mercury was studied, another problem was found: The
calculations of precession of Mercury's perihelion was in contradiction
with the observed value. The calculations gave larger value than the
observed ones. The first attempt to solve this problem was again to
invoke the {}``dark matter'' idea. It was thought that there must
be a planet, Vulcan, in the solar system \cite{Goldhaber,Verrier}.
However, this did not solve the problem since Vulcan was never found.
The solution came with modifying gravity: replacing Newtonian gravity
with the Einsteinian one \cite{Goldhaber}. 

In GR the spacetime geometry is determined by the matter, energy,
pressure etc. and the dynamics of matter or light is determined by
the geometry of spacetime. GR is constructed with the idea of equivalence
principle and the general coordinate invariance (and with the added
assumption that the equations are wave-type equations with second
derivatives on the basic fields). The Einstein equation is 
\begin{equation}
R_{\mu\nu}-\frac{1}{2}g_{\mu\nu}R=\kappa T_{\mu\nu},\label{Einstein_eq}
\end{equation}
where the left hand side is the geometry part and it is known as the
Einstein tensor and the right hand side is the matter part. In (\ref{Einstein_eq}),
$R_{\mu\nu}$ and $R$ are the Ricci tensor and Ricci scalar, respectively.
These quantities can be thought as a measure of the curvature of spacetime.
They depend on the spacetime metric $g_{\mu\nu}$ and its derivatives.
The connection is taken to be metric compatible ($\nabla g=0$). In
the right hand side, there is a coupling constant related to the Newton's
constant which in four dimensions is $\kappa=8\pi G$, and $T_{\mu\nu}$
is the energy-momentum tensor coming from the matter sector.

Einstein's equation, right after was introduced, was immediately applied
to solve the then more-than fifty year old problem of Mercury's orbit.
After the Schwarzschild solution was found, which is the unique static
spherically symmetric time-independent solution to (\ref{Einstein_eq}),
the perihelion precession of Mercury was calculated and the answer
was found to be compatible with the observed value \cite{Goldhaber,Weinberg3,Inverno,Carroll}.
There were also other predictions that came out from GR such as the
bending of light passing by the sun and the gravitational redshift
of light which were also supported by observations. Inspite of these
successful solutions and predictions there are some observations that
GR in its basic form as given in (\ref{Einstein_eq}) cannot explain.
The data taken from supernova explosions \cite{Risse,Perlmutter,Hinterbichler}
show that the universe has an accelerated expansion. GR cannot explain
this phenomenon in its pure form%
\footnote{$T_{\mu\nu}$ is taken as the standard energy-momentum tensor of the
matter.%
}. 

If GR is taken as the correct theory, there must exist a dark energy
component in the matter-energy budget of the universe which can be
represented as a constant term, the cosmological constant, $\Lambda$.
The density of the dark energy is estimated to be $75\%$ of the total
energy density of the universe and numerically it reads
\begin{equation}
\rho_{\Lambda}=4\times10^{-6}\frac{GeV}{cm^{3}},\label{vacuum_dens}
\end{equation}
which is very small compared to the vacuum energy density that comes
from the energy density of the vacuum of quantum field theory (QFT);
\begin{equation}
\rho_{\Lambda}\sim10^{-123}\rho_{QFT},\label{cos_cons_prob}
\end{equation}
\cite{Hinterbichler,Rubakov,Weinberg}. This mismatch of the {}``experimental''
and {}``theoretical'' values of the energy density of the vacuum
is known as the cosmological constant problem. Therefore, it can be
thought that GR is not the full theory and it must be modified in
the infrared (IR), namely at ultra-large distances or ultra-weak interacting
regimes, such that without a need of dark energy the accelerating
expansion of the universe happens. {[}Admittedly, this alone will
not solve the question of why QFT vacuum is almost empty.{]} One of
the suitable candidates for this modification is to give a tiny mass
to the graviton in the theory, in such a way that massive GR still
passes the solar system tests.

From the perspective of QFT, at low energies, GR can be thought as
a weakly interacting massless spin-2 field. The story starts with
the classification of particles with respect to their spins and masses,
as degrees of freedom \cite{Hinterbichler,Weinberg1}. The representations
of these degrees of freedom are fields. For spin-0 mode of massless
particles the representations is a scalar field, and for spin-1 it
is a vector field. For spin-2 mode the symmetry is the general coordinate
invariance when the interactions are taken into account, and around
flat and maximally symmetric spaces, one has a massless helicity-2
particle. This points to GR \cite{Hinterbichler,Gupta,Kraichnan,Weinberg2,Deser1,Boulware,Fang,Wald}
with the action 
\begin{equation}
I=\frac{1}{\kappa}\int d^{4}x\sqrt{-g}R\,,\label{Ein-Hilb}
\end{equation}
where $g$ is the determinant of the spacetime metric and the rest
are defined in section \ref{The-Curved-Spacetime}. Therefore, it
is also natural to combine the quantum theory with GR.

There are other motivations to try to modify GR and perhaps construct
a quantum gravity theory. The Schwarzschild solution of (\ref{Einstein_eq})
is a black hole with a curvature singularity, to explain or resolve
this type of singularities a quantum gravity theory is needed. Also,
to understand the very early universe when both the quantum effects
and gravitational effects are dominant at the same time, a unification
of gravity and quantum mechanics is necessary. There are attempts
to construct quantum gravity theories. String theory is one such candidate.
Higher curvature gravity theories also are candidates. With this motivation
it is also possible to add higher curvature terms to Einstein gravity
(\ref{Ein-Hilb}). The higher curvature terms are negligible at low
energies, but they dominate at high energy domains. Another reason
to introduce the higher curvature terms is that the Einstein-Hilbert
action (\ref{Ein-Hilb}) is not renormalizable. Higher curvature terms
make the theory renormalizable but ruin the unitarity, namely yields
negative norm states in the scattering matrix \cite{Stelle1,Stelle2}. 

To summarize the argument, we can say that GR needs modification at
both the IR and ultraviolet (UV) regimes. Therefore, in this thesis
Einstein gravity will be modified both by adding mass and higher curvature
terms. Hence we modify the theory at both the UV and IR regimes.

The outline of the thesis is as follows: In the next sections of this
chapter the massive gravity and higher derivative gravity models are
discussed separately. In the massive gravity model mainly the Pauli-Fierz
\cite{Pauli1,Pauli2} model is studied and also the nature of the
van Dam-Veltman-Zakharov \cite{vanDam,Zakharov} discontinuity between
the strictly massless theory and the arbitrarily small mass theory
is discussed. Then some specific three dimensional massive theories
are studied. After that, some technical details will be given about
the relevant spacetimes, the cosmological constant and the higher
dimensional gravity models. At the end of this Chapter the higher
derivative Pais-Uhlenbeck oscillators are briefly reviewed. In the
second chapter the general higher curvature massive gravity theory
is analyzed. Its unitarity structure is studied by computing the tree
level scattering amplitude in generic $D$ dimensions. Also, the Newtonian
potentials are calculated. The third chapter is devoted to the analysis
of the three dimensional unitary theory. These two chapters are based
on the papers {}``\emph{Massive Higher Derivative Gravity in D-dimensional
Anti-de Sitter Spacetimes}'' \cite{Gullu1} and {}``\emph{Canonical
Structure of Higher Derivative Gravity in 3D }'' \cite{Gullu2} respectively.
Then, the conclusion part comes. Also, an appendix part is added so
that some of the calculations can be followed easily.

\section{Modifying Gravity}

In this part the modification of the Einstein theory is discussed.
By modifying gravity we mean that we change the degrees of freedom
of the theory in some way. First the mass term is added to the theory
which changes the degrees of freedom from two to five in $3+1$ dimensions.
Also, adding higher derivative terms change the degrees of freedom.
The massive gravity theory is discussed first. Then the higher derivative
gravity theory is considered. At the end of this part, two 3-dimensional
massive higher derivative gravity theories are introduced.

\subsection{Modifying Gravity with Mass Terms:}

There are two ways to give mass to a gravity theory. One is to add
directly the mass to the action. The other way is to introduce scalar
fields. When these fields are evaluated at the background the general
covariance is broken and the graviton gets mass. Both ways end up
in the same class of theories. In this discussion we will follow the
first way. The mass is added to the Einstein-Hilbert theory in a Lorentz
invariant way as follows 
\begin{equation}
I=\int d^{4}x\sqrt{-g}\left[\frac{1}{\kappa}R-\frac{1}{2}m^{2}h^{\mu\nu}\left(h_{\mu\nu}-\eta_{\mu\nu}h\right)\right],\label{PF_action}
\end{equation}
which is known as the Pauli-Fierz (PF) action, the second part is
the PF mass term \cite{Pauli1,Pauli2} and $m^{2}$ is the mass parameter.
We first consider this theory in a flat background. In (\ref{PF_action})
$\eta_{\mu\nu}$ is the flat spacetime metric and $h_{\mu\nu}$ is
the linear part of the metric perturbation, $g_{\mu\nu}=\eta_{\mu\nu}-h_{\mu\nu}$,
and $h=\eta_{\mu\nu}h^{\mu\nu}$ is trace of this metric perturbation.
To see more explicitly how the added PF part gives mass to the graviton,
we should linearize the theory around the background spacetime. From
both (\ref{Einstein_eq}) and (\ref{Ein-Hilb}), the linearized Einstein
tensor can be written in the absence of sources as
\begin{equation}
\mathcal{G}_{\mu\nu}^{L}=\frac{1}{2}\left(\partial^{\sigma}\partial_{\mu}h_{\nu\sigma}+\partial^{\sigma}\partial_{\nu}h_{\mu\sigma}-\partial^{2}h_{\mu\nu}-\partial_{\mu}\partial_{\nu}h\right)-\frac{1}{2}\eta_{\mu\nu}\left(-\partial^{2}h+\partial^{\sigma}\partial^{\rho}h_{\sigma\rho}\right)=0,\label{lineer_Einstein}
\end{equation}
where $\partial^{2}\equiv\partial_{\mu}\partial^{\mu}$ and the linear
parts of the Ricci tensor and Ricci scalar are 
\begin{equation}
R_{\mu\nu}^{L}=\frac{1}{2}\left(\partial^{\sigma}\partial_{\mu}h_{\nu\sigma}+\partial^{\sigma}\partial_{\nu}h_{\mu\sigma}-\partial^{2}h_{\mu\nu}-\partial_{\mu}\partial_{\nu}h\right),\,\, R^{L}=-\partial^{2}h+\partial^{\sigma}\partial^{\mu}h_{\sigma\mu}.\label{Ricci_tens}
\end{equation}
If the metric perturbation is constrained to be transverse and traceless,
that are $\partial^{\mu}h_{\mu\nu}=0$ and $h=0$, then (\ref{lineer_Einstein})
yields 
\begin{equation}
\partial^{2}\left(h_{\mu\nu}-\eta_{\mu\nu}h\right)=0,\label{box_pf}
\end{equation}
and for massive particles it is known that the Klein-Gordon equation
must hold that is 
\begin{equation}
\left(\partial^{2}-m^{2}\right)\phi=0,\label{KG_equation}
\end{equation}
 for the mostly plus signature of the metric. To write (\ref{box_pf})
as a massive field equation, the PF term must be added:
\begin{equation}
\left(\partial^{2}-m^{2}\right)\left(h_{\mu\nu}-\eta_{\mu\nu}h\right)=0.\label{KG_PF}
\end{equation}
Since this term cannot be obtained from the curvature terms, it is
added to the action (\ref{Ein-Hilb}) by brute force. Later on, it
will be seen that modifying GR with higher curvature terms will generate
this mass term automatically when the coupling constants of the higher
curvature terms have a special combination. From the transverse and
traceless conditions it can be seen easily that the PF theory describes
a massive spin-2 field. The metric perturbation is a symmetric rank-2
tensor, so that it has ten independent degrees of freedom in $3+1$
dimensions. However, from the transversality condition there are four
constraints which eliminate four degrees of freedom leaving six. The
tracelessness condition also eliminates one degree of freedom and
the theory has five degrees of freedom. We must also note that this
mass term is the unique ghost and tachyon-free Lorentz-invariant combination.
When the sign is changed in the middle of the PF mass term, then it
produces tachyons in its propagator structure \cite{Hinterbichler}. 

The massless limit of the PF theory must tend to Einstein gravity
as the usual continuity arguments in physics dictate. However, the
$m^{2}\rightarrow0$ limit does not give the results of the $m^{2}=0$
theory. For two point sources the interaction potential, in the Newtonian
limit of GR, in four dimensions is 
\begin{equation}
U=-\frac{Gm_{1}m_{2}}{r}.\label{newt_pot}
\end{equation}
However, the massive theory gives 
\begin{equation}
U=-\frac{4}{3}\frac{Gm_{1}m_{2}}{r},\label{vdvz_pot}
\end{equation}
where the Newtonian potential is greater than the usual one. This
effect is known as the van Dam-Veltman-Zakharov (vDVZ) discontinuity%
\footnote{Redefinition of the Newton's constant does not solve the problem,
since then the deflection of light changes by $25\%$.%
} \cite{vanDam,Zakharov}. The linearized PF theory is different from
the linearized GR in the massless limit. This discontinuity appears
in flat backgrounds \cite{higuchi,Porrati1,kogan}. However, in curved
spacetime this discontinuity does not appear. From the general amplitude
equation (\ref{mainresult}) it can be seen more explicitly. There
is a $\frac{M^{2}}{\Lambda}$ fraction by which the flat spacetime
limit and massless limit do not commute. Going to flat spacetime limit
first the vDVZ discontinuity appears in the massless limit, but taking
the massless limit first the discontinuity disappears in four dimensions
\cite{Gullu1,Porrati1}. Hence, the introduction of a cosmological
constant to the theory solves this problem since the smallness of
the mass can be compared with another measurable quantity. {[}One
could argue that discontinuity has been mainly replaced by the non-commutativity
of the limits, which is a valid objection.{]} From the pole structure
of the general amplitude equation (\ref{mainresult}), it can be seen
that for de Sitter spacetime ($\Lambda>0$) a pole produces a tachyon
and it is absent for anti-de Sitter ($\Lambda<0$) spacetime%
\footnote{We should also note that Vainshtein claimed that \cite{vainshtein}
the discontinuity disappears at the nonlinear level once the finite
Schwarzschild radius of one of the scattering particles is taken into
account. %
}.

\subsection{Modifying Gravity with Higher Curvature Terms:}

To make GR perturbatively renormalizable, quadratic curvature terms
$\alpha R^{2}+\beta R_{\mu\nu}R^{\mu\nu}+\gamma R_{\mu\sigma\nu\rho}R^{\mu\sigma\nu\rho}$
must be added to the Einstein-Hilbert action (\ref{Ein-Hilb}) \cite{Stelle1}.
But adding these terms makes the theory nonunitary because of the
non-decoupling ghost introduced by the middle term $\beta R_{\mu\nu}R^{\mu\nu}$
\cite{Stelle2}. When this term is excluded the unitarity of the theory
is regained but the theory becomes nonrenormalizable. To have a consistent
quantum gravity, both of these properties must be provided. 

Before discussing further, the notion of unitarity must be explained.
With unitarity, it is meant that the theory must be both ghost and
tachyon free. Ghost is a particle that has a negative kinetic energy
and tachyon is a particle that has negative mass-squared (in flat
space). In a more technical language the signs of the propagators
must be correct that is (for the $\left(-,+,+,+\right)$ signature)
the (scalar propagator) 
\begin{equation}
D\left(p\right)\sim\frac{1}{p^{2}+m^{2}}\label{propagator}
\end{equation}
where $p$ is the four-momentum. In the canonical form, the Lagrangian
should be 
\begin{equation}
\mathcal{L}=\frac{1}{2}\psi\left(t,\vec{x}\right)\left(\square-m^{2}\right)\psi\left(t,\vec{x}\right),\label{can_propagator}
\end{equation}
the kinetic term is the d'Alembertian operator and it must be positive.
The mass-squared must again be greater than zero. Note that only the
tree-level unitarity is considered here. The Feynman diagram of the
tree-level scattering amplitude is shown in the figure 
\begin{figure}
\begin{centering}
\includegraphics{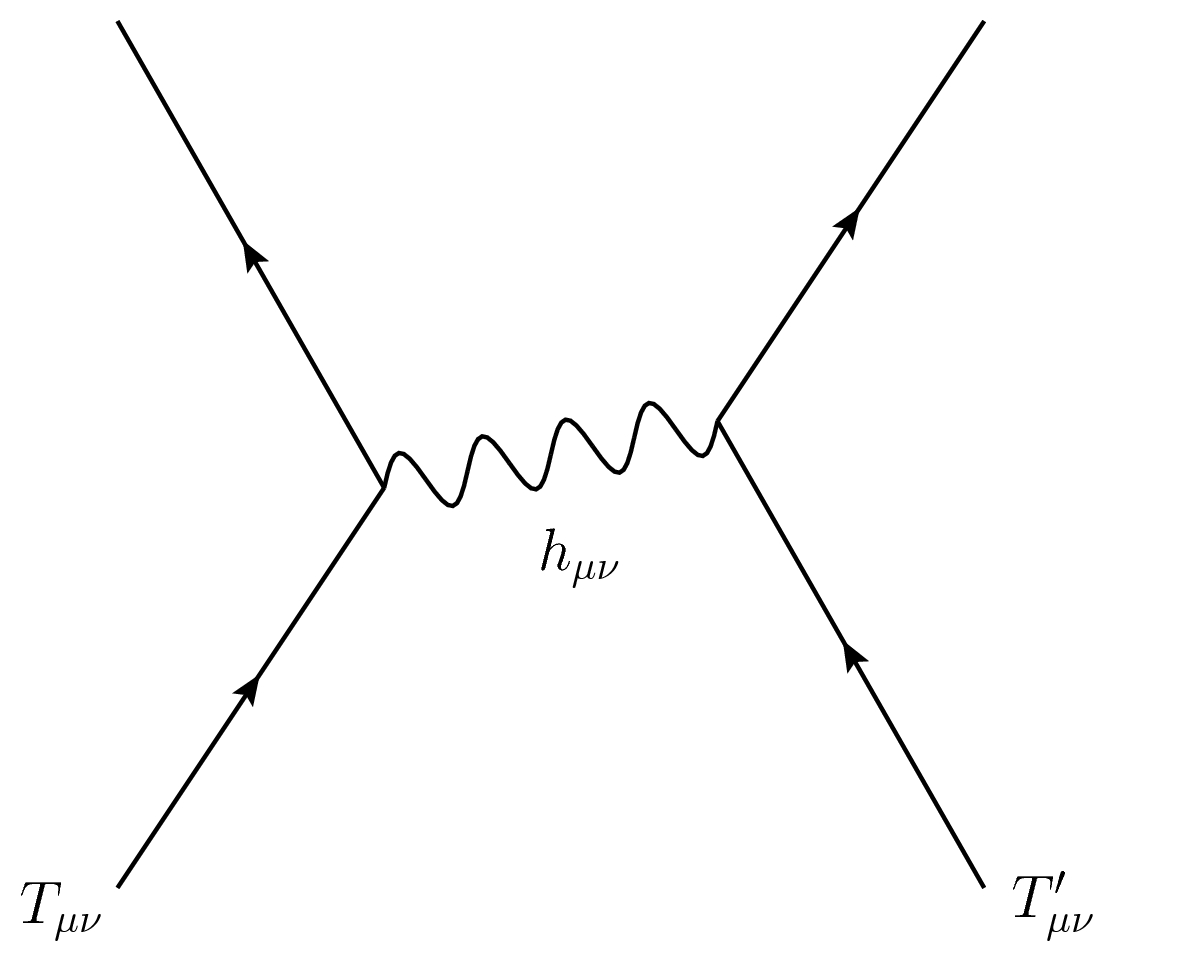}
\par\end{centering}

\caption{The tree-level one-particle scattering amplitude between two background
covariantly conserved sources mediated by a graviton exchange.}
\end{figure}
.

\subsection{Topologically Massive Gravity}

In the discussion to follow, we will need a specific 3D theory that
has been studied a lot in the literature. This is the topologically
massive gravity (TMG) theory introduced in \cite{Djt1,Djt2} with
the action
\begin{equation}
I=\int d^{3}x\sqrt{-g}\left[\frac{1}{\kappa}R-\frac{1}{2\mu}\epsilon^{\lambda\mu\nu}\Gamma_{\phantom{\rho}\lambda\sigma}^{\rho}\left(\partial_{\mu}\Gamma_{\phantom{\sigma}\rho\nu}^{\sigma}+\frac{2}{3}\Gamma_{\phantom{\sigma}\mu\beta}^{\sigma}\Gamma_{\phantom{\beta}\nu\rho}^{\beta}\right)\right].\label{grav_chern_simon}
\end{equation}
Here $\mu$ is the coupling constant of the gravitational Chern-Simons
term. The sign of $\mu$ is not fixed. $\epsilon^{\lambda\mu\nu}$
is the three dimensional anti-symmetric tensor which is connected
to the Levi-Civita symbol as $\epsilon^{\lambda\mu\nu}\equiv-\frac{1}{\sqrt{-g}}\tilde{\epsilon}^{\lambda\mu\nu}$.
The pure Einstein-Hilbert action does not propagate any degrees of
freedom in 3D. After the gravitational Chern-Simons term is introduced,
TMG propagates a single massive particle with helicity $+2$ or $-2$
(not both since the theory is not parity-invariant). The action for
this theory can be written in terms of the gauge-invariant functions%
\footnote{Note that the gauge-invariant functions will be discussed later in
the third chapter.%
} (\ref{gen_cs_action}) as 
\begin{equation}
I=\frac{1}{2}\int d^{3}x\,\left[\frac{1}{\kappa}\left(\phi q+\sigma^{2}\right)+\frac{1}{\mu}\sigma\left(q+\Box\phi\right)\right].\label{grav_chern_siom_gauge}
\end{equation}
Taking the variation with respect to the $q$ field one of the other
fields can be eliminated. This variation yields
\begin{equation}
\delta q:\;\frac{1}{\kappa}\phi+\frac{1}{\mu}\sigma=0,\label{var_q}
\end{equation}
from which $\phi=-\frac{\kappa}{\mu}\sigma$ follows. Putting this
result back into (\ref{grav_chern_siom_gauge}) yields the action
\begin{equation}
I=-\frac{\kappa}{2\mu^{2}}\int d^{3}x\,\left[\sigma\Box\sigma-\frac{\mu^{2}}{\kappa^{2}}\sigma^{2}\right].\label{cs_can_form2}
\end{equation}
From (\ref{cs_can_form2}) it can be seen that this theory propagates
only a single massive degree of freedom with mass $m^{2}=\frac{\mu^{2}}{\kappa^{2}}$.
Not to have a negative kinetic energy term, $\kappa$ must be chosen
negative which yields the {}``wrong-sign'' Einstein-Hilbert term.
In fact this model propagates a single massive scalar spin-2 mode
\cite{Djt1}. Since the spin of the particle depends on the sign of
$\mu$, this model is a parity violating theory, as noted above. The
detailed discussion and calculations for this model is given in Chapter
3.

\subsection{New Massive Gravity and Critical Gravity}

Another interesting higher derivative massive gravity theory is the
recently introduced {}``New Massive Gravity'' \cite{Bht1,Bht2}.
The action for this model is 
\begin{equation}
I=-\frac{1}{\kappa}\int d^{3}x\,\sqrt{-g}\left\{ -R+\frac{1}{m^{2}}\left(R_{\mu\nu}^{2}-\frac{3}{8}R^{2}\right)\right\} ,\label{nmg}
\end{equation}
where there is a relation between the coupling constants of the higher
curvature terms as $8\alpha+3\beta=0$ and the mass reads in terms
of the coupling constants as $m^{2}=-\frac{1}{\kappa\beta}$. For
the unitarity of the theory, $\kappa$ must be negative, and $\beta>0$.
This model also implies the wrong-sign Einstein-Hilbert term. 

This model is not a higher derivative theory at the linearized level.
The canonical form of (\ref{nmg}) is (see more details in section
\ref{Higher-derivative-spin-2-in-flat}) 
\begin{equation}
I=\int d^{3}x\,\left\{ -\frac{1}{2\kappa}\left(\phi\square\phi+\frac{1}{\kappa\beta}\phi^{2}\right)+\frac{\beta}{2}\left(\sigma\Box\sigma+\frac{1}{\kappa\beta}\sigma^{2}\right)\right\} ,\label{can_nmg}
\end{equation}
where it can be seen that this model has two modes propagating with
the same masses. Therefore, this theory is parity invariant. Also,
in this form the necessity of negative $\kappa$ can be seen explicitly.

After this theory was introduced \cite{Bht1} a research activity
started in massive 3D gravity theories. Especially in Anti-de Sitter
(AdS) backgrounds some classical solutions and conserved charges of
this theory was given \cite{Gullu1,Bht2,Clement1,Sun1,Sun2,Giribet1,Oliva,gurses,Aliev_Ahmedov1,Aliev_Ahmedov2,Aliev_Ahmedov3,Gurses_Sisman}.
The question of how the theory gains mass was answered in \cite{Suat,Suat_Tanhayi}.
In this work Higgs mechanism of the general quadratic theory in three
dimensions was given by writing the theory in a Weyl invariant form.
Breaking the Weyl invariance in the background, graviton gains mass,
scalar field remains massless and the Weyl gauge field gains mass
or can remain massless depending on the parameters.

The four and higher dimensional extensions of this theory, called
{}``the critical gravity'', was also investigated in \cite{Lu,Sisman,Porrati2}.
For these cases the special points are again set to zero $3\alpha+\beta=0$
and $4\alpha(D-1)+D\beta=0$ for four and the generic $D$ dimensional
cases respectively, for which the massive scalar mode vanishes \cite{Gullu1}.
In the critical gravity, the mass of spin-2 is set to zero and the
energy of black holes become zero for the AdS vacuum. It is also found
that the energy of the massless spin-2 modes vanishes on-shell. There
remains only spin-2 logarithmic modes with positive energy.

\section{Constant Curvature Spacetimes}

In our discussion we will need the maximally symmetric backgrounds
which are the flat, Anti-de Sitter (AdS) and de Sitter (dS) geometries.
Maximally symmetric means that the curvature of the spacetime is constant
everywhere and in any direction. Therefore, if the curvature is known
in one point it is enough to know the curvature at every point. There
are only three maximally symmetric spacetimes that are Minkowski,
which is flat so that the curvature is zero everywhere, dS and AdS.
These spacetimes are classified with respect to the curvature scalar
$R$, the dimensionality of spacetime and the signature of the metric.
De Sitter space is a $\left(D-1\right)$-dimensional surface embedded
into a $D$ dimensional flat spacetime for which the signature of
the metric reads $\left(-,+,+,...,+\right)$. For the AdS spacetime
the signature of the embedding metric reads $\left(-,+,+,...,-\right)$.
For AdS spacetime there is no horizon but for dS there is a horizon
\cite{Abbott,Witten,Strominger}. Examples of the maximally symmetric
flat spaces are planes with Euclidean signature $(+,+,+)$ and its
higher dimensional generalizations. Spheres $\left(S^{D}\right)$
are dS or positively curved spacetimes, and hyperboloids $\left(H^{D}\right)$
are negatively curved or AdS spacetimes.

The Riemann tensor of a maximally symmetric $D$-dimensional spacetime
is 
\begin{equation}
R_{\mu\sigma\nu\rho}=\frac{R}{D\left(D-1\right)}\left(g_{\mu\nu}g_{\sigma\rho}-g_{\mu\rho}g_{\sigma\nu}\right),\label{Riemann}
\end{equation}
where the curvature scalar $R$ is constant and $D$ is the dimensionality
of the spacetime \cite{Carroll}. For AdS or dS the curvature scalar
is $R=\frac{2D\Lambda}{\left(D-2\right)}$. Here, $\Lambda$ is the
cosmological constant and for $\Lambda<0$ the spacetime becomes AdS,
negatively curved spacetime, and for $\Lambda>0$ it is dS, positively
curved spacetime. If $\Lambda=0$ then the spacetime is flat. 

The cosmological constant is related to the vacuum energy density
in $D=4$ as 
\begin{equation}
\rho_{\text{vac}}=\frac{\Lambda}{8\pi G}.\label{vac_ener}
\end{equation}
Here $\rho$ is the energy density of the vacuum which is proportional
to the pressure of a perfect fluid $T_{\mu\nu}=\left(\rho+p\right)U_{\mu}U_{\nu}+pg_{\mu\nu}$.
To see this proportionality the energy-momentum tensor in the Einstein
equation is divided into matter and vacuum pieces,
\begin{equation}
R_{\mu\nu}-\frac{1}{2}g_{\mu\nu}R=8\pi G\left(T_{\mu\nu}^{\text{M}}+T_{\mu\nu}^{\text{vac}}\right),\label{Einstein_mat_eng}
\end{equation}
where the vacuum energy-momentum tensor is $T_{\mu\nu}^{\text{vac}}=-\rho_{\text{vac}}g_{\mu\nu}$.
The vacuum energy-momentum tensor becomes a perfect if $-\rho_{\text{vac}}=p_{\text{vac}}$.
The Einstein equation can be written with a cosmological constant
term as 
\begin{equation}
R_{\mu\nu}-\frac{1}{2}g_{\mu\nu}R+\Lambda g_{\mu\nu}=8\pi GT_{\mu\nu}.\label{cos_Einstein}
\end{equation}
 The relation between the vacuum energy density and the cosmological
constant can be seen from these equations \cite{Carroll}. The maximally
symmetric solutions of this equation are dS and AdS (and also Minkowski)
spacetimes, for the pure vacuum case.

In the first part of the analysis the higher curvature massive gravity
is studied in a $D$ dimensional (A)dS spacetime $\left(\text{AdS}_{D}\right)$.
In this way the effects of higher dimensions can be seen. Also, with
this general approach a unitary theory may be found in $D\geq5$.

\section{Canonical Analysis}

In this part some technical details of the canonical analysis are
given. To put a Lagrangian into the canonical form means to write
it in terms of scalar fields that are in the form of harmonic oscillator
Lagrangians. The field equations of these Lagrangians are Klein-Gordon
equation. So that, KG equation can be seen as simple harmonic oscillators.
From these oscillators the particle spectrum can be studied. Also,
for investigation of the unitarity of a theory, this form is useful.
Therefore, first the KG equation for free particle is reviewed. Then
the generalization to higher derivative of this equation known as
Pais-Uhlenbeck oscillator is discussed.

\subsection{The Klein-Gordon Equation}

The story starts from the quantization of the non-relativistic energy
equation for a free particle. The energy of a free particle is 
\begin{equation}
E=\frac{p^{2}}{2m},\label{energy_free_part}
\end{equation}
where $E$ is the energy and $p$ is the momentum of the particle.
To quantize this, the variables are promoted to being operators by
taking $E\rightarrow H=i\hbar\partial_{t}$ and $\vec{p}\rightarrow-i\hbar\vec{\nabla}$,
where $\hbar$ is the reduced Planck constant, and $\frac{\partial}{\partial t}\equiv\partial_{t}=\partial_{0}$,
$\nabla\equiv\frac{\partial}{\partial x}$ for one dimensional space%
\footnote{Here the derivative is $\partial_{\mu}=\left(\frac{\partial}{\partial t},\,\nabla\right)$
and the four momentum is $P^{\mu}=\left(E,\,\vec{p}\right)=i\partial^{\mu}$
in the natural units.%
}. With these substitutions (\ref{energy_free_part}) turns to the
well known Schrödinger equation
\begin{equation}
i\hbar\partial_{t}\psi\left(t,\vec{x}\right)=-\frac{\hbar^{2}}{2m}\nabla^{2}\psi\left(t,\vec{x}\right),\label{Schrodinger_eq}
\end{equation}
where $\psi\left(t,\vec{x}\right)$ is a complex function. A relativistic
version of this equation can be obtained by upgrading the relativistic
dispersion relation%
\footnote{If the equation is not squared the operators will be inside the square
root $E=\sqrt{p^{2}c^{2}+m^{2}c^{4}}\rightarrow i\hbar\partial_{t}\psi=\sqrt{\left(-i\hbar\nabla\right)^{2}c^{2}+m^{2}c^{4}}\psi$.
This form has some problems in evaluating the operators. Also, it
is nonlocal. %
} \cite{Griffiths}
\begin{equation}
E^{2}=p^{2}c^{2}+m^{2}c^{4},\label{rela_ener_eq}
\end{equation}
to an operator equation 
\begin{equation}
\left(\frac{1}{c^{2}}\partial_{t}^{2}-\nabla^{2}+\frac{m^{2}c^{2}}{\hbar^{2}}\right)\psi\left(t,\vec{x}\right)=0,\label{rel_sch}
\end{equation}
where $c$ is the speed of light. Going to natural units $c=1,\;\hbar=1$
and defining the d'Alembertian operator $\square\equiv\partial_{\mu}\partial^{\mu}$,
where the components of the partial derivative is $\partial^{\mu}=\left(\partial^{0},\,-\vec{\nabla}\right)$
(in Minkowski background with the signature $\left(-,+,+,+\right)$)
yields%
\footnote{Note that for the signature $\left(+,-,-,-\right)$ the KG equation
reads $\left(\square+m^{2}\right)\psi=0$.%
}
\begin{equation}
\left(\square-m^{2}\right)\psi\left(t,\vec{x}\right)=0,\label{KG_eq}
\end{equation}
which is the classical relativistic wave equation of a massive particle.
Taking the Fourier transform of $\psi\left(t,\vec{x}\right)=\int\frac{d^{3}x}{2\pi}e^{i\vec{p}.\vec{x}}\psi\left(t,\vec{p}\right)$
and putting it into (\ref{rel_sch}), the equation of motion for a
harmonic oscillator is found with the frequency $\omega_{p}=\sqrt{|p|^{2}+m^{2}}$,
that is \cite{Peskin,Reinhardt} 
\begin{equation}
\left(\partial_{t}^{2}+\omega_{p}^{2}\right)\psi\left(t,\vec{p}\right)=0.\label{har_osc}
\end{equation}
In the field theoretical approach the same result can be found by
taking the real spin-0 particle Lagrangian density with mass $m$
\begin{equation}
\mathcal{L}\left(t,x\right)=-\frac{1}{2}\left(\partial_{\mu}\psi\partial^{\mu}\psi+m^{2}\psi^{2}\right),\label{spin0_lag}
\end{equation}
where the field depends both on time and space $\psi=\psi\left(t,x\right)$.
The Euler-Lagrange equations $\frac{\partial\mathcal{L}}{\partial\psi}=\partial_{\mu}\frac{\partial\mathcal{L}}{\mbox{\ensuremath{\partial\left(\ensuremath{\partial^{\mu}\psi}\right)}}}$
of (\ref{spin0_lag}) gives (\ref{KG_eq}) \cite{Peskin,Reinhardt}. 

After getting some basic informations about the relativistic wave
equation, the higher derivative terms can be added to the Lagrangian
to generalize this discussion. The field equations of the higher derivative
Lagrangians are known as the Pais-Uhlenbeck oscillators \cite{Pais}.

\subsection{Pais-Uhlenbeck Oscillators}

The Lagrangian density of the higher derivative real scalar field
is \cite{Sarioglu} 
\begin{equation}
\mathcal{L}=-\frac{1}{2}\psi\left(\prod_{i=1}^{N}\left(\square+m_{i}^{2}\right)\right)\psi.\label{pu_oscl_lag}
\end{equation}
For simplicity $N$ is set to two, i. e. $N=2$, and the examination
is done in the nonrelativistic limit, that is all the space derivatives
are dropped. For these conditions the Lagrangian density becomes 
\begin{equation}
\mathcal{L}=-\frac{1}{2}\left\{ \ddot{q}^{2}-\left(\omega_{1}^{2}+\omega_{2}^{2}\right)\dot{q}^{2}+\omega_{1}^{2}\omega_{2}^{2}q^{2}\right\} ,\label{sec_pu_lag}
\end{equation}
where $\omega_{i}$ is used instead of $m_{i}$. Also, note that the
$\omega_{1}=\omega_{2}$ case will be different from the $\omega_{1}\ne\omega_{2}$
case, where the first case is the degenerate case and the Lagrangian
density becomes purely quadratic
\begin{equation}
\mathcal{L}=-\frac{1}{2}\left\{ \ddot{q}+\omega^{2}q\right\} ^{2}.\label{pur_quad}
\end{equation}
The Euler-Lagrange equation for (\ref{sec_pu_lag}) is 
\begin{equation}
\frac{d^{2}}{dt^{2}}\left(\frac{\partial\mathcal{L}}{\partial\ddot{q}}\right)-\frac{d}{dt}\left(\frac{\partial\mathcal{L}}{\partial\dot{q}}\right)+\frac{\partial\mathcal{L}}{\partial q}=0,\label{eul_lag_eq}
\end{equation}
which can be obtained by taking the variation of (\ref{sec_pu_lag})
with respect to the variable $q=q\left(t\right)$. For n-th order
variational problem (\ref{eul_lag_eq}) takes the form 
\begin{equation}
\sum_{n=0}^{N}\left(-1\right)^{n}\frac{d^{n}}{dt^{n}}\left(\frac{\partial\mathcal{L}}{\partial q^{\left(n\right)}}\right)=0\label{gen_eul_lag_eq}
\end{equation}
in the non-relativistic limit. This equation (\ref{gen_eul_lag_eq})
can be obtained by taking the variation of the general Lagrangian
$L=L\left(q,\dot{q},\ddot{q},\dots,q^{\left(n\right)}\right)$, which
is \cite{Ostrogradski,Simon} 
\begin{equation}
\delta S=\int_{t_{i}}^{t_{f}}dt\delta q\left[-\sum_{n=0}^{N}\left(-\frac{d}{dt}\right)^{n}\frac{\partial L}{\partial q^{\left(n\right)}}\right]+\left[\sum_{n=0}^{N-1}P_{q^{\left(n\right)}}\delta q^{\left(n\right)}\right]_{t_{i}}^{t_{f}},\label{gen_var_eq}
\end{equation}
where $P_{q^{\left(n\right)}}$ is defined as 
\begin{equation}
P_{q^{\left(n\right)}}\equiv\sum_{k=n+1}^{N}\left(-\frac{d}{dt}\right)^{k-n-1}\frac{\partial L}{\partial q^{\left(k\right)}}.\label{momentum}
\end{equation}
The Hamiltonian for the generic case is written \cite{Ostrogradski,Simon}
as
\begin{equation}
H=\sum_{n=0}^{N-1}P_{q^{\left(n\right)}}q^{\left(n+1\right)}-L.\label{hamiltonian}
\end{equation}
From (\ref{gen_eul_lag_eq}) the equations of motion can be found
for $N=2$ as 
\begin{equation}
q^{\left(4\right)}+\left(\omega_{1}^{2}+\omega_{2}^{2}\right)\ddot{q}+\omega_{1}^{2}\omega_{2}^{2}q=0,\label{eqn_of_mot}
\end{equation}
where $q^{\left(4\right)}$ is the fourth time derivative of $q$.
The Hamiltonian (\ref{hamiltonian}) and the momentum (\ref{momentum})
of the system for $N=2$ reads
\begin{equation}
H=-\frac{1}{2}\left\{ \ddot{q}^{2}-2q^{\left(3\right)}\dot{q}-\left(\omega_{1}^{2}+\omega_{2}^{2}\right)\dot{q}^{2}-\omega_{1}^{2}\omega_{2}^{2}q^{2}\right\} ,\label{ham_n2}
\end{equation}
and 
\begin{equation}
P_{q^{\left(0\right)}}=\left(\omega_{1}^{2}+\omega_{2}^{2}\right)\dot{q}+q^{\left(3\right)}\quad\text{and}\quad P_{q^{\left(1\right)}}=-\ddot{q},\label{momentums_n2}
\end{equation}
respectively. For the non-degenerate case $\omega_{1}\ne\omega_{2}$
with the Pais-Uhlenbeck variables $Q_{1}\equiv q+\frac{\ddot{q}}{\omega_{2}^{2}}$
and $Q_{2}\equiv q+\frac{\ddot{q}}{\omega_{1}^{2}}$ can be diagonalized
as 
\begin{equation}
H=\frac{\omega_{1}^{4}}{2\left(\omega_{2}^{2}-\omega_{1}^{2}\right)}\left(\dot{Q}_{2}^{2}+\omega_{2}^{2}\dot{Q}_{2}^{2}\right)-\frac{\omega_{2}^{4}}{2\left(\omega_{2}^{2}-\omega_{1}^{2}\right)}\left(\dot{Q}_{1}^{2}+\omega_{1}^{2}\dot{Q}_{1}^{2}\right),\label{diag_ham_pu}
\end{equation}
where the second term is a ghost term. However, for the degenerate
case there can be a unitary region for proper constants \cite{Sarioglu}.

\section{Conventions}

In what follows we will use the following conventions:

\subsection{Flat Spacetime:}

The flat spacetime metric has the mostly positive signature $\text{diag}\left(\eta_{\mu\nu}\right)=\left(-,+,+,+,\dots\right)$
in its diagonal components. Since the spacetime is flat, the curvature
tensors and scalars are zero $R_{\phantom{\sigma}\mu\rho\nu}^{\sigma}=R_{\mu\nu}=R=0$.
The partial derivative reads $\partial^{\mu}=\left(\partial^{t},\,-\partial^{i}\right)=\left(\partial^{0},\,-\vec{\nabla}\right)$
and $\partial_{\mu}=\left(-\partial_{t},\,-\partial_{i}\right)=\left(-\partial_{0},\,-\vec{\nabla}\right)$.
The d'Alembertian operator becomes $\square=\partial^{2}=\partial_{\mu}\partial^{\mu}=-\partial_{0}^{2}+\partial_{i}^{2}=-\partial_{0}^{2}+\nabla^{2}$.
The Greek indices run as $\mu=0,1,2,\dots$ that counts all the coordinates
of the spacetime. The Latin indices run as $i=1,2,3,\dots$ and they
denote the spatial coordinates.

\subsection{The Curved Spacetime\label{The-Curved-Spacetime}:}

The metric has the signature $\left(-,+,+,+,\dots\right)$ and the
Christoffel connection reads 
\begin{equation}
\Gamma_{\mu\nu}^{\sigma}=\frac{1}{2}g^{\sigma\lambda}\left(\partial_{\mu}g_{\nu\lambda}+\partial_{\nu}g_{\mu\lambda}-\partial_{\lambda}g_{\mu\nu}\right),\label{Christoffel}
\end{equation}
and the commutation relation of the covariant derivatives are 
\begin{equation}
\left[\nabla_{\mu},\nabla_{\nu}\right]V^{\sigma}=R_{\mu\nu\phantom{\sigma}\lambda}^{\phantom{\mu\nu}\sigma}V^{\lambda},\label{Riemann1}
\end{equation}
where
\begin{equation}
R_{\mu\nu\phantom{\sigma}\lambda}^{\phantom{\mu\nu}\sigma}=\partial_{\mu}\Gamma_{\nu\lambda}^{\sigma}-\partial_{\nu}\Gamma_{\mu\lambda}^{\sigma}+\Gamma_{\mu\rho}^{\sigma}\Gamma_{\nu\lambda}^{\rho}-\Gamma_{\nu\rho}^{\sigma}\Gamma_{\mu\lambda}^{\rho}\label{Riemann2}
\end{equation}
is Riemann tensor. The covariant derivative acts on the covariant
and contravariant vectors as 
\begin{equation}
\nabla_{\mu}\omega_{\nu}=\partial_{\mu}\omega_{\nu}-\Gamma_{\mu\nu}^{\lambda}\omega_{\lambda},\quad\nabla_{\mu}V^{\nu}=\partial_{\mu}V^{\nu}+\Gamma_{\mu\lambda}^{\nu}V^{\lambda}.\label{cov_der}
\end{equation}
The Ricci tensor and Ricci scalar are defined as 
\begin{equation}
R_{\mu\nu}=g^{\rho\sigma}R_{\rho\mu\sigma\nu}\quad R=g^{\mu\nu}R_{\mu\nu},\label{Ricci_ten_scal}
\end{equation}
respectively. The d'Alembertian is $\square=\nabla^{\mu}\nabla_{\mu}$.
The Greek and Latin indices have the same meaning as in the flat spacetime
case.

\part{Massive Higher Derivative Gravity in $D$-dimensional Anti-de Sitter
Spacetimes }

\section{Introduction}

In gravity there is no theory which is unitary and renormalizable
at the same time. To get a renormalizable theory in four dimensions,
higher derivative terms, $\alpha R^{2}+\beta R_{\mu\nu}^{2}$, are
added to Einstein-Hilbert action \cite{Stelle1}. In doing so one
can gain renormalizability but loses unitary. The coupling constant
of the square of the Ricci tensor introduces a non-decoupling ghost
term. However, if we omit this term then we gain unitary yet lose
renormalizability. If a theory is non-unitary it shows itself as a
repulsive force in the Newtonian limit between static sources. Therefore
the theory has a better UV behaviour because of this repulsive force.
In field theory this is what usually happens: To have a better behaved
theory, ghosts are introduced during the renormalization process but
they decouple at the end if the theory is unitary. Therefore, bartering
unitarity with renormalizability can not be accepted. 

In three dimensions there is a perturbatively renormalizable and tree-level
unitary theory \cite{Bht1,Bht2} in flat background for a special
choice of coupling constants of higher derivative terms that is $8\alpha+3\beta=0$
and for a reversed sign of Einstein-Hilbert term \cite{nakason,oda,nakasonoda,Deser2}
and this theory is named as {}``New Massive Gravity'' (NMG), that
is a parity-preserving spin-2 theory. However, we do not know that
if this special ratio between $\alpha$ and $\beta$ will survive
renormalization at a given loop level. Another interesting thing about
this theory is that at the linearized level the theory has a massive
graviton with helicities $\pm2$ in its spectrum. Therefore, the theory
has a non-linear extension to the Pauli-Fierz (PF) mass term. For
this reason this theory solves an old problem of finding a non-linear
extension of PF mass term. The equivalence of NMG and PF was shown
in \cite{Bht1}%
\footnote{Note that, beside this parity-preserving theory, there is also the
parity violating Topologically Massive Gravity (TMG) \cite{Djt1,Djt2}.%
}. The physical meaning of this equivalence must be well understood
since NMG theory is background diffeomorphism invariant not only at
non-linear level but also at the linear level. However, PF theory
is invariant under Killing symmetries of the $2+1$ dimensional Minkowski
spacetime. To have a better understanding about these symmetries a
quite interesting approach was put forward in \cite{Deser2}: In the
absence of the Einstein-Hilbert term, the Weyl invariant form of the
linearized NMG is written. Therefore, introducing the Einstein-Hilbert
term at the linearized level breaks this symmetry and produces the
mass of the graviton. As a result one can think that Einstein-Hilbert
term provides the mass and the higher derivative terms gives the kinetic
energy. This perspective explains the sign change of Einstein-Hilbert
term. Looking back, this result is expected since pure Einstein gravity
is non-dynamical and does not give any propagation in three dimensions.
It is like the mass term in a scalar field theory at the linearized
level. When a kinetic energy is introduced to the theory then it plays
a role in the dynamics. In the case of NMG the kinetic energy comes
from the higher derivative terms. Consequently, the mass of the graviton
comes from Einstein-Hilbert term by breaking Weyl invariance. This
point may be important for constructing massive gravity theories in
other dimensions.

In this chapter, the most general quadratic curvature gravity theory
will be considered in a $D$ dimensional (anti)-de Sitter background.
We also augment this theory with a PF mass term. We will study the
propagator structure of this theory by finding its one-particle exchange
amplitude between two covariantly conserved sources. For this reason
we first linearize the action 

\begin{eqnarray}
I & = & \int d^{{D}}x\,\sqrt{-g}\left\{ \frac{1}{\kappa}R-\frac{2\Lambda_{0}}{\kappa}+\alpha R^{2}+\beta R_{\mu\nu}^{^{2}}+\gamma\left(R_{\mu\nu\sigma\rho}^{2}-4R_{\mu\nu}^{2}+R^{2}\right)\right\} \nonumber \\
 &  & +\int d^{D}x\,\sqrt{-g}\left\{ -\frac{M^{2}}{4\kappa}\left(h_{\mu\nu}^{2}-h^{2}\right)+{\cal {L}}_{\mbox{matter}}\right\} ,\label{action}
\end{eqnarray}
where $\Lambda_{0}$ is the bare cosmological constant. $\kappa$
is related to the $D$-dimensional Newton's constant as $\kappa\equiv2\Omega_{D-2}G_{D}$
where $\Omega_{D-2}$ is the $D-2$ dimensional solid angle. The other
parameters are the coupling constants $\alpha$, $\beta$, $\gamma$
and $M^{2}$ is the mass parameter. In total we have a seven parameter
theory which is the most general quadratic model including the dimensions
and cosmological constants%
\footnote{In the absence of the source terms and at the linearized level, one
can reduce the number of parameters in the action \cite{vas1,vas2},
but here for the sake of generality we shall work with (\ref{action}).%
}. Because of these parameters the theory potentially has various interesting
limits and discontinuities. At this level, there are no constraints
on the parameters. They may be positive, negative or zero. After computing
the tree-level amplitude we will constrain the parameters not to have
any ghosts or tachyons in the theory. Also, some terms do not contribute
to the equations of motion for some specific dimensions. For example,
for $D=4$ the Gauss-Bonnet term becomes a surface term and for $D=3$
it vanishes identically. Therefore, in three dimensions all the information
is carried by the Ricci tensor and the Riemann tensor has no more
information. In three dimensions the model can be extended by adding
the Chern-Simons term $\mu\left(\Gamma\partial\Gamma+\frac{2}{3}\Gamma^{3}\right)$
to (\ref{action}). But here we will stick to (\ref{action}). In
the next chapter we study this case in detail%
\footnote{See \cite{Sarioglu,dest} for this case, without the higher curvature
terms.%
}. The theory reduces to $R^{2}$ model with the PF term for two dimensions.
In this chapter we consider $D\geq3$. The theory that we consider
has general covariance except for the PF mass term.

Various limits of the spin-2 model which is defined by the linearization
of (\ref{action}) have been studied in the literature. Though, for
some certain limits of the above action there may still be some interesting
new models. One of them is NMG which is the case for $D=3$ and in
flat background without the PF mass term and $8\alpha+3\beta=0$.
Apart from finding such new models we will also explore discontinuities
of the full seven parameter theory. The discontinuities come out while
the order of limits are changed when some of the parameters approach
zero. One such discontinuity is the so called van vDVZ discontinuity.
The resolution of this discontinuity comes from the introduction of
the cosmological constant. Then taking $\frac{M^{2}}{\Lambda}\rightarrow0$
limit GR results are recovered at tree-level \cite{higuchi,Porrati1,kogan,vainshtein,deserwaldron2}%
\footnote{Another resolution of the discontinuity may follow even in flat space
if the Schwarzschild radius of the scattering objects is taken as
a second mass scale in the theory \cite{vainshtein}.%
}. However, the discontinuity reappears when the quantum corrections
are taken into account \cite{duff}. Up to now the linear theory has
been considered. Once, we consider the non-linear theory a ghost arises
in massive gravity and this effect is known as the Boulware-Deser
instability \cite{bouldeser}.

The ingredients of this chapter is as follows: In the second section
we will analyze the linear equations of motion which is obtained around
an AdS background. We also discuss certain special limits. In the
next section we calculate the one-particle scattering amplitude and
the following sections will be devoted to the discussions of some
limits and discontinuities. At the last section the conclusions and
discussions will be given.

\section{Linearized Equations}

The field equations can be found by taking the variation of the action
(\ref{action}) with respect to the metric. After variation we get
the field equations as follows:

\begin{eqnarray}
\frac{1}{\kappa}\left(R_{\mu\nu}-\frac{1}{2}g_{\mu\nu}R+\Lambda_{0}g_{\mu\nu}\right)+2\alpha R\left(R_{\mu\nu}-\frac{1}{4}g_{\mu\nu}R\right)+\left(2\alpha+\beta\right)\left(g_{\mu\nu}\square-\nabla_{\mu}\nabla_{\nu}\right)R\nonumber \\
+2\gamma\left[RR_{\mu\nu}-2R_{\mu\sigma\nu\rho}R^{\sigma\rho}+R_{\mu\sigma\rho\tau}R_{\nu}^{\;\;\sigma\rho\tau}-2R_{\mu\sigma}R_{\nu}^{\;\;\sigma}-\frac{1}{4}g_{\mu\nu}\left(R_{\tau\lambda\sigma\rho}^{2}-4R_{\sigma\rho}^{2}+R^{2}\right)\right]\nonumber \\
+\beta\square\left(R_{\mu\nu}-\frac{1}{2}g_{\mu\nu}R\right)+2\beta\left(R_{\mu\sigma\nu\rho}-\frac{1}{4}g_{\mu\nu}R_{\sigma\rho}\right)R^{\sigma\rho}+\frac{M^{2}}{2\kappa}\left(h_{\mu\nu}-\bar{g}_{\mu\nu}h\right)=\tau_{\mu\nu}.\label{fieldequations}
\end{eqnarray}

Here $\tau_{\mu\nu}$ is energy-momentum tensor coming from the source
terms, etc... . The background metric $\bar{g}_{\mu\nu}$ namely the
vacuum, is a non-singular solution to the field equations in the absence
of the matter terms. The vacuum is the (anti)-de Sitter space which
is the maximally symmetric vacuum with the Riemann, Ricci tensors
and Ricci scalar curvature given respectively as 
\begin{eqnarray}
\bar{R}_{\mu\rho\nu\sigma}=\frac{2\Lambda}{(D-1)(D-2)}\left(\bar{g}_{\mu\nu}\bar{g}_{\rho\sigma}-\bar{g}_{\mu\sigma}\bar{g}_{\nu\rho}\right),\hskip0.5cm\bar{R}_{\mu\nu}=\frac{2\Lambda}{D-2}\bar{g}_{\mu\nu},\hskip0.5cm\bar{R}=\frac{2D\Lambda}{D-2}.\label{background}
\end{eqnarray}
We made all the contractions with respect to the background metric.
If we put the background metric (\ref{background}) in the field equations
(\ref{fieldequations}), the effective cosmological constant $\Lambda$
can be found in terms of $\alpha$, $\beta$, $\gamma$, and $\Lambda_{0}$:
\begin{equation}
\frac{\Lambda-\Lambda_{0}}{2\kappa}+f(\alpha,\beta,\gamma)\Lambda^{2}=0,\label{effectivecosmo}
\end{equation}
where $f(\alpha,\beta,\gamma)\equiv\left(\alpha D+\beta\right)\frac{\left(D-4\right)}{\left(D-2\right)^{2}}+\gamma\frac{\left(D-3\right)\left(D-4\right)}{\left(D-1\right)\left(D-2\right)}$.
This is a second order equation with respect to $\Lambda$. Therefore
it has two solutions which are 

\begin{equation}
\Lambda=-\frac{1}{4\kappa f}\left[1\pm\sqrt{1+8\kappa f\Lambda_{0}}\right],\label{cosmogen}
\end{equation}
and for reality of the vacuum $8\kappa\Lambda_{0}f\geq-1$. Hence
the spacetime can be both dS or AdS. One of these solutions vanishes
in the absence of $\Lambda_{0}$. In this case the non-vanishing root
becomes $\Lambda=-\frac{1}{2\kappa f}$. There are some exceptional
points of this equation of motion for which the effective cosmological
constant becomes equal to the bare cosmological constant. In four
dimensions $\Lambda=\Lambda_{0}$ since $f(\alpha,\beta,\gamma)$
becomes zero. Also, in three dimensions if we set $3\alpha+\beta=0$
then we again have $f(\alpha,\beta,\gamma)=0$ and $\Lambda=\Lambda_{0}$.
Moreover, if we set $\gamma=0$ and $\alpha D+\beta=0$, which gives
$f(\alpha,\beta,\gamma)=0$, the theory again has $\Lambda_{0}=\Lambda$.
In three dimensions without setting $3\alpha+\beta=0$, we have $\Lambda_{\pm}=\frac{1\pm\sqrt{1-8\kappa\left(3\alpha+\beta\right)\Lambda_{0}}}{4\kappa\left(3\alpha+\beta\right)}$
and for $\Lambda$ to be real $1\geq8\kappa\left(3\alpha+\beta\right)\Lambda_{0}$.

The next step is to find the linear equation of motion of (\ref{fieldequations})
around the constant curvature background, $g_{\mu\nu}=\bar{g}_{\mu\nu}+h_{\mu\nu}$.
By using (\ref{cosmogen}) we eliminate the bare cosmological constant
and after linearization the equation of motion becomes \cite{dt2}

\begin{eqnarray}
T_{\mu\nu}\left(h\right) & = & a\,{\mathcal{G}}_{\mu\nu}^{L}+\left(2\alpha+\beta\right)\left(\bar{g}_{\mu\nu}\bar{\square}-\bar{\nabla}_{\mu}\bar{\nabla}_{\nu}+\frac{2\Lambda}{D-2}\bar{g}_{\mu\nu}\right)R^{L}\nonumber \\
 & + & \beta\left(\bar{\square}{\mathcal{G}}_{\mu\nu}^{L}-\frac{2\Lambda}{D-1}\bar{g}_{\mu\nu}R^{L}\right)+\frac{M^{2}}{2\kappa}\left(h_{\mu\nu}-\bar{g}_{\mu\nu}h\right),\label{linearizedfirst}
\end{eqnarray}
Note that the PF term is already linear. Here we define a new constant
in terms of the other coupling constants as

\begin{equation}
a\equiv\frac{1}{\kappa}+\frac{4\Lambda D}{D-2}\alpha+\frac{4\Lambda}{D-1}\beta+\frac{4\Lambda(D-3)(D-4)}{(D-1)(D-2)}\gamma.\label{eq:a_eqn}
\end{equation}

$T_{\mu\nu}\left(h\right)$ is the energy-momentum tensor which contains
all higher order terms and the source $\tau_{\mu\nu}$. ${\mathcal{G}}_{\mu\nu}^{L}$
is the linear cosmological Einstein tensor 

\begin{equation}
{\mathcal{G}}_{\mu\nu}^{L}=R_{\mu\nu}^{L}-\frac{1}{2}\bar{g}_{\mu\nu}R^{L}-\frac{2\Lambda}{D-2}h_{\mu\nu}.\label{einstein}
\end{equation}
$R_{\mu\nu}^{L}$ and $R^{L}$ are the linearized Ricci tensor and
the linearized scalar curvature $R^{L}\equiv\left(g^{\mu\nu}R_{\mu\nu}\right)^{L}$,
respectively. They read 
\begin{equation}
R_{\mu\nu}^{L}=\frac{1}{2}\left(\bar{\nabla}^{\sigma}\bar{\nabla}_{\mu}h_{\nu\sigma}+\bar{\nabla}^{\sigma}\bar{\nabla}_{\nu}h_{\mu\sigma}-\bar{\square}h_{\mu\nu}-\bar{\nabla}_{\mu}\bar{\nabla}_{\nu}h\right),\,\, R^{L}=-\bar{\square}h+\bar{\nabla}^{\sigma}\bar{\nabla}^{\mu}h_{\sigma\mu}-\frac{2\Lambda}{D-2}h.\label{ricci}
\end{equation}

To calculate the scattering amplitude we need the trace of (\ref{linearizedfirst}),
that is $T=\bar{g}_{\mu\nu}T^{\mu\nu}$ and it reads

\begin{equation}
\left[\left(4\alpha(D-1)+D\beta\right)\bar{\square}-(D-2)\left(\frac{1}{\kappa}+4f\Lambda\right)\right]R^{L}-\frac{M^{2}}{\kappa}(D-1)h=2T.\label{tracedenk}
\end{equation}
This equation is a wave equation, and for $4\alpha(D-1)+D\beta=0$
something special happens, since the dynamical part of this equation
vanishes. Actually, for $D=3$ this choice will lead us to the NMG.
After computing the tree level scattering amplitude between two sources,
and constraining the theory to have no ghosts or tachyons will give
us this special point. Before moving on we can still analyze some
limits of the theory at the linearized level. For this purpose we
drop the sources. Since there are no sources we can not get the unitarity
regions but we can capture the ranges of the parameters in which tachyons
drop out. The discussion bifurcates whether $M^{2}$ is zero or not.
First we will consider $M^{2}\ne0$.

\subsection{Massive case:}

We take the divergence and double divergence of (\ref{linearizedfirst})
to find the constraints on the deviation part of the metric, $h_{\mu\nu}$.
Since, (\ref{linearizedfirst}) is divergence free without the mass
term, the constraints come from the divergence of the mass term as 

\begin{equation}
\bar{\nabla}^{\mu}h_{\mu\nu}-\bar{\nabla}_{\nu}h=0,\hskip1cm\bar{\nabla}^{\mu}\bar{\nabla}^{\nu}h_{\mu\nu}-\bar{\square}h=0.\label{divdenk}
\end{equation}
Replacing the covariant derivatives that appear in the background
Ricci scalar in (\ref{ricci}) which is 
\begin{align*}
\bar{\nabla}^{\sigma}\bar{\nabla}^{\mu}h_{\sigma\mu} & =\left[\bar{\nabla}^{\sigma},\bar{\nabla}^{\mu}\right]h_{\sigma\mu}+\bar{\nabla}^{\sigma}\bar{\nabla}^{\mu}h_{\sigma\mu},\\
 & =\bar{R}_{\phantom{\sigma\mu}\sigma}^{\sigma\mu\phantom{\sigma}\lambda}h_{\lambda\mu}+\bar{R}_{\phantom{\sigma\mu}\mu}^{\sigma\mu\phantom{\mu}\lambda}h_{\sigma\lambda}+\bar{\nabla}^{\sigma}\bar{\nabla}_{\sigma}h,\\
 & =\frac{2\Lambda}{(D-2)}\bar{g}^{\mu\lambda}h_{\lambda\mu}-\frac{2\Lambda}{(D-2)}\bar{g}^{\sigma\lambda}h_{\sigma\lambda}+\bar{\nabla}^{\sigma}\bar{\nabla}_{\sigma}h,\\
 & =\bar{\nabla}^{\sigma}\bar{\nabla}_{\sigma}h,
\end{align*}
and using the first equation in (\ref{divdenk}) lead to $R^{L}=-\frac{2\Lambda}{D-2}h$.
For the flat space case we have $R^{L}=0$. This choice forces the
field to be traceless (\ref{tracedenk}), and from the first equation
of (\ref{divdenk}) it becomes transverse. The field $h_{\mu\nu}$
has $\frac{D(D+1)}{2}$ independent components. The transverse-traceless
condition gives $D+1$ equations that eliminates $D+1$ of the independent
components. Thus we are left with the remaining $(D+1)(D-2)/2$ independent
components. The linear Einstein tensor becomes $\mathcal{G}_{\mu\nu}^{L}=R_{\mu\nu}^{L}=-\partial^{2}h_{\mu\nu}$,
where we have changed the places of the covariant derivatives in (\ref{ricci}).
With these constraints the field equation for the remaining independent
components of the field becomes 
\begin{equation}
\left(\beta\partial^{4}+\frac{1}{\kappa}\partial^{2}-\frac{M^{2}}{\kappa}\right)h_{\mu\nu}=0,
\end{equation}
which is a quadratic equation that describes two massive excitations
with masses 
\begin{equation}
m_{\pm}^{2}=-\frac{1}{2\kappa\beta}\pm\frac{1}{2|\kappa\beta|}\sqrt{1+4\kappa\beta M^{2}}.\label{2kutle}
\end{equation}
To have real masses $M^{2}\geq-\frac{1}{4\beta\kappa}$, and at the
saturation point the two masses become equal. These masses become
non-tachyonic when the parameters are chosen properly. If $\kappa$
and $\beta$ have the same sign, one of the excitation becomes negative
and produces a tachyon. If they have opposite signs, the term inside
the square root must be $0<4\mid\kappa\beta\mid M^{2}<1$. In this
case both excitations become non-tachyonic. However, this theory is
not unitary as we will see. 

For the generic $\Lambda$ case the trace equation reads 
\begin{equation}
\left[(4\alpha(D-1)+D\beta)\square-(D-2)\left(\frac{1}{\kappa}+4\Lambda f\right)+\frac{M^{2}}{2\kappa\Lambda}(D-1)(D-2)\right]h=0.\label{trace_field_eq}
\end{equation}
 In this equation we can see that $h$ becomes a dynamical scalar
field. The dynamical part can be eliminated by setting $4\alpha(D-1)+D\beta=0$.
After choosing this special point we still have two options to satisfy
(\ref{trace_field_eq}). We may set $h=0$ and the field becomes again
transverse and traceless, and the field equations read 
\begin{equation}
\left[\left(a+\beta\bar{\square}\right)\left(\frac{2\Lambda}{\left(D-1\right)\left(D-2\right)}-\frac{1}{2}\bar{\square}\right)+\frac{M^{2}}{2\kappa}\right]h_{\mu\nu}=0,\label{field_eq_l_neq_zero}
\end{equation}
and it has generically two excitations 
\begin{equation}
m_{1,2}=\frac{2\Lambda}{\left(D-1\right)\left(D-2\right)}-\frac{a}{2\beta}\pm\sqrt{\left(\frac{2\Lambda}{\left(D-1\right)\left(D-2\right)}+\frac{a}{2\beta}\right)^{2}+\frac{M^{2}}{\kappa}}.\label{massive_excitation_l_neq_z}
\end{equation}
After computing the one-particle exchange amplitude we will again
discuss this theory and see that for special points, this theory will
give us a ghost and tachyon free theory. The other option to satisfy
(\ref{trace_field_eq}) is tuning the mass as 
\begin{equation}
M^{2}=\frac{2\Lambda\kappa}{D-1}\left(\frac{1}{\kappa}-\frac{\Lambda\beta(D-4)}{D-1}+4\Lambda\gamma\frac{(D-3)(D-4)}{(D-1)(D-2)}\right),\label{partially_massless}
\end{equation}
which is the partially massless point, vanishes for flat background
and for curved background a higher derivative gauge invariance appears
\cite{deserwaldron}. Therefore, the field has one less degree of
freedom (DOF) compared to the massive one. Moreover, the mass is allowed
to be negative in AdS as long as it satisfies the Breitenlohner-Freedman
type bound. In four dimensions higher derivative terms vanish and
the mass only depends on the cosmological constant, $M^{2}=\frac{2\Lambda}{3}$.
In higher dimensions the partially massless theory depends on the
higher derivative terms.

\subsection{Massless case:}

Without the Pauli-Fierz mass term the theory becomes invariant under
background diffeomorphisms $\delta_{\xi}h_{\mu\nu}=\bar{\nabla}_{\mu}\xi_{\nu}+\bar{\nabla}_{\nu}\xi_{\mu}$,
since $\delta_{\xi}{\mathcal{G}}_{\mu\nu}^{L}=0$ and $\delta_{\xi}R^{L}=0$.
As we mentioned above the divergence and double divergence of (\ref{linearizedfirst})
do not give any constraint on $h_{\mu\nu}$. Therefore, $R^{L}$ becomes
a dynamical variable for $T=0$. If the coefficient in front of the
d'Alembertian operator is fixed to zero than the dynamics of $R^{L}$
vanishes. The equation (\ref{tracedenk}) is satisfied for two conditions:
$R^{L}$ can be zero or the coefficient, $\frac{1}{\kappa}+4\Lambda f$,
can be set to zero in which case $R^{L}$ need not vanish. The later
one is not acceptable because the field equations leave a gauge invariant
object undetermined.

\section{Tree-level Amplitude}

To find the Newton potentials and to analyze the particle spectrum
of (\ref{action}), we are going to find the tree-level scattering
amplitude between two covariantly conserved sources which is defined
as 
\begin{equation}
A\equiv\frac{1}{4}\int d^{D}x\:\sqrt{-\bar{g}}T{}_{\mu\nu}^{\prime}\left(x\right)h^{\mu\nu}\left(x\right),\label{amplitude1}
\end{equation}
by considering the full theory (\ref{linearizedfirst}). In this equation
$T{}_{\mu\nu}^{\prime}$ is one of the sources and $h^{\mu\nu}$ is
the deviation which is produced by the unprimed source. The factor
$\frac{1}{4}$ is put to get the correct Newton's constant. Since
the field has independent components we need to decompose $h^{\mu\nu}$
to eliminate the unphysical parts of it. Therefore, the decomposition
must be done in such a way that the physical components of $h_{\mu\nu}$
will be determined by $T^{\mu\nu}$. The usual choice is to define
\begin{equation}
h_{\mu\nu}\equiv h_{\mu\nu}^{TT}+\bar{\nabla}_{(\mu}V_{\nu)}+\bar{\nabla}_{\mu}\bar{\nabla}_{\nu}\phi+\bar{g}_{\mu\nu}\psi,\label{decompose}
\end{equation}
where $h_{\mu\nu}^{TT}$ is the transverse and traceless part of the
deviation. $V_{\mu}$ is the vector part with a symmetrization that
is defined with a $\frac{1}{2}$ factor and it is divergence free.
$\phi$ and $\psi$ are scalar functions. With this definition (\ref{decompose}),
the amplitude equation becomes
\begin{equation}
A=\frac{1}{4}\int d^{D}x\:\sqrt{-\bar{g}}\left(T_{\mu\nu}^{\prime}h^{TT\mu\nu}+T^{\prime}\psi\right),\label{amplitude2}
\end{equation}
where the terms in the middle of (\ref{decompose}) vanish since we
have covariantly conserved sources and the total derivative vanishes
at the boundary by use of Stoke's theorem. In (\ref{amplitude2})
the tensorial quantities and $\psi$ must be written in terms of the
trace of energy momentum tensor. In order to get this we take the
trace, divergence and double divergence of (\ref{decompose}) 
\begin{equation}
h=\bar{\square}\phi+D\psi,\hskip1cm\bar{\square}h=\bar{\square}^{2}\phi+\frac{2\Lambda}{\left(D-2\right)}\bar{\square}\phi+\bar{\square}\psi,\label{handboxh}
\end{equation}
where we used $\bar{\nabla}^{\nu}\bar{\nabla}^{\mu}h_{\mu\nu}=\bar{\square}h$.
This condition comes from the nonzero mass term and it is not a gauge
condition. From these two equations we can write $\bar{\square}\phi$
in terms of $\bar{\square}\psi$. We hit the first equation with the
$\bar{\square}$ operator and subtract it from the second one which
gives us 
\begin{equation}
\bar{\square}\phi=\frac{\left(D-1\right)\left(D-2\right)}{2\Lambda}\bar{\square}\psi,\label{phidenk}
\end{equation}
which can be put into the first equation in (\ref{handboxh}) yielding
\begin{equation}
h=\left(\frac{\left(D-1\right)\left(D-2\right)}{2\Lambda}\bar{\square}+D\right)\psi.\label{kdenk}
\end{equation}
This equation can be inserted into (\ref{tracedenk}) so that $\psi$
can be written in terms of the trace of the energy momentum tensor
as follows 
\begin{equation}
\psi=\left\{ \frac{\Lambda}{\kappa}+4\Lambda f-c\Lambda\bar{\square}-\frac{M^{2}}{2\kappa}\left(D-1\right)\right\} ^{-1}\left(\frac{\left(D-1\right)\left(D-2\right)}{2\Lambda}\bar{\square}+D\right)^{-1}T,\label{psi_and_T}
\end{equation}
where $c\equiv\frac{4(D-1)\alpha}{D-2}+\frac{D\beta}{D-2}$. Now we
will write $h_{\mu\nu}^{TT}$ such that it is determined by the trace
of the energy momentum tensor. First of all, we will find the transverse
and traceless part of the field equations by using the Lichnerowicz
operator $\triangle_{L}^{(2)}$ acting on spin-2 symmetric tensors
\begin{equation}
\triangle_{L}^{(2)}h_{\mu\nu}=-\bar{\square}h_{\mu\nu}-2\bar{R}_{\mu\rho\nu\sigma}h^{\rho\sigma}+2\bar{R}^{\rho}\,_{(\mu}h_{\nu)\rho}.\label{lich}
\end{equation}
Some properties of this operator (that we need) were collected in
\cite{Porrati1} 
\begin{eqnarray}
\triangle_{L}^{(2)}\nabla_{(\mu}V_{\nu)} & = & \nabla_{(\mu}\triangle_{L}^{(1)}V_{\nu)},\hskip0.5cm\triangle_{L}^{(1)}V_{\mu}=\left(-\square+\Lambda\right)V_{\mu},\hskip0.5cm\nabla^{\mu}\triangle_{L}^{(2)}h_{\mu\nu}=\triangle_{L}^{(1)}\nabla^{\mu}h_{\mu\nu},\nonumber \\
\triangle_{L}^{(2)}g_{\mu\nu}\phi & = & g_{\mu\nu}\triangle_{L}^{(0)}\phi,\hskip0.5cm\triangle_{L}^{(0)}\phi=-\square\phi,\hskip0.5cm\nabla^{\mu}\triangle_{L}^{(1)}V_{\mu}=\triangle_{L}^{(0)}\nabla^{\mu}V_{\mu}.\label{lichpro}
\end{eqnarray}
 Using these properties we have 
\begin{equation}
{\mathcal{G}}_{\mu\nu}^{LTT}=\frac{1}{2}\triangle_{L}^{(2)}h_{\mu\nu}^{TT}-\frac{2\Lambda}{\left(D-2\right)}h_{\mu\nu}^{TT}.\label{transverse_traceless_G}
\end{equation}
 With this equation the transverse traceless part of the deviation
can be written in terms of the transverse traceless part of the energy
momentum tensor as 
\begin{equation}
h_{\mu\nu}^{TT}=2\left\{ (\beta\bar{\square}+a)(\triangle_{L}^{(2)}-\frac{4\Lambda}{D-2})+\frac{M^{2}}{\kappa}\right\} ^{-1}T_{\mu\nu}^{TT}.\label{tracelessh}
\end{equation}
The only thing that is left is to decompose the energy momentum tensor
so that we can connect $h_{\mu\nu}^{TT}$ with the energy momentum
tensor. To find this we first decompose the energy momentum tensor
as (\ref{decompose}). Then we take the double divergence of this
equation, keeping in mind that we have covariantly conserved sources.
Also taking the trace, one can find 
\begin{eqnarray}
T_{\mu\nu}^{TT}=T_{\mu\nu}-\frac{\bar{g}_{\mu\nu}}{D-1}T & + & \frac{1}{D-1}\left(\bar{\nabla}_{\mu}\bar{\nabla}_{\nu}+\frac{2\Lambda\bar{g}_{\mu\nu}}{\left(D-1\right)\left(D-2\right)}\right)\nonumber \\
 &  & \phantom{\frac{1}{D-1}}\times\left(\bar{\square}+\frac{2\Lambda D}{\left(D-1\right)\left(D-2\right)}\right)^{-1}T.\label{denkTT}
\end{eqnarray}
 We are ready to compute the scattering amplitude by putting (\ref{psi_and_T},
\ref{tracelessh}, \ref{denkTT}) into (\ref{amplitude2}). After
doing manipulations by using (\ref{lichpro}) and suppressing the
integral sign the amplitude equation becomes 
\begin{eqnarray}
4A & = & 2T{}_{\mu\nu}^{\prime}\left\{ (\beta\bar{\square}+a)(\triangle_{L}^{(2)}-\frac{4\Lambda}{D-2})+\frac{M^{2}}{\kappa}\right\} ^{-1}T^{\mu\nu}\nonumber \\
 & + & \frac{2}{D-1}T^{\prime}\left\{ (\beta\bar{\square}+a)(\bar{\square}+\frac{4\Lambda}{D-2})-\frac{M^{2}}{\kappa}\right\} ^{-1}T\label{mainresult}\\
 & - & \frac{4\Lambda}{(D-2)(D-1)^{2}}T^{\prime}\left\{ (\beta\bar{\square}+a)(\bar{\square}+\frac{4\Lambda}{D-2})-\frac{M^{2}}{\kappa}\right\} ^{-1}\left\{ \bar{\square}+\frac{2\Lambda D}{(D-2)(D-1)}\right\} ^{-1}T\nonumber \\
 & + & \frac{2}{(D-2)(D-1)}T^{\prime}\left\{ \frac{1}{\kappa}+4\Lambda f-c\bar{\square}-\frac{M^{2}}{2\kappa\Lambda}(D-1)\right\} ^{-1}\left\{ \bar{\square}+\frac{2\Lambda D}{(D-2)(D-1)}\right\} ^{-1}T.\nonumber 
\end{eqnarray}
 From this main result we can consider various limits. Solving this
integral for non zero cosmological constant is highly nontrivial.
However, the particle spectrum of the theory can be considered by
analyzing the pole structure of the amplitude. For the general case
there are four poles which read as 
\begin{eqnarray}
\bar{\square}_{1} & = & -\frac{2\Lambda D}{\left(D-1\right)\left(D-2\right)},\nonumber \\
\bar{\square}_{2,3} & = & \frac{1}{\beta}\left\{ -\left(\frac{a}{2}+\frac{2\Lambda\beta}{\left(D-2\right)}\right)\pm\sqrt{\left(\frac{a}{2}+\frac{2\Lambda\beta}{\left(D-2\right)}\right)^{2}-\beta\left(\frac{4\Lambda a}{\left(D-2\right)}-\frac{M^{2}}{\kappa}\right)}\right\} ,\label{eq:poles}\\
\bar{\square}_{4} & = & \frac{1}{c}\left(\kappa^{-1}+4\Lambda f-\frac{M^{2}}{2\kappa\Lambda}(D-1)\right).\nonumber 
\end{eqnarray}
 To have tachyon free theory these poles must be positive. By choosing
the constants properly such a theory can be found. However, this is
not the only restriction on the parameters that appear in the general
quadratic theory. With these poles the residues can be found. These
residues must be negative in order not to have any ghost terms. This
is the other restriction that we use while finding a ghost and tachyon
free model. However, in the most general form the residues are cumbersome.
Therefore, we will restrict the theory and compute the poles and residues
for some interesting limits. Also, we will compute the Newtonian potentials
for these limits.

\section{Massive theory in flat spacetime:}

Looking at (\ref{mainresult}), we can see that the massless limit
and the flat space limit do not commute because of the $\frac{M^{2}}{\Lambda}$
term. Therefore, if we want to get the usual Newtonian potentials
in flat space and without mass term, the flat space limit must be
taken before the massless limit. Otherwise, we encounter the well
known vDVZ discontinuity. Let us first look at the limits which produce
the vDVZ discontinuity that means we first take the flat space limit
$\Lambda\rightarrow0$ and go to massless limit $M^{2}\rightarrow0$.
With the flat space limit (\ref{mainresult}) becomes; 
\begin{equation}
4A=-2T{}_{\mu\nu}^{\prime}\left\{ \beta\partial^{4}+\frac{1}{\kappa}\partial^{2}-\frac{M^{2}}{\kappa}\right\} ^{-1}T^{\mu\nu}+\frac{2}{D-1}T'\left\{ \beta\partial^{4}+\frac{1}{\kappa}\partial^{2}-\frac{M^{2}}{\kappa}\right\} ^{-1}T,\label{flatvdvz-1}
\end{equation}
here the Lichnerowicz operator goes to $\triangle_{L}^{(2)}=-\partial^{4}$
and this can be seen from (\ref{lich}). The spectrum of (\ref{flatvdvz-1})
has two massive excitations which can be found by solving the second
order equation for $\partial^{2}$ and the masses are the same as
(\ref{2kutle}). Modifying (\ref{flatvdvz-1}) with the mass terms
give 
\begin{equation}
4A=-\frac{2}{\beta(m_{-}^{2}-m_{+}^{2})}\left\{ T{}_{\mu\nu}^{\prime}\left(\frac{1}{\partial^{2}-m_{+}^{2}}-\frac{1}{\partial^{2}-m_{-}^{2}}\right)T^{\mu\nu}-\frac{1}{(D-1)}T^{\prime}\left(\frac{1}{\partial^{2}-m_{+}^{2}}-\frac{1}{\partial^{2}-m_{-}^{2}}\right)T\right\} .\label{flatvdvz-2}
\end{equation}
For $\beta<0$ and $\beta>0$, (\ref{flatvdvz-2}) produces a massive
ghost. Therefore, $\beta$ must be set to zero to avoid ghost terms.
The Newtonian potential energy (U) can be calculated by use of Green's
function technique. For simplicity we take the sources as $T{}_{00}^{\prime}\equiv m_{1}\delta(x-x_{1})$,
$T_{00}\equiv m_{2}\delta(x^{\prime}-x_{2})$, where $m_{1}$ and
$m_{2}$ are masses of the sources, and the other terms of the energy-momentum
tensor are taken to be zero. Moreover, we calculate these potentials
in four and three dimensions. The Green's function for modified Helmholtz
equation $\nabla^{2}-k^{2}$ reads $\frac{1}{2\pi}K_{0}\left(k|r_{1}-r_{2}|\right)$
and $\frac{e^{-k|r_{1}-r_{2}|}}{4\pi|r_{1}-r_{2}|}$ \cite{Arfken}
for two and three space dimensions, respectively. Then the potential
energies read 
\begin{equation}
U=\begin{cases}
\frac{1}{2\beta(m_{+}^{2}-m_{-}^{2})}\frac{m_{1}m_{2}}{4\pi}[K_{0}(m_{-}r)-K_{0}(m_{+}r)]\hspace{2cm}D=3,\\
\frac{m_{1}m_{2}}{3\beta(m_{+}^{2}-m_{-}^{2})}\frac{1}{4\pi r}[e^{-m_{-}r}-e^{-m_{+}r}]\hspace{3.5cm}D=4,
\end{cases}\label{potential_energy_vdvz}
\end{equation}
where $r\equiv|\vec{x_{1}}-\vec{x_{2}}|$ and $m_{\pm}^{2}$ are defined
in (\ref{2kutle}). If we take the $\beta\rightarrow0$ limit to take
care of the ghost term, (\ref{potential_energy_vdvz}) becomes 
\begin{equation}
U=\begin{cases}
-\frac{\kappa}{8\pi}m_{1}m_{2}K_{0}(Mr)\hspace{2cm}D=3,\\
-\frac{4}{3}\frac{Gm_{1}m_{2}}{r}e^{-Mr}\hspace{2cm}D=4.
\end{cases}\label{massive3d_and_4d}
\end{equation}
The first equation of (\ref{massive3d_and_4d}) is the Newtonian limit
of massive gravity in three dimensions and it gives attractive force
when $\kappa$ is positive. For this case the $M\rightarrow0$ limit
does not exist. As $x\to0$, $K_{0}(x)\to-\ln(x/2)+\gamma_{E}$, which
gives the expected $\frac{1}{r}$ force for small separation of the
sources. The second equation of (\ref{massive3d_and_4d}) is the Newtonian
limit of massive gravity in four dimensions. Unlike the three dimensional
case, $M\rightarrow0$ limit exists. However, the $\frac{4}{3}$ term
indicates the famous vDVZ discontinuity. $G$ is the Newton constant
and it is taken as $\frac{\kappa}{16\pi}$. Therefore, in three dimensions
massive gravity gives the correct Newtonian limit despite that in
four dimensions it does not give the expected limit.

\section{Massless theory in flat spacetime:}

In the above discussion we see that first taking the flat space limit
and then going to massless limit does not give us the expected Newtonian
limit. In this section we first take the massless limit and then go
to the flat space limit. When we set $M^{2}=0$ and $\Lambda\rightarrow0$
in (\ref{mainresult}), we get
\begin{eqnarray}
 & 4A= & -2T{}_{\mu\nu}^{\prime}\left\{ \beta\partial^{4}+\frac{1}{\kappa}\partial^{2}\right\} ^{-1}T^{\mu\nu}+\frac{2}{(D-1)}T^{\prime}\left\{ \beta\partial^{4}+\frac{1}{\kappa}\partial^{2}\right\} ^{-1}T\nonumber \\
 &  & -\frac{2}{(D-1)(D-2)}T^{\prime}\left\{ c\partial^{4}-\frac{1}{\kappa}\partial^{2}\right\} ^{-1}T.\label{ozel}
\end{eqnarray}
Unlike (\ref{flatvdvz-1}), (\ref{ozel}) has three poles that are
\begin{equation}
\partial_{1}^{2}=0,\hskip1cm\partial_{2}^{2}=-\frac{1}{\kappa\beta},\hskip1cm\partial_{3}^{2}=\frac{1}{\kappa c}.\label{poles_3d}
\end{equation}
To make a full analysis we need the residues of these poles which
read 
\[
Res(\partial_{1}^{2})=\frac{2\kappa\left(3-D\right)}{\left(D-2\right)}\hskip1cmRes(\partial_{2}^{2})=\frac{2\kappa\left(D-2\right)}{\left(D-1\right)}\hskip1cmRes\left(\partial_{3}^{2}\right)=-\frac{2\kappa}{\left(D-1\right)\left(D-2\right)}.
\]
From the second pole we see that $\kappa\beta<0$ for not having tachyon
and form the residue of this pole $\kappa<0$ not to have a ghost.
For negative $\kappa$ the residue of the massless pole fixes the
dimension to three, since for $D>3$ makes the residue positive in
which case it produces a massless ghost. The residue of the third
pole becomes positive for negative $\kappa$. To eliminate this residue
one has to set $c=0$. For these conditions the theory becomes unitary
and has its special name as New Massive Gravity (NMG). If we look
to the Newtonian potential which is written as 
\begin{equation}
U=\frac{\kappa}{8\pi}m_{1}m_{2}\left(K_{0}(m_{g}r)-K_{0}(m_{0}r)\right)\hspace{2cm}D=3,\label{potential_3d}
\end{equation}
where $m_{g}^{2}\equiv-\frac{1}{\kappa\beta}$ and $m_{0}^{2}\equiv\frac{1}{\kappa(8\alpha+3\beta)}$.
For negative $\kappa$, the second term gives a massive ghost which
decouples in the case of $8\alpha+3\beta=0$. In this case only the
first term lives and gives an attractive force. Also, if we choose
the Pauli-Fierz mass term as $M=m_{g}$ the NMG has the same Newtonian
limit as the usual massive gravity. For $D>3$, a massive ghost comes
out and does not decouple unless $\beta=0$. For example we can look
at $D=4$: 
\begin{equation}
U=-\frac{Gm_{1}m_{2}}{r}\left(1-\frac{4}{3}e^{-m_{g}r}+\frac{1}{3}e^{-m_{a}r}\right),\label{potential_4d}
\end{equation}
where $m_{a}^{2}\equiv\frac{1}{2\kappa(3\alpha+\beta)}$. The term
in the middle is the ghost term \cite{Stelle1}. When we take $\beta\rightarrow0$
the ghost term vanishes and taking $m_{a}\rightarrow\infty$ the last
term drops out and we are left with the usual Newtonian potential
in which case the vDVZ discontinuity does not appear. Therefore, vDVZ
discontinuity does not appear in a curved background, since we can
compare the smallness of the mass of graviton with another scale that
is the cosmological constant. We can also generalize the Newtonian
potentials to $D$ dimensions. For this we need the general Green's
function for the operators $\frac{1}{\partial^{2}}$ and $\frac{1}{\partial^{2}-m_{i}^{2}}$.
They are 
\[
G\left(x,x^{\prime}\right)=\frac{1}{\left(2\pi\right)^{\frac{D-1}{2}}}\frac{1}{r^{D-3}}\left[2^{\frac{\left(D-5\right)}{2}}\Gamma\left(\frac{D-3}{2}\right)\right]
\]
and 
\[
G\left(x,x^{\prime}\right)=\frac{1}{\left(2\pi\right)^{\frac{D-1}{2}}}\frac{1}{r^{\frac{D-3}{2}}}\left[\left(\frac{1}{m_{i}^{2}}\right)^{\frac{3-D}{4}}K_{\frac{D-3}{2}}\left(r\, m_{i}^{2}\right)\right],
\]
respectively. $\Gamma$ is the gamma function and $K_{\nu}$ is the
modified Bessel function. We first write the amplitude equation (\ref{ozel})
as 
\begin{align}
2A & =\frac{\left(D-2\right)\kappa\, m_{1}\, m_{2}}{\left(D-1\right)}\delta(x-x_{1})\frac{1}{\partial^{2}-m_{g}^{2}}\delta(x-x_{2})\nonumber \\
 & -\frac{\left(D^{2}-4D+3\right)\kappa\, m_{1}m_{2}}{(D-1)\left(D-2\right)}\delta(x-x_{1})\frac{1}{\partial^{2}}\delta(x-x_{2})\nonumber \\
 & -\frac{\kappa\, m_{1}\, m_{2}}{(D-1)(D-2)}\delta(x-x_{1})\frac{1}{\partial^{2}-m_{a}^{2}}\delta(x-x_{2}).\label{amplitude_of_general}
\end{align}
 Taking again static sources, we find the potential as 
\begin{align}
U & =\frac{\left(D-2\right)\kappa\, m_{1}\, m_{2}}{\left(D-1\right)}\frac{1}{\left(2\pi\right)^{\frac{D-1}{2}}}\frac{1}{r^{\frac{D-3}{2}}}\left[\left(\frac{1}{m_{g}^{2}}\right)^{\frac{3-D}{4}}K_{\frac{D-3}{2}}\left(r\, m_{g}^{2}\right)\right]\nonumber \\
 & -\frac{\kappa\, m_{1}\, m_{2}}{(D-1)(D-2)}\frac{1}{\left(2\pi\right)^{\frac{D-1}{2}}}\frac{1}{r^{\frac{D-3}{2}}}\left[\left(\frac{1}{m_{a}^{2}}\right)^{\frac{3-D}{4}}K_{\frac{D-3}{2}}\left(r\, m_{a}^{2}\right)\right]\nonumber \\
 & -\frac{\left(D^{2}-4D+3\right)\kappa\, m_{1}m_{2}}{(D-1)\left(D-2\right)}\frac{1}{\left(2\pi\right)^{\frac{D-1}{2}}}\frac{1}{r^{D-3}}\left[2^{\frac{\left(D-5\right)}{2}}\Gamma\left(\frac{D-3}{2}\right)\right].\label{gen_new_pot}
\end{align}
For any desired dimension this equation gives the Newtonian potential
in flat spacetime and without any mass term. For $\kappa>0$ the first
term signals the ghost problem for all dimensions and the last two
terms give attractive forces. We can confirm \cite{Stelle1} that
in any dimension the quadratic curvature gravity theory is not unitary
except for the NMG point.

For curved backgrounds the calculation of the Newtonian potential
is complicated because of the Lichnerowicz operator. Nonetheless,
we can say that in curved backgrounds there will be an additional
term to the Newtonian potentials. Apart from the term that contains
Lichnerowicz operator, other terms can be calculated as above.

\section{Conclusion and Discussion for Chapter-2}

The tree level scattering amplitude of the general quadratic gravity
theory with a Pauli-Fierz mass term in $D$ dimensions is found and
various limits of this amplitude equation is studied. For these limits
the Newtonian potentials for three and four dimensions are calculated.
First the flat space limit is taken and then the massless limit. Looking
at the potential energies it is seen that the theory is not unitary.
For taking care of the ghost term the $\beta\rightarrow0$ limit is
taken. In this limit the theory becomes unitary but the theory suffers
from the vDVZ discontinuity in four dimensions. Massive gravity gives
the correct Newtonian potentials in three dimensions since the $M^{2}\rightarrow0$
limit does not exist. The second limit that is taken is the massless
limit and then the flat space limit. Analyzing the pole structure
and the residues of the amplitude and calculating the Newtonian potential
energies it is seen that the theory is unitary for a special condition
for the coupling constants of the theory that is $8\alpha+3\beta=0$
in three dimensions. Also, the coupling constant of Hilbert-Einstein
term becomes negative. Apart from three dimensions the general quadratic
theory is non-unitary. To make the theory unitary again the $\beta\rightarrow0$
limit is taken and for this case it becomes unitary. The usual Newtonian
potential is obtained in four dimensions by taking all coupling constants
of higher derivative terms to zero. Also, the Newtonian potentials
for generic dimensions in flat space time are calculated.

\part{CANONICAL STRUCTURE OF HIGHER DERIVATIVE GRAVITY IN 3D}

\section{Introduction}

In the previous chapter we saw that the general quadratic gravity
theory has a special case in three dimensions. This special case is
$8\alpha+3\beta=0$ and $\kappa<0$ for the Lagrangian $\kappa^{-1}R+\alpha R^{2}+\beta R_{\mu\nu}^{2}$
and named as New Massive Gravity (NMG) \cite{Bht1,Bht2}. Also, the
parity violating extension of NMG is found \cite{Bht1,Bht2} by adding
a gravitational Chern-Simons term to the theory. These theories have
massive ghost-free spin-2 particles in their free spectrum around
both flat and (anti)-de Sitter (A)dS. The most interesting property
of NMG is that it gives a non linear extension to the Pauli-Fierz
mass term for spin-2 particles. It is the only known example which
provides this nonlinear extension. Thus, NMG is a suitable candidate
to be a perturbatively well defined quantum gravity theory in three
dimensions if it is unitary beyond tree level.

The theory is studied in many directions. Its ghost-freedom and tree
level unitarity \cite{Deser1,Gullu1,Bht1,Bht2,nakason} and Newtonian
limits \cite{Gullu1} have been studied out. Furthermore, its classical
solutions and issues related to the classical solutions were studied
\cite{Bht1,Bht2,Clement1,Sun1,Giribet1,Oliva,gurses}. Also, its supergravity
extension were given in \cite{andringa}. 

In this chapter we study the canonical structure of the general quadratic
curvature theory and give an explicitly gauge invariant analysis of
it in three dimensions. The analysis is done for both flat and de
Sitter (dS) spacetimes. In the flat space analysis the gravitational
Chern-Simons term is added to the theory. By writing the canonical
form of the general quadratic action we can easily see how NMG is
singled out among all other higher curvature theories as a regular
{}``harmonic oscillator'' which can be thought as massive free field.
These type oscillators do not have the Ostragradskian instability
which spoil every higher time derivative theory \cite{woodard}%
\footnote{It was claimed that adding interactions might yield stable higher-time
derivative theories \cite{smilga}.%
}. Except of NMG, at the linearized level, all the other quadratic
theories are ghost-ridden Pais-Uhlenbeck oscillators. Moreover, the
discussion is extended by discussing the Newtonian limits, weak fields,
the scattering of particles with mass and spin at the tree level.

The content of this chapter is as follows: In the first section the
flat spacetime analysis is done by writing the canonical structure
for both quadratic curvature theory with and without the gravitational
Chern-Simons term. Also, the effects of static sources are analyzed
in detail. Moreover, the weak field solutions are obtained in circularly
symmetric case. In the second section the discussion is extended to
curved spacetimes namely to de Sitter spacetime. Some of the useful
calculations are given in the Appendix.

\section{Higher-derivative spin-2 in flat spacetime\label{Higher-derivative-spin-2-in-flat}}

We start the analysis with the flat space case since it is simpler
than the curved space background which is discussed in the next section.
Without any constraints on the parameters the general action is

\begin{equation}
I=\int d^{3}x\,\sqrt{-g}\left(\frac{1}{\kappa}R+\alpha R^{2}+\beta R_{\mu\nu}^{2}\right).\label{Non-linear_action}
\end{equation}
To get the desired spin-2 model the action must be expanded around
flat background $g_{\mu\nu}=\eta_{\mu\nu}+h_{\mu\nu}$, where $\eta_{\mu\nu}$
is the usual flat spacetime metric with signature $\left(-,+,+\right)$
and $h_{\mu\nu}$ is the spin-2 field which actually is a symmetric
rank-2 tensor and without any constraints has not only spin-2 component
but also spin-1 and spin-0 components. In this section we do not consider
the gravitational Chern-Simons term. However, in the next section
we add it to (\ref{Non-linear_action}). To get the action for $h_{\mu\nu}$
we first linearize the field equations that comes from the variation
of (\ref{Non-linear_action}) and then integrate the linear field
equations. While integrating them, one must pay attention to the overall
sign factor since the signs will become important in the discussion
of unitarity. Up to boundary terms (\ref{Non-linear_action}) becomes
\begin{equation}
I=-\frac{1}{2}\int d^{3}x\, h_{\mu\nu}\left[\frac{1}{\kappa}\mathcal{G}_{L}^{\mu\nu}+\left(2\alpha+\beta\right)\left(\eta^{\mu\nu}\Box-\partial^{\mu}\partial^{\nu}\right)R_{L}+\beta\Box\mathcal{G}_{L}^{\mu\nu}\right].\label{flat_sec_ord_action}
\end{equation}
Here, $\mathcal{G}_{L}^{\mu\nu}$ is the linearized Einstein tensor
and $R_{L}$ is the linearized Ricci scalar which is defined as $R_{L}\equiv\left(g_{\mu\nu}R^{\mu\nu}\right)_{L}$.
The linearized Einstein and Ricci tensors, and linearized curvature
scalar are as follows
\begin{gather}
\mathcal{G}_{L}^{\mu\nu}=R_{L}^{\mu\nu}-\frac{1}{2}\eta^{\mu\nu}R_{L},\qquad R_{L}=\partial_{\alpha}\partial_{\beta}h^{\alpha\beta}-\Box h,\nonumber \\
R_{L}^{\mu\nu}=\frac{1}{2}\left(\partial_{\sigma}\partial^{\mu}h^{\nu\sigma}+\partial_{\sigma}\partial^{\nu}h^{\mu\sigma}-\Box h^{\mu\nu}-\partial^{\mu}\partial^{\nu}h\right),\qquad h=\eta^{\mu\nu}h_{\mu\nu},\label{flat_lin_tensors}
\end{gather}
 where $\square$ is the D'Alembertian operator that is $\Box=\partial_{\mu}\partial^{\mu}=-\partial_{0}^{2}+\nabla^{2}$.
Since the metric is perturbed around flat background, we raise and
lower the indices with the flat metric $\eta_{\mu\nu}$. To analyze
the canonical structure and explore the free fields $h_{\mu\nu}$
must be decomposed. This is done as follows:
\begin{align}
h_{ij} & \equiv\left(\delta_{ij}+\hat{\partial}_{i}\hat{\partial}_{j}\right)\phi-\hat{\partial}_{i}\hat{\partial}_{j}\chi+\left(\epsilon_{ik}\hat{\partial}_{k}\hat{\partial}_{j}+\epsilon_{jk}\hat{\partial}_{k}\hat{\partial}_{i}\right)\xi,\nonumber \\
h_{0i} & \equiv-\epsilon_{ij}\partial_{j}\eta+\partial_{i}N_{L},\qquad h_{00}\equiv N,\label{Metric_decomposition}
\end{align}
 where $\phi,\,\chi,\,\xi,\,$$\eta,\, N_{L},\, N$ are free functions
of time and space $\left(t,\,\vec{x}\right)$ and $\hat{\partial}_{i}\equiv\partial_{i}/\sqrt{-\nabla^{2}}$.

After decomposing the spin-2 field, the components of the linearized
Einstein tensor must be written in terms of these free functions.
Let us give some details of these calculations.

We first write the trace of the spin-2 field:
\begin{equation}
\eta^{\mu\nu}h_{\mu\nu}=h=\eta^{00}h_{00}+\eta^{ij}h_{ij}.\label{trace_h}
\end{equation}
Putting (\ref{Metric_decomposition}) in (\ref{trace_h}) gives us
\[
h=-N+h_{i}^{i}=-N+\phi+\chi,
\]
where we have used $\hat{\partial}_{i}\hat{\partial}^{i}=-1$ and
note that $h_{i}^{i}=\phi+\chi$. The linear Einstein tensor has three
components that are $\mathcal{G}_{00}^{L},\,\mathcal{G}_{0i}^{L}$
and $\mathcal{G}_{ij}^{L}$. Using (\ref{flat_lin_tensors}) and summing
the repeated indices these components can be written as
\[
\mathcal{G}_{00}^{L}=\frac{1}{2}\left(\partial^{i}\partial^{j}h_{ij}-\partial^{i}\partial_{i}h_{k}^{\phantom{k}k}\right),
\]
\[
\mathcal{G}_{0i}^{L}=\frac{1}{2}\left(\partial_{j}\partial_{0}h_{i}^{\phantom{i}j}+\partial_{j}\partial_{i}h_{0}^{\phantom{0}j}-\partial_{j}\partial^{j}h_{0i}-\partial_{0}\partial_{i}h_{k}^{\phantom{k}k}\right),
\]
\begin{align*}
\mathcal{G}_{ij}^{L}= & \frac{1}{2}\left(\partial_{0}\partial_{i}h_{j}^{\phantom{j}0}+\partial_{0}\partial_{j}h_{i}^{\phantom{i}0}-\partial_{0}\partial^{0}h_{ij}-\partial_{k}\partial^{k}h_{ij}+\partial_{i}\partial_{j}h_{0}^{\phantom{0}0}-\partial_{i}\partial_{j}h_{k}^{\phantom{k}k}\right)\\
 & -\frac{1}{2}\delta_{ij}\left(\partial^{0}\partial^{0}h_{00}+2\partial^{0}\partial^{k}h_{0k}+\partial^{l}\partial^{k}h_{lk}+\partial_{0}\partial^{0}h_{00}-\delta^{kl}\partial_{0}\partial^{0}h_{kl}+\partial_{k}\partial^{k}h_{00}-\delta^{kl}\partial_{n}\partial^{n}h_{kl}\right).
\end{align*}
After using (\ref{Metric_decomposition}) these components can be
written in terms of gauge invariant objects as follows
\begin{align}
\mathcal{G}_{00}^{L} & =-\frac{1}{2}\nabla^{2}\phi,\qquad\mathcal{G}_{0i}^{L}=-\frac{1}{2}\left(\epsilon_{ik}\partial_{k}\sigma+\partial_{i}\dot{\phi}\right),\nonumber \\
\mathcal{G}_{ij}^{L} & =-\frac{1}{2}\left[\left(\delta_{ij}+\hat{\partial}_{i}\hat{\partial}_{j}\right)q-\hat{\partial}_{i}\hat{\partial}_{j}\ddot{\phi}-\left(\epsilon_{ik}\hat{\partial}_{k}\hat{\partial}_{j}+\epsilon_{jk}\hat{\partial}_{k}\hat{\partial}_{i}\right)\dot{\sigma}\right],\label{Einstein_tensor_gauge_inv}
\end{align}
where {}`` $\dot{}$ '' means differentiation with respect to time
$t$ and $q,\,\sigma,\,\text{and}\,\phi$ are defined as 
\begin{equation}
q\equiv\nabla^{2}N-2\nabla^{2}\dot{N_{L}}+\ddot{\chi},\qquad\sigma\equiv\dot{\xi}-\nabla^{2}\eta,\label{flat_gauge_inv_comb}
\end{equation}
which are gauge invariant under the transformation $\delta_{\zeta}h_{\mu\nu}=\partial_{\mu}\zeta_{\nu}+\partial_{\nu}\zeta_{\mu}$
and the Laplacian operator is $\nabla^{2}=\partial_{i}^{2}=\partial_{i}\partial^{i}=\partial_{i}\partial_{i}$
. Note that, $\phi$ is already a gauge invariant component of $h_{\mu\nu}$. 

Let us show this invariance of these objects: First we define the
three vector as $\xi_{\mu}=\left(\Lambda_{0},\,\epsilon_{ij}\partial_{j}+\partial_{i}K\right)$
and then we transform the components of the spin-2 fields. From the
transformations of $h_{00}$ and $h_{0i}$ we get 
\begin{equation}
\eta^{\prime}=\eta-\dot{\Lambda},\qquad N_{L}^{\prime}=N_{L}+\dot{K}+\Lambda_{0},\qquad N^{\prime}=N+2\dot{\Lambda}_{0}.\label{gauge_inv_1}
\end{equation}
From the last component we get three more equations by multiplying
$h_{ij}^{\prime}=h_{ij}+\partial_{i}\xi_{j}+\partial_{j}\xi_{i}$
with $\partial_{j}\partial_{i}$, $\epsilon_{il}\partial_{l}$ and
$\delta_{ij}$ that are 
\begin{equation}
\chi^{\prime}=\chi+2\nabla^{2}K,\qquad\xi^{\prime}=\xi-\nabla^{2}\Lambda,\qquad\phi^{\prime}=\phi,\label{gauge_inv_2}
\end{equation}
respectively. From here we can see explicitly the gauge invariance
of $\phi$. Using (\ref{gauge_inv_1}) and (\ref{gauge_inv_2}), one
can get (\ref{flat_gauge_inv_comb}). The linearized Ricci scalar
can be written in terms of the gauge invariant functions:
\begin{align*}
R^{L} & =-\left(\partial_{\alpha}\partial^{\alpha}\eta^{\mu\nu}-\partial^{\mu}\partial^{\nu}\right)h_{\mu\nu}\\
 & =\partial_{0}^{2}h_{i}^{\phantom{i}i}+\partial_{i}^{2}\left(h_{00}-h_{i}^{\phantom{i}i}\right)-2\partial_{0}\partial_{i}h_{0i}+\partial_{i}\partial_{j}h_{ij},
\end{align*}
again using (\ref{Metric_decomposition}) the scalar curvature becomes
\[
R^{L}=\ddot{\phi}+\ddot{\chi}-\nabla^{2}\phi+\nabla^{2}N-2\nabla^{2}N_{L}.
\]
Putting the first equation of (\ref{flat_gauge_inv_comb}) finally
we get
\begin{equation}
R_{L}=q-\Box\phi.\label{Ricci_scalar_gauge_inv}
\end{equation}
Although we have six free functions, the Bianchi identity $\partial_{\mu}\mathcal{G}_{L}^{\mu\nu}=0$
reduces the number of arbitrary functions to three. Therefore, there
are two sets of functions that can be used which are $\phi,\,\sigma,\, q$
and $\phi,\,\sigma,\, R_{L}$. We can write the total action in terms
of these gauge invariant free functions. Let us start with the Einstein-Hilbert
part of the action that is 
\[
I_{EH}=-\frac{1}{2\kappa}\int d^{3}x\, h_{\mu\nu}\mathcal{G}_{L}^{\mu\nu},
\]
then doing the summation about the repeated indices and using (\ref{Metric_decomposition}),
(\ref{Einstein_tensor_gauge_inv}) and (\ref{flat_gauge_inv_comb})
we get 
\begin{equation}
I_{EH}=-\frac{1}{2\kappa}\int d^{3}x\, h_{\mu\nu}\mathcal{G}_{L}^{\mu\nu}=\frac{1}{2\kappa}\int d^{3}x\,\left(\phi q+\sigma^{2}\right).\label{EH_action}
\end{equation}
Here we can see that the pure Einstein theory does not have any propagating
degrees of freedom since there is not a dynamical part which shows
itself as a D'Alembertian operator. With the same route we can find
the other two parts of the action. Instead of using this technique
one can also use the self-adjointness of the operators. In both ways
one can write the quadratic parts of the action in explicitly gauge
invariant forms. They are 

\begin{equation}
I_{2\alpha+\beta}=-\frac{2\alpha+\beta}{2}\int d^{3}x\, h_{\mu\nu}\left(\eta^{\mu\nu}\Box-\partial^{\mu}\partial^{\nu}\right)R_{L}=\frac{2\alpha+\beta}{2}\int d^{3}x\, R_{L}^{2},\label{two_alpha_plus_beta_action}
\end{equation}

\begin{align}
I_{\beta} & =-\frac{\beta}{2}\int d^{3}x\, h_{\mu\nu}\Box\mathcal{G}_{L}^{\mu\nu}=-\frac{\beta}{2}\int d^{3}x\,\left(-2\mathcal{G}_{\mu\nu}^{L}\mathcal{G}_{L}^{\mu\nu}+\frac{1}{2}R_{L}^{2}\right)=\frac{\beta}{2}\int d^{3}x\,\left(q\Box\phi+\sigma\Box\sigma\right).\label{beta_action}
\end{align}
In the first action we move all the derivatives on $h_{\mu\nu}$ and
that combination gives us again the linearized Ricci scalar. In the
second action $I_{\beta}$, first the derivatives are moved on the
spin-2 field and then (\ref{flat_lin_tensors}) is used. Finally using
the Bianchi identity one can achieve (\ref{beta_action}). Adding
(\ref{EH_action}), (\ref{two_alpha_plus_beta_action}) and (\ref{beta_action})
the total action comes out in gauge invariant combinations as 

\begin{align}
I & =\frac{1}{2}\int d^{3}x\,\left[\frac{1}{\kappa}\phi q+\left(2\alpha+\beta\right)\left(q-\Box\phi\right)^{2}+\beta q\Box\phi\right]+\frac{\beta}{2}\int d^{3}x\,\left(\sigma\Box\sigma+\frac{1}{\kappa\beta}\sigma^{2}\right).\label{flat_total_action}
\end{align}
From this equation, it can be seen immediately that $\sigma$ shows
itself as a single massive scalar field with mass $m_{g}^{2}=-\frac{1}{\kappa\beta}$.
Not to have negative mass $\kappa\beta$ must be negative and not
to have ghost $\beta$ must be positive for the $\sigma$ field. For
these signs $\kappa$ must be chosen negative. For the remaining part
of the action the discussion bifurcates whether $2\alpha+\beta=0$,
or not. Let us discuss these cases separately:

\subsection{$2\alpha+\beta\ne0$ case:}

In this case the $q$ field can be eliminated by taking the variation
with respect to it. Doing so gives 
\begin{equation}
q=-\frac{1}{2\left(2\alpha+\beta\right)}\left(\frac{\phi}{\kappa}+\beta\square\phi\right)+\square\phi,\label{q_field}
\end{equation}
and inserting the $q$ field into (\ref{flat_total_action}), the
action for the $\phi$ field becomes

\begin{equation}
I_{\phi}=\frac{1}{2}\int d^{3}x\,\left[\frac{\beta\left(8\alpha+3\beta\right)}{4\left(2\alpha+\beta\right)}\left(\Box\phi\right)^{2}+\frac{\left(4\alpha+\beta\right)}{2\kappa\left(2\alpha+\beta\right)}\phi\Box\phi-\frac{1}{4\kappa^{2}\left(2\alpha+\beta\right)}\phi^{2}\right].\label{phi_action}
\end{equation}
In this action there are some special limits. The first limit is the
NMG limit $8\alpha+3\beta=0$. For this limit the higher derivative
term drops out and the $\phi$ field describes a single massive degree
of freedom
\begin{equation}
I_{NMG,\phi}=-\frac{1}{2\kappa}\int d^{3}x\,\left(\phi\square\phi+\frac{1}{\kappa\beta}\phi^{2}\right),\label{phi_field_nmg}
\end{equation}
which has the same mass as the $\sigma$ field. Since both degrees
of freedom have the same masses in the NMG limit, the theory is parity-invariant.
Note that, not to have a ghost $\kappa$ must again be negative. Also,
NMG is not a higher derivative theory at the linearized level. The
theory does not have the Ostragradski instability which appears in
theories that have higher order derivatives. 

Another interesting limit is $4\alpha+\beta=0$. The middle term in
(\ref{phi_action}) drops out and the action (\ref{phi_action}) becomes
\begin{equation}
I_{\phi}=\frac{1}{4\kappa}\int d^{3}x\,\left(\kappa\beta\left(\square\phi\right)^{2}-\frac{1}{\kappa\beta}\phi^{2}\right).\label{phi_field_other_limit}
\end{equation}
This theory is tachyonic for $\kappa<0$ and gives a ghost term for
$\kappa>0$, keeping in mind that $\kappa\beta<0$. 

Taking $\beta=0$ also drops the higher derivative term. For this
case the action reads 
\begin{equation}
I_{\phi}=\frac{1}{2\kappa}\int d^{3}x\,\left(\phi\square\phi-\frac{1}{8\kappa\alpha}\phi^{2}\right).\label{phi_field_beta_0}
\end{equation}
In this case, to avoid ghost term $\kappa>0$ and not to have tachyon
$\alpha>0$, where the mass can be defined as $m^{2}=\frac{1}{8\kappa\alpha}$. 

For the general coupling constants, the above action (\ref{phi_action})
is a higher derivative Pais-Uhlenbeck oscillator. In order to decouple
the fields we define new fields that are linear in terms of old fields.
By inspection the fields can be defined as follows 

\begin{equation}
\varphi_{1}\equiv\phi-\frac{\Box\phi}{m_{g}^{2}},\qquad\varphi_{2}\equiv\phi-\frac{\Box\phi}{m_{s}^{2}},\label{redefined_fields}
\end{equation}
and with these definitions the uncoupled Lagrangian must be as follows
\[
K_{1}\varphi_{1}\left(\Box-m_{s}^{2}\right)\varphi_{1}+K_{2}\varphi_{2}\left(\Box-m_{g}^{2}\right)\varphi_{2}.
\]
Here $K_{1}$ and $K_{2}$ are unknowns. To find these unknown terms
we put (\ref{redefined_fields}) into above Lagrangian and compare
that result with (\ref{phi_action}). After getting the unknown factors
the action can be written in terms of simple oscillators and (\ref{phi_action})
becomes

\begin{equation}
I_{\phi}=\frac{1}{64\kappa\left(2\alpha+\beta\right)^{2}}\int d^{3}x\,\left[\left(8\alpha+3\beta\right)^{2}\varphi_{1}\left(\Box-m_{s}^{2}\right)\varphi_{1}-\beta^{2}\varphi_{2}\left(\Box-m_{g}^{2}\right)\varphi_{2}\right],\label{decoupled_phi_action}
\end{equation}
and the two fields $\varphi_{1},\,\varphi_{2}$ have masses $m_{s}$
and $m_{g}$ respectively. $m_{g}$ is defined above and $m_{s}$
is defined as

\[
m_{s}^{2}=\frac{1}{\kappa\left(8\alpha+3\beta\right)}.
\]
Note that for the general case the only restriction on the coupling
constants is $2\alpha+\beta\ne0$. For $\kappa<0$ the $\varphi_{1}$
field gives ghost and for $\kappa>0$ the $\varphi_{2}$ has negative
kinetic energy. Also, not to have negative mass term $8\alpha+3\beta$
must be positive for positive $\kappa$ and negative for negative
$\kappa$.

\subsection{$2\alpha+\beta=0$ case:}

In this case we can follow two routes: we either start from (\ref{flat_total_action})
or from (\ref{decoupled_phi_action}). These two actions give different
results and as we have seen above in the second action this limit
provides a singular theory, while in the first action it does not. 

We start with (\ref{decoupled_phi_action}) and take $\epsilon\equiv2\alpha+\beta\rightarrow0$
limit. By this limit $m_{s}=\frac{m_{g}^{2}}{\left(1-\frac{4\epsilon}{\beta}\right)}\approx m_{g}^{2}\left(1+\frac{4\epsilon}{\beta}\right)$
where in the last step we have taken the Taylor expansion around small
$\epsilon$. Note that $m_{g}^{2}$ does not depend on $\epsilon$.
With the same limit we can write the decoupled fields in terms of
$\epsilon$ as $\varphi_{2}=\phi-\frac{\square\phi}{m_{g}^{2}\left(1+\frac{4\epsilon}{\beta}\right)}\approx\varphi_{1}+\frac{4\epsilon}{\beta m_{g}^{2}}$
and again $\varphi_{1}$ does not depend on $\epsilon$. With these
expansions, (\ref{decoupled_phi_action}) gives us up to second order
\begin{align}
I_{\phi} & =\frac{1}{8\kappa\epsilon}\int d^{3}x\,\left\{ \frac{\beta}{m_{g}^{2}}\left[\left(\Box-m_{g}^{2}\right)\phi\right]^{2}-4\epsilon\phi\left(\Box-m_{g}^{2}\right)\phi+O\left(\epsilon^{2}\right)\right\} .\label{phi_action_limit_2a+b_zero}
\end{align}
This action is a degenerate Pais-Uhlenbeck oscillator which is known
as ghost free for some parameter ranges. However, if we start from
(\ref{flat_total_action}) and set $2\alpha+\beta=0$, we get
\begin{align}
I_{\phi} & =\frac{\beta}{2}\int d^{3}x\,\left(q\Box\phi-m_{g}^{2}q\phi\right).\label{phi_action_2a+b_zero}
\end{align}
Taking variation with respect to $q$ or $\phi$ gives us a massive
wave equation. However, these equations do not say anything about
the ghost structure or tachyonic behaviour of the theory. So that,
we must again separate the fields by redefining the fields such that
$q\equiv m_{g}^{2}\left(\Psi_{1}+\Psi_{2}\right)$ and $\phi\equiv\Psi_{1}-\Psi_{2}$.
With these definitions (\ref{phi_action_2a+b_zero}) turns into 
\begin{align}
I & =\frac{m_{g}^{2}\beta}{2}\int d^{3}x\,\left[\left(\Psi_{1}\Box\Psi_{1}-m_{g}^{2}\Psi_{1}^{2}\right)-\left(\Psi_{2}\Box\Psi_{2}-m_{g}^{2}\Psi_{2}^{2}\right)\right].\label{2a+b_zero_seperate_fields}
\end{align}
From the above discussion we know that $\beta$ must be positive not
to have ghost term, but in this case if it is taken positive, $\Psi_{2}$
becomes a ghost excitation. As we discussed in the previous chapter
the Newtonian limit of this theory is interesting \cite{Gullu1}.
The Newtonian potential of this theory becomes zero when two static
sources are taken into account, since the ghost excitation that is
the repulsive component cancels the spin-2 part which is the attractive
component. This is the same situation that happens in the pure Einstein
gravity.

\subsection{Adding static and spinning sources}

Up to now the analysis depended on theories without any interaction.
However, when interaction enters into the picture it may change the
particle spectrum of the theory. So that, we turn our analysis to
source dependent higher derivative gravity. First we add static sources
to our analysis and then we generalize this analysis by adding spinning
masses in flat spacetime.

\subsubsection{Static Sources:}

The matter can be added to the theory by the usual gravity-matter
coupling that is 

\begin{equation}
I_{\text{source}}=\frac{1}{2}\int d^{3}x\, h_{\mu\nu}T^{\mu\nu}.\label{source_eq}
\end{equation}
For the static source case we take $T^{00}=\rho\left(\vec{x}\right)$,
$T^{0i}=0$, $T^{ij}=0,$ and the above action becomes

\begin{equation}
I_{\text{source}}=\frac{1}{2}\int d^{3}x\, N\rho\left(\vec{x}\right)=\frac{1}{2}\int d^{3}x\,\left(\frac{1}{\nabla^{2}}q+2\dot{N}_{L}-\frac{1}{\nabla^{2}}\ddot{\chi}\right)\rho\left(\vec{x}\right),\label{source_eq1}
\end{equation}
where we used $q\equiv\nabla^{2}N-2\nabla^{2}\dot{N_{L}}+\ddot{\chi}\Rightarrow N=\frac{1}{\nabla^{2}}q+2\dot{N}_{L}-\frac{1}{\nabla^{2}}\ddot{\chi}$.
After dropping the boundary terms and taking the source terms static,
that is $\dot{\rho}\left(\vec{x}\right)=0$, and using the symmetry
property of the Green's function in the first term of (\ref{source_eq1}),
the source action becomes
\[
I_{\text{source}}=\frac{1}{2}\int d^{3}x\, q\frac{1}{\nabla^{2}}\rho.
\]
We add this action to the general total action (\ref{flat_total_action})
\begin{align}
I & =\frac{1}{2}\int d^{3}x\,\left[\frac{1}{\kappa}\phi q+\left(2\alpha+\beta\right)\left(q-\Box\phi\right)^{2}+\beta q\Box\phi+q\frac{1}{\nabla^{2}}\rho\right]+\frac{\beta}{2}I_{\sigma},\label{flat_total_source_action}
\end{align}
where $I_{\sigma}$ is the action that is only constructed from the
$\sigma$ field and is already in the simple harmonic oscillator form.
To eliminate the last term in the parenthesis of (\ref{flat_total_source_action})
we define new fields so that we can decouple the fields easily. The
last and the first terms can be combined and redefined as a new field
$\varphi\equiv\phi+\kappa\frac{1}{\nabla^{2}}\rho$ and by inspection
the $q$ field can be redefined as $\tilde{q}\equiv q+\kappa\rho$.
Putting these new fields into (\ref{flat_total_source_action}) and
taking $\square=-\partial_{t}^{2}+\nabla^{2}$, we get the total action
as
\begin{align}
I & =\frac{1}{2}\int d^{3}x\,\left[\frac{1}{\kappa}\left(\varphi\tilde{q}-\kappa\varphi\rho+\sigma^{2}\right)+\left(2\alpha+\beta\right)\left(\tilde{q}-\Box\varphi\right)^{2}\right.\nonumber \\
 & \phantom{=\frac{1}{2}\int d^{3}x\,}\left.+\beta\left(\tilde{q}\Box\varphi-\kappa\rho\Box\varphi-\kappa\tilde{q}\rho+\kappa^{2}\rho^{2}+\sigma\Box\sigma\right)\right].\label{flat_total_source_action1}
\end{align}
 Let us concentrate on the NMG case, that is $8\alpha+3\beta=0$.
For this case $2\alpha+\beta=\frac{\beta}{4}$. Taking variation with
respect to $\tilde{q}$ field, the following equation comes out 
\[
\frac{1}{\kappa}\varphi+\frac{\beta}{2}\left(\tilde{q}-\square\varphi\right)+\beta\left(\square\varphi-\kappa\rho\right)=0.
\]
From this equation $\tilde{q}$ can be defined in terms of the other
fields as 
\begin{equation}
\tilde{q}=2\kappa\rho-\square\varphi-\frac{2}{\kappa\beta}\varphi.\label{q_tilde}
\end{equation}
Putting (\ref{q_tilde}) into (\ref{flat_total_source_action1}) the
full action becomes
\begin{equation}
I=\frac{1}{2}\int d^{3}x\,\left[\beta\left(\sigma\Box\sigma-m_{g}^{2}\sigma^{2}\right)-\frac{1}{\kappa}\left(\varphi\Box\varphi-m_{g}^{2}\varphi^{2}\right)+\varphi\rho\right].\label{interaction_action}
\end{equation}
The last term in the action is the interaction part and it gives an
attractive potential energy for negative $\kappa$. In order to find
the potential we take the variation of (\ref{interaction_action})
with respect to $\varphi$ field that gives 
\begin{equation}
\frac{2}{\kappa}\left(\square-m_{g}^{2}\right)\varphi=\rho\Rightarrow\frac{2}{\kappa}\left(\square-m_{g}^{2}\right)\varphi=\frac{\kappa}{2}\frac{1}{\square-m_{g}^{2}}\rho.\label{inter_potential}
\end{equation}
Putting the $\varphi$ into the interaction part of (\ref{interaction_action})
the potential energy reads

\begin{equation}
U=\frac{\kappa}{4}\int d^{2}x\,\rho_{1}\frac{1}{\nabla^{2}-m_{g}^{2}}\rho_{2}=\frac{\kappa}{8\pi}m_{1}m_{2}\text{\ensuremath{K_{0}}}\left(m_{g}r\right).\label{Potential_source}
\end{equation}
From this equation it can be seen easily that for negative $\kappa$
the interaction part gives an attractive potential energy, that means
in NMG case adding static sources does not spoil the unitarity structure
or the tachyonic behaviour. In (\ref{Potential_source}) the sources
are defined as $\rho_{1}\left(\vec{x}\right)=m_{1}\delta^{\left(2\right)}\left(\vec{x}-\vec{x}_{1}\right)$,
$\rho_{2}\left(\vec{x}\right)=m_{2}\delta^{\left(2\right)}\left(\vec{x}-\vec{x}_{2}\right)$,
and $K_{0}$ is the modified Bessel function of the second kind. This
result also matches with the result that we found in the previous
chapter \cite{Gullu1}.

\subsubsection{Spinning masses:}

For this case the energy-momentum tensor must be written as
\[
T_{00}=m\delta^{\left(2\right)}\left(\vec{r}-\vec{r}_{1}\right),\qquad T_{\phantom{i}0}^{i}=\frac{1}{2}j\,\epsilon^{ij}\partial_{j}\delta^{\left(2\right)}\left(\vec{r}-\vec{r}_{1}\right),\qquad T_{ij}=0,
\]
where $m$ is the mass and $j$ is the spin of the point-like source.
The general amplitude and Newtonian potential was calculated for static
sources at flat spacetime in $D$-dimensions in the previous chapter
\cite{Gullu1}. They become
\begin{eqnarray*}
4A & = & \int d^{3}x\,\left\{ -2T_{\mu\nu}^{\prime}\left[\beta\partial^{4}+\frac{1}{\kappa}\partial^{2}\right]^{-1}T^{\mu\nu}+T^{\prime}\left[\beta\partial^{4}+\frac{1}{\kappa}\partial^{2}\right]^{-1}T\right.\\
 &  & \phantom{\int d^{3}x\,}\left.-T^{\prime}\left[\left(8\alpha+3\beta\right)\partial^{4}-\frac{1}{\kappa}\partial^{2}\right]^{-1}T\right\} ,
\end{eqnarray*}
and 
\[
U=\frac{\kappa\, m_{1}\, m_{2}}{2}\frac{1}{\left(2\pi\right)}\left[K_{0}\left(r\, m_{g}\right)-K_{0}\left(r\, m_{a}\right)\right]
\]
in three dimensions. The only added part from the spin will come from
the $T_{0i}$ components of the energy-momentum tensor. It will read
as 

\[
-4T_{i0}^{\prime}\left(\beta\partial^{4}+\frac{1}{\kappa}\partial^{2}\right)^{-1}T^{i0}=-\frac{j_{1}j_{2}}{\beta m_{g}^{2}}\partial_{i}\delta^{\left(2\right)}\left(\vec{r}-\vec{r}_{1}\right)\left(\frac{1}{\partial^{2}}-\frac{1}{\partial^{2}-m_{g}^{2}}\right)\partial_{i}\delta^{\left(2\right)}\left(\vec{r}-\vec{r}_{2}\right).
\]
The Newtonian potential of these operators was calculated in the previous
chapter. Using that result the potential energy reads
\begin{eqnarray*}
-4T'_{i0}\left\{ \beta\partial^{4}+\frac{1}{\kappa}\partial^{2}\right\} ^{-1}T^{i0} & = & -\frac{4}{\beta}T'_{i0}\frac{1}{\partial^{2}(\partial^{2}+\frac{1}{\beta\kappa})}T_{i0}\\
 & = & \frac{4}{\beta m_{g}^{2}}T'_{i0}\frac{1}{\partial^{2}}T_{i0}-\frac{4}{\beta m_{g}^{2}}T'_{i0}\frac{1}{\partial^{2}-m_{g}^{2}}T_{i0}\\
 & = & +\frac{j_{1}j_{2}}{\beta m_{g}^{2}}\epsilon^{ij}\partial_{j}\delta^{2}(\vec{r}-\vec{r}_{1})\frac{1}{\partial^{2}}\epsilon^{ik}\partial_{k}\delta^{2}(\vec{r}-\vec{r}_{2})\\
 & \phantom{=} & -\frac{j_{1}j_{2}}{\beta m_{g}^{2}}\epsilon^{ij}\partial_{j}\delta^{2}(\vec{r}-\vec{r}_{1})\frac{1}{\partial^{2}-m_{g}^{2}}\epsilon^{ik}\partial_{k}\delta^{2}(\vec{r}-\vec{r}_{2}).
\end{eqnarray*}
Using $\epsilon^{ij}\epsilon^{ik}=\delta^{jk}$ the spinning part
of the amplitude becomes
\begin{eqnarray*}
-4T'_{i0}\left\{ \beta\partial^{4}+\frac{1}{\kappa}\partial^{2}\right\} ^{-1}T^{i0} & = & \frac{j_{1}j_{2}}{\beta m_{g}^{2}}\partial_{i}\delta^{2}(\vec{r}-\vec{r}_{1})\frac{1}{\partial^{2}}\partial_{i}\delta^{2}(\vec{r}-\vec{r}_{2})\\
 & \phantom{=} & -\frac{j_{1}j_{2}}{\beta m_{g}^{2}}\partial_{i}\delta^{2}(\vec{r}-\vec{r}_{1})\frac{1}{\partial^{2}-m_{g}^{2}}\partial_{i}\delta^{2}(\vec{r}-\vec{r}_{2}).
\end{eqnarray*}
Using the Green's function of these operators and taking the differentiations
in front of the functions and carrying the space integrations, we
get
\begin{eqnarray*}
-4T'_{i0}\left\{ \beta\partial^{4}+\frac{1}{\kappa}\partial^{2}\right\} ^{-1}T^{i0} & = & -\frac{j_{1}j_{2}}{2\pi\beta m_{g}^{2}}\vec{\nabla}^{2}ln|\vec{r}_{1}-\vec{r}_{2}|\\
 & \phantom{=} & -\frac{j_{1}j_{2}}{2\pi\beta m_{g}^{2}}\vec{\nabla}^{2}K_{0}\left(m_{g}|\vec{r}_{1}-\vec{r}_{2}|\right).
\end{eqnarray*}
Note that $\vec{\nabla}^{2}ln|\vec{r}_{1}-\vec{r}_{2}|=-2\pi\delta^{2}\left(\vec{r}_{1}-\vec{r}_{2}\right)$
and if we assume that the sources are in separate positions, that
is $\vec{r}_{1}\ne\vec{r}_{2}$ the first term in the above equation
vanishes and the second term becomes $\left(\vec{\nabla}^{2}-m_{g}^{2}\right)K_{0}\left(m_{g}|\vec{r}_{1}-\vec{r}_{2}|\right)=2\pi\delta^{2}\left(\vec{r}_{1}-\vec{r}_{2}\right)=0\Rightarrow\vec{\nabla}^{2}K_{0}\left(m_{g}|\vec{r}_{1}-\vec{r}_{2}|\right)=m_{g}^{2}K_{0}\left(m_{g}|\vec{r}_{1}-\vec{r}_{2}|\right)$
and the amplitude takes the form 
\begin{equation}
-4T_{i0}^{\prime}\left(\beta\partial^{4}+\frac{1}{\kappa}\partial^{2}\right)^{-1}T^{i0}=-\frac{j_{1}j_{2}}{2\pi\beta}\text{\ensuremath{K_{0}}}\left(m_{g}\left|\vec{r}_{1}-\vec{r}_{2}\right|\right).\label{ampl_spin}
\end{equation}
With this spinning part, the total Newtonian potential energy,$U=\text{Amplitude}/\text{time}$,
can be written as

\textbf{
\begin{equation}
U=\frac{\kappa}{8\pi}\left(m_{1}m_{2}+4m_{g}^{2}j_{1}j_{2}\right)\text{\ensuremath{K_{0}}}\left(m_{g}\left|\vec{r}_{1}-\vec{r}_{2}\right|\right)-\frac{\kappa}{8\pi}m_{1}m_{2}\text{\ensuremath{K_{0}}}\left(m_{s}\left|\vec{r}_{1}-\vec{r}_{2}\right|\right).\label{pot_energy_spin}
\end{equation}
}The signs of $j_{1}$ and $j_{2}$ are not fixed so that they can
be both positive or negative. Therefore, potential energy coming from
this part could be both attractive or repulsive. In the NMG case this
situation does not change since for this case only the last term in
(\ref{pot_energy_spin}) drops out.

\subsection{Weak field approximation}

Up to this point we have found some results from the linearized theory.
In this section we will try to get some of the above results from
the nonlinear theory (\ref{Non-linear_action}). We start with the
ansatz\textbf{
\begin{equation}
ds^{2}=-f\left(r\right)dt^{2}+\frac{b^{2}\left(r\right)}{f\left(r\right)}dr^{2}+r^{2}d\theta^{2},\label{ansatz}
\end{equation}
}and insert it to (\ref{Non-linear_action}). Here $f\left(r\right)$
and $b\left(r\right)$ are functions and the action will be varied
with respect to these functions%
\footnote{See the details of this Weyl trick in \cite{dt3}%
}. We will consider only the NMG theory. To find an approximate solution
the functions can be defined as follows: $f\left(r\right)=1+\int^{r}dr\, a\left(r\right)$,
$b\left(r\right)=1+\int^{r}dr\, v\left(r\right)$. Here, $v\left(r\right)$
and $a\left(r\right)$ are small functions. Putting the ansatz into
(\ref{Non-linear_action}) with the defined functions we get
\begin{equation}
\frac{4}{\kappa}v+2\beta v^{\prime\prime}+2\beta a^{\prime\prime}+r\beta a^{\prime\prime\prime}=0,\label{w_A_eqn}
\end{equation}
\begin{equation}
\beta r^{2}a^{\prime\prime}+\frac{2}{\kappa}r^{2}a+2r\beta v'-2\beta v=0,\label{dif_of_w_A_eqns}
\end{equation}
up to first order in $v\left(r\right)$ and $a\left(r\right)$. Here,
$^{\prime}$ denotes differentiation with respect to $r$. From (\ref{w_A_eqn})
and (\ref{dif_of_w_A_eqns}), $v\left(r\right)$ can be written in
terms of $a\left(r\right)$ as $v\left(r\right)=a\left(r\right)+\frac{r}{2}a^{\prime}\left(r\right)$.
To have an ordinary differential equation we put $v\left(r\right)$
into (\ref{dif_of_w_A_eqns}) and we get 
\begin{equation}
r^{2}a^{\prime\prime}+ra^{\prime}-a\left(m_{g}^{2}r^{2}+1\right)=0.\label{A_eqn2}
\end{equation}
By solving this equation we can find $a\left(r\right)$ as $a\left(r\right)=c_{1}\text{\ensuremath{I_{1}}}\left(m_{g}r\right)+c_{2}\text{\ensuremath{K_{1}}}\left(m_{g}r\right)$,
where $c_{1}$ and $c_{2}$ are constants. From this equation also
$v\left(r\right)$ can be determined using $v\left(r\right)=a\left(r\right)+\frac{r}{2}a^{\prime}\left(r\right)$.
From the above ansatz we know that $g_{00}\approx-1-\int^{r}dr\, a\left(r\right)$.
For $g_{rr}$ we first take the square of $b\left(r\right)$ and drop
the second order term. Then we expand $\frac{1}{f\left(r\right)}$
up to first order that is $\frac{1}{1+\epsilon}\approx1-\epsilon$.
Combining these approximations we get $g_{rr}\approx1+\int^{r}dr\,\left[2v\left(r\right)-a\left(r\right)\right]$.
We know that for decaying fields that is for $r\rightarrow\infty$,
$a\left(r\right)\rightarrow0$. Since $I_{1}$ diverges to infinity
for decaying fields $c_{1}$ must vanish, and the metric components
become 
\begin{equation}
g_{00}\approx-1+c\text{\ensuremath{K_{0}}}\left(m_{g}r\right),\qquad g_{rr}\approx1+d\text{\ensuremath{K_{1}}}\left(m_{g}r\right).\label{compo_ansatz}
\end{equation}
Here $c$ and $d$ are constants. When we compare these components
with Schwarzschild solution we see that the constants must be related
to the mass of the sources. This is consistent with our earlier result
(\ref{Potential_source}).

\subsection{Higher-derivative gravity plus a Chern-Simons term}

In this part the above discussion will be extended by adding the gravitational
Chern-Simons term to the quadratic gravity theory in flat space. We
will follow the same route; first we will decompose the linear form
of the action that comes from the gravitational Chern-Simons term.
Then the coupled field will be decomposed by taking the Fourier transform
of the Lagrangian and writing it in a matrix form and diagonalizing
this matrix. The general quadratic action with gravitational Chern-Simons
term is \cite{Djt1,Djt2}

\begin{equation}
I=\int d^{3}x\,\sqrt{-g}\left[\frac{1}{\kappa}R+\alpha R^{2}+\beta R_{\mu\nu}^{2}-\frac{1}{2\mu}\epsilon^{\lambda\mu\nu}\Gamma_{\phantom{\rho}\lambda\sigma}^{\rho}\left(\partial_{\mu}\Gamma_{\phantom{\sigma}\rho\nu}^{\sigma}+\frac{2}{3}\Gamma_{\phantom{\sigma}\mu\beta}^{\sigma}\Gamma_{\phantom{\sigma}\nu\rho}^{\beta}\right)\right],\label{cs_non-linear_action}
\end{equation}
where $\epsilon_{012}=1$, and $\mu$ is the Chern-Simons coupling
with an arbitrary sign%
\footnote{Without the $\alpha$, $\beta$ terms, but with a Pauli-Fierz mass
term, canonical analysis was carried out in \cite{Sarioglu,dest}.%
}. 

The linearization of the gravitational Chern-Simons bit yields
\[
I_{CS}=-\frac{1}{2\mu}\int d^{3}x\,\epsilon_{\mu\alpha\beta}\mathcal{G}_{L}^{\alpha\nu}\partial^{\mu}h_{\phantom{\beta}\nu}^{\beta}+O\left(h^{3}\right).
\]
First we write all the summations that is 

\begin{align}
I_{CS} & =-\frac{1}{2\mu}\int d^{3}x\,\epsilon_{\mu\alpha\beta}\mathcal{G}_{L}^{\alpha\nu}\partial^{\mu}h_{\phantom{\beta}\nu}^{\beta}=-\frac{1}{2\mu}\int d^{3}x\,\eta^{\rho\nu}\epsilon^{\mu\alpha\beta}\mathcal{G}_{\alpha\rho}^{L}\partial_{\mu}h_{\beta\nu}\nonumber \\
 & =\frac{1}{2\mu}\int d^{3}x\,\epsilon_{ij}\left(-\mathcal{G}_{i0}^{L}\partial_{0}h_{j0}+\mathcal{G}_{00}^{L}\partial_{i}h_{j0}-\mathcal{G}_{j0}^{L}\partial_{i}h_{00}-\mathcal{G}_{jk}^{L}\partial_{0}h_{ik}+\mathcal{G}_{0k}^{L}\partial_{j}h_{ik}-\mathcal{G}_{ik}^{L}\partial_{j}h_{0k}\right),\label{cs_summed_action}
\end{align}
where $\epsilon_{ij}\equiv\epsilon_{0ij}$. Using (\ref{Metric_decomposition})
and (\ref{Einstein_tensor_gauge_inv}) the terms that appear in the
last line in the above equation are 
\begin{align}
 & -\epsilon_{ij}\mathcal{G}_{0i}^{L}\partial_{0}h_{0j}=\frac{1}{2}\left(-\sigma\nabla^{2}\dot{N}_{L}+\ddot{\phi}\nabla^{2}\eta\right),\qquad\epsilon_{ij}\mathcal{G}_{00}^{L}\partial_{i}h_{0j}=-\frac{1}{2}\nabla^{2}\phi\nabla^{2}\eta,\nonumber \\
 & -\epsilon_{ij}\mathcal{G}_{0j}^{L}\partial_{i}h_{00}=\frac{1}{2}\sigma\nabla^{2}N,\qquad-\epsilon_{ij}\mathcal{G}_{jk}^{L}\partial_{0}h_{ik}=\frac{1}{2}\left(q\dot{\xi}-\ddot{\phi}\dot{\xi}+\sigma\ddot{\chi}-\sigma\ddot{\phi}\right),\nonumber \\
 & \epsilon_{ij}\mathcal{G}_{0k}^{L}\partial_{j}h_{ik}=\frac{1}{2}\left(\sigma\nabla^{2}\phi+\dot{\xi}\nabla^{2}\phi\right),\qquad-\epsilon_{ij}\mathcal{G}_{ik}^{L}\partial_{j}h_{0k}=\frac{1}{2}\left(-q\nabla^{2}\eta-\sigma\nabla^{2}\dot{N}_{L}\right),\label{cs_comp}
\end{align}
where we have taken out total derivatives in the needed steps and
dropped the boundary terms. With these equations the action of the
gravitational Chern-Simons term takes the form 
\begin{align}
I_{CS} & =\frac{1}{2\mu}\int d^{3}x\,\left[\sigma\left(q+\Box\phi\right)\right],\label{cs_flat_action}
\end{align}
where we have defined the $q$ and $\sigma$ terms in the previous
section. With these, action (\ref{cs_non-linear_action}) becomes
\begin{align}
I & =\frac{1}{2}\int d^{3}x\,\left[\frac{1}{\kappa}\left(\phi q+\sigma^{2}\right)+\left(2\alpha+\beta\right)\left(q-\Box\phi\right)^{2}+\beta\left(q\Box\phi+\sigma\Box\sigma\right)+\frac{1}{\mu}\sigma\left(q+\Box\phi\right)\right],\label{gen_cs_action}
\end{align}
where all the fields are gauge invariant. For the general case, that
is $2\alpha+\beta\ne0$ and both $\alpha\ne0,\;\beta\ne0$, we eliminate
the $q$ field by taking the variation of (\ref{gen_cs_action}) with
respect to the $q$ field. From this equation of motion we get 
\begin{equation}
0=\frac{1}{\kappa}\phi+2\left(2\alpha+\beta\right)\left(q-\Box\phi\right)+\beta\Box\phi+\frac{1}{\mu}\sigma,\label{cs_flat_act_var_wrt_q_filed}
\end{equation}
and 
\begin{equation}
q=\left(4\alpha+\beta\right)\Box\phi-\frac{1}{\kappa}\phi-\frac{1}{\mu}\sigma.\label{q_field_in_flat_cs_case}
\end{equation}
Putting (\ref{q_field_in_flat_cs_case}) in (\ref{gen_cs_action}),
the general action takes the following form 

\begin{align}
I & =\frac{1}{2}\int d^{3}x\,\left\{ \beta\left[\sigma\Box\sigma+\left(\frac{1}{\kappa\beta}-\frac{1}{4\mu^{2}\beta\left(2\alpha+\beta\right)}\right)\sigma^{2}\right]+\left[\frac{1}{\mu}+\frac{\left(4\alpha+\beta\right)}{2\mu\left(2\alpha+\beta\right)}\right]\sigma\Box\phi\right.\nonumber \\
 & \phantom{=\,}\left.-\frac{1}{2\kappa\mu\left(2\alpha+\beta\right)}\sigma\phi+\frac{1}{\kappa}\left[\frac{\beta\kappa\left(8\alpha+3\beta\right)}{4\left(2\alpha+\beta\right)}\left(\Box\phi\right)^{2}+\frac{\left(4\alpha+\beta\right)}{2\left(2\alpha+\beta\right)}\phi\Box\phi-\frac{1}{4\kappa\left(2\alpha+\beta\right)}\phi^{2}\right]\right\} .\label{without_q_cs}
\end{align}
For a proper analysis we must decouple the $\sigma$ and the $\phi$
fields. This decoupling procedure can be done mathematically for generic
$\alpha$ and $\beta$ but since we are interested in the NMG case
we take the $8\alpha+3\beta=0$ point. For this case the action reads
\begin{align*}
I_{NMG-CS} & =\frac{\beta}{2}\int d^{3}x\,\left\{ \left[\sigma\Box\sigma-\left(m_{g}^{2}+\frac{1}{\mu^{2}\beta^{2}}\right)\sigma^{2}\right]+\frac{2m_{g}^{2}}{\beta\mu}\sigma\phi+m_{g}^{2}\left(\phi\Box\phi-m_{g}^{2}\phi^{2}\right)\right\} .
\end{align*}
To decouple these fields we will follow a different route: First we
take the Fourier transform of the fields and put the Lagrangian in
a matrix form then we diagonalize this matrix. 

Taking the Fourier transform of the Lagrangian and putting in the
matrix form yields
\[
L_{FT}=\left(\begin{array}{cc}
\tilde{\sigma} & m_{g}\tilde{\phi}\end{array}\right)\left[\begin{array}{cc}
-\left(k^{2}+m_{g}^{2}+\frac{1}{\mu^{2}\beta^{2}}\right) & \frac{m_{g}}{\beta\mu}\\
\frac{m_{g}}{\beta\mu} & -\left(k^{2}+m_{g}^{2}\right)
\end{array}\right]\left(\begin{array}{c}
\tilde{\sigma}\\
m_{g}\tilde{\phi}
\end{array}\right).
\]
Using the eigenvalue equation $det\left(\lambda\, I-A\right)=0$ the
eigenvalues of the $A$ matrix can be found where $I$ is $2\times2$
identity matrix, $A$ is the above matrix and $\lambda$ is the eigenvalues
of the $A$ matrix. Solving the eigenvalue equation for $\lambda$
we get 
\begin{align*}
\lambda^{2}+\left[2\left(k^{2}+m_{g}^{2}\right)+\frac{1}{\mu^{2}\beta^{2}}\right]\lambda+\left(k^{2}+m_{g}^{2}\right)^{2}+\frac{k^{2}}{\mu^{2}\beta^{2}} & =0,
\end{align*}
with roots 
\begin{equation}
\lambda_{\pm}=-k^{2}-m_{g}^{2}-\frac{1}{2\mu^{2}\beta^{2}}\pm\frac{1}{\mu\beta}\sqrt{m_{g}^{2}+\frac{1}{4\mu^{2}\beta^{2}}}.\label{eigenvalue_eq}
\end{equation}
Now, let's find eigenvectors. For $\lambda_{\pm}$;
\begin{equation}
\left[\begin{array}{cc}
-\frac{1}{2\mu^{2}\beta^{2}}\mp\frac{1}{\mu\beta}\sqrt{m_{g}^{2}+\frac{1}{4\mu^{2}\beta^{2}}} & \frac{m_{g}}{\beta\mu}\\
\frac{m_{g}}{\beta\mu} & \frac{1}{2\mu^{2}\beta^{2}}\mp\frac{1}{\mu\beta}\sqrt{m_{g}^{2}+\frac{1}{4\mu^{2}\beta^{2}}}
\end{array}\right]\left[\begin{array}{c}
u_{1}^{\pm}\\
u_{2}^{\pm}
\end{array}\right]=\left[\begin{array}{c}
0\\
0
\end{array}\right],\label{eigenvector_eq}
\end{equation}
\begin{equation}
\left[\begin{array}{c}
u_{1}^{\pm}\\
u_{2}^{\pm}
\end{array}\right]=N_{\pm}\left[\begin{array}{c}
1\\
\frac{1}{2\mu\beta m_{g}}\pm\frac{1}{m_{g}}\sqrt{m_{g}^{2}+\frac{1}{4\mu^{2}\beta^{2}}}
\end{array}\right],\; N_{\pm}=\left[1+\left(\frac{1}{2\mu\beta m_{g}}\pm\frac{1}{m_{g}}\sqrt{m_{g}^{2}+\frac{1}{4\mu^{2}\beta^{2}}}\right)^{2}\right]^{-\frac{1}{2}}\label{eigenvectors_n}
\end{equation}
where $N_{\pm}$ are the normalization factors. Then we construct
a modal matrix from the components of the eigenvectors such that
\begin{equation}
P=\left[\begin{array}{cc}
N_{+} & N_{-}\\
N_{+}\left(\frac{1}{2\mu\beta m_{g}}+\frac{1}{m_{g}}\sqrt{m_{g}^{2}+\frac{1}{4\mu^{2}\beta^{2}}}\right) & N_{-}\left(\frac{1}{2\mu\beta m_{g}}-\frac{1}{m_{g}}\sqrt{m_{g}^{2}+\frac{1}{4\mu^{2}\beta^{2}}}\right)
\end{array}\right],\label{modal_matrix}
\end{equation}
and since the eigenvectors are orthogonal to each other the inverse
of the modal matrix is equal to its transpose. With the inverse of
this modal matrix we can define new fields such that $\Psi=P^{-1}X$,
where $X$ is the original field, 
\begin{equation}
\Psi=\left(\begin{array}{c}
\tilde{\Psi}_{+}\\
\tilde{\Psi}_{-}
\end{array}\right)=\left[\begin{array}{cc}
N_{+} & N_{+}\left(\frac{1}{2\mu\beta m_{g}}+\frac{1}{m_{g}}\sqrt{m_{g}^{2}+\frac{1}{4\mu^{2}\beta^{2}}}\right)\\
N_{-} & N_{-}\left(\frac{1}{2\mu\beta m_{g}}-\frac{1}{m_{g}}\sqrt{m_{g}^{2}+\frac{1}{4\mu^{2}\beta^{2}}}\right)
\end{array}\right]\left(\begin{array}{c}
\tilde{\sigma}\\
m_{g}\tilde{\phi}
\end{array}\right).\label{new_fields}
\end{equation}
Then, the diagonalized $A$ matrix can be constructed as $D=P^{-1}\, A\, P$
which is just $D=\left[\begin{array}{cc}
\lambda_{+} & 0\\
0 & \lambda_{-}
\end{array}\right]$, from which the transformed Lagrangian reads $L_{FT}=\Psi\, D\,\Psi$.
Then, the action in the decoupled form becomes
\begin{align}
I_{NMG-CS} & =\frac{\beta}{2}\int d^{3}x\,\left[\Psi_{+}\Box\Psi_{+}-\left(m_{g}^{2}+\frac{1}{2\mu^{2}\beta^{2}}-\frac{1}{\mu\beta}\sqrt{m_{g}^{2}+\frac{1}{4\mu^{2}\beta^{2}}}\right)\Psi_{+}^{2}\right.\nonumber \\
 & \phantom{=\frac{\beta}{2}\int d^{3}x\,}\left.\Psi_{-}\Box\Psi_{-}-\left(m_{g}^{2}+\frac{1}{2\mu^{2}\beta^{2}}+\frac{1}{\mu\beta}\sqrt{m_{g}^{2}+\frac{1}{4\mu^{2}\beta^{2}}}\right)\Psi_{-}^{2}\right],\label{nmg_cs_dec_action}
\end{align}
or in a more concrete form 
\begin{equation}
I_{NMG-CS}=\frac{\beta}{2}\int d^{3}x\,\left(\Psi_{+}\Box\Psi_{+}-m_{+}^{2}\Psi_{+}^{2}+\Psi_{-}\Box\Psi_{-}-m_{-}^{2}\Psi_{-}^{2}\right),\label{nmg_cs_concrete_action}
\end{equation}
where the masses read
\begin{equation}
m_{\pm}^{2}=m_{g}^{2}+\frac{1}{2\mu^{2}\beta^{2}}\pm\frac{1}{\mu\beta}\sqrt{m_{g}^{2}+\frac{1}{4\mu^{2}\beta^{2}}},\label{masses_nmg_cs}
\end{equation}
and the fields are inverse Fourier transformed. The masses are same
with those of \cite{Bht1,Bht2,andringa}. Since masses of the helicity
modes are different this theory is a parity violating theory. To check
the results of topologically massive gravity we take $\beta\rightarrow0$
limit. In this limit $m_{+}$ diverges and drops out, therefore we
are left with a single degree of freedom that has a mass $m_{-}=-\left|\mu\right|/\kappa$
\cite{Djt1,Djt2}. This result can be seen by Taylor expanding the
square root part of (\ref{masses_nmg_cs}) up to third order.

\section{Higher-derivative spin-2 in a de Sitter background}

Up to now our analysis was based on the flat space time. Now, we will
change the background to a constant curvature background. Specifically
we will study the canonical structure of higher derivative gravity
whose action is defined as

\begin{equation}
I=\int d^{3}x\,\sqrt{-g}\left[\frac{1}{\kappa}\left(R-2\Lambda_{0}\right)+\alpha R^{2}+\beta R_{\mu\nu}^{2}\right],\label{nonlineer_act_cosmo}
\end{equation}
in an (anti)-de Sitter background. Here $\Lambda_{0}$ is the bare
cosmological constant. The linearization of (\ref{nonlineer_act_cosmo})
yields
\begin{equation}
I=-\frac{1}{2}\int d^{3}x\,\sqrt{-\bar{g}}\, h_{\mu\nu}\left[a\mathcal{G}_{L}^{\mu\nu}+\left(2\alpha+\beta\right)\left(\bar{g}^{\mu\nu}\Box-\nabla^{\mu}\nabla^{\nu}+\frac{2}{\ell^{2}}\bar{g}^{\mu\nu}\right)R_{L}+\beta\left(\Box\mathcal{G}_{L}^{\mu\nu}-\frac{1}{\ell^{2}}\bar{g}^{\mu\nu}R_{L}\right)\right],\label{generic_act}
\end{equation}
where $1/\ell^{2}$ is the cosmological constant and $a\equiv\frac{1}{\kappa}+\frac{12}{\ell^{2}}\alpha+\frac{2}{\ell^{2}}\beta$.
The cosmological constant can be related to the coupling constants
$\alpha$, $\beta$, $\kappa$ and the bare cosmological constant
as $\frac{1}{\ell^{2}}=\frac{1}{4\kappa\left(3\alpha+\beta\right)}\left[1\pm\sqrt{1-8\kappa\Lambda_{0}\left(3\alpha+\beta\right)}\right]$
\cite{Gullu1,dt2,dt1}. We will carry the analysis in a de Sitter
(dS) background since it is simpler than an anti-de Sitter (AdS) background.
Nevertheless, from these expressions one can find the results in AdS
spacetime by taking $\ell\rightarrow i\ell$ transformation. This
transformation can be taken since the results that come in the dS
background are analytic in $\ell$%
\footnote{To keep the signature intact, one also needs to Wick rotate a space
coordinate%
}. For dS, we define the background metric $\bar{g}_{\mu\nu}$ in the
Poincaré form as 

\begin{equation}
ds^{2}=\frac{\ell^{2}}{t^{2}}\left(-dt^{2}+dx^{2}+dy^{2}\right),\label{ds_metric}
\end{equation}
by which all the raising and lowering operations and covariant derivatives
will be made. The perturbation can be defined as 

\begin{equation}
g_{\mu\nu}=\frac{\ell^{2}}{t^{2}}\eta_{\mu\nu}+h_{\mu\nu},\label{perturbation_ds}
\end{equation}
where $\frac{\ell^{2}}{t^{2}}\eta_{\mu\nu}=\bar{g}_{\mu\nu}$ and
$\eta_{\mu\nu}$ is the flat spacetime metric. With this perturbation
the linearized forms of Einstein and Ricci tensors and scalar curvature
can be written as

\begin{gather}
\mathcal{G}_{\mu\nu}^{L}=R_{\mu\nu}^{L}-\frac{1}{2}\bar{g}_{\mu\nu}R_{L}-\frac{2}{\ell^{2}}h_{\mu\nu},\nonumber \\
R_{\mu\nu}^{L}=\frac{1}{2}\left(\nabla^{\sigma}\nabla_{\mu}h_{\nu\sigma}+\nabla^{\sigma}\nabla_{\nu}h_{\mu\sigma}-\Box h_{\mu\nu}-\nabla_{\mu}\nabla_{\nu}h\right),\qquad R_{L}=\nabla_{\alpha}\nabla_{\beta}h^{\alpha\beta}-\Box h-\frac{2}{\ell^{2}}h,\label{ds_lin_tensors}
\end{gather}
where D'Alembertian operator defined as $\Box\equiv\nabla_{\mu}\nabla^{\mu}=\frac{t^{2}}{\ell^{2}}\eta^{\mu\nu}\nabla_{\mu}\nabla_{\nu}$.
The deviation part of the metric $h_{\mu\nu}$ can be decomposed into
its {}``spatial'' tensor $h_{ij}$, vector $h_{0i}$ and {}``scalar''
$h_{00}$ parts as follows

\begin{align}
h_{ij} & \equiv\frac{\ell^{2}}{t^{2}}\left[\left(\delta_{ij}+\hat{\nabla}_{i}\hat{\nabla}_{j}\right)\phi-\hat{\nabla}_{i}\hat{\nabla}_{j}\chi+\left(\tilde{\epsilon}_{i}^{\phantom{i}k}\hat{\nabla}_{k}\hat{\nabla}_{j}+\tilde{\epsilon}_{j}^{\phantom{j}k}\hat{\nabla}_{k}\hat{\nabla}_{i}\right)\xi\right]\nonumber \\
 & =\frac{\ell^{2}}{t^{2}}\left[\left(\delta_{ij}+\hat{\nabla}_{i}\hat{\nabla}_{j}\right)\phi-\hat{\nabla}_{i}\hat{\nabla}_{j}\chi+\frac{t^{2}}{\ell^{2}}\left(\tilde{\epsilon}_{ik}\hat{\nabla}_{k}\hat{\nabla}_{j}+\tilde{\epsilon}_{jk}\hat{\nabla}_{k}\hat{\nabla}_{i}\right)\xi\right],\label{curve_comp_h_1}\\
h_{0i} & \equiv\frac{\ell^{2}}{t^{2}}\left(-\tilde{\epsilon}_{i}^{\phantom{i}k}\nabla_{k}\eta+\partial_{i}N_{L}\right)=\frac{\ell^{2}}{t^{2}}\left(-\frac{t^{2}}{\ell^{2}}\tilde{\epsilon}_{ij}\nabla_{j}\eta+\partial_{i}N_{L}\right),\nonumber \\
h_{00} & \equiv\frac{\ell^{2}}{t^{2}}N,\nonumber 
\end{align}
where $\hat{\nabla}_{i}\equiv\nabla_{i}/\sqrt{-\nabla_{k}^{2}}$ and
the covariant derivative is defined for the two-dimensional space
metric $\gamma_{ij}=\frac{\ell^{2}}{t^{2}}\delta_{ij}$. The Latin
indices are $i=1,2$ for space dimensions. Since the components of
the two dimensional metric has no space dependence and is flat, the
covariant derivative reduces to partial derivative, $\nabla_{i}\rightarrow\partial_{i}$,
and $\hat{\partial}_{i}\equiv\partial_{i}/\sqrt{-\partial_{k}^{2}}$.
$\tilde{\epsilon}_{ik}$ is the Levi-Civita tensor which is related
to the corresponding tensor density as

\begin{equation}
\tilde{\epsilon}_{ik}=\sqrt{\gamma}\epsilon_{ik}\quad\Rightarrow\quad\tilde{\epsilon}_{ik}=\frac{\ell^{2}}{t^{2}}\epsilon_{ik}.\label{ten_den_eq}
\end{equation}
The convention for $\epsilon_{ik}$ is $\epsilon_{12}=1$ (the convention
for Levi-Civita tensor density for the upper indices is $\epsilon^{12}=1$
naturally with the induced metric). Therefore, the final result of
the above decomposition becomes 
\begin{align}
h_{ij} & =\frac{\ell^{2}}{t^{2}}\left[\left(\delta_{ij}+\hat{\partial}_{i}\hat{\partial}_{j}\right)\phi-\hat{\partial}_{i}\hat{\partial}_{j}\chi+\left(\epsilon_{ik}\hat{\partial}_{k}\hat{\partial}_{j}+\epsilon_{jk}\hat{\partial}_{k}\hat{\partial}_{i}\right)\xi\right],\nonumber \\
h_{0i} & =\frac{\ell^{2}}{t^{2}}\left(-\epsilon_{ij}\partial_{j}\eta+\partial_{i}N_{L}\right),\qquad h_{00}=\frac{\ell^{2}}{t^{2}}N.\label{curve_comp_h_2}
\end{align}
 The raising and lowering operations for spatial indices are done
with $\delta_{ij}$. Note that, the specific choice of decomposition
involves $\frac{\ell^{2}}{t^{2}}$ coefficients and these coefficients
help us to check the flat space limit, $\ell\rightarrow\infty$, $\frac{\ell}{t}\rightarrow1$
at every step in our calculations. 

At this stage we will find the gauge invariant combinations that will
be constructed from the six scalar functions. In flat space case $\phi$
is gauge invariant. However, for curved background it is not gauge
invariant anymore. The components of $h_{\mu\nu}$ transforms under
the gauge transformations $\delta_{\zeta}h_{\mu\nu}=\nabla_{\mu}\zeta_{\nu}+\nabla_{\nu}\zeta_{\mu}$
as 
\begin{align}
\delta_{\zeta}\phi & =2\frac{t}{\ell^{2}}\zeta_{0},\qquad\delta_{\zeta}\chi=2\frac{t^{2}}{\ell^{2}}\left(\partial_{i}^{2}\kappa+\frac{1}{t}\zeta_{0}\right),\qquad\delta_{\zeta}\xi=\frac{t^{2}}{\ell^{2}}\partial_{i}^{2}\zeta,\nonumber \\
\delta_{\zeta}\eta & =\frac{t^{2}}{\ell^{2}}\left(\dot{\zeta}+\frac{2}{t}\zeta\right),\qquad\delta_{\zeta}N_{L}=\frac{t^{2}}{\ell^{2}}\left(\dot{\kappa}+\zeta_{0}+\frac{2}{t}\kappa\right),\qquad\delta_{\zeta}N=2\frac{t^{2}}{\ell^{2}}\left(\dot{\zeta}_{0}+\frac{1}{t}\zeta_{0}\right),\label{gauge_tran_h}
\end{align}
where the components of $\zeta_{\mu}$ are defined as $\zeta_{\mu}\equiv\left(\zeta_{0},-\epsilon_{ij}\partial_{j}\zeta+\partial_{i}\kappa\right)$.
Looking at the linearized Bianchi identity, $\nabla_{\mu}\mathcal{G}^{\mu\nu}=0$,
there should be at least three independent gauge invariant combinations
which are constructed out of the six scalar fields and their derivatives.
These combinations can be found by inspections. However, looking at
the independent components of the gauge-invariant tensor $\mathcal{G}_{L}^{\mu\nu}$
the gauge-invariant combinations can be found easily. Following this
route four gauge-invariant functions can be found: 
\begin{align}
f & \equiv\frac{\ell}{t}\left[\phi-\frac{2}{t}N_{L}+\frac{1}{t}\frac{1}{\nabla^{2}}\left(\dot{\phi}+\dot{\chi}-\frac{2}{t}N\right)\right],\qquad p\equiv\frac{\ell}{t}\left(\dot{\phi}-\frac{1}{t}N\right),\nonumber \\
q & \equiv\frac{\ell}{t}\left[\nabla^{2}N+\ddot{\chi}-2\nabla^{2}\dot{N}_{L}-\frac{1}{t}\left(\dot{N}-2\nabla^{2}N_{L}+\dot{\chi}\right)+\frac{2}{t^{2}}N\right],\qquad\sigma\equiv\frac{\ell}{t}\left(\dot{\xi}-\nabla^{2}\eta\right),\label{gauge_inv_comb}
\end{align}
and from the Bianchi identity we get a relation between these functions
as 
\begin{equation}
t\nabla^{2}\left(\dot{f}-p+\frac{f}{t}\right)-\dot{p}-q=0.\label{Bianchi_id}
\end{equation}
 We can find the components of the linearized Einstein tensor in terms
of these gauge invariant fields, that are 
\begin{gather}
\mathcal{G}_{00}^{L}=-\frac{t}{2\ell}\nabla^{2}f,\qquad\mathcal{G}_{0i}^{L}=-\frac{t}{2\ell}\left(\partial_{i}p+\epsilon_{ik}\partial_{k}\sigma\right),\nonumber \\
\mathcal{G}_{ij}^{L}=-\frac{t}{2\ell}\left[\left(\delta_{ij}+\hat{\partial}_{i}\hat{\partial}_{j}\right)q-\hat{\partial}_{i}\hat{\partial}_{j}\dot{p}-\left(\epsilon_{ik}\hat{\partial}_{k}\hat{\partial}_{j}+\epsilon_{jk}\hat{\partial}_{k}\hat{\partial_{i}}\right)\dot{\sigma}\right].\label{lin_ein_ten_gauge_inv_fun}
\end{gather}
 We can also write the linearized curvature scalar in terms of the
gauge-invariant fields 
\begin{equation}
R_{L}=\frac{t^{3}}{\ell^{3}}\left(q-\nabla^{2}f+\dot{p}\right)=\frac{t^{4}}{\ell^{3}}\nabla^{2}\left(\dot{f}-p\right).\label{lin_curve_scal_gauge_inv_fun}
\end{equation}
 In the second equality we used the Bianchi identity (\ref{Bianchi_id}). 

Using the above relations (\ref{generic_act}) can be written in terms
of the gauge-invariant functions. The Einstein-Hilbert part of the
action takes the following form: 
\begin{equation}
I_{EH}=-\frac{a}{2}\int d^{3}x\,\sqrt{-\bar{g}}\, h_{\mu\nu}\mathcal{G}_{L}^{\mu\nu}=\frac{a}{2}\int d^{3}x\,\left[\frac{\ell^{2}}{t^{2}}fR_{L}+\frac{t}{\ell}\left(f\nabla^{2}f+p^{2}+\sigma^{2}\right)\right].\label{eh_gauge_inv_func}
\end{equation}
For the $2\alpha+\beta$ part of the action, using the self-adjointness
of the involved operators, $h_{\mu\nu}$ can be replaced by some gauge-invariant
combinations. Also, doing so makes the computations simpler as in
the flat space case. With this trick, the $2\alpha+\beta$ part of
the action becomes 
\begin{equation}
I_{2\alpha+\beta}=-\frac{\left(2\alpha+\beta\right)}{2}\int d^{3}x\,\sqrt{-\bar{g}}h_{\mu\nu}\left(\bar{g}^{\mu\nu}\Box-\nabla^{\mu}\nabla^{\nu}+\frac{2}{\ell^{2}}\bar{g}^{\mu\nu}\right)R_{L}=\frac{\left(2\alpha+\beta\right)}{2}\int d^{3}x\,\sqrt{-\bar{g}}R_{L}^{2}.\label{2alpha_beta_gauge_inv_fun}
\end{equation}
 For the $\beta$ part of the general action, we have 
\begin{align*}
I_{\beta} & =-\frac{\beta}{2}\int d^{3}x\,\sqrt{-\bar{g}}h_{\mu\nu}\left(\Box\mathcal{G}_{L}^{\mu\nu}-\frac{1}{\ell^{2}}\bar{g}^{\mu\nu}R_{L}\right)=-\frac{\beta}{2}\int d^{3}x\,\sqrt{-\bar{g}}\left[\left(\Box h_{\mu\nu}\right)\mathcal{G}_{L}^{\mu\nu}-\frac{1}{\ell^{2}}hR_{L}\right].
\end{align*}
First, writing $R_{\mu\nu}^{L}$ (\ref{ds_lin_tensors}) such that
the indices of covariant derivatives $\mu$ and $\nu$ stay at the
left, and then using the Bianchi identity, the following action can
be found for the $\beta$ part 
\[
I_{\beta}=-\frac{\beta}{2}\int d^{3}x\,\sqrt{-\bar{g}}\left(-2\mathcal{G}_{\mu\nu}^{L}\mathcal{G}_{L}^{\mu\nu}+\frac{1}{2}R_{L}^{2}+\frac{2}{\ell^{2}}h_{\mu\nu}\mathcal{G}_{L}^{\mu\nu}\right).
\]
If we had not used this method and computed $h_{\mu\nu}\Box\mathcal{G}_{L}^{\mu\nu}$
directly, to put the result into an explicitly gauge-invariant form
would have been more difficult and taken more time. Finally, the general
action can be written in terms of the gauge-invariant form by collecting
all the parts that are computed above as follows 
\begin{align}
I & =\frac{1}{2}\int d^{3}x\,\left\{ \left(a+\frac{2\beta}{\ell^{2}}\right)\left[\frac{\ell^{2}}{t^{2}}fR_{L}+\frac{t}{\ell}\left(f\nabla^{2}f+p^{2}+\sigma^{2}\right)\right]+\left(2\alpha+\beta\right)\frac{\ell^{3}}{t^{3}}R_{L}^{2}\right.\nonumber \\
 & \phantom{=\frac{1}{2}\int d^{3}x\,}\left.+\beta\frac{t^{3}}{\ell^{3}}\left[\dot{\sigma}^{2}+\sigma\nabla^{2}\sigma+\dot{p}^{2}+p\nabla^{2}p+\left(\nabla^{2}f\right)^{2}\right.\right.\label{total_act_gauge_inv}\\
 & \phantom{=\frac{1}{2}\int d^{3}x\,+\beta\frac{t^{3}}{\ell^{3}}}\left.\left.+\frac{\ell^{3}}{t^{3}}R_{L}\nabla^{2}f-\frac{\ell^{3}}{t^{3}}R_{L}\dot{p}-\dot{p}\nabla^{2}f\right]\right\} .\nonumber 
\end{align}
The above action gives (\ref{flat_total_action}) when the flat space
limit is taken. Also, the fields that appear in (\ref{total_act_gauge_inv})
are not independent. However, defining new field such that $\varphi\equiv\nabla^{2}f$
, and after using the Bianchi identity (\ref{Bianchi_id}), the above
action (\ref{total_act_gauge_inv}) takes a rather simple form in
which the $\sigma$ field decouples
\begin{align}
I & =\frac{1}{2}\int d^{3}x\,\left\{ \left(a+\frac{2\beta}{\ell^{2}}\right)\frac{t}{\ell}\left(-tp\varphi+p^{2}\right)+\left(2\alpha+\beta\right)\frac{t^{5}}{\ell^{3}}\left(\dot{\varphi}-\nabla^{2}p\right)^{2}\right.\nonumber \\
 & \phantom{=\frac{1}{2}\int d^{3}x\,}\left.+\beta\frac{t^{3}}{\ell^{3}}\left(\dot{p}^{2}-p\nabla^{2}p-\varphi^{2}-t\varphi\nabla^{2}p-t\dot{p}\dot{\varphi}-\varphi\dot{p}\right)\right\} +I_{\sigma},\label{total_action_in_3_variable}
\end{align}
 where the $\sigma$ action is 
\begin{equation}
I_{\sigma}=\frac{1}{2}\int d^{3}x\,\left[\beta\frac{t^{3}}{\ell^{3}}\left(\dot{\sigma}^{2}+\sigma\nabla^{2}\sigma\right)+\left(a+\frac{2\beta}{\ell^{2}}\right)\frac{t}{\ell}\sigma^{2}\right].\label{sigma_action}
\end{equation}
In flat spacetime, cosmological Einstein theory does not have any
propagating degrees of freedom for vanishing $\alpha$ and $\beta$
in $D=3$. Just like the flat spacetime, without higher curvature
terms the total action does not have any propagating degrees of freedom.
For non-vanishing $\alpha$ and $\beta$, (\ref{total_action_in_3_variable})
has three degrees of freedom. From the decoupled field we can read
the mass if (\ref{sigma_action}) is put in a correct canonical form.
For a minimally coupled scalar field the correct canonical form is
\begin{equation}
I=-\frac{1}{2}\int d^{3}x\,\sqrt{-g}\left(\partial_{\mu}\Phi\partial^{\mu}\Phi+m^{2}\Phi^{2}\right)=-\frac{1}{2}\int d^{3}x\,\left\{ \frac{\ell}{t}\left[-\dot{\Phi}^{2}+\left(\partial_{i}\Phi\right)^{2}\right]+\frac{\ell^{3}}{t^{3}}m^{2}\Phi^{2}\right\} .\label{correct_can_form}
\end{equation}
To get the correct canonical dimension the $\sigma$ field is rescaled
as $\sigma\rightarrow\frac{\ell^{2}}{t^{2}}\sigma$. After this rescaling,
(\ref{sigma_action}) becomes 
\begin{equation}
I_{\sigma}=-\frac{\beta}{2}\int d^{3}x\,\left[\frac{t}{\ell}\left(-\dot{\sigma}^{2}+\left(\nabla\sigma\right)^{2}\right)-\frac{\ell^{3}}{t^{3}}\left(\frac{a}{\beta}+\frac{2}{\ell^{2}}\right)\sigma^{2}\right],\label{rescaled_sigma_act}
\end{equation}
and for the $\sigma$ field the mass reads 
\begin{equation}
m_{g}^{2}=-\frac{a}{\beta}-\frac{2}{\ell^{2}}=-\frac{1}{\kappa\beta}-\frac{12\alpha}{\ell^{2}\beta}-\frac{4}{\ell^{2}}=-\frac{1}{\kappa\beta}-\frac{4\left(3\alpha-\beta\right)}{\beta\ell^{2}}.\label{eq:dS_m_g}
\end{equation}
Unlike the flat space case, diagonalizing the $\varphi$, $p$ action
is complicated for the general coupling constants. Since we want to
see the oscillators of this theory we can use other methods. One of
the methods is to Fourier transform the fields in the $\vec{x}$ space
and then computing the zero-momentum limit, that is dropping the Laplacians
in the action. The action does not change the number of degrees of
freedom by doing this manipulation, since, $\nabla^{2}\left(\text{field}\right)$
is not the lowest order term anymore. Another method is to directly
work on the equations of motion. Both methods will be studied separately
below.

\subsection{Masses from the nonrelativistic limit}

In this part we take the nonrelativistic limit of the general action
(\ref{total_action_in_3_variable}) by dropping the $\nabla^{2}$
terms. Since the $\sigma$ part of the action is already in an oscillator
form, it is not taken into account. Then the action becomes 
\begin{equation}
I=\frac{1}{2}\int d^{3}x\,\left[\left(a+\frac{2\beta}{\ell^{2}}\right)\frac{t}{\ell}\left(-tp\varphi+p^{2}\right)+\left(2\alpha+\beta\right)\frac{t^{5}}{\ell^{3}}\dot{\varphi}^{2}+\beta\frac{t^{3}}{\ell^{3}}\left(\dot{p}^{2}-\varphi^{2}-t\dot{p}\dot{\varphi}-\varphi\dot{p}\right)\right].\label{nonrelativ_action}
\end{equation}
At this step we must decouple the $\varphi$ and $p$ fields. For
this reason we rescale the $\varphi$ field as $\varphi\rightarrow\frac{1}{t}\varphi$.
Also, we separate the constant $2\alpha+\beta=\frac{\beta}{4}+\frac{8\alpha+3\beta}{4}$
to see clearly the NMG limit. With these (\ref{nonrelativ_action})
becomes
\begin{align}
I & =\frac{1}{2}\int d^{3}x\,\left[\left(a+\frac{2\beta}{\ell^{2}}\right)\frac{t}{\ell}\left(-p\varphi+p^{2}\right)+\frac{\beta}{4}\frac{t^{3}}{\ell^{3}}\left(\dot{\varphi}^{2}-\frac{\varphi^{2}}{t^{2}}+4\dot{p}^{2}-4\dot{p}\dot{\varphi}\right)\right.\nonumber \\
 & \phantom{=\frac{1}{2}\int d^{3}x\,}+\left.\frac{\left(8\alpha+3\beta\right)}{4}\frac{t^{3}}{\ell^{3}}\left(\dot{\varphi}^{2}+\frac{3\varphi^{2}}{t^{2}}\right)\right].\label{nonrelativ_action_res}
\end{align}
However, this rescaling is not enough to separate the fields. Therefore,
we define a new field $\Phi\equiv\varphi-2p$. With this definition
the $\varphi$ and $\Phi$ fields decouple. The $\Phi$ field action
can be read as 
\begin{equation}
I_{\Phi}=\frac{\beta}{8}\int d^{3}x\,\left[\frac{t^{3}}{\ell^{3}}\dot{\Phi}^{2}+\frac{t}{\ell}\left(\frac{a}{\beta}+\frac{2}{\ell^{2}}\right)\Phi^{2}\right],\label{phi_act}
\end{equation}
and gives the same mass as the $\sigma$ field (\ref{eq:dS_m_g}).
Therefore, the $\Phi$ field becomes the spin-2 helicity partner of
the $\sigma$ field. The action for the $\varphi$ field reads 
\begin{equation}
I_{\varphi}=\frac{\left(8\alpha+3\beta\right)}{8}\int d^{3}x\,\left[\frac{t^{3}}{\ell^{3}}\dot{\varphi}^{2}-\frac{1}{\left(8\alpha+3\beta\right)}\frac{t}{\ell}\left(a-\frac{24\alpha}{\ell^{2}}-\frac{6\beta}{\ell^{2}}\right)\varphi^{2}\right],\label{varphi_act}
\end{equation}
which is the spin-0 mode. To read the mass of this mode, we must put
it into the canonical form. For this reason we again rescale the $\varphi$
field $\varphi\rightarrow\frac{\ell^{2}}{t^{2}}\varphi$. After this
rescaling the mass can be read as 
\begin{equation}
m_{s}^{2}=\frac{1}{\kappa\left(8\alpha+3\beta\right)}-\frac{4}{\ell^{2}}\left(\frac{3\alpha+\beta}{8\alpha+3\beta}\right).\label{mass_spin0}
\end{equation}
For the NMG case this mode drops out and we are left with the $m_{g}^{2}$
and this result matches with \cite{Bht1}. Note that in \cite{Bht1}
the analysis was carried out by introducing auxiliary fields, and
these fields can be eliminated in such a way that the action gives
spin-2 field with a Pauli-Fierz mass. However, we reach the same results
by using canonical analysis. For general coupling constants, to make
the same analysis as in \cite{Bht1} we must introduce two auxiliary
fields and rewrite the action (\ref{Non-linear_action}) in terms
of these fields: 
\begin{equation}
\mathcal{L}=\frac{1}{\kappa}\sqrt{-g}{\mathcal{G}}_{\mu\nu}^{L}\left[R-f^{\mu\nu}G_{\mu\nu}-\phi R+\frac{m_{1}^{2}}{2}\phi^{2}+\frac{m_{2}^{2}}{4}\left(f^{\mu\nu}f_{\mu\nu}-f^{2}\right)\right],\label{action_aux_fields}
\end{equation}
where $\phi$ and $f_{\mu\nu}$ are auxiliary fields and the masses
are $m_{1}^{2}=-\frac{4}{\kappa\left(8\alpha+3\beta\right)}$ and
$m_{2}^{2}=-\frac{1}{\kappa\beta}$. Then we linearize (\ref{action_aux_fields})
around flat background, that is 
\begin{align}
\kappa\mathcal{L}_{linearized} & =-\left(\frac{1}{2}h^{\mu\nu}+f^{\mu\nu}\right)\mathcal{G}_{\mu\nu}^{L}-\phi R_{L}-\frac{2}{\kappa\left(8\alpha+3\beta\right)}\phi^{2}-\frac{1}{4\kappa\beta}\left(f^{\mu\nu}f_{\mu\nu}-f^{2}\right).\label{lineer_aux}
\end{align}
Here we can see explicitly that for the NMG limit the $\phi$ field
drops out and $f_{\mu\nu}$ can be chosen as $f_{\mu\nu}=-h_{\mu\nu}$.
However, for the generic case it is not clear how the fields $\phi$,
$f_{\mu\nu}$ and $h_{\mu\nu}$ decouples. One possible way is to
rescale the $h_{\mu\nu}$ field \cite{hindawi}.

In this section we have discussed the canonical structure of the generic
action in the nonrelativistic limit. It would also be interesting
to get the same results from the analysis of the relativistic equations.
The next section is devoted to the discussion of the relativistic
equations for the NMG limit.

\subsection{Equations of motions in the NMG case}

The action (\ref{total_action_in_3_variable}) for the $8\alpha+3\beta=0$
case becomes 
\begin{align}
I & =\frac{\beta}{2}\int d^{3}x\,\left\{ m_{g}^{2}\frac{t}{\ell}\left(tp\varphi-p^{2}\right)+\frac{t^{5}}{4\ell^{3}}\left(\dot{\varphi}-\nabla^{2}p\right)^{2}\right.\nonumber \\
 & \phantom{=\frac{1}{2}\int d^{3}x\,}\left.+\frac{t^{3}}{\ell^{3}}\left(\dot{p}^{2}-p\nabla^{2}p-\varphi^{2}-t\varphi\nabla^{2}p-t\dot{p}\dot{\varphi}-\varphi\dot{p}\right)\right\} ,\label{action_nmg_rel}
\end{align}
where we dropped the $\sigma$ field. From this action one can claim
that there are two degrees of freedom, which is in conflict with our
earlier results \cite{Bht1,Bht2}. However, if we look at the Hessian
matrix $\mathcal{H}=\frac{\partial^{2}\mathcal{L}}{\partial\dot{q}_{i}\partial\dot{q}_{j}}$,
that is 
\begin{equation}
\mathcal{H}=\frac{\beta t^{3}}{4\ell^{3}}\left(\begin{array}{cc}
t^{2} & -2t\\
-2t & 4
\end{array}\right),\label{hessian}
\end{equation}
and compute its determinant, $\det\mathcal{H}=0$, we see that there
should be a constraint in this model. Therefore the time derivatives
of the fields cannot be separated in terms of the canonical momenta
\[
\Pi_{\varphi}\equiv\frac{\partial\mathcal{L}}{\partial\dot{\varphi}}=\frac{\beta t^{5}}{4\ell^{3}}\left(\dot{\varphi}-\nabla^{2}p-\frac{2}{t}\dot{p}\right),\qquad\Pi_{p}\equiv\frac{\partial\mathcal{L}}{\partial\dot{p}}=\frac{\beta t^{3}}{2\ell^{3}}\left(2\dot{p}-t\dot{\varphi}-\varphi\right).
\]
Since the equations of motion are needed we take the variations of
(\ref{action_nmg_rel}) with respect to $\varphi$ and $p$ fields
which yield
\begin{equation}
\delta\varphi:\quad\frac{m_{g}^{2}t^{2}}{\ell}p-\frac{t^{3}}{\ell^{3}}\left(2\varphi+t\nabla^{2}p+\dot{p}\right)-\frac{1}{2\ell^{3}}\partial_{0}\left[t^{5}\left(\dot{\varphi}-\nabla^{2}p\right)-2t^{4}\dot{p}\right]=0,\label{eom_1}
\end{equation}
 and
\begin{equation}
\delta p:\quad\frac{m_{g}^{2}t}{\ell}\left(t\varphi-2p\right)-\frac{t^{5}}{2\ell^{3}}\left(\dot{\varphi}-\nabla^{2}p+\frac{4}{t^{2}}p+\frac{2}{t}\varphi\right)-\frac{1}{\ell^{3}}\partial_{0}\left[t^{3}\left(2\dot{p}-t\dot{\varphi}-\varphi\right)\right]=0.\label{eom_2}
\end{equation}
To decouple the fields, we define $\dot{\varphi}=\nabla^{2}p$ with
a hint that this choice makes $R_{L}=0$. With this definition, the
other equation becomes 
\begin{equation}
\frac{\ell}{t}\left(-\ddot{\varphi}-\frac{1}{t}\dot{\varphi}+\nabla^{2}\varphi\right)-\frac{\ell^{3}}{t^{3}}\left(m_{g}^{2}-\frac{1}{\ell^{2}}\right)\varphi=0.\label{eom_phi}
\end{equation}
This equation is not in the canonical wave equation form in dS. We
again rescale the field $\varphi\rightarrow\varphi/t$ to have canonical
wave equation and this rescaling yields
\begin{equation}
\frac{\ell}{t}\left(-\ddot{\varphi}+\frac{1}{t}\dot{\varphi}+\nabla^{2}\varphi\right)-\frac{\ell^{3}}{t^{3}}m_{g}^{2}\varphi=0,\Rightarrow\left(\Box-m_{g}^{2}\right)\varphi=0,\label{eom_phi_can}
\end{equation}
which is the same as the $\sigma$ field. From the relativistic equations,
we again see that for NMG limit the quadratic gravity theory is parity
preserving theory in three dimensions.

\section{Conclusions and Discussion for Chapter-3}

In this chapter the canonical structure of the linearized quadratic
gravity theories have been studied. The analysis is done in an explicitly
gauge-invariant way. Moreover, the analysis is made both for flat
and dS backgrounds in three dimensions. In flat background case, the
general quadratic action has been written in canonical wave equation
form. After the fields are decoupled they generate three harmonic
oscillators. The coupling constants $\kappa$, $\alpha$, $\beta$
are chosen to have a ghost-free and non-tachyonic theory. When the
coupling constants are fixed in this way, the NMG theory is singled
out as the unique unitary higher derivative massive gravity (but not
a higher-time derivative theory). Apart from this theory, all other
higher derivative gravity models are higher-derivative Pais-Uhlenbeck
oscillators which have ghost modes.

The analysis is also extended by adding static sources and spinning
masses to the theory. The effects of the sources are described by
computing the Newtonian potentials for both cases. In addition, the
weak field limit of the theory is computed at the nonlinear level
by using the circularly symmetric ansatz. Another extension that is
done for the flat spacetime is adding the gravitational Chern-Simons
term to the general quadratic theory. With this addition the NMG theory
is investigated and in this limit it is found that the oscillators
decouple with different masses. Therefore, this model is parity violating
theory as expected.

In dS spacetime case, the general action is written in terms of three
gauge-invariant functions. These functions are constructed from the
derivatives of the components of metric perturbation. The fields are
decoupled by two different ways. The first way to decouple fields
is to go to the nonrelativistic limit by dropping out the two dimensional
Laplacians in the action. The second decoupling way is to get the
field equations in the relativistic form. 

These gauge-invariant actions that are constructed in this chapter
may be useful when nonlinearities and interactions are introduced
to the theory. Apart from this usefulness there are other interesting
points about the theories that we discussed in this chapter. Tuned
values of the parameters give rise to uncommon phenomena, for instance
partial masslessness or chiral gravity, which especially arise in
(anti)-de Sitter spacetimes. These matters are open to study.

\part{Conclusion}

In this dissertation, the most general quadratic curvature massive
gravity theory is analyzed to find a specific unitary theory. For
this aim the one-particle scattering amplitude is calculated. Also,
to understand the found unitary theory, that is NMG, the most quadratic
curvature theory is written in canonical form in three dimensions
for both flat and constant curvature spacetimes.

In the second chapter we have studied the $D$ dimensional quadratic
gravity theory augmented with a Pauli-Fierz mass term. First the linear
theory without sources is studied for both massive and massless limits.
For the massive case, a wave equation is found with two massive excitations
in flat spacetime. In curved spacetime the trace of the metric perturbation
becomes a dynamical scalar field. When the dynamical part is dropped
by fixing the coupling constants one solution of this equation is
the partially massless point which comes out only in curved backgrounds.
For the massless limit both the linearized Einstein tensor and the
linearized Ricci scalar are background diffeomorphism invariant. Also,
the scalar curvature must be zero in flat spacetime, but need not
be zero in curved background when the dynamical part of the equation
is eliminated by fixing the constants. 

After the analysis in the linearized level, we have moved on to compute
the one-particle scattering amplitude between two covariantly conserved
sources. To find the exchange amplitude the perturbation part must
be expressed in terms of the energy-momentum tensor. However, the
components of the perturbation part of the metric are not independent.
Hence, it is decomposed into its parts to obtain the independent components.
After computing the necessary elements, the scattering amplitude is
calculated. From this general amplitude equation, the poles and residues
can be calculated to have an idea about the particle spectrum of the
most general theory. Moreover, the Newtonian potentials can be computed
from this amplitude equation. Nevertheless, the equations are very
complicated for the most general case. The calculations are done only
for some interesting limits. First the massive theory is considered
in flat spacetime. For this case it is seen that the theory suffers
from ghosts, and to get rid of the ghost term the coupling constant
of the square of Ricci tensor must be taken to zero. Also, the Newtonian
potentials are calculated for static point sources in four and three
dimensions. From the four dimensional Newtonian potential, the vDVZ
discontinuity is obtained. The massive gravity theory does not fit
in with the pure Einstein theory. But in three dimensions it works
well with the pure Einstein theory for small separations of the sources
since the massless limit does not exist. The other interesting limit
is the massless theory in flat spacetime. In this case first mass
is set to zero and then the flat spacetime limit is taken. For this
case the poles, corresponding residues and Newtonian potential energies
are calculated. When the theory is constrained to ghost and tachyon
freedom, from the poles, residues and Newtonian potentials, it is
seen that the dimension must be set to three, the coupling constant
of the Einstein-Hilbert term must taken to be negative and there must
be a relation between the coupling constants of the higher curvature
terms as $8\alpha+3\beta=0$. This theory is known as NMG. Therefore,
this theory is obtained from a different perspective than that of
\cite{Bht1,Bht2}. In three dimensions the Newtonian potential energy
again has the same potential energy as the Einstein gravity at the
NMG limit. The Newtonian potential energy in four dimensions has a
ghost term and to get rid of this term $\beta\rightarrow0$ limit
must be taken. Also, for this limit the general Newtonian potential
energy is calculated for $D$ dimensions. From this equation it can
be seen that for all dimensions there is a ghost term.

In the third chapter, the three dimensional quadratic curvature theory
is studied in more detail. To understand how the NMG theory is singled
out among other three dimensional theories, the most general action
is written in canonical form in both flat and dS spacetimes. First
the flat spacetime case is studied. To get the canonical form, the
metric perturbation is decomposed in terms of six scalar functions.
From these functions three gauge-invariant functions are constructed
by the help of Bianchi identity. The components of linearized Einstein
tensor is written in terms of these gauge-invariant combinations.
With the help of these components and the linearized Ricci scalar,
the most general quadratic action is written in the canonical form
with these three gauge-invariant fields, one of which is automatically
decoupled from the others. The decoupling of the other two fields
depends on whether $2\alpha+\beta$ is zero or not. These two cases
are discussed separately. When this constant is set to a generic value
apart from zero, one of the fields that has no dynamics in the action
can be written in terms of the other dynamical field. After this manipulation,
a decoupled action is computed. This action is analyzed for some special
points and it is found that for NMG case, the higher-derivative term
disappears and we are left with a parity-invariant theory. Except
for the NMG case, this theory describes a ghost like excitation. When
we set $2\alpha+\beta=0$ at the action level and define a new field,
a decoupled action can be written which has a ghost excitation. Then
the discussion is extended by adding static sources and spinning masses
to the action separately. For the static case an interaction part
comes to the theory. The Newtonian potential energy of this interaction
part becomes attractive for negative $\kappa$. By adding spinning
masses, the spin-spin interaction part is also found. Unfortunately,
this interaction can be repulsive or attractive depending on the signs
of the spins. Moreover, some of the found results are also obtained
from the nonlinear theory for the NMG limit. The last part for the
flat spacetime discussion is to add the gravitational Chern-Simons
term to the general quadratic action. For this case the masses of
the excitations are found. Since, these masses are different, this
is a parity-violating theory. Also, for the topologically massive
gravity limit, the expected result is found.

In the second part of the third chapter, the discussion is extended
to curved backgrounds, namely to dS backgrounds. In this part the
calculations are repeated for this case and again three gauge invariant
functions are found. With the help of these functions the generic
action is written with one decoupled field and two coupled fields.
For the decoupled field the mass is computed, and for the coupled
fields the mass is obtained from both the nonrelativistic limit and
the equations of motion.

In the Appendix A, we give some details of the computation which
can be helpful to follow the discussion.

\appendix

\section{Metric Decomposition and The Higher Curvature Action in dS Background}

\subsection{Introduction}

In this chapter some of the calculations are given to follow the bulk
of the dissertation. Here the gauge invariant action is calculated
in dS spacetime. For this purpose, first the gauge invariance of the
metric perturbation is written, then from the components of the Einstein
tensor the gauge-invariant functions are defined. Then the action
is written in terms of these gauge-invariant functions. In this part
all the raising and lowering operations are made by the background
metric and all covariant derivatives are defined with the background
metric.

The metric is taken as 
\begin{equation}
g_{\mu\nu}=\bar{g}_{\mu\nu}+h_{\mu\nu},\label{metric_ds_1}
\end{equation}
where $\bar{g}_{\mu\nu}=\frac{\ell^{2}}{t^{2}}\eta_{\mu\nu}$ is the
background metric and $\eta_{\mu\nu}$ is the flat spacetime metric.
The perturbation part of the metric $h_{\mu\nu}$ is decomposed as
\begin{align}
h_{ij} & \equiv\left(\delta_{ij}+\hat{\partial}_{i}\hat{\partial}_{j}\right)\phi-\hat{\partial}_{i}\hat{\partial}_{j}\chi+\left(\epsilon_{ik}\hat{\partial}_{k}\hat{\partial}_{j}+\epsilon_{jk}\hat{\partial}_{k}\hat{\partial}_{i}\right)\xi,\nonumber \\
h_{0i} & \equiv\left(-\epsilon_{ij}\partial_{j}\eta+\partial_{i}N_{L}\right),\qquad h_{00}\equiv N.\label{met_per_dec}
\end{align}
Here $\epsilon_{ij}$ is the tensor density with the convention $\epsilon_{12}=1$.

The trace of the perturbation metric is 
\begin{align}
h & =\bar{g}^{\mu\nu}h_{\mu\nu}=\frac{t^{2}}{a^{2}}\eta^{\mu\nu}h_{\mu\nu}=\frac{t^{2}}{a^{2}}\left(-h_{00}+h_{ii}\right),\nonumber \\
 & =\frac{t^{2}}{a^{2}}\left(-N+\phi+\chi\right).\label{trace_of_h_mu_nu}
\end{align}
Some useful identities are as follows:
\begin{align}
\partial_{i}h_{0i} & =\partial_{i}\left(-\epsilon_{ik}\partial_{k}\eta+\partial_{i}N_{L}\right)=\partial_{i}^{2}N_{L},\nonumber \\
\partial_{i}\partial_{j}h_{ij} & =\partial_{i}\partial_{j}\left[\left(\delta_{ij}+\hat{\partial}_{i}\hat{\partial}_{j}\right)\phi-\hat{\partial}_{i}\hat{\partial}_{j}\chi+\left(\epsilon_{ik}\hat{\partial}_{k}\hat{\partial}_{j}+\epsilon_{jk}\hat{\partial}_{k}\hat{\partial}_{i}\right)\xi\right],\nonumber \\
 & =\partial_{i}\left[\left(\partial_{i}-\partial_{i}\right)\phi+\partial_{i}\chi-\epsilon_{ik}\partial_{k}\xi\right]=\partial_{i}^{2}\chi,\label{useful_identities}\\
h_{ii} & =\left(\delta_{ii}+\hat{\partial}_{i}\hat{\partial}_{i}\right)\phi-\hat{\partial}_{i}\hat{\partial}_{i}\chi+\left(\epsilon_{ik}\hat{\partial}_{k}\hat{\partial}_{i}+\epsilon_{ik}\hat{\partial}_{k}\hat{\partial}_{i}\right)\xi=\phi+\chi.\nonumber 
\end{align}

Later the six free functions are redefined as $\tilde{\psi}\equiv\frac{\ell^{2}}{t^{2}}\psi$,
to get simple form of the results. With this redefinition, the metric
(\ref{metric_ds_1}) takes the form 
\begin{equation}
g_{\mu\nu}=\frac{\ell^{2}}{t^{2}}\left(\eta_{\mu\nu}+h_{\mu\nu}\right).\label{metric_ds_2}
\end{equation}

In this appendix only the calculations of the dS part is given but
the flat spacetime part is given at the needed steps by taking the
limits $\ell\rightarrow\infty,\,\ell/t\rightarrow1$.

\subsection{The Gauge-Invariance of Metric Decomposition:}

The gauge transformation is $\delta_{\zeta}h_{\mu\nu}=\nabla_{\mu}\zeta_{\nu}+\nabla_{\nu}\zeta_{\mu}$
where 
\begin{equation}
\zeta_{\mu}=\left(\zeta_{0},\,-\epsilon_{ij}\partial_{j}\zeta+\partial_{i}\kappa\right)\label{gauge_vec}
\end{equation}
 and $\epsilon_{ij}$ is the Levi-Civita tensor density with the convention
$\epsilon_{12}=1$. To calculate the gauge-invariance of the metric
components the connection coefficients are needed. For (\ref{metric_ds_1})
the coefficients can be calculated by use of 
\begin{align}
\Gamma_{\sigma\nu}^{\mu}= & \frac{1}{2}\bar{g}^{\mu\lambda}\left(\partial_{\sigma}\bar{g}_{\nu\lambda}+\partial_{\nu}\bar{g}_{\sigma\lambda}-\partial_{\lambda}\bar{g}_{\nu\sigma}\right),\label{connection}
\end{align}
where for notational simplicity we omit the bar sign on $\Gamma_{\sigma\nu}^{\mu}$
and in what follows we shall also omit the bar on the covariant derivatives.
There are only three non-trivial components of (\ref{connection})
\begin{equation}
\Gamma_{00}^{0}=-\frac{1}{t},\;\Gamma_{ij}^{0}=-\frac{1}{t}\delta_{ij},\;\Gamma_{0j}^{i}=-\frac{1}{t}\delta_{ij}.\label{comp_conec}
\end{equation}
All other components of the connection coefficients are zero. For
the full tensor the gauge invariance can be written as 
\begin{equation}
h_{\mu\nu}^{\prime}-h_{\mu\nu}=\partial_{\mu}\zeta_{\nu}+\partial_{\nu}\zeta_{\mu}-2\Gamma_{\mu\nu}^{\lambda}\zeta_{\lambda},\label{gauge}
\end{equation}
after decomposing the covariant derivative.

\subsubsection{The $h_{00}$ Component:}

Let start with the $h_{00}$ component. Using (\ref{gauge});
\begin{align}
h_{00}^{\prime}-h_{00} & =2\partial_{0}\zeta_{0}-2\Gamma_{00}^{0}\zeta_{0}-2\Gamma_{00}^{i}\zeta_{i}\nonumber \\
N^{\prime}-N & =2\left(\dot{\zeta}_{0}+\frac{1}{t}\zeta_{0}\right),\label{gauge_N}
\end{align}
where in the second line, (\ref{comp_conec}) is used. Taking $h_{\mu\nu}\rightarrow\frac{\ell^{2}}{t^{2}}h_{\mu\nu}$
\begin{equation}
\delta_{\zeta}N=2\frac{t^{2}}{\ell^{2}}\left(\dot{\zeta}_{0}+\frac{1}{t}\zeta_{0}\right).\label{gauge_h_00}
\end{equation}

\subsubsection{The $h_{i0}$ Component:}

The $h_{i0}$ component can be transformed again using (\ref{gauge})
as 
\begin{align}
h_{i0}^{\prime}-h_{i0} & =\partial_{0}\zeta_{i}+\partial_{i}\zeta_{0}-2\Gamma_{0i}^{j}\zeta_{j}-2\Gamma_{0i}^{0}\zeta_{0}\nonumber \\
-\epsilon_{ij}\partial_{j}\eta^{\prime}+\partial_{i}N_{L}^{\prime}+\epsilon_{ij}\partial_{j}\eta-\partial_{i}N_{L} & =\partial_{0}\left(-\epsilon_{ij}\partial_{j}\zeta+\partial_{i}\kappa\right)+\partial_{i}\zeta_{0}+\frac{2}{t}\delta_{ij}\left(-\epsilon_{kj}\partial_{k}\zeta+\partial_{j}\kappa\right)\nonumber \\
-\epsilon_{ij}\partial_{j}\left(\eta^{\prime}-\eta\right)+\partial_{i}\left(N_{L}^{\prime}-N_{L}\right) & =-\partial_{j}\epsilon_{ij}\left(\dot{\zeta}-\frac{2}{t}\zeta\right)+\partial_{i}\left(\dot{\kappa}+\zeta_{0}+\frac{2}{t}\kappa\right),\label{gauge_h_0i1}
\end{align}
where we have used (\ref{met_per_dec}) and (\ref{comp_conec}) in
the left and right hand sides of the second line, respectively. Equating
both sides with respect to their coefficients, we found 
\begin{align}
\delta_{\zeta}N_{L} & =\left(\dot{\kappa}+\zeta_{0}+\frac{2}{t}\kappa\right),\quad\delta_{\zeta}\eta=\left(\dot{\zeta}-\frac{2}{t}\zeta\right),\label{gauge_N_L_and_eta}
\end{align}
 and for the redefinition of the metric perturbation, (\ref{gauge_N_L_and_eta})
take the following form 
\begin{equation}
\delta_{\zeta}N_{L}=\frac{t^{2}}{\ell^{2}}\left(\dot{\kappa}+\zeta_{0}+\frac{2}{t}\kappa\right),\;\delta_{\zeta}\eta=\frac{t^{2}}{\ell^{2}}\left(\dot{\zeta}-\frac{2}{t}\zeta\right).\label{gauge_h_0i}
\end{equation}

\subsubsection{The $h_{ij}$ Component:}

From (\ref{gauge}) the last term is can be written as 
\begin{align}
h_{ij}^{\prime}-h_{ij}= & \partial_{j}\zeta_{i}+\partial_{i}\zeta_{j}-2\Gamma_{ji}^{0}\zeta_{0}-2\Gamma_{ji}^{k}\zeta_{k}\nonumber \\
= & \left(\delta_{ij}+\hat{\partial}_{i}\hat{\partial}_{j}\right)\phi^{\prime}-\hat{\partial}_{i}\hat{\partial}_{j}\chi\prime+\left(\epsilon_{ik}\hat{\partial}_{k}\hat{\partial}_{j}+\epsilon_{jk}\hat{\partial}_{k}\hat{\partial}_{i}\right)\xi^{\prime}\nonumber \\
 & -\left(\delta_{ij}+\hat{\partial}_{i}\hat{\partial}_{j}\right)\phi+\hat{\partial}_{i}\hat{\partial}_{j}\chi-\left(\epsilon_{ik}\hat{\partial}_{k}\hat{\partial}_{j}+\epsilon_{jk}\hat{\partial}_{k}\hat{\partial}_{i}\right)\xi\nonumber \\
= & \partial_{j}\left(-\epsilon_{ik}\partial_{k}\zeta+\partial_{i}\kappa\right)+\partial_{i}\left(-\epsilon_{jk}\partial_{k}\zeta+\partial_{j}\kappa\right)+\frac{2}{t}\delta_{ij}\zeta_{0},\label{gauge_h_ij}
\end{align}
where we have again used (\ref{met_per_dec}),(\ref{gauge_vec}) and
(\ref{comp_conec}). The (\ref{gauge_h_ij}) is multiplied with $\partial_{j}$
in order to eliminate $\phi$ field. Note that $\partial_{j}\hat{\partial}_{i}\hat{\partial}_{j}=-\partial_{i}$
and $\hat{\partial}_{i}\hat{\partial}_{i}=-1$. Then 
\begin{equation}
\partial_{i}\chi\prime-\epsilon_{ik}\partial_{k}\xi^{\prime}-\partial_{i}\chi+\epsilon_{ik}\partial_{k}\xi=-\epsilon_{ik}\partial_{j}^{2}\partial_{k}\zeta+2\partial_{i}\partial_{j}^{2}\kappa+\frac{2}{t}\partial_{i}\zeta_{0},\label{gauge_h_ij_partial_j}
\end{equation}
and equating both sides with respect to their coefficients, we get
\begin{align}
\partial_{i}\left(\chi\prime-\chi\right) & =2\partial_{j}^{2}\kappa+\frac{2}{t}\zeta_{0},\nonumber \\
\delta_{\zeta}\chi & =2\left(\partial_{j}^{2}\kappa+\frac{1}{t}\zeta_{0}\right),\label{gauge_chi}
\end{align}
and 
\begin{align}
-\epsilon_{ik}\partial_{k}\left(\xi^{\prime}-\xi\right)= & -\epsilon_{ik}\partial_{j}^{2}\partial_{k}\zeta\nonumber \\
\delta_{\zeta}\xi= & \partial_{j}^{2}\zeta.\label{gauge_xi}
\end{align}
For the redefined metric, (\ref{gauge_chi}) and (\ref{gauge_xi})
becomes 
\begin{equation}
\delta_{\zeta}\chi=2\frac{t^{2}}{\ell^{2}}\left(\partial_{j}^{2}\kappa+\frac{1}{t}\zeta_{0}\right).\label{gauge_h_ij_chi}
\end{equation}
and 
\begin{equation}
\delta_{\zeta}\xi=\frac{t^{2}}{\ell^{2}}\partial_{j}^{2}\zeta.\label{gauge_h_ij_ksi}
\end{equation}
We can eliminate $\xi$ field by multiplying (\ref{gauge_h_ij}) with
$\delta_{ij}$ and using (\ref{gauge_chi}) we get 
\begin{equation}
\phi^{\prime}-\phi=\frac{2}{t}\zeta_{0},\label{gauge_phi}
\end{equation}
and for the redefined metric perturbation (\ref{gauge_phi}) reads
\begin{equation}
\delta_{\zeta}\phi=\frac{2t}{\ell^{2}}\zeta_{0}.\label{gauge_h_ij_phi}
\end{equation}
For the flat spacetime case the non-gauge-invariant functions become
\begin{align}
 & \delta_{\zeta}\phi=0,\;\delta_{\zeta}\xi=\partial_{j}^{2}\zeta,\;\delta_{\zeta}\chi=2\partial_{j}^{2}\kappa,\nonumber \\
 & \delta_{\zeta}N_{L}=\dot{\kappa}+\zeta_{0},\;\delta_{\zeta}\eta=\dot{\zeta},\delta_{\zeta}N=2\dot{\zeta}_{0}.\label{gauge_flat}
\end{align}
From (\ref{gauge_flat}) it can be seen that apart from $\phi$ function
all other functions are not gauge-invariant. 

In this part we see that the functions are not gauge-invariant. Therefore,
the next section is devoted to find gauge invariant functions. For
this purpose, we need to find the components of the linear Einstein
tensor in terms of the free functions since it is known that linear
Einstein tensor must satisfy the Bianchi identity. Therefore, its
components must be gauge invariant. 

In the next section we first find the components of Ricci tensor and
scalar in terms of these six free functions. Then using these components,
the components of the linear Einstein tensor are written. Then we
define the gauge invariant functions and also check their invariance.
Moreover, we see that Bianchi identity give us a relation between
these gauge invariant functions.

\subsection{Gauge-Invariant Combinations:}

The Bianchi identity gives a clue to finding these gauge invariant
combinations, constructed from the six free functions, since $\nabla_{\mu}\mathcal{G}_{L}^{\mu\nu}=0$.
Therefore, the components of the linear Einstein tensor must be composed
of such combinations that are invariant under a gauge transformation.
The components of the linear Einstein tensor are found one by one.

Before calculating the components of Einstein tensor the components
of the linear Ricci tensor and the Scalar curvature must be calculated.

\subsubsection{Ricci Tensor:}

The Ricci tensor is 
\begin{equation}
R_{\mu\nu}^{L}=\frac{1}{2}\left(\nabla^{\sigma}\nabla_{\mu}h_{\nu\sigma}+\nabla^{\sigma}\nabla_{\nu}h_{\mu\sigma}-\Box h_{\mu\nu}-\nabla_{\mu}\nabla_{\nu}h\right).\label{Linear_Ricci_ten}
\end{equation}
Extracting the covariant derivatives, the Ricci tensor becomes
\begin{align}
R_{\mu\nu}^{L} & =\frac{1}{2}\left(\nabla^{\sigma}\nabla_{\mu}h_{\nu\sigma}+\nabla^{\sigma}\nabla_{\nu}h_{\mu\sigma}-\Box h_{\mu\nu}-\nabla_{\mu}\nabla_{\nu}h\right),\nonumber \\
 & =\frac{1}{2}\left(\bar{g}^{\sigma\rho}\nabla_{\rho}\nabla_{\mu}h_{\nu\sigma}+\bar{g}^{\sigma\rho}\nabla_{\rho}\nabla_{\nu}h_{\mu\sigma}-\bar{g}^{\sigma\rho}\nabla_{\rho}\nabla_{\sigma}h_{\mu\nu}\right)-\frac{1}{2}\nabla_{\mu}\left(\partial_{\nu}h\right),\nonumber \\
 & =\frac{1}{2}\bar{g}^{\sigma\rho}\nabla_{\rho}\left(\nabla_{\mu}h_{\nu\sigma}+\nabla_{\nu}h_{\mu\sigma}-\nabla_{\sigma}h_{\mu\nu}\right)-\frac{1}{2}\left(\partial_{\mu}\partial_{\nu}h-\Gamma_{\mu\nu}^{\lambda}\partial_{\lambda}h\right),\nonumber \\
 & =\frac{1}{2}\bar{g}^{\sigma\rho}\nabla_{\rho}\left[\left(\partial_{\mu}h_{\nu\sigma}-\Gamma_{\mu\nu}^{\lambda}h_{\lambda\sigma}-\Gamma_{\mu\sigma}^{\lambda}h_{\nu\lambda}\right)+\left(\partial_{\nu}h_{\mu\sigma}-\Gamma_{\nu\mu}^{\lambda}h_{\lambda\sigma}-\Gamma_{\nu\sigma}^{\lambda}h_{\mu\lambda}\right)\right.\nonumber \\
 & \phantom{=}\left.-\left(\partial_{\sigma}h_{\mu\nu}-\Gamma_{\sigma\mu}^{\lambda}h_{\lambda\nu}-\Gamma_{\sigma\nu}^{\lambda}h_{\mu\lambda}\right)\right]-\frac{1}{2}\left(\partial_{\mu}\partial_{\nu}h-\Gamma_{\mu\nu}^{\lambda}\partial_{\lambda}h\right),\label{eq:Ricci_ten_05}
\end{align}
and 
\begin{align}
R_{\mu\nu}^{L} & =\frac{1}{2}\bar{g}^{\sigma\rho}\nabla_{\rho}\left[\partial_{\mu}h_{\nu\sigma}-2\Gamma_{\mu\nu}^{\lambda}h_{\lambda\sigma}+\partial_{\nu}h_{\mu\sigma}-\partial_{\sigma}h_{\mu\nu}\right]-\frac{1}{2}\left(\partial_{\mu}\partial_{\nu}h-\Gamma_{\mu\nu}^{\lambda}\partial_{\lambda}h\right),\nonumber \\
 & =\frac{1}{2}\bar{g}^{\sigma\rho}\left\{ \left[\partial_{\rho}\left(\partial_{\mu}h_{\nu\sigma}\right)-\Gamma_{\rho\mu}^{\alpha}\left(\partial_{\alpha}h_{\nu\sigma}\right)-\Gamma_{\rho\nu}^{\alpha}\left(\partial_{\mu}h_{\alpha\sigma}\right)-\Gamma_{\rho\sigma}^{\alpha}\left(\partial_{\mu}h_{\nu\alpha}\right)\right]\right.\nonumber \\
 & \phantom{=\frac{1}{2}\bar{g}^{\sigma\rho}}+\left[\partial_{\rho}\left(\partial_{\nu}h_{\mu\sigma}\right)-\Gamma_{\rho\nu}^{\alpha}\left(\partial_{\alpha}h_{\mu\sigma}\right)-\Gamma_{\rho\mu}^{\alpha}\left(\partial_{\nu}h_{\alpha\sigma}\right)-\Gamma_{\rho\sigma}^{\alpha}\left(\partial_{\nu}h_{\mu\alpha}\right)\right]\nonumber \\
 & \phantom{=\frac{1}{2}\bar{g}^{\sigma\rho}}\left.-\left[\partial_{\rho}\left(\partial_{\sigma}h_{\mu\nu}\right)-\Gamma_{\rho\sigma}^{\alpha}\left(\partial_{\alpha}h_{\mu\nu}\right)-\Gamma_{\rho\mu}^{\alpha}\left(\partial_{\sigma}h_{\alpha\nu}\right)-\Gamma_{\rho\nu}^{\alpha}\left(\partial_{\sigma}h_{\mu\alpha}\right)\right]\right\} \label{Ricci_ten1}\\
 & \phantom{=\frac{1}{2}\bar{g}^{\sigma\rho}}-\bar{g}^{\sigma\rho}\left[\partial_{\rho}\left(\Gamma_{\mu\nu}^{\lambda}h_{\lambda\sigma}\right)-\Gamma_{\rho\mu}^{\alpha}\Gamma_{\alpha\nu}^{\lambda}h_{\lambda\sigma}-\Gamma_{\rho\nu}^{\alpha}\Gamma_{\mu\alpha}^{\lambda}h_{\lambda\sigma}-\Gamma_{\rho\sigma}^{\alpha}\Gamma_{\mu\nu}^{\lambda}h_{\lambda\alpha}\right]\nonumber \\
 & \phantom{=}-\frac{1}{2}\left(\partial_{\mu}\partial_{\nu}h-\Gamma_{\mu\nu}^{\lambda}\partial_{\lambda}h\right),\nonumber 
\end{align}
after collecting terms in parenthesis we obtain,
\begin{align}
2R_{\mu\nu}^{L} & =\bar{g}^{\sigma\rho}\left\{ \partial_{\rho}\left(\partial_{\mu}h_{\nu\sigma}+\partial_{\nu}h_{\mu\sigma}-\partial_{\sigma}h_{\mu\nu}\right)-\Gamma_{\rho\mu}^{\alpha}\left(\partial_{\alpha}h_{\nu\sigma}+\partial_{\nu}h_{\alpha\sigma}-\partial_{\sigma}h_{\alpha\nu}\right)\phantom{\left[\left(\Gamma_{\mu}^{\lambda}\right)\right]}\right.\nonumber \\
 & \phantom{=\bar{g}^{\sigma\rho}}-\Gamma_{\rho\nu}^{\alpha}\left(\partial_{\mu}h_{\alpha\sigma}+\partial_{\alpha}h_{\mu\sigma}-\partial_{\sigma}h_{\mu\alpha}\right)-\Gamma_{\rho\sigma}^{\alpha}\left(\partial_{\mu}h_{\nu\alpha}+\partial_{\nu}h_{\mu\alpha}-\partial_{\alpha}h_{\mu\nu}\right)\nonumber \\
 & \phantom{=\bar{g}^{\sigma\rho}}\left.-2\left[\partial_{\rho}\left(\Gamma_{\mu\nu}^{\lambda}h_{\lambda\sigma}\right)-\Gamma_{\rho\mu}^{\alpha}\Gamma_{\alpha\nu}^{\lambda}h_{\lambda\sigma}-\Gamma_{\rho\nu}^{\alpha}\Gamma_{\mu\alpha}^{\lambda}h_{\lambda\sigma}-\Gamma_{\rho\sigma}^{\alpha}\Gamma_{\mu\nu}^{\lambda}h_{\lambda\alpha}\right]\right\} \label{Ricci_ten2}\\
 & \phantom{=}-\left(\partial_{\mu}\partial_{\nu}h-\Gamma_{\mu\nu}^{\lambda}\partial_{\lambda}h\right).\nonumber 
\end{align}
Putting $\bar{g}^{\sigma\rho}=\frac{t^{2}}{\ell^{2}}\eta^{\sigma\rho}$
and summing $\sigma$ and $\rho$ indices we get
\begin{align}
R_{\mu\nu}^{L} & =-\frac{t^{2}}{2\ell^{2}}\left\{ \partial_{0}\left(\partial_{\mu}h_{\nu0}+\partial_{\nu}h_{\mu0}-\partial_{0}h_{\mu\nu}\right)-\Gamma_{0\mu}^{\alpha}\left(\partial_{\alpha}h_{\nu0}+\partial_{\nu}h_{\alpha0}-\partial_{0}h_{\alpha\nu}\right)\phantom{\left[\left(\Gamma_{\mu}^{\lambda}\right)\right]}\right.\nonumber \\
 & \phantom{=-\frac{t^{2}}{2\ell^{2}}}-\Gamma_{0\nu}^{\alpha}\left(\partial_{\mu}h_{\alpha0}+\partial_{\alpha}h_{\mu0}-\partial_{0}h_{\mu\alpha}\right)-\Gamma_{00}^{\alpha}\left(\partial_{\mu}h_{\nu\alpha}+\partial_{\nu}h_{\mu\alpha}-\partial_{\alpha}h_{\mu\nu}\right)\nonumber \\
 & \phantom{=-\frac{t^{2}}{2\ell^{2}}}\left.-2\left[\partial_{0}\left(\Gamma_{\mu\nu}^{\lambda}h_{\lambda0}\right)-\Gamma_{0\mu}^{\alpha}\Gamma_{\alpha\nu}^{\lambda}h_{\lambda0}-\Gamma_{0\nu}^{\alpha}\Gamma_{\mu\alpha}^{\lambda}h_{\lambda0}-\Gamma_{00}^{\alpha}\Gamma_{\mu\nu}^{\lambda}h_{\lambda\alpha}\right]\right\} \nonumber \\
 & \phantom{=}+\frac{t^{2}}{2\ell^{2}}\left\{ \partial_{i}\left(\partial_{\mu}h_{\nu i}+\partial_{\nu}h_{\mu i}-\partial_{i}h_{\mu\nu}\right)-\Gamma_{i\mu}^{\alpha}\left(\partial_{\alpha}h_{\nu i}+\partial_{\nu}h_{\alpha i}-\partial_{i}h_{\alpha\nu}\right)\phantom{\phantom{\left[\left(\Gamma_{\mu}^{\lambda}\right)\right]}}\right.\label{Ricci_ten3}\\
 & \phantom{=+\frac{t^{2}}{2\ell^{2}}}-\Gamma_{i\nu}^{\alpha}\left(\partial_{\mu}h_{\alpha i}+\partial_{\alpha}h_{\mu i}-\partial_{i}h_{\mu\alpha}\right)-\Gamma_{ii}^{\alpha}\left(\partial_{\mu}h_{\nu\alpha}+\partial_{\nu}h_{\mu\alpha}-\partial_{\alpha}h_{\mu\nu}\right)\nonumber \\
 & \phantom{=+\frac{t^{2}}{2\ell^{2}}}\left.-2\left[\partial_{i}\left(\Gamma_{\mu\nu}^{\lambda}h_{\lambda i}\right)-\Gamma_{i\mu}^{\alpha}\Gamma_{\alpha\nu}^{\lambda}h_{\lambda i}-\Gamma_{i\nu}^{\alpha}\Gamma_{\mu\alpha}^{\lambda}h_{\lambda i}-\Gamma_{ii}^{\alpha}\Gamma_{\mu\nu}^{\lambda}h_{\lambda\alpha}\right]\right\} \nonumber \\
 & \phantom{=}-\frac{1}{2}\left(\partial_{\mu}\partial_{\nu}h-\Gamma_{\mu\nu}^{\lambda}\partial_{\lambda}h\right).\nonumber 
\end{align}
With the help of (\ref{comp_conec}), (\ref{Ricci_ten3}) becomes
\begin{align}
R_{\mu\nu}^{L} & =-\frac{t^{2}}{2\ell^{2}}\left\{ \partial_{0}\left(\partial_{\mu}h_{\nu0}+\partial_{\nu}h_{\mu0}-\partial_{0}h_{\mu\nu}\right)+\frac{2}{t}\left(\partial_{\mu}h_{\nu0}+\partial_{\nu}h_{\mu0}-\partial_{0}h_{\mu\nu}\right)\phantom{\left[\frac{1}{t}\right]}\right.\nonumber \\
 & \phantom{=-\frac{t^{2}}{2\ell^{2}}}+\frac{1}{t}\left(\partial_{\mu}h_{\nu0}+\partial_{\nu}h_{\mu0}-\partial_{0}h_{\mu\nu}\right)\nonumber \\
 & \phantom{=-\frac{t^{2}}{2\ell^{2}}}\left.-2\left[\partial_{0}\left(\Gamma_{\mu\nu}^{\lambda}h_{\lambda0}\right)+\frac{1}{t}\left(2\Gamma_{\mu\nu}^{\lambda}h_{\lambda0}+\Gamma_{\mu\nu}^{\lambda}h_{\lambda0}\right)\right]\right\} \nonumber \\
 & \phantom{=}+\frac{t^{2}}{2\ell^{2}}\left\{ \partial_{i}\left(\partial_{\mu}h_{\nu i}+\partial_{\nu}h_{\mu i}-\partial_{i}h_{\mu\nu}\right)-\Gamma_{i\mu}^{\alpha}\left(\partial_{\alpha}h_{\nu i}+\partial_{\nu}h_{\alpha i}-\partial_{i}h_{\alpha\nu}\right)\phantom{\left[\frac{1}{t}\right]}\right.\label{Ricci_ten4}\\
 & \phantom{=+\frac{t^{2}}{2\ell^{2}}}-\Gamma_{i\nu}^{\alpha}\left(\partial_{\mu}h_{\alpha i}+\partial_{\alpha}h_{\mu i}-\partial_{i}h_{\mu\alpha}\right)+\frac{2}{t}\left(\partial_{\mu}h_{\nu0}+\partial_{\nu}h_{\mu0}-\partial_{0}h_{\mu\nu}\right)\nonumber \\
 & \phantom{=+\frac{t^{2}}{2\ell^{2}}}\left.-2\left[\partial_{i}\left(\Gamma_{\mu\nu}^{\lambda}h_{\lambda i}\right)-\Gamma_{i\mu}^{\alpha}\Gamma_{\alpha\nu}^{\lambda}h_{\lambda i}-\Gamma_{i\nu}^{\alpha}\Gamma_{\mu\alpha}^{\lambda}h_{\lambda i}+\frac{2}{t}\Gamma_{\mu\nu}^{\lambda}h_{\lambda0}\right]\right\} \nonumber \\
 & \phantom{=}-\frac{1}{2}\left(\partial_{\mu}\partial_{\nu}h-\Gamma_{\mu\nu}^{\lambda}\partial_{\lambda}h\right),\nonumber 
\end{align}
finally, we end up with
\begin{align}
R_{\mu\nu}^{L} & =-\frac{t^{2}}{2\ell^{2}}\left\{ \partial_{0}\left(\partial_{\mu}h_{\nu0}+\partial_{\nu}h_{\mu0}-\partial_{0}h_{\mu\nu}\right)-\partial_{i}\left(\partial_{\mu}h_{\nu i}+\partial_{\nu}h_{\mu i}-\partial_{i}h_{\mu\nu}\right)\phantom{\left[\frac{1}{t}\right]}\right.\nonumber \\
 & \phantom{=-\frac{t^{2}}{2\ell^{2}}}+\frac{1}{t}\left(\partial_{\mu}h_{\nu0}+\partial_{\nu}h_{\mu0}-\partial_{0}h_{\mu\nu}\right)+\Gamma_{i\mu}^{\alpha}\left(\partial_{\alpha}h_{\nu i}+\partial_{\nu}h_{\alpha i}-\partial_{i}h_{\alpha\nu}\right)\nonumber \\
 & \phantom{=-\frac{t^{2}}{2\ell^{2}}}+\Gamma_{i\nu}^{\alpha}\left(\partial_{\mu}h_{\alpha i}+\partial_{\alpha}h_{\mu i}-\partial_{i}h_{\mu\alpha}\right)\nonumber \\
 & \phantom{=-\frac{t^{2}}{2\ell^{2}}}\left.-2\left[\partial_{0}\left(\Gamma_{\mu\nu}^{\lambda}h_{\lambda0}\right)-\partial_{i}\left(\Gamma_{\mu\nu}^{\lambda}h_{\lambda i}\right)+\frac{1}{t}\Gamma_{\mu\nu}^{\lambda}h_{\lambda0}+\Gamma_{i\mu}^{\alpha}\Gamma_{\alpha\nu}^{\lambda}h_{\lambda i}+\Gamma_{i\nu}^{\alpha}\Gamma_{\mu\alpha}^{\lambda}h_{\lambda i}\right]\right\} \label{R_mn_gen_form}\\
 & \phantom{=}-\frac{1}{2}\left(\partial_{\mu}\partial_{\nu}h-\Gamma_{\mu\nu}^{\lambda}\partial_{\lambda}h\right).\nonumber 
\end{align}

With this result the components of the Ricci tensor can be calculated.

\subsubsection{The $R_{00}^{L}$ Component:}

Using (\ref{R_mn_gen_form}), $R_{00}^{L}$ component becomes
\begin{align}
R_{00}^{L} & =-\frac{t^{2}}{2\ell^{2}}\left\{ \partial_{0}\left(\partial_{0}h_{00}+\partial_{0}h_{00}-\partial_{0}h_{00}\right)-\partial_{i}\left(\partial_{0}h_{0i}+\partial_{0}h_{0i}-\partial_{i}h_{00}\right)\right.\nonumber \\
 & \phantom{=-\frac{t^{2}}{2\ell^{2}}}+\frac{1}{t}\left(\partial_{0}h_{00}+\partial_{0}h_{00}-\partial_{0}h_{00}\right)\nonumber \\
 & \phantom{=-\frac{t^{2}}{2\ell^{2}}}+\Gamma_{i0}^{\alpha}\left(\partial_{\alpha}h_{0i}+\partial_{0}h_{\alpha i}-\partial_{i}h_{\alpha0}\right)+\Gamma_{i0}^{\alpha}\left(\partial_{0}h_{\alpha i}+\partial_{\alpha}h_{0i}-\partial_{i}h_{0\alpha}\right)\label{R_00_comp1}\\
 & \phantom{=-\frac{t^{2}}{2\ell^{2}}}\left.-2\left[\partial_{0}\left(\Gamma_{00}^{\lambda}h_{\lambda0}\right)-\partial_{i}\left(\Gamma_{00}^{\lambda}h_{\lambda i}\right)+\frac{1}{t}\Gamma_{00}^{\lambda}h_{\lambda0}+\Gamma_{i0}^{\alpha}\Gamma_{\alpha0}^{\lambda}h_{\lambda i}+\Gamma_{i0}^{\alpha}\Gamma_{0\alpha}^{\lambda}h_{\lambda i}\right]\right\} \nonumber \\
 & \phantom{=}-\frac{1}{2}\left(\partial_{0}\partial_{0}h-\Gamma_{00}^{\lambda}\partial_{\lambda}h\right),\nonumber 
\end{align}
rearranging and doing some cancellations yield
\begin{align}
R_{00}^{L} & =-\frac{t^{2}}{2\ell^{2}}\left\{ \partial_{0}^{2}h_{00}-2\partial_{0}\partial_{i}h_{0i}+\partial_{i}^{2}h_{00}+\frac{1}{t}\partial_{0}h_{00}+2\Gamma_{i0}^{\alpha}\left(\partial_{\alpha}h_{0i}+\partial_{0}h_{\alpha i}-\partial_{i}h_{\alpha0}\right)\right.\nonumber \\
 & \phantom{=-\frac{t^{2}}{2\ell^{2}}}\left.-2\left[\partial_{0}\left(\Gamma_{00}^{\lambda}h_{\lambda0}\right)-\partial_{i}\left(\Gamma_{00}^{\lambda}h_{\lambda i}\right)+\frac{1}{t}\Gamma_{00}^{\lambda}h_{\lambda0}+2\Gamma_{i0}^{\alpha}\Gamma_{\alpha0}^{\lambda}h_{\lambda i}\right]\right\} \label{R_00_comp2}\\
 & \phantom{=}-\frac{1}{2}\left(\partial_{0}\partial_{0}h-\Gamma_{00}^{\lambda}\partial_{\lambda}h\right).\nonumber 
\end{align}
Summing the repeated indices and using (\ref{comp_conec}) gives
\begin{align}
R_{00}^{L} & =-\frac{t^{2}}{2\ell^{2}}\left(\partial_{0}^{2}h_{00}-2\partial_{0}\partial_{i}h_{0i}+\partial_{i}^{2}h_{00}\right)-\frac{t}{2\ell^{2}}\left(3\partial_{0}h_{00}-2\partial_{0}h_{ii}-2\partial_{i}h_{0i}\right)\nonumber \\
 & \phantom{=}+\frac{2}{\ell^{2}}h_{ii}-\frac{1}{2}\left(\partial_{0}\partial_{0}+\frac{1}{t}\partial_{0}\right)h.\label{R_00_comp3}
\end{align}
Putting (\ref{met_per_dec}) in (\ref{R_00_comp3}) yields,
\begin{align}
R_{00}^{L} & =-\frac{t^{2}}{2\ell^{2}}\left(\ddot{N}-2\partial_{i}^{2}\dot{N}_{L}+\partial_{i}^{2}N\right)-\frac{t}{2\ell^{2}}\left(3\dot{N}-2\dot{\phi}-2\dot{\chi}-2\partial_{i}^{2}N_{L}\right)\nonumber \\
 & \phantom{=}+\frac{2}{\ell^{2}}\left(\phi+\chi\right)-\frac{1}{2}\left(\partial_{0}\partial_{0}+\frac{1}{t}\partial_{0}\right)\left[\frac{t^{2}}{\ell^{2}}\left(-N+\phi+\chi\right)\right],\nonumber \\
 & =-\frac{t^{2}}{2\ell^{2}}\left(\ddot{N}+\partial_{i}^{2}N-2\partial_{i}^{2}\dot{N}_{L}\right)-\frac{t}{2\ell^{2}}\left(3\dot{N}-2\dot{\phi}-2\dot{\chi}-2\partial_{i}^{2}N_{L}\right)+\frac{2}{\ell^{2}}\left(\phi+\chi\right)\nonumber \\
 & \phantom{=}+\frac{t^{2}}{2\ell^{2}}\left(\ddot{N}-\ddot{\phi}-\ddot{\chi}\right)+\frac{5t}{2\ell^{2}}\left(\dot{N}-\dot{\phi}-\dot{\chi}\right)+\frac{2}{\ell^{2}}\left(N-\phi-\chi\right),\label{R_00_comp4}
\end{align}
and after the cancellations the final answer of $R_{00}^{L}$ component
becomes 
\begin{equation}
R_{00}^{L}=-\frac{t^{2}}{2\ell^{2}}\left(\partial_{i}^{2}N-2\partial_{i}^{2}\dot{N}_{L}+\ddot{\phi}+\ddot{\chi}\right)+\frac{t}{2\ell^{2}}\left(2\dot{N}-3\dot{\phi}-3\dot{\chi}+2\partial_{i}^{2}N_{L}\right)+\frac{2}{\ell^{2}}N.\label{R_00_gen_func}
\end{equation}
Also, taking the $t/\ell\rightarrow1$ and $\ell\rightarrow\infty$
limits the flat spacetime case of this component can be got 
\begin{equation}
R_{00}^{L}=-\frac{1}{2}\left(\partial_{i}^{2}N-2\partial_{i}^{2}\dot{N}_{L}+\ddot{\phi}+\ddot{\chi}\right).\label{R_00_gen_func_flat}
\end{equation}

\subsubsection{The $R_{0i}^{L}$ Component:}

From (\ref{R_mn_gen_form}) $R_{0j}^{L}$ term can be written as 
\begin{align}
R_{0j}^{L} & =-\frac{t^{2}}{2\ell^{2}}\left\{ \partial_{0}\left(\partial_{0}h_{j0}+\partial_{j}h_{00}-\partial_{0}h_{0j}\right)-\partial_{i}\left(\partial_{0}h_{ji}+\partial_{j}h_{0i}-\partial_{i}h_{0j}\right)\right.\nonumber \\
 & \phantom{=-\frac{t^{2}}{2\ell^{2}}}+\frac{1}{t}\left(\partial_{0}h_{j0}+\partial_{j}h_{00}-\partial_{0}h_{0j}\right)\nonumber \\
 & \phantom{=-\frac{t^{2}}{2\ell^{2}}}+\Gamma_{i0}^{\alpha}\left(\partial_{\alpha}h_{ji}+\partial_{j}h_{\alpha i}-\partial_{i}h_{\alpha j}\right)+\Gamma_{ij}^{\alpha}\left(\partial_{0}h_{\alpha i}+\partial_{\alpha}h_{0i}-\partial_{i}h_{0\alpha}\right)\label{R_0j_comp1}\\
 & \phantom{=-\frac{t^{2}}{2\ell^{2}}}\left.-2\left[\partial_{0}\left(\Gamma_{0j}^{\lambda}h_{\lambda0}\right)-\partial_{i}\left(\Gamma_{0j}^{\lambda}h_{\lambda i}\right)+\frac{1}{t}\Gamma_{0j}^{\lambda}h_{\lambda0}+\Gamma_{i0}^{\alpha}\Gamma_{\alpha j}^{\lambda}h_{\lambda i}+\Gamma_{ij}^{\alpha}\Gamma_{0\alpha}^{\lambda}h_{\lambda i}\right]\right\} \nonumber \\
 & \phantom{=}-\frac{1}{2}\left(\partial_{0}\partial_{j}h-\Gamma_{0j}^{\lambda}\partial_{\lambda}h\right),\nonumber 
\end{align}
rearranging and doing some cancellations we get
\begin{align}
R_{0j}^{L} & =-\frac{t^{2}}{2\ell^{2}}\left\{ \partial_{0}\partial_{j}h_{00}-\partial_{0}\partial_{i}h_{ij}-\partial_{j}\partial_{i}h_{0i}+\partial_{i}^{2}h_{0j}+\frac{1}{t}\partial_{j}h_{00}\right.\nonumber \\
 & \phantom{=-\frac{t^{2}}{2\ell^{2}}}+\Gamma_{i0}^{\alpha}\left(\partial_{\alpha}h_{ji}+\partial_{j}h_{\alpha i}-\partial_{i}h_{\alpha j}\right)+\Gamma_{ij}^{\alpha}\left(\partial_{0}h_{\alpha i}+\partial_{\alpha}h_{0i}-\partial_{i}h_{0\alpha}\right)\nonumber \\
 & \phantom{=-\frac{t^{2}}{2\ell^{2}}}\left.-2\left[\partial_{0}\left(\Gamma_{0j}^{\lambda}h_{\lambda0}\right)-\partial_{i}\left(\Gamma_{0j}^{\lambda}h_{\lambda i}\right)+\frac{1}{t}\Gamma_{0j}^{\lambda}h_{\lambda0}+\Gamma_{i0}^{\alpha}\Gamma_{\alpha j}^{\lambda}h_{\lambda i}+\Gamma_{ij}^{\alpha}\Gamma_{0\alpha}^{\lambda}h_{\lambda i}\right]\right\} \nonumber \\
 & \phantom{=}-\frac{1}{2}\left(\partial_{0}\partial_{j}h-\Gamma_{0j}^{\lambda}\partial_{\lambda}h\right).\label{R_0j_comp2}
\end{align}
Again, using (\ref{comp_conec}) in (\ref{R_0j_comp2}) gives us 
\begin{align}
R_{0j}^{L} & =-\frac{t^{2}}{2\ell^{2}}\left\{ \partial_{0}\partial_{j}h_{00}-\partial_{0}\partial_{i}h_{ij}-\partial_{j}\partial_{i}h_{0i}+\partial_{i}^{2}h_{0j}\right.\nonumber \\
 & \phantom{=-\frac{t^{2}}{2\ell^{2}}}+\frac{1}{t}\partial_{j}h_{00}-\frac{1}{t}\left(\partial_{i}h_{ji}+\partial_{j}h_{ii}-\partial_{i}h_{ij}\right)-\frac{1}{t}\left(\partial_{0}h_{0j}+\partial_{0}h_{0j}-\partial_{j}h_{00}\right)\nonumber \\
 & \phantom{=-\frac{t^{2}}{2\ell^{2}}}\left.-2\left[-\frac{1}{t}\partial_{0}h_{j0}+\frac{1}{t^{2}}h_{j0}+\frac{1}{t}\partial_{i}h_{ji}-\frac{1}{t^{2}}h_{j0}+\frac{1}{t^{2}}h_{0j}+\frac{1}{t^{2}}h_{0j}\right]\right\} \nonumber \\
 & \phantom{=}-\frac{1}{2}\left(\partial_{0}\partial_{j}h+\frac{1}{t}\partial_{j}h\right),\label{R_0j_comp3}
\end{align}
doing cancellations and summing the same terms yields
\begin{align}
R_{0j}^{L} & =-\frac{t^{2}}{2\ell^{2}}\left(\partial_{0}\partial_{j}h_{00}-\partial_{0}\partial_{i}h_{ij}-\partial_{j}\partial_{i}h_{0i}+\partial_{i}^{2}h_{0j}\right)-\frac{t}{2\ell^{2}}\left(2\partial_{j}h_{00}-2\partial_{i}h_{ij}-\partial_{j}h_{ii}\right)\nonumber \\
 & \phantom{=}+\frac{2}{\ell^{2}}h_{0j}-\frac{1}{2}\left(\partial_{0}+\frac{1}{t}\right)\partial_{j}h.\label{R_0j_comp4}
\end{align}
Putting (\ref{met_per_dec}) in (\ref{R_0j_comp4}) bring forth
\begin{align}
R_{0j}^{L} & =-\frac{t^{2}}{2\ell^{2}}\left(\partial_{j}\dot{N}-\partial_{j}\dot{\chi}+\epsilon_{ji}\partial_{i}\dot{\xi}-\partial_{j}\partial_{i}^{2}N_{L}-\epsilon_{jk}\partial_{k}\partial_{i}^{2}\eta+\partial_{j}\partial_{i}^{2}N_{L}\right)\nonumber \\
 & \phantom{=}-\frac{t}{2\ell^{2}}\left(2\partial_{j}N-2\partial_{j}\chi+2\epsilon_{ji}\partial_{i}\xi-\partial_{j}\phi-\partial_{j}\chi\right)+\frac{2}{\ell^{2}}\left(-\epsilon_{ji}\partial_{i}\eta+\partial_{j}N_{L}\right)\label{R_0j_comp5}\\
 & \phantom{=}-\frac{1}{2}\left(\partial_{0}+\frac{1}{t}\right)\left[\frac{t^{2}}{\ell^{2}}\partial_{j}\left(-N+\phi+\chi\right)\right].\nonumber 
\end{align}
Rearranging and opening the parenthesis in the last line in (\ref{R_0j_comp5})
gives us
\begin{align}
R_{0j}^{L} & =-\frac{t^{2}}{2\ell^{2}}\left[\partial_{j}\left(\dot{N}-\dot{\chi}\right)+\epsilon_{jk}\partial_{k}\left(\dot{\xi}-\partial_{i}^{2}\eta\right)\right]-\frac{t}{2\ell^{2}}\left[\partial_{j}\left(2N-3\chi-\phi\right)+2\epsilon_{ji}\partial_{i}\xi\right]\nonumber \\
 & \phantom{=}+\frac{2}{\ell^{2}}\left(-\epsilon_{ji}\partial_{i}\eta+\partial_{j}N_{L}\right)+\frac{t^{2}}{2\ell^{2}}\partial_{j}\left(\dot{N}-\dot{\chi}-\dot{\phi}\right)+\frac{t}{2\ell^{2}}\partial_{j}\left(3N-3\chi-3\phi\right),\label{R_0j_comp_6}
\end{align}
and the overall result comes out as 
\begin{align}
R_{0j}^{L} & =-\frac{t^{2}}{2\ell^{2}}\left[\partial_{j}\dot{\phi}+\epsilon_{jk}\partial_{k}\left(\dot{\xi}-\partial_{i}^{2}\eta\right)\right]+\frac{t}{2\ell^{2}}\left[\partial_{j}\left(N-2\phi\right)-2\epsilon_{ji}\partial_{i}\xi\right]\nonumber \\
 & \phantom{=}+\frac{2}{\ell^{2}}\left(-\epsilon_{ji}\partial_{i}\eta+\partial_{j}N_{L}\right).\label{R_0j_gen_fun}
\end{align}
The flat spacetime limit, $t/\ell\rightarrow1$ and $\ell\rightarrow\infty$,
of this expression is 
\begin{align}
R_{0j}^{L} & =-\frac{1}{2}\left[\partial_{j}\dot{\phi}+\epsilon_{jk}\partial_{k}\left(\dot{\xi}-\partial_{i}^{2}\eta\right)\right].\label{R_0j_gen_func_flat}
\end{align}

\subsubsection{The $R_{jk}^{L}$ Component:}

The last term is $R_{jk}^{L}$ and it can be written from (\ref{R_mn_gen_form})
as 
\begin{align}
R_{jk}^{L} & =-\frac{t^{2}}{2\ell^{2}}\left\{ \partial_{0}\left(\partial_{j}h_{k0}+\partial_{k}h_{j0}-\partial_{0}h_{jk}\right)-\partial_{i}\left(\partial_{j}h_{ki}+\partial_{k}h_{ji}-\partial_{i}h_{jk}\right)\right.\nonumber \\
 & \phantom{=-\frac{t^{2}}{2\ell^{2}}}+\frac{1}{t}\left(\partial_{j}h_{k0}+\partial_{k}h_{j0}-\partial_{0}h_{jk}\right)\nonumber \\
 & \phantom{=-\frac{t^{2}}{2\ell^{2}}}+\Gamma_{ij}^{\alpha}\left(\partial_{\alpha}h_{ki}+\partial_{k}h_{\alpha i}-\partial_{i}h_{\alpha k}\right)+\Gamma_{ik}^{\alpha}\left(\partial_{j}h_{\alpha i}+\partial_{\alpha}h_{ji}-\partial_{i}h_{j\alpha}\right)\label{R_jk_comp1}\\
 & \phantom{=-\frac{t^{2}}{2\ell^{2}}}\left.-2\left[\partial_{0}\left(\Gamma_{jk}^{\lambda}h_{\lambda0}\right)-\partial_{i}\left(\Gamma_{jk}^{\lambda}h_{\lambda i}\right)+\frac{1}{t}\Gamma_{jk}^{\lambda}h_{\lambda0}+\Gamma_{ij}^{\alpha}\Gamma_{\alpha k}^{\lambda}h_{\lambda i}+\Gamma_{ik}^{\alpha}\Gamma_{j\alpha}^{\lambda}h_{\lambda i}\right]\right\} \nonumber \\
 & \phantom{=}-\frac{1}{2}\left(\partial_{j}\partial_{k}h-\Gamma_{jk}^{\lambda}\partial_{\lambda}h\right),\nonumber 
\end{align}
and using (\ref{comp_conec}) yields
\begin{align}
R_{jk}^{L} & =-\frac{t^{2}}{2\ell^{2}}\left(2\partial_{0}\partial_{j}h_{k0}-2\partial_{j}\partial_{i}h_{ki}-\partial_{0}^{2}h_{jk}+\partial_{i}^{2}h_{jk}\right)\nonumber \\
 & \phantom{=}-\frac{t}{2\ell^{2}}\left(2\partial_{j}h_{k0}-3\partial_{0}h_{jk}+2\eta_{jk}\partial_{0}h_{00}-2\eta_{jk}\partial_{i}h_{0i}\right)\label{R_jk_comp2}\\
 & \phantom{=}+\frac{2}{\ell^{2}}h_{jk}-\frac{1}{2}\left(\partial_{j}\partial_{k}h+\frac{1}{t}\eta_{jk}\partial_{0}h\right).\nonumber 
\end{align}
Putting (\ref{met_per_dec}) in (\ref{R_jk_comp2}) gives us 
\begin{align}
R_{jk}^{L} & =\left(-\frac{t^{2}}{2\ell^{2}}\right)\left\{ \partial_{j}\left[-2\epsilon_{kl}\partial_{l}\left(\dot{\eta}+\frac{1}{t}\eta\right)+\partial_{k}\left(2\dot{N}_{L}-N+\frac{2}{t}N_{L}\right)\right]\right.\nonumber \\
 & \phantom{=\left(-\frac{t^{2}}{2\ell^{2}}\right)}\left.-\left(\partial_{0}^{2}+\frac{3}{t}\partial_{0}+\frac{4}{t^{2}}\right)\left[\left(\delta_{jk}+\hat{\partial}_{j}\hat{\partial}_{k}\right)\phi-\hat{\partial}_{j}\hat{\partial}_{k}\chi+\left(\epsilon_{jl}\hat{\partial}_{l}\hat{\partial}_{k}+\epsilon_{kl}\hat{\partial}_{l}\hat{\partial}_{j}\right)\xi\right]\right.\nonumber \\
 & \phantom{\ell=\left(-\frac{t^{2}}{2\ell^{2}}\right)}\left.+\left[\delta_{jk}\partial_{i}^{2}\phi-\left(\epsilon_{jl}\partial_{l}\partial_{k}-\epsilon_{kl}\partial_{l}\partial_{j}\right)\xi\right]\right.\nonumber \\
 & \phantom{=\left(-\frac{t^{2}}{2\ell^{2}}\right)}\left.+\frac{1}{t}\delta_{jk}\left[2\left(\dot{N}-\partial_{i}^{2}N_{L}\right)+\left(-\dot{N}+\dot{\phi}+\dot{\chi}\right)+\frac{2}{t}\left(-N+\phi+\chi\right)\right]\right\} ,\label{R_jk_comp3}
\end{align}
where $\left(\epsilon_{jl}\partial_{l}\partial_{k}-\epsilon_{kl}\partial_{l}\partial_{j}\right)\xi$
should be eliminated due to the fact that $j$ and $k$ are symmetric.
Then the final form for $R_{jk}^{L}$ becomes
\begin{align}
R_{jk}^{L} & =\left(-\frac{t^{2}}{2\ell^{2}}\right)\left\{ \partial_{j}\left[-2\epsilon_{kl}\partial_{l}\left(\dot{\eta}+\frac{1}{t}\eta\right)+\partial_{k}\left(2\dot{N}_{L}-N+\frac{2}{t}N_{L}\right)\right]\right.\nonumber \\
 & \phantom{=\left(-\frac{t^{2}}{2\ell^{2}}\right)}\left.-\left(\partial_{0}^{2}+\frac{3}{t}\partial_{0}+\frac{4}{t^{2}}\right)\left[\left(\delta_{jk}+\hat{\partial}_{j}\hat{\partial}_{k}\right)\phi-\hat{\partial}_{j}\hat{\partial}_{k}\chi+\left(\epsilon_{jl}\hat{\partial}_{l}\hat{\partial}_{k}+\epsilon_{kl}\hat{\partial}_{l}\hat{\partial}_{j}\right)\xi\right]\right.\label{R_jk_gen_func}\\
 & \phantom{=\left(-\frac{t^{2}}{2\ell^{2}}\right)}\left.+\delta_{jk}\partial_{i}^{2}\phi+\frac{1}{t}\delta_{jk}\left[2\left(\dot{N}-\partial_{i}^{2}N_{L}\right)+\left(-\dot{N}+\dot{\phi}+\dot{\chi}\right)+\frac{2}{t}\left(-N+\phi+\chi\right)\right]\right\} .\nonumber 
\end{align}
The flat spacetime result of this component is 
\begin{align}
R_{jk}^{L} & =\left(-\frac{1}{2}\right)\left\{ \partial_{j}\left[-2\epsilon_{kl}\partial_{l}\left(\dot{\eta}+\frac{1}{t}\eta\right)+\partial_{k}\left(2\dot{N}_{L}-N+\frac{2}{t}N_{L}\right)\right]\right.\nonumber \\
 & \phantom{=\left(-\frac{1}{2}\right)}\left.-\left(\partial_{0}^{2}+\frac{3}{t}\partial_{0}+\frac{4}{t^{2}}\right)\left[\left(\delta_{jk}+\hat{\partial}_{j}\hat{\partial}_{k}\right)\phi-\hat{\partial}_{j}\hat{\partial}_{k}\chi+\left(\epsilon_{jl}\hat{\partial}_{l}\hat{\partial}_{k}+\epsilon_{kl}\hat{\partial}_{l}\hat{\partial}_{j}\right)\xi\right]\right.\nonumber \\
 & \phantom{=\left(-\frac{1}{2}\right)}\left.+\delta_{jk}\partial_{i}^{2}\phi+\frac{1}{t}\delta_{jk}\left[2\left(\dot{N}-\partial_{i}^{2}N_{L}\right)+\left(-\dot{N}+\dot{\phi}+\dot{\chi}\right)+\frac{2}{t}\left(-N+\phi+\chi\right)\right]\right\} ,\label{R_jk_gen_func_flat}
\end{align}
where the $t/\ell\rightarrow1$ and $\ell\rightarrow\infty$ limits
are used.

In order to calculate the components of Einstein tensor the Ricci
scalar is also needed. The next section is devoted to the calculation
of Ricci scalar.

\subsection{Ricci Scalar}

Now, let us work on the $R_{L}$ term: It is defined as 
\begin{align}
R_{L} & \equiv\nabla_{\mu}\nabla_{\nu}h^{\mu\nu}-\Box h-2\Lambda h,\nonumber \\
 & =\frac{t^{4}}{\ell^{4}}\eta^{\rho\mu}\eta^{\sigma\nu}\nabla_{\rho}\nabla_{\sigma}h_{\mu\nu}-\frac{t^{2}}{\ell^{2}}\eta^{\mu\nu}\nabla_{\mu}\nabla_{\nu}h-\frac{2}{\ell^{2}}h,\label{R_scalar1}
\end{align}
where in the second line the upper indices are lowered. In (\ref{R_scalar1})
there are two terms that must be written in terms of the free functions.
These are
\begin{equation}
\nabla_{\mu}\nabla_{\nu}h=\partial_{\mu}\partial_{\nu}h-\Gamma_{\mu\nu}^{\lambda}\partial_{\lambda}h,\label{R_scalar_iden1}
\end{equation}
and
\begin{align}
\nabla_{\rho}\nabla_{\sigma}h_{\mu\nu} & =\partial_{\rho}\left(\nabla_{\sigma}h_{\mu\nu}\right)-\Gamma_{\rho\sigma}^{\alpha}\left(\nabla_{\alpha}h_{\mu\nu}\right)-\Gamma_{\rho\mu}^{\alpha}\left(\nabla_{\sigma}h_{\alpha\nu}\right)-\Gamma_{\rho\nu}^{\alpha}\left(\nabla_{\sigma}h_{\mu\alpha}\right),\nonumber \\
 & =\partial_{\rho}\left(\partial_{\sigma}h_{\mu\nu}-\Gamma_{\sigma\mu}^{\lambda}h_{\lambda\nu}-\Gamma_{\sigma\nu}^{\lambda}h_{\mu\lambda}\right)-\Gamma_{\rho\sigma}^{\alpha}\left(\partial_{\alpha}h_{\mu\nu}-\Gamma_{\alpha\mu}^{\lambda}h_{\lambda\nu}-\Gamma_{\alpha\nu}^{\lambda}h_{\mu\lambda}\right)\label{R_scalar_iden2}\\
 & \phantom{=}-\Gamma_{\rho\mu}^{\alpha}\left(\partial_{\sigma}h_{\alpha\nu}-\Gamma_{\sigma\alpha}^{\lambda}h_{\lambda\nu}-\Gamma_{\sigma\nu}^{\lambda}h_{\alpha\lambda}\right)-\Gamma_{\rho\nu}^{\alpha}\left(\partial_{\sigma}h_{\mu\alpha}-\Gamma_{\sigma\mu}^{\lambda}h_{\lambda\alpha}-\Gamma_{\sigma\alpha}^{\lambda}h_{\mu\lambda}\right).\nonumber 
\end{align}
After the application of contractions, using (\ref{comp_conec}) and
doing some simple manipulations these terms become
\begin{align}
-\frac{t^{2}}{\ell^{2}}\eta^{\mu\nu}\nabla_{\mu}\nabla_{\nu}h & =-\frac{t^{2}}{\ell^{2}}\eta^{\mu\nu}\left(\partial_{\mu}\partial_{\nu}h-\Gamma_{\mu\nu}^{\lambda}\partial_{\lambda}h\right),\nonumber \\
 & =\frac{t^{2}}{\ell^{2}}\left[\left(\partial_{0}\partial_{0}h-\Gamma_{00}^{\lambda}\partial_{\lambda}h\right)-\left(\partial_{i}\partial_{i}h-\Gamma_{ii}^{\lambda}\partial_{\lambda}h\right)\right],\nonumber \\
 & =\frac{t^{2}}{\ell^{2}}\left[\partial_{0}\partial_{0}h-\partial_{i}\partial_{i}h+\frac{1}{t}\left(\partial_{0}h-2\partial_{0}h\right)\right],\nonumber \\
 & =\frac{t^{2}}{\ell^{2}}\left[\partial_{0}\partial_{0}h-\partial_{i}\partial_{i}h-\frac{1}{t}\partial_{0}h\right],\nonumber \\
 & =\frac{t^{2}}{\ell^{2}}\left(\partial_{0}\partial_{0}h-\partial_{i}\partial_{i}h\right)-\frac{t}{\ell^{2}}\partial_{0}h,\label{R_scalar_iden3}
\end{align}
and
\begin{align}
\frac{t^{4}}{\ell^{4}}\eta^{\rho\mu}\eta^{\sigma\nu}\nabla_{\rho}\nabla_{\sigma}h_{\mu\nu} & =\frac{t^{4}}{\ell^{4}}\left[\partial_{0}\left(\partial_{0}h_{00}-\Gamma_{00}^{\lambda}h_{\lambda0}-\Gamma_{00}^{\lambda}h_{0\lambda}\right)-\Gamma_{00}^{\alpha}\left(\partial_{\alpha}h_{00}-\Gamma_{\alpha0}^{\lambda}h_{\lambda0}-\Gamma_{\alpha0}^{\lambda}h_{0\lambda}\right)\right.\nonumber \\
 & \phantom{=}\left.-\Gamma_{00}^{\alpha}\left(\partial_{0}h_{\alpha0}-\Gamma_{0\alpha}^{\lambda}h_{\lambda0}-\Gamma_{00}^{\lambda}h_{\alpha\lambda}\right)-\Gamma_{00}^{\alpha}\left(\partial_{0}h_{0\alpha}-\Gamma_{00}^{\lambda}h_{\lambda\alpha}-\Gamma_{0\alpha}^{\lambda}h_{0\lambda}\right)\right]\nonumber \\
 & \phantom{=}-\frac{t^{4}}{\ell^{4}}\left[\partial_{0}\left(\partial_{i}h_{0i}-\Gamma_{i0}^{\lambda}h_{\lambda i}-\Gamma_{ii}^{\lambda}h_{0\lambda}\right)-\Gamma_{0i}^{\alpha}\left(\partial_{\alpha}h_{0i}-\Gamma_{\alpha0}^{\lambda}h_{\lambda i}-\Gamma_{\alpha i}^{\lambda}h_{0\lambda}\right)\right.\nonumber \\
 & \phantom{=}\left.-\Gamma_{00}^{\alpha}\left(\partial_{i}h_{\alpha i}-\Gamma_{i\alpha}^{\lambda}h_{\lambda i}-\Gamma_{ii}^{\lambda}h_{\alpha\lambda}\right)-\Gamma_{0i}^{\alpha}\left(\partial_{i}h_{0\alpha}-\Gamma_{i0}^{\lambda}h_{\lambda\alpha}-\Gamma_{i\alpha}^{\lambda}h_{0\lambda}\right)\right]\nonumber \\
 & \phantom{=}-\frac{t^{4}}{\ell^{4}}\left[\partial_{i}\left(\partial_{0}h_{i0}-\Gamma_{0i}^{\lambda}h_{\lambda0}-\Gamma_{00}^{\lambda}h_{i\lambda}\right)-\Gamma_{i0}^{\alpha}\left(\partial_{\alpha}h_{i0}-\Gamma_{\alpha i}^{\lambda}h_{\lambda0}-\Gamma_{\alpha0}^{\lambda}h_{i\lambda}\right)\right.\nonumber \\
 & \phantom{=}\left.-\Gamma_{ii}^{\alpha}\left(\partial_{0}h_{\alpha0}-\Gamma_{0\alpha}^{\lambda}h_{\lambda0}-\Gamma_{00}^{\lambda}h_{\alpha\lambda}\right)-\Gamma_{i0}^{\alpha}\left(\partial_{0}h_{i\alpha}-\Gamma_{0i}^{\lambda}h_{\lambda\alpha}-\Gamma_{0\alpha}^{\lambda}h_{i\lambda}\right)\right]\nonumber \\
 & \phantom{=}+\frac{t^{4}}{\ell^{4}}\left[\partial_{i}\left(\partial_{j}h_{ij}-\Gamma_{ji}^{\lambda}h_{\lambda j}-\Gamma_{jj}^{\lambda}h_{i\lambda}\right)-\Gamma_{ij}^{\alpha}\left(\partial_{\alpha}h_{ij}-\Gamma_{\alpha i}^{\lambda}h_{\lambda j}-\Gamma_{\alpha j}^{\lambda}h_{i\lambda}\right)\right.\nonumber \\
 & \phantom{=}\left.-\Gamma_{ii}^{\alpha}\left(\partial_{j}h_{\alpha j}-\Gamma_{j\alpha}^{\lambda}h_{\lambda j}-\Gamma_{jj}^{\lambda}h_{\alpha\lambda}\right)-\Gamma_{ij}^{\alpha}\left(\partial_{j}h_{i\alpha}-\Gamma_{ji}^{\lambda}h_{\lambda\alpha}-\Gamma_{j\alpha}^{\lambda}h_{i\lambda}\right)\right].\label{R_scalar_iden4}
\end{align}
Summing the repeated indices in (\ref{R_scalar_iden4}) and again
using (\ref{comp_conec}) yields
\begin{equation}
\nabla_{\mu}\nabla_{\nu}h^{\mu\nu}=\frac{t^{4}}{\ell^{4}}\left(\partial_{0}\partial_{0}h_{00}-2\partial_{0}\partial_{i}h_{0i}+\partial_{i}\partial_{j}h_{ij}\right)+\frac{t^{3}}{\ell^{4}}\left(\partial_{0}h_{00}-\partial_{0}h_{ii}\right).\label{R_scalar_iden5}
\end{equation}
Putting (\ref{R_scalar_iden3}) and (\ref{R_scalar_iden5}) into (\ref{R_scalar1})
and using $\Lambda=1/\ell^{2}$ gives us
\begin{align}
R_{L} & =\left[\frac{t^{4}}{\ell^{4}}\left(\partial_{0}\partial_{0}h_{00}-2\partial_{0}\partial_{i}h_{0i}+\partial_{i}\partial_{j}h_{ij}\right)+\frac{t^{3}}{\ell^{4}}\left(\partial_{0}h_{00}-\partial_{0}h_{ii}\right)\right]\nonumber \\
 & \phantom{=}+\left[\frac{t^{2}}{\ell^{2}}\left(\partial_{0}\partial_{0}-\partial_{i}\partial_{i}\right)-\frac{t}{\ell^{2}}\partial_{0}-\frac{2}{\ell^{2}}\right]h.\label{R_scalar1.5}
\end{align}
Putting (\ref{met_per_dec}) in (\ref{R_scalar1}) yields
\begin{align}
R_{L} & =\left[\frac{t^{4}}{\ell^{4}}\left(\ddot{N}-2\partial_{i}^{2}\dot{N}_{L}+\partial_{i}^{2}\chi\right)+\frac{t^{3}}{\ell^{4}}\left(\dot{N}-\dot{\phi}-\dot{\chi}\right)\right]\nonumber \\
 & \phantom{=}+\left[\frac{t^{2}}{\ell^{2}}\left(\partial_{0}\partial_{0}-\partial_{i}\partial_{i}\right)-\frac{t}{\ell^{2}}\partial_{0}-\frac{2}{\ell^{2}}\right]\left[\frac{t^{2}}{\ell^{2}}\left(-N+\phi+\chi\right)\right],\nonumber \\
 & =\left[\frac{t^{4}}{\ell^{4}}\left(\ddot{N}-2\partial_{i}^{2}\dot{N}_{L}+\partial_{i}^{2}\chi\right)+\frac{t^{3}}{\ell^{4}}\left(\dot{N}-\dot{\phi}-\dot{\chi}\right)\right]\nonumber \\
 & \phantom{=}+\left[\frac{t^{4}}{\ell^{4}}\left(\partial_{0}\partial_{0}-\partial_{i}\partial_{i}\right)+\frac{4t^{3}}{\ell^{4}}\partial_{0}+\frac{2t^{2}}{\ell^{4}}-\frac{t^{3}}{\ell^{4}}\partial_{0}-\frac{2t^{2}}{\ell^{4}}-\frac{2t^{2}}{\ell^{4}}\right]\left(-N+\phi+\chi\right),\nonumber \\
 & =\left[\left(\frac{t^{4}}{\ell^{4}}\partial_{0}\partial_{0}+\frac{t^{3}}{\ell^{4}}\partial_{0}\right)N+\left(\frac{t^{4}}{\ell^{4}}\partial_{i}^{2}-\frac{t^{3}}{\ell^{4}}\partial_{0}\right)\chi-\frac{t^{4}}{\ell^{4}}2\partial_{i}^{2}\dot{N}_{L}-\frac{t^{3}}{\ell^{4}}\partial_{0}\phi\right]\nonumber \\
 & \phantom{=}+\left[\frac{t^{4}}{\ell^{4}}\left(\partial_{0}\partial_{0}-\partial_{i}\partial_{i}\right)+\frac{3t^{3}}{\ell^{4}}\partial_{0}-\frac{2t^{2}}{\ell^{4}}\right]\left(-N+\phi+\chi\right),\nonumber \\
 & =\left[\left(\frac{t^{4}}{\ell^{4}}\partial_{i}\partial_{i}-\frac{2t^{3}}{\ell^{4}}\partial_{0}+\frac{2t^{2}}{\ell^{4}}\right)N+\left(\frac{t^{4}}{\ell^{4}}\partial_{0}^{2}+\frac{2t^{3}}{\ell^{4}}\partial_{0}-\frac{2t^{2}}{\ell^{4}}\right)\chi-\frac{t^{4}}{\ell^{4}}2\partial_{i}^{2}\dot{N}_{L}\right]\nonumber \\
 & \phantom{=}+\left[\frac{t^{4}}{\ell^{4}}\left(\partial_{0}\partial_{0}-\partial_{i}\partial_{i}\right)+\frac{2t^{3}}{\ell^{4}}\partial_{0}-\frac{2t^{2}}{\ell^{4}}\right]\phi,\label{R_scalar2}
\end{align}
and finally the scalar curvature read
\begin{equation}
R_{L}=\frac{t^{4}}{\ell^{4}}\left(\partial_{i}^{2}N+\ddot{\chi}-2\partial_{i}^{2}\dot{N}_{L}+\ddot{\phi}-\partial_{i}^{2}\phi\right)+\frac{2t^{3}}{\ell^{4}}\left(-\dot{N}+\dot{\phi}+\dot{\chi}\right)-\frac{2t^{2}}{\ell^{4}}\left(-N+\phi+\chi\right).\label{R_scalar3}
\end{equation}
In the flat spacetime limit (\ref{R_scalar3}) becomes
\begin{equation}
R_{L}=\partial_{j}^{2}N+\ddot{\chi}-2\partial_{i}^{2}\dot{N}_{L}+\left(\partial_{0}^{2}-\partial_{j}^{2}\right)\phi,\label{R_scalar_flat}
\end{equation}
which is the flat spacetime result for the scalar curvature. 

With the results that we found in the above calculations, (\ref{R_00_gen_func}),
(\ref{R_0j_gen_fun}), (\ref{R_jk_gen_func}) and (\ref{R_scalar3}),
we are ready to find the components of the Einstein tensor in terms
of the free functions of metric perturbation.

\subsection{The Einstein Tensor }

The linear Einstein tensor is defined as follows

\begin{equation}
\mathcal{G}_{\mu\nu}^{L}=R_{\mu\nu}^{L}-\frac{1}{2}\bar{g}_{\mu\nu}R_{L}-2\Lambda h_{\mu\nu}.\label{Lin_Einstein_ten}
\end{equation}
The components of the Einstein tensor is investigated term by term.

\subsubsection{The $\mathcal{G}_{00}^{L}$ Term:}

The $00$-component of linear Einstein tensor,$\mathcal{G}_{00}^{L}$,
can be written by using (\ref{R_00_gen_func}), (\ref{R_scalar3})
and (\ref{met_per_dec}) as follows,
\begin{align}
\mathcal{G}_{00}^{L} & =R_{00}^{L}-\frac{1}{2}\bar{g}_{00}R_{L}-2\Lambda h_{00},\nonumber \\
 & =\left[-\frac{t^{2}}{2\ell^{2}}\left(\partial_{i}^{2}N-2\partial_{i}^{2}\dot{N}_{L}+\ddot{\phi}+\ddot{\chi}\right)+\frac{t}{2\ell^{2}}\left(2\dot{N}-3\dot{\phi}-3\dot{\chi}+2\partial_{i}^{2}N_{L}\right)+\frac{2}{\ell^{2}}N\right]-\frac{2}{\ell^{2}}N\nonumber \\
 & \phantom{=}-\frac{1}{2}\frac{\ell^{2}}{t^{2}}\eta_{00}\left\{ \frac{t^{4}}{\ell^{4}}\left[\partial_{i}^{2}N+\ddot{\chi}-2\partial_{i}^{2}\dot{N}_{L}+\left(\partial_{0}^{2}-\partial_{i}^{2}\right)\phi\right]\right.\nonumber \\
 & \phantom{=}\left.+\frac{2t^{3}}{\ell^{4}}\left(-\dot{N}+\dot{\phi}+\dot{\chi}\right)-\frac{2t^{2}}{\ell^{4}}\left(-N+\phi+\chi\right)\right\} ,\nonumber \\
 & =\left[-\frac{t^{2}}{2\ell^{2}}\left(\partial_{i}^{2}N-2\partial_{i}^{2}\dot{N}_{L}+\ddot{\phi}+\ddot{\chi}\right)+\frac{t}{2\ell^{2}}\left(2\dot{N}-3\dot{\phi}-3\dot{\chi}+2\partial_{i}^{2}N_{L}\right)\right]\nonumber \\
 & \phantom{=}+\left\{ \frac{t^{2}}{2\ell^{2}}\left[\partial_{i}^{2}N+\ddot{\chi}-2\partial_{i}^{2}\dot{N}_{L}+\left(\partial_{0}^{2}-\partial_{i}^{2}\right)\phi\right]+\frac{t}{\ell^{2}}\left(-\dot{N}+\dot{\phi}+\dot{\chi}\right)-\frac{1}{\ell^{2}}\left(-N+\phi+\chi\right)\right\} ,\label{G_00_comp1}
\end{align}
and doing cancellations and summations we get
\begin{equation}
\mathcal{G}_{00}^{L}=-\frac{t^{2}}{2\ell^{2}}\partial_{i}^{2}\phi-\frac{t}{2\ell^{2}}\left(\dot{\phi}+\dot{\chi}-2\partial_{i}^{2}N_{L}\right)-\frac{1}{\ell^{2}}\left(-N+\phi+\chi\right).\label{G_00_comp}
\end{equation}
To simplify (\ref{G_00_comp}), the metric perturbation functions
are redefined as $\phi\rightarrow\frac{\ell^{2}}{t^{2}}\phi$ and
the time derivative transforms as$\dot{\phi}\rightarrow-2\frac{\ell^{2}}{t^{3}}\phi+\frac{\ell^{2}}{t^{2}}\dot{\phi}$.
With these redefinitions the $00$-component of the Einstein tensor
becomes 
\begin{align}
\mathcal{G}_{00}^{L} & =-\frac{1}{2}\partial_{i}^{2}\phi-\frac{t}{2\ell^{2}}\left(-2\frac{\ell^{2}}{t^{3}}\phi+\frac{\ell^{2}}{t^{2}}\dot{\phi}-2\frac{\ell^{2}}{t^{3}}\chi+\frac{\ell^{2}}{t^{2}}\dot{\chi}-2\frac{\ell^{2}}{t^{2}}\partial_{i}^{2}N_{L}\right)-\frac{1}{t^{2}}\left(-N+\phi+\chi\right)\nonumber \\
 & =-\frac{1}{2}\partial_{i}^{2}\phi-\left(-\frac{1}{t^{2}}\phi+\frac{1}{2t}\dot{\phi}-\frac{1}{t^{2}}\chi+\frac{1}{2t}\dot{\chi}-\frac{1}{t}\partial_{i}^{2}N_{L}\right)-\frac{1}{t^{2}}\left(-N+\phi+\chi\right)\nonumber \\
 & =-\frac{1}{2}\partial_{i}^{2}\phi-\left(\frac{1}{2t}\dot{\phi}+\frac{1}{2t}\dot{\chi}-\frac{1}{t^{2}}N\right)+\frac{1}{t}\partial_{i}^{2}N_{L}\nonumber \\
 & =-\frac{1}{2}\left[\partial_{i}^{2}\phi+\frac{1}{t}\left(\dot{\phi}+\dot{\chi}-\frac{2}{t}N\right)-\frac{2}{t}\partial_{i}^{2}N_{L}\right].\label{G_00_comp_trans}
\end{align}
From the last line of (\ref{G_00_comp_trans}) a gauge invariant function
can be defined: 
\begin{equation}
f\equiv\frac{\ell}{t}\left[\phi+\frac{1}{t\nabla^{2}}\left(\dot{\phi}+\dot{\chi}-\frac{2}{t}N\right)-\frac{2}{t}N_{L}\right],\label{gauge_f}
\end{equation}
then 
\begin{equation}
\mathcal{G}_{00}^{L}=-\frac{t}{2\ell}\nabla^{2}f.\label{G_00_gauge}
\end{equation}
Also, for flat spacetime this term becomes 
\begin{equation}
\mathcal{G}_{00}^{L}=-\frac{1}{2}\nabla^{2}\phi,\label{G_00_gauge_flat}
\end{equation}
where $f=\phi$ for the flat spacetime background.

\subsubsection{The $\mathcal{G}_{0j}^{L}$ Term:}

The $0j$-component of linear Einstein tensor,$\mathcal{G}_{0j}^{L}=R_{0j}^{L}-2\Lambda h_{0j}$,
can be written by using (\ref{R_0j_gen_fun}) and (\ref{met_per_dec})
as follows,
\begin{align}
\mathcal{G}_{0j}^{L} & =\left\{ -\frac{t^{2}}{2\ell^{2}}\left[\partial_{j}\dot{\phi}+\epsilon_{jk}\partial_{k}\left(\dot{\xi}-\partial_{i}^{2}\eta\right)\right]+\frac{t}{2\ell^{2}}\left[\partial_{j}\left(N-2\phi\right)-2\epsilon_{ji}\partial_{i}\xi\right]+\frac{2}{\ell^{2}}h_{0j}\right\} -\frac{2}{\ell^{2}}h_{0j},\nonumber \\
 & =-\frac{t^{2}}{2\ell^{2}}\left[\partial_{j}\dot{\phi}+\epsilon_{jk}\partial_{k}\left(\dot{\xi}-\partial_{i}^{2}\eta\right)\right]+\frac{t}{2\ell^{2}}\left[\partial_{j}\left(N-2\phi\right)-2\epsilon_{ji}\partial_{i}\xi\right].\label{G_0j_comp}
\end{align}
Again doing the same redefinitions, $\phi\rightarrow\frac{\ell^{2}}{t^{2}}\phi$
and $\dot{\phi}\rightarrow-2\frac{\ell^{2}}{t^{3}}\phi+\frac{\ell^{2}}{t^{2}}\dot{\phi}$,
(\ref{G_0j_comp}) becomes 
\begin{align}
\mathcal{G}_{0j}^{L} & =-\frac{1}{2}\left[\partial_{j}\left(-\frac{2}{t}\phi+\dot{\phi}\right)+\epsilon_{jk}\partial_{k}\left(-\frac{2}{t}\xi+\dot{\xi}-\partial_{i}^{2}\eta\right)\right]+\frac{1}{2t}\left[\partial_{j}\left(N-2\phi\right)-2\epsilon_{ji}\partial_{i}\xi\right]\nonumber \\
 & =-\frac{1}{2}\left[\partial_{j}\dot{\phi}+\epsilon_{jk}\partial_{k}\left(\dot{\xi}-\partial_{i}^{2}\eta\right)-\frac{1}{t}\partial_{j}N\right].\label{G_0j_comp1}
\end{align}
From (\ref{G_0j_comp1}) two gauge-invariant combinations can be defined
as 
\begin{equation}
p\equiv\frac{\ell}{t}\left(\dot{\phi}-\frac{1}{t}N\right),\quad\sigma\equiv\frac{\ell}{t}\left(\dot{\xi}-\partial_{i}^{2}\eta\right).\label{gauge_p_sigma}
\end{equation}
With these definitions, (\ref{gauge_p_sigma}), the $0j$-component
of the linear Einstein tensor reads 
\begin{equation}
\mathcal{G}_{0j}^{L}=-\frac{t}{2\ell}\left[\partial_{j}p+\epsilon_{jk}\partial_{k}\sigma\right].\label{G_0j_gauge}
\end{equation}
For the flat spacetime this component yields 
\begin{equation}
\mathcal{G}_{0j}^{L}=-\frac{1}{2}\left[\partial_{j}\dot{\phi}+\epsilon_{jk}\partial_{k}\sigma\right],\label{G_0j_gauge_flat}
\end{equation}
with the flat spacetime version of the gauge-invariant functions,
that are 
\begin{equation}
p\equiv\dot{\phi},\quad\sigma\equiv\left(\dot{\xi}-\partial_{i}^{2}\eta\right).\label{gauge_p_sigma_flat}
\end{equation}

\subsubsection{The $\mathcal{G}_{jk}^{L}$ Term:}

The $jk$-component of linear Einstein tensor,$\mathcal{G}_{jk}^{L}$,
can be written by putting (\ref{R_jk_gen_func}), (\ref{R_scalar3})
and (\ref{met_per_dec}) into the following equation

\begin{equation}
\mathcal{G}_{jk}^{L}=R_{jk}^{L}-\frac{1}{2}\frac{\ell^{2}}{t^{2}}\delta_{jk}R_{L}-\frac{2}{\ell^{2}}h_{jk},\label{G_jk_comp1}
\end{equation}
that yields 
\begin{align}
\mathcal{G}_{jk}^{L} & =\left(-\frac{t^{2}}{2\ell^{2}}\right)\left\{ \partial_{j}\left[-2\epsilon_{kl}\partial_{l}\left(\dot{\eta}+\frac{1}{t}\eta\right)+\partial_{k}\left(2\dot{N}_{L}-N+\frac{2}{t}N_{L}\right)\right]\right.\nonumber \\
 & \phantom{=\left(-\frac{t^{2}}{2\ell^{2}}\right)}\left.-\left(\partial_{0}^{2}+\frac{3}{t}\partial_{0}+\frac{4}{t^{2}}\right)\left[\left(\delta_{jk}+\hat{\partial}_{j}\hat{\partial}_{k}\right)\phi-\hat{\partial}_{j}\hat{\partial}_{k}\chi+\left(\epsilon_{jl}\hat{\partial}_{l}\hat{\partial}_{k}+\epsilon_{kl}\hat{\partial}_{l}\hat{\partial}_{j}\right)\xi\right]\right.\nonumber \\
 & \phantom{=\left(-\frac{t^{2}}{2\ell^{2}}\right)}\left.+\delta_{jk}\partial_{i}^{2}\phi+\frac{1}{t}\delta_{jk}\left[2\left(\dot{N}-\partial_{i}^{2}N_{L}\right)+\left(-\dot{N}+\dot{\phi}+\dot{\chi}\right)+\frac{2}{t}\left(-N+\phi+\chi\right)\right]\right\} \label{G_jk_comp2}\\
 & \phantom{=}-\frac{1}{2}\delta_{jk}\left\{ \frac{t^{2}}{\ell^{2}}\left[\partial_{i}^{2}N+\ddot{\chi}-2\partial_{i}^{2}\dot{N}_{L}+\left(\partial_{0}^{2}-\partial_{i}^{2}\right)\phi\right]+\frac{2t}{\ell^{2}}\left(-\dot{N}+\dot{\phi}+\dot{\chi}\right)-\frac{2}{\ell^{2}}\left(-N+\phi+\chi\right)\right\} \nonumber \\
 & \phantom{=}-\frac{2}{\ell^{2}}h_{jk},\nonumber 
\end{align}
and with some cancellations it takes the following form
\begin{align}
\mathcal{G}_{jk}^{L} & =\left(-\frac{t^{2}}{2\ell^{2}}\right)\left\{ \partial_{j}\left[-2\epsilon_{kl}\partial_{l}\left(\dot{\eta}+\frac{1}{t}\eta\right)+\partial_{k}\left(2\dot{N}_{L}-N+\frac{2}{t}N_{L}\right)\right]\right.\nonumber \\
 & \phantom{=\left(-\frac{t^{2}}{2\ell^{2}}\right)}\left.-\left(\partial_{0}^{2}+\frac{3}{t}\partial_{0}\right)\left[\hat{\partial}_{j}\hat{\partial}_{k}\phi-\hat{\partial}_{j}\hat{\partial}_{k}\chi+\left(\epsilon_{jl}\hat{\partial}_{l}\hat{\partial}_{k}+\epsilon_{kl}\hat{\partial}_{l}\hat{\partial}_{j}\right)\xi\right]\right\} \nonumber \\
 & \phantom{=}-\frac{1}{2}\delta_{jk}\left\{ \frac{t^{2}}{\ell^{2}}\left(\partial_{i}^{2}N+\ddot{\chi}-2\partial_{i}^{2}\dot{N}_{L}\right)-\frac{t}{\ell^{2}}\left(\dot{N}+2\partial_{i}^{2}N_{L}\right)+\frac{3t}{\ell^{2}}\dot{\chi}\right\} .\label{G_jk_comp3}
\end{align}
After redefining all the functions as $\phi\rightarrow\frac{\ell^{2}}{t^{2}}\phi$
for which the first second derivatives are $\dot{\phi}\rightarrow-\frac{2\ell^{2}}{t^{3}}\phi+\frac{\ell^{2}}{t^{2}}\dot{\phi}$,
and $\ddot{\phi}\rightarrow\frac{6\ell^{2}}{t^{4}}\phi-\frac{4\ell^{2}}{t^{3}}\dot{\phi}+\frac{\ell^{2}}{t^{2}}\ddot{\phi}$
respectively, (\ref{G_jk_comp3}) becomes
\begin{align}
-2\mathcal{G}_{jk}^{L} & =2\epsilon_{kl}\partial_{j}\partial_{l}\left(\frac{1}{t}\eta-\dot{\eta}\right)+\partial_{j}\partial_{k}\left(-\frac{2}{t}N_{L}+2\dot{N}_{L}-N\right)\nonumber \\
 & \phantom{=}+\hat{\partial}_{j}\hat{\partial}_{k}\left(\frac{1}{t}\dot{\phi}-\ddot{\phi}\right)-\hat{\partial}_{j}\hat{\partial}_{k}\left(\frac{1}{t}\dot{\chi}-\ddot{\chi}\right)+\left(\epsilon_{jl}\hat{\partial}_{l}\hat{\partial}_{k}+\epsilon_{kl}\hat{\partial}_{l}\hat{\partial}_{j}\right)\left(\frac{1}{t}\dot{\xi}-\ddot{\xi}\right)\label{G_jk_comp4}\\
 & \phantom{=}+\delta_{jk}\left(\partial_{i}^{2}N-\frac{1}{t}\dot{\chi}+\ddot{\chi}-2\partial_{i}^{2}\dot{N}_{L}+\frac{2}{t}\partial_{i}^{2}N_{L}+\frac{2}{t^{2}}N-\frac{1}{t}\dot{N}\right).\nonumber 
\end{align}
Since the $j$ and $k$ indices are symmetric, the first term can
be manipulated such that 
\begin{align}
2\epsilon_{kl}\partial_{j}\partial_{l}\left(\frac{1}{t}\eta-\dot{\eta}\right) & =\left(\epsilon_{kl}\partial_{j}\partial_{l}+\epsilon_{jl}\partial_{k}\partial_{l}\right)\left(\frac{1}{t}\eta-\dot{\eta}\right)\nonumber \\
 & =-\left(\epsilon_{kl}\hat{\partial}_{j}\hat{\partial}_{l}+\epsilon_{jl}\hat{\partial}_{k}\hat{\partial}_{l}\right)\left(\frac{1}{t}\partial_{i}^{2}\eta-\partial_{i}^{2}\dot{\eta}\right)\label{G_jk_iden}
\end{align}
where in the second line the unit derivatives are introduced. After
this manipulation (\ref{G_jk_comp4}) becomes 
\begin{align}
-2\mathcal{G}_{jk}^{L} & =-\left(\epsilon_{kl}\hat{\partial}_{j}\hat{\partial}_{l}+\epsilon_{jl}\hat{\partial}_{k}\hat{\partial}_{l}\right)\left(\frac{1}{t}\partial_{i}^{2}\eta-\partial_{i}^{2}\dot{\eta}-\frac{1}{t}\dot{\xi}+\ddot{\xi}\right)\nonumber \\
 & \phantom{=}-\hat{\partial}_{j}\hat{\partial}_{k}\left(-\frac{1}{t}\dot{\phi}+\ddot{\phi}+\frac{2}{t^{2}}N-\frac{1}{t}\dot{N}\right)\label{G_jk_comp5}\\
 & \phantom{=}+\left(\delta_{jk}+\hat{\partial}_{j}\hat{\partial}_{k}\right)\left(\partial_{i}^{2}N-\frac{1}{t}\dot{\chi}+\ddot{\chi}-2\partial_{i}^{2}\dot{N}_{L}+\frac{2}{t}\partial_{i}^{2}N_{L}+\frac{2}{t^{2}}N-\frac{1}{t}\dot{N}\right),\nonumber 
\end{align}
where the terms $\frac{2}{t^{2}}\hat{\partial}_{k}\hat{\partial}_{j}N$
and $\frac{1}{t}\hat{\partial}_{k}\hat{\partial}_{j}\dot{N}$ are
added and subtracted. The second parenthesis in the first term is
$\dot{\sigma}$ and the second term is $\dot{p}$. The last term can
be defined as 
\begin{equation}
q\equiv\frac{\ell}{t}\left(\partial_{i}^{2}N-\frac{1}{t}\dot{\chi}+\ddot{\chi}-2\partial_{i}^{2}\dot{N}_{L}+\frac{2}{t}\partial_{i}^{2}N_{L}+\frac{2}{t^{2}}N-\frac{1}{t}\dot{N}\right),\label{gauge_q}
\end{equation}
which is also gauge-invariant function. Finally, the $jk$-component
of the linear Einstein tensor becomes 
\begin{align}
\mathcal{G}_{jk}^{L} & =-\frac{t}{2\ell}\left\{ -\left(\epsilon_{kl}\hat{\partial}_{j}\hat{\partial}_{l}+\epsilon_{jl}\hat{\partial}_{k}\hat{\partial}_{l}\right)\dot{\sigma}-\hat{\partial}_{j}\hat{\partial}_{k}\dot{p}+\left(\delta_{jk}+\hat{\partial}_{j}\hat{\partial}_{k}\right)q\right\} .\label{G_jk_gauge}
\end{align}
The flat spacetime version of this component is 
\begin{equation}
\mathcal{G}_{jk}^{L}=-\frac{1}{2}\left\{ -\left(\epsilon_{kl}\hat{\partial}_{j}\hat{\partial}_{l}+\epsilon_{jl}\hat{\partial}_{k}\hat{\partial}_{l}\right)\dot{\sigma}-\hat{\partial}_{j}\hat{\partial}_{k}\ddot{\phi}+\left(\delta_{jk}+\hat{\partial}_{j}\hat{\partial}_{k}\right)q\right\} ,\label{G_jk_gauge_flat}
\end{equation}
where the function $q$ reads 
\begin{equation}
q=\partial_{i}^{2}N-\frac{1}{t}\dot{\chi}+\ddot{\chi}-2\partial_{i}^{2}\dot{N}_{L}.\label{gauge_q_flat}
\end{equation}

\subsection{Gauge-Invariance of the Defined Functions}

The functions can be checked whether they are gauge-invariant or not.
Let us start with,
\begin{equation}
\delta_{\zeta}\sigma=\frac{\ell}{t}\left(\delta_{\zeta}\dot{\xi}-\nabla^{2}\delta_{\zeta}\eta\right),\label{gauge_check_sigma1}
\end{equation}
and using (\ref{gauge_h_00}, \ref{gauge_h_0i}, \ref{gauge_h_ij_chi},
\ref{gauge_h_ij_ksi}, \ref{gauge_h_ij_phi}) and their derivatives
(\ref{gauge_check_sigma1}) becomes
\begin{equation}
\delta_{\zeta}\sigma=\frac{\ell}{t}\left(\frac{2t}{\ell^{2}}\nabla^{2}\zeta+\frac{t^{2}}{\ell^{2}}\nabla^{2}\dot{\zeta}-\frac{t^{2}}{\ell^{2}}\nabla^{2}\dot{\zeta}-\frac{2t}{\ell^{2}}\nabla^{2}\zeta\right)=0,\label{gauge_check_sigma2}
\end{equation}
where we have used $\delta_{\zeta}\dot{\xi}=\frac{2t}{\ell^{2}}\partial_{j}^{2}\zeta+\frac{t^{2}}{\ell^{2}}\partial_{j}^{2}\dot{\zeta}.$
Again using (\ref{gauge_h_00}, \ref{gauge_h_0i}, \ref{gauge_h_ij_chi},
\ref{gauge_h_ij_ksi}, \ref{gauge_h_ij_phi}) the $p$ function reads
\begin{align}
\delta_{\zeta}p & =\frac{\ell}{t}\left(\delta_{\zeta}\dot{\phi}-2\frac{t}{\ell^{2}}\dot{\zeta}_{0}-2\frac{1}{\ell^{2}}\zeta_{0}\right),\label{gauge_check_p1}\\
\delta_{\zeta}p & =\frac{\ell}{t}\left(\frac{2}{\ell^{2}}\zeta_{0}+\frac{2t}{\ell^{2}}\dot{\zeta_{0}}-2\frac{t}{\ell^{2}}\dot{\zeta}_{0}-2\frac{1}{\ell^{2}}\zeta_{0}\right)=0,\label{gauge_check_p2}
\end{align}
where we have again used the derivative of fields, $\delta_{\zeta}\dot{\phi}=\frac{2}{\ell^{2}}\zeta_{0}+\frac{2t}{\ell^{2}}\dot{\zeta_{0}}$.
The other two functions can be handled in the same way. 
\begin{align}
\delta_{\zeta}f & =\frac{\ell}{t}\left[\delta_{\zeta}\phi+\frac{1}{t\nabla^{2}}\left(\delta_{\zeta}\dot{\phi}+\delta_{\zeta}\dot{\chi}-\frac{2}{t}\delta_{\zeta}N\right)-\frac{2}{t}\delta_{\zeta}N_{L}\right],\label{gauge_check_f1}\\
\frac{t}{\ell}\delta_{\zeta}f & =\frac{1}{t\nabla^{2}}\left[\frac{2}{\ell^{2}}\zeta_{0}+\frac{2t}{\ell^{2}}\dot{\zeta_{0}}+4\frac{t}{\ell^{2}}\left(\nabla^{2}\kappa+\frac{1}{t}\zeta_{0}\right)+2\frac{t^{2}}{\ell^{2}}\left(\nabla^{2}\dot{\kappa}+\frac{1}{t}\dot{\zeta_{0}}-\frac{1}{t^{2}}\zeta_{0}\right)-4\frac{t}{\ell^{2}}\left(\dot{\zeta}_{0}+\frac{1}{t}\zeta_{0}\right)\right]\nonumber \\
 & \phantom{=}+\frac{2t}{\ell^{2}}\zeta_{0}-\frac{2t}{\ell^{2}}\left(\dot{\kappa}+\zeta_{0}+\frac{2}{t}\kappa\right)=0,\label{gauge_check_f2}
\end{align}
where we have used $\delta_{\zeta}\dot{\chi}=4\frac{t}{\ell^{2}}\left(\partial_{j}^{2}\kappa+\frac{1}{t}\zeta_{0}\right)+2\frac{t^{2}}{\ell^{2}}\left(\partial_{j}^{2}\dot{\kappa}+\frac{1}{t}\dot{\zeta_{0}}-\frac{1}{t^{2}}\zeta_{0}\right)$.
The $q$ function is 
\begin{align}
\frac{t}{\ell}\delta_{\zeta}q & =\nabla^{2}\delta_{\zeta}N-\frac{1}{t}\delta_{\zeta}\dot{\chi}+\delta_{\zeta}\ddot{\chi}-2\partial_{i}^{2}\delta_{\zeta}\dot{N}_{L}+\frac{2}{t}\nabla^{2}\delta_{\zeta}N_{L}+\frac{2}{t^{2}}\delta_{\zeta}N-\frac{1}{t}\delta_{\zeta}\dot{N}\label{gauge_check_q1}\\
\frac{t}{\ell}\delta_{\zeta}q & =\frac{2t^{2}}{\ell^{2}}\nabla^{2}\left(\dot{\zeta}_{0}+\frac{1}{t}\zeta_{0}\right)-\frac{4}{\ell^{2}}\left(\nabla^{2}\kappa+\frac{1}{t}\zeta_{0}\right)-\frac{2t}{\ell^{2}}\left(\nabla^{2}\dot{\kappa}+\frac{1}{t}\dot{\zeta_{0}}-\frac{1}{t^{2}}\zeta_{0}\right)\nonumber \\
 & \phantom{=}+\frac{4}{\ell^{2}}\left(\nabla^{2}\kappa+\frac{1}{t}\zeta_{0}\right)+\frac{4t}{\ell^{2}}\left(\nabla^{2}\dot{\kappa}+\frac{1}{t}\dot{\zeta_{0}}-\frac{1}{t^{2}}\zeta_{0}\right)\nonumber \\
 & \phantom{=}+\frac{4t}{\ell^{2}}\left(\nabla^{2}\dot{\kappa}+\frac{1}{t}\dot{\zeta_{0}}-\frac{1}{t^{2}}\zeta_{0}\right)+\frac{2t^{2}}{\ell^{2}}\left(\nabla^{2}\ddot{\kappa}+\frac{1}{t}\ddot{\zeta_{0}}-\frac{1}{t^{2}}\dot{\zeta_{0}}-\frac{1}{t^{2}}\dot{\zeta_{0}}+\frac{2}{t^{3}}\zeta_{0}\right)\nonumber \\
 & \phantom{=}-\frac{4t}{\ell^{2}}\nabla^{2}\left(\dot{\kappa}+\zeta_{0}+\frac{2}{t}\kappa\right)-\frac{2t^{2}}{\ell^{2}}\nabla^{2}\left(\ddot{\kappa}+\dot{\zeta_{0}}+\frac{2}{t}\dot{\kappa}-\frac{2}{t^{2}}\kappa\right)\nonumber \\
 & \phantom{=}+\frac{2t}{\ell^{2}}\nabla^{2}\left(\dot{\kappa}+\zeta_{0}+\frac{2}{t}\kappa\right)+\frac{4}{\ell^{2}}\left(\dot{\zeta}_{0}+\frac{1}{t}\zeta_{0}\right)\nonumber \\
 & \phantom{=}-\frac{4}{\ell^{2}}\left(\dot{\zeta}_{0}+\frac{1}{t}\zeta_{0}\right)-\frac{2t}{\ell^{2}}\left(\ddot{\zeta_{0}}+\frac{1}{t}\dot{\zeta_{0}}-\frac{1}{t^{2}}\zeta_{0}\right)=0.\label{gauge_check_q2}
\end{align}
Therefore, these functions are indeed invariant under gauge transformations.

\subsection{Ricci Scalar in Terms of Gauge Invariant functions}

The Ricci scalar can also be written in terms of the gauge-invariant
functions. We first write (\ref{R_scalar3}) with the modified functions,
$\phi\rightarrow\frac{\ell^{2}}{t^{2}}\phi$. With these functions
the Ricci scalar takes the form 
\begin{equation}
R_{L}=\frac{t^{2}}{\ell^{2}}\left(\partial_{i}^{2}N+\ddot{\chi}-2\partial_{i}^{2}\dot{N}_{L}+\ddot{\phi}-\partial_{i}^{2}\phi-\frac{2}{t}\dot{\chi}+\frac{4}{t}\nabla^{2}N_{L}-\frac{2}{t}\dot{\phi}+\frac{6}{t^{2}}N-\frac{2}{t}\dot{N}\right).\label{R_scalar_gauge1}
\end{equation}
Using the definitions of the gauge-invariant functions, (\ref{gauge_f}),
(\ref{gauge_p_sigma}) and (\ref{gauge_q}), the Ricci scalar can
be written as 
\begin{equation}
R_{L}=\frac{t^{3}}{\ell^{3}}\left(q-\nabla^{2}f+\dot{p}\right).\label{gauge_inv_R_scalar}
\end{equation}
For the flat spacetime case 
\begin{equation}
R_{L}=q-\nabla^{2}\phi+\ddot{\phi},\label{gauge_inv_R_flat1}
\end{equation}
and using $\square=-\partial_{0}+\nabla^{2}$ the Ricci scalar becomes
\begin{equation}
R_{L}=q-\square\phi.\label{gauge_inv_R_flat2}
\end{equation}

\subsection{The Bianchi Identity}

The Bianchi identity can also be written in terms of the gauge-invariant
functions. First the identity is decomposed into its components by
doing the summation in the repeated indices:
\begin{align}
\nabla^{\mu}\mathcal{G}_{\mu\nu}^{L} & =\bar{g}^{\mu\sigma}\nabla_{\sigma}\mathcal{G}_{\mu\nu}^{L}=0\label{Bianchi_iden1}\\
 & =\bar{g}^{\mu\sigma}\left(\partial_{\sigma}\mathcal{G}_{\mu\nu}^{L}-\Gamma_{\sigma\mu}^{\lambda}\mathcal{G}_{\lambda\nu}^{L}-\Gamma_{\sigma\nu}^{\lambda}\mathcal{G}_{\lambda\mu}^{L}\right),\nonumber \\
 & =\bar{g}^{0\sigma}\left(\partial_{\sigma}\mathcal{G}_{0\nu}^{L}-\Gamma_{\sigma0}^{\lambda}\mathcal{G}_{\lambda\nu}^{L}-\Gamma_{\sigma\nu}^{\lambda}\mathcal{G}_{\lambda0}^{L}\right)\nonumber \\
 & \phantom{=}+\bar{g}^{i\sigma}\left(\partial_{\sigma}\mathcal{G}_{i\nu}^{L}-\Gamma_{\sigma i}^{\lambda}\mathcal{G}_{\lambda\nu}^{L}-\Gamma_{\sigma\nu}^{\lambda}\mathcal{G}_{\lambda i}^{L}\right),\nonumber \\
 & 0=-\frac{t^{2}}{\ell^{2}}\left(\partial_{0}\mathcal{G}_{0\nu}^{L}-\Gamma_{00}^{0}\mathcal{G}_{0\nu}^{L}-\Gamma_{0\nu}^{0}\mathcal{G}_{00}^{L}-\Gamma_{0\nu}^{k}\mathcal{G}_{k0}^{L}\right)\nonumber \\
 & \phantom{=}+\frac{t^{2}}{\ell^{2}}\delta^{ij}\left(\partial_{j}\mathcal{G}_{i\nu}^{L}-\Gamma_{ji}^{0}\mathcal{G}_{0\nu}^{L}-\Gamma_{j\nu}^{0}\mathcal{G}_{0i}^{L}-\Gamma_{j\nu}^{k}\mathcal{G}_{ki}^{L}\right).\label{Bianchi_iden2}
\end{align}
There is two equations depending on the $\nu$ index. For $\nu=0$,
(\ref{Bianchi_iden2}) yields
\begin{align}
0 & =-\frac{t^{2}}{\ell^{2}}\left(\partial_{0}\mathcal{G}_{00}^{L}-\Gamma_{00}^{0}\mathcal{G}_{00}^{L}-\Gamma_{00}^{0}\mathcal{G}_{00}^{L}-\Gamma_{00}^{k}\mathcal{G}_{k0}^{L}\right)\nonumber \\
 & \phantom{=}+\frac{t^{2}}{\ell^{2}}\delta^{ij}\left(\partial_{j}\mathcal{G}_{i0}^{L}-\Gamma_{ji}^{0}\mathcal{G}_{00}^{L}-\Gamma_{j0}^{0}\mathcal{G}_{0i}^{L}-\Gamma_{j0}^{k}\mathcal{G}_{ki}^{L}\right),\label{Bianchi_iden3}
\end{align}
and using (\ref{comp_conec}),
\begin{align}
0 & =-\frac{t^{2}}{\ell^{2}}\left(\partial_{0}\mathcal{G}_{00}^{L}+\frac{2}{t}\mathcal{G}_{00}^{L}\right)\nonumber \\
 & \phantom{=}+\frac{t^{2}}{\ell^{2}}\left(\partial_{i}\mathcal{G}_{i0}^{L}+\frac{2}{t}\mathcal{G}_{00}^{L}+\frac{1}{t}\mathcal{G}_{ii}^{L}\right),\nonumber \\
0 & =-\frac{t^{2}}{\ell^{2}}\left(\partial_{0}\mathcal{G}_{00}^{L}-\partial_{i}\mathcal{G}_{i0}^{L}-\frac{1}{t}\mathcal{G}_{ii}^{L}\right).\label{Bianchi_iden4}
\end{align}
Using the gauge-invariant form of the components of the Einstein tensor
and their derivatives, that are
\begin{align}
\mathcal{G}_{00}^{L} & =-\frac{t}{2\ell}\nabla^{2}f\Rightarrow\partial_{0}\mathcal{G}_{00}^{L}=-\frac{1}{2\ell}\nabla^{2}f-\frac{t}{2\ell}\nabla^{2}\dot{f},\label{G_00_der}\\
\mathcal{G}_{i0}^{L} & =-\frac{t}{2\ell}\left[\partial_{i}p+\epsilon_{ik}\partial_{k}\sigma\right]\Rightarrow\partial_{i}\mathcal{G}_{i0}^{L}=-\frac{t}{2\ell}\partial_{i}^{2}p=-\frac{t}{2\ell}\nabla^{2}p,\label{G_0j_der}
\end{align}
and 
\begin{align}
\mathcal{G}_{ii}^{L} & =-\frac{t}{2\ell}\delta_{ij}\left\{ -\left(\epsilon_{il}\hat{\partial}_{j}\hat{\partial}_{l}+\epsilon_{jl}\hat{\partial}_{i}\hat{\partial}_{l}\right)\dot{\sigma}-\hat{\partial}_{j}\hat{\partial}_{i}\dot{p}+\left(\delta_{ji}+\hat{\partial}_{j}\hat{\partial}_{i}\right)q\right\} \nonumber \\
 & =-\frac{t}{2\ell}\left(\dot{p}+q\right).\label{G_ii_contr}
\end{align}
With these equations, (\ref{G_00_der}), (\ref{G_0j_der}) and (\ref{G_ii_contr}),
the Bianchi identity yields
\begin{align}
0 & =\left(-\frac{1}{2\ell}\nabla^{2}f-\frac{t}{2\ell}\nabla^{2}\dot{f}+\frac{t}{2\ell}\nabla^{2}p+\frac{1}{2\ell}\left(\dot{p}+q\right)\right),\nonumber \\
 & =\left(-\nabla^{2}f-t\nabla^{2}\dot{f}+t\nabla^{2}p+\dot{p}+q\right),\label{Bianchi_iden5}
\end{align}
and finally
\begin{equation}
t\nabla^{2}\left(\frac{1}{t}f+\dot{f}-p\right)-\dot{p}-q=0.\label{Bianchi_identity_gauge1}
\end{equation}
Also, $\nu$ can be taken as $\nu=n$. For this choice (\ref{Bianchi_iden2})
reads 
\begin{align}
0 & =-\frac{t^{2}}{\ell^{2}}\left(\partial_{0}\mathcal{G}_{0n}^{L}-\Gamma_{00}^{0}\mathcal{G}_{0n}^{L}-\Gamma_{0n}^{0}\mathcal{G}_{00}^{L}-\Gamma_{0n}^{k}\mathcal{G}_{k0}^{L}\right)\nonumber \\
 & \phantom{=}+\frac{t^{2}}{\ell^{2}}\delta^{ij}\left(\partial_{j}\mathcal{G}_{in}^{L}-\Gamma_{ji}^{0}\mathcal{G}_{0n}^{L}-\Gamma_{jn}^{0}\mathcal{G}_{0i}^{L}-\Gamma_{jn}^{k}\mathcal{G}_{ki}^{L}\right),\nonumber \\
0 & =-\left(\partial_{0}\mathcal{G}_{0n}^{L}+\frac{1}{t}\mathcal{G}_{0n}^{L}+\frac{1}{t}\mathcal{G}_{n0}^{L}\right)\nonumber \\
 & \phantom{=}+\left(\partial_{i}\mathcal{G}_{in}^{L}+\frac{2}{t}\mathcal{G}_{0n}^{L}+\frac{1}{t}\mathcal{G}_{0n}^{L}\right),\nonumber \\
0 & =\left(\partial_{0}\mathcal{G}_{0n}^{L}-\frac{1}{t}\mathcal{G}_{0n}^{L}-\partial_{i}\mathcal{G}_{in}^{L}\right).\label{Bianchi_iden6}
\end{align}
Again using (\ref{G_0j_der}) and the derivative of (\ref{G_jk_gauge}),
that is 
\begin{equation}
\partial_{i}\mathcal{G}_{in}^{L}=-\frac{t}{2\ell}\left(-\epsilon_{nl}\partial_{l}\dot{\sigma}+\partial_{n}\dot{p}\right),\label{G_jk_der}
\end{equation}
(\ref{Bianchi_iden6}) yields
\begin{align}
0 & =-\frac{t}{2\ell}\left[\partial_{n}\dot{p}+\epsilon_{nk}\partial_{k}\dot{\sigma}\right]+\frac{1}{2\ell}\left[\partial_{n}p+\epsilon_{nk}\partial_{k}\sigma\right]+\frac{t}{2\ell}\left(-\epsilon_{nl}\hat{\partial}_{i}\hat{\partial}_{l}\dot{\sigma}+\partial_{n}\dot{p}\right),\nonumber \\
0 & =-\partial_{i}\dot{p}-\epsilon_{ik}\partial_{k}\dot{\sigma}+\frac{1}{t}\left[\partial_{i}p+\epsilon_{ik}\partial_{k}\sigma\right]-\epsilon_{nl}\partial_{l}\dot{\sigma}+\partial_{n}\dot{p},\nonumber \\
0 & =-2\epsilon_{nk}\partial_{k}\dot{\sigma}+\frac{1}{t}\left[\partial_{n}p+\epsilon_{nk}\partial_{k}\sigma\right],\label{Bianchi_iden_7}
\end{align}
and 
\begin{equation}
0=\epsilon_{nk}\partial_{k}\left(-2\dot{\sigma}+\frac{1}{t}\sigma\right)+\frac{1}{t}\partial_{n}p.\label{Bianchi_identity_gauge2}
\end{equation}
Using (\ref{Bianchi_identity_gauge1}), (\ref{gauge_inv_R_scalar})
becomes
\begin{equation}
R_{L}=\frac{t^{3}}{\ell^{3}}\left(q-\nabla^{2}f+\dot{p}\right)=\frac{t^{4}}{\ell^{3}}\nabla^{2}\left(\dot{f}-p\right).\label{Ricci_gauge_Bianchi}
\end{equation}

Up to now, we write the components of linearized Ricci tensor and
linear Einstein tensor in terms of gauge invariant functions. Moreover,
the linear Ricci scalar is also written in a gauge invariant form.
Now we can write (\ref{generic_act}) in terms of gauge invariant
functions by using these identities. The next section is devoted to
find a gauge invariant action.

\subsection{Metric Decomposition and Gauge-Invariant Form Of The Higher Derivative
Action}

Linearized action is
\begin{equation}
I=-\frac{1}{2}\int d^{3}x\,\sqrt{\bar{g}}\, h_{\mu\nu}\left[a\mathcal{G}_{L}^{\mu\nu}+\left(2\alpha+\beta\right)\left(\bar{g}^{\mu\nu}\Box-\nabla^{\mu}\nabla^{\nu}+2\Lambda\bar{g}^{\mu\nu}\right)R_{L}+\beta\left(\Box\mathcal{G}_{L}^{\mu\nu}-\Lambda\bar{g}^{\mu\nu}R_{L}\right)\right]\label{Linearized_action}
\end{equation}
 where background metric is
\begin{equation}
ds^{2}=\frac{\ell^{2}}{t^{2}}\left[-dt^{2}+dx^{2}+dy^{2}\right]\quad\Rightarrow\quad\bar{g}_{\mu\nu}=\frac{\ell^{2}}{t^{2}}\eta_{\mu\nu}.\label{metric_app}
\end{equation}
Note that the term in the parenthesis is the field equations of the
quadratic curvature theory. The constant $a$ in front of $\mathcal{G}_{L}^{\mu\nu}$
is \cite{Gullu1}
\begin{equation}
a\equiv\frac{1}{\kappa}+12\Lambda\alpha+2\Lambda\beta.\label{coeff_a}
\end{equation}
Linearized form of Einstein and Ricci tensors, and Ricci scalar are
given in (\ref{Lin_Einstein_ten}, \ref{Linear_Ricci_ten}, \ref{R_scalar1})
with the definition $\Box\equiv\nabla_{\mu}\nabla^{\mu}=\frac{t^{2}}{\ell^{2}}\eta^{\mu\nu}\nabla_{\mu}\nabla_{\nu}$.

\subsubsection{The Einstein-Hilbert Part}

Let us work on the Einstein-Hilbert part first:
\begin{align}
I_{E} & =-\frac{a}{2}\int d^{3}x\,\sqrt{\bar{g}}h_{\mu\nu}\mathcal{G}_{L}^{\mu\nu}=-\frac{a}{2}\int d^{3}x\,\frac{\ell^{3}}{t^{3}}\frac{t^{4}}{\ell^{4}}\eta^{\rho\mu}\eta^{\sigma\nu}h_{\rho\sigma}\mathcal{G}_{\mu\nu}^{L},\nonumber \\
 & =-\frac{a}{2}\int d^{3}x\,\frac{t}{\ell}\eta^{\rho\mu}\eta^{\sigma\nu}h_{\rho\sigma}\mathcal{G}_{\mu\nu}^{L}.\label{Einstein_Hilbert}
\end{align}
Writing Einstein-Hilbert part in terms of tensor components yields
\begin{align}
I_{E} & =-\frac{a}{2}\int d^{3}x\,\frac{t}{\ell}\left(h_{00}\mathcal{G}_{00}^{L}-2h_{0i}\mathcal{G}_{0i}^{L}+h_{ij}\mathcal{G}_{ij}^{L}\right).\label{EH_comp}
\end{align}
Let us calculate term by term this action:

\subsubsection{Decomposition of $h_{00}\mathcal{G}_{00}^{L}$ term}

Let's start with the first term, that is $h_{00}\mathcal{G}_{00}^{L}$
where
\begin{equation}
\mathcal{G}_{00}^{L}=R_{00}^{L}-\frac{1}{2}\bar{g}_{00}R_{L}-2\Lambda h_{00}.\label{G_00}
\end{equation}
Using (\ref{G_00_comp}) and (\ref{met_per_dec}), (\ref{h_00_G_00})
becomes

\begin{equation}
h_{00}\mathcal{G}_{00}^{L}=N\left[-\frac{t^{2}}{2\ell^{2}}\partial_{i}^{2}\phi-\frac{t}{2\ell^{2}}\left(\dot{\phi}+\dot{\chi}-2\partial_{i}^{2}N_{L}\right)-\frac{1}{\ell^{2}}\left(-N+\phi+\chi\right)\right].\label{h_00_G_00}
\end{equation}
For the flat spacetime case (\ref{h_00_G_00}) becomes
\begin{equation}
h_{00}\mathcal{G}_{00}^{L}=-\frac{1}{2}N\partial_{i}^{2}\phi.\label{h_00_G_00_flat}
\end{equation}

\subsubsection{Decomposition of $h_{0j}\mathcal{G}_{0j}^{L}$ term}

Let us move to second term in Einstein-Hilbert action which is $-2h_{0j}\mathcal{G}_{0j}^{L}$.
Using (\ref{G_0j_comp}) and (\ref{met_per_dec}) $h_{0j}\mathcal{G}_{0j}^{L}$
becomes

\begin{align}
h_{0j}\mathcal{G}_{0j}^{L} & =\left(-2h_{0j}\left\{ -\frac{t^{2}}{2\ell^{2}}\left[\partial_{j}\dot{\phi}+\epsilon_{jk}\partial_{k}\left(\dot{\xi}-\partial_{i}^{2}\eta\right)\right]+\frac{t}{2\ell^{2}}\left[\partial_{j}\left(N-2\phi\right)-2\epsilon_{jk}\partial_{k}\xi\right]\right\} \right),\nonumber \\
 & =\frac{t^{2}}{\ell^{2}}\left(h_{0j}\left\{ \partial_{j}\left[\dot{\phi}-\frac{1}{t}\left(N-2\phi\right)\right]+\epsilon_{jk}\partial_{k}\left[\left(\dot{\xi}-\partial_{i}^{2}\eta\right)+\frac{2}{t}\xi\right]\right\} \right),\nonumber \\
 & =\frac{t^{2}}{\ell^{2}}\left\{ -\left(\partial_{j}h_{0j}\right)\left[\dot{\phi}-\frac{1}{t}\left(N-2\phi\right)\right]-\left(\epsilon_{jk}\partial_{k}h_{0j}\right)\left[\left(\dot{\xi}-\partial_{i}^{2}\eta\right)+\frac{2}{t}\xi\right]\right\} ,\label{h_0j_G_0j1}
\end{align}
where the integral sign is suppressed and at the last line the boundary
terms dropped. Also note that,\foreignlanguage{turkish}{
\begin{align}
\partial_{j}h_{0j} & =\partial_{j}^{2}N_{L},\label{delh_0j}
\end{align}
}and
\begin{align}
\epsilon_{jk}\partial_{k}h_{0j} & =\epsilon_{jk}\partial_{k}\left(-\epsilon_{ji}\partial_{i}\eta+\partial_{j}N_{L}\right)=-\epsilon_{ji}\epsilon_{jk}\partial_{k}\partial_{i}\eta+\epsilon_{jk}\partial_{k}\partial_{j}N_{L},\nonumber \\
 & =-\delta_{ik}\partial_{k}\partial_{i}\eta=-\partial_{i}^{2}\eta.\label{ep_delh_oj}
\end{align}
Then (\ref{h_0j_G_0j1}) yields,
\begin{align}
\left(-2h_{0j}\mathcal{G}_{0j}^{L}\right) & =\frac{t^{2}}{\ell^{2}}\left\{ \partial_{i}^{2}\eta\left[\left(\dot{\xi}-\partial_{i}^{2}\eta\right)+\frac{2}{t}\xi\right]-\partial_{j}^{2}N_{L}\left[\dot{\phi}-\frac{1}{t}\left(N-2\phi\right)\right]\right\} ,\nonumber \\
 & =\frac{t^{2}}{\ell^{2}}\left\{ \left(\dot{\xi}-\partial_{i}^{2}\eta\right)\partial_{i}^{2}\eta-\dot{\phi}\left(\partial_{j}^{2}N_{L}\right)+\frac{1}{t}\left[2\xi\partial_{i}^{2}\eta+\left(N-2\phi\right)\left(\partial_{j}^{2}N_{L}\right)\right]\right\} .\label{h_0j_G_0j}
\end{align}
In $t/\ell\rightarrow1$ and $\ell\rightarrow\infty$ limits (\ref{h_0j_G_0j})
becomes
\begin{align}
-2h_{0j}\mathcal{G}_{0j}^{L} & =\left(\dot{\xi}-\partial_{i}^{2}\eta\right)\partial_{i}^{2}\eta-\dot{\phi}\left(\partial_{j}^{2}N_{L}\right),\label{h_0j_G_0j_flat}
\end{align}
which is the result for the flat spacetime case.

\subsubsection{Decomposition of $h_{jk}\mathcal{G}_{jk}^{L}$ term}

Using (\ref{G_jk_comp3}) and the first equation in (\ref{met_per_dec}),
$h_{jk}\mathcal{G}_{jk}^{L}$ takes the following form
\begin{align}
h_{jk}\mathcal{G}_{jk}^{L} & =\left(-\frac{t^{2}}{2\ell^{2}}\right)\left\{ h_{jk}\partial_{j}\left[-2\epsilon_{kl}\partial_{l}\left(\dot{\eta}+\frac{1}{t}\eta\right)+\partial_{k}\left(2\dot{N}_{L}-N+\frac{2}{t}N_{L}\right)\right]\right.\nonumber \\
 & \phantom{=\left(-\frac{t^{2}}{2\ell^{2}}\right)}\left.-h_{jk}\left(\partial_{0}^{2}+\frac{3}{t}\partial_{0}\right)\left[\hat{\partial}_{j}\hat{\partial}_{k}\phi-\hat{\partial}_{j}\hat{\partial}_{k}\chi+\left(\epsilon_{jl}\hat{\partial}_{l}\hat{\partial}_{k}+\epsilon_{kl}\hat{\partial}_{l}\hat{\partial}_{j}\right)\xi\right]\right\} \nonumber \\
 & \phantom{=}-\frac{1}{2}h_{jk}\delta_{jk}\left\{ \frac{t^{2}}{\ell^{2}}\left(\partial_{i}^{2}N+\ddot{\chi}-2\partial_{i}^{2}\dot{N}_{L}\right)-\frac{t}{\ell^{2}}\left(\dot{N}+2\partial_{i}^{2}N_{L}\right)+\frac{3t}{\ell^{2}}\dot{\chi}\right\} .\label{h_jk_G_jk1}
\end{align}
Note that $\partial_{j}h_{jk}=\partial_{k}\chi-\epsilon_{kl}\partial_{l}\xi$
and $\delta_{jk}h_{jk}=\phi+\chi$ and again the integral sign is
suppressed. Taking the space derivative in front of the metric perturbation
in the first term with dropping the boundary terms the equation becomes
\begin{align}
h_{jk}\mathcal{G}_{jk}^{L} & =\left(-\frac{t^{2}}{2a\ell^{2}}\right)\left\{ -\left(\partial_{k}\chi-\epsilon_{kl}\partial_{l}\xi\right)\left[-2\epsilon_{kl}\partial_{l}\left(\dot{\eta}+\frac{1}{t}\eta\right)+\partial_{k}\left(2\dot{N}_{L}-N+\frac{2}{t}N_{L}\right)\right]\right.\nonumber \\
 & \phantom{=\left(-\frac{t^{2}}{2\ell^{2}}\right)}\left.-h_{jk}\left(\partial_{0}^{2}+\frac{3}{t}\partial_{0}\right)\left[\hat{\partial}_{j}\hat{\partial}_{k}\phi-\hat{\partial}_{j}\hat{\partial}_{k}\chi+\left(\epsilon_{jl}\hat{\partial}_{l}\hat{\partial}_{k}+\epsilon_{kl}\hat{\partial}_{l}\hat{\partial}_{j}\right)\xi\right]\right\} \nonumber \\
 & \phantom{=}-\frac{1}{2}\left(\phi+\chi\right)\left\{ \frac{t^{2}}{\ell^{2}}\left(\partial_{i}^{2}N+\ddot{\chi}-2\partial_{i}^{2}\dot{N}_{L}\right)-\frac{t}{\ell^{2}}\left(\dot{N}+2\partial_{i}^{2}N_{L}\right)+\frac{3t}{\ell^{2}}\dot{\chi}\right\} .\label{h_jk_G_jk2}
\end{align}
The multiplication of the metric perturbation with the middle term
is 
\begin{align}
\left[\left(\delta_{jk}+\hat{\partial}_{k}\hat{\partial}_{j}\right)\phi-\hat{\partial}_{k}\hat{\partial}_{j}\chi+\left(\epsilon_{kn}\hat{\partial}_{n}\hat{\partial}_{j}+\epsilon_{jn}\hat{\partial}_{n}\hat{\partial}_{k}\right)\xi\right]\times\nonumber \\
\left[\hat{\partial}_{j}\hat{\partial}_{k}\phi^{\left(i\right)}-\hat{\partial}_{j}\hat{\partial}_{k}\chi^{\left(i\right)}+\left(\epsilon_{jl}\hat{\partial}_{l}\hat{\partial}_{k}+\epsilon_{kl}\hat{\partial}_{l}\hat{\partial}_{j}\right)\xi^{\left(i\right)}\right] & =-\chi\phi^{\left(i\right)}+\chi\chi^{\left(i\right)}+2\xi\xi^{\left(i\right)},\label{h_jk_G_jk_iden}
\end{align}
where $^{\left(i\right)}$ denotes time derivative and $i=1,2$. Putting
this equation with suitable derivative signs and doing the cancellations
(\ref{h_jk_G_jk2}) turns to be

\begin{align}
h_{jk}\mathcal{G}_{jk}^{L} & =\left(-\frac{t^{2}}{2\ell^{2}}\right)\left\{ \left(-2\phi\partial_{i}^{2}\dot{N}_{L}+2\xi\partial_{i}^{2}\dot{\eta}+\phi\partial_{i}^{2}N+\phi\ddot{\chi}+\chi\ddot{\phi}-2\xi\ddot{\xi}\right)\right.\nonumber \\
 & \phantom{\phantom{=\left(-\frac{t^{2}}{2\ell^{2}}\right)}}\left.+\frac{1}{t}\left(2\xi\partial_{i}^{2}\eta-6\xi\dot{\xi}-2\phi\partial_{i}^{2}N_{L}-\phi\dot{N}-\chi\dot{N}+3\phi\dot{\chi}+3\chi\dot{\phi}\right)\right\} .\label{h_jk_G_jk}
\end{align}
In the flat spacetime limit (\ref{h_jk_G_jk}) becomes
\begin{align}
h_{jk}\mathcal{G}_{jk}^{L} & =-\frac{1}{2}\left(-2\phi\partial_{i}^{2}\dot{N}_{L}+2\xi\partial_{i}^{2}\dot{\eta}+\phi\partial_{i}^{2}N+\phi\ddot{\chi}+\chi\ddot{\phi}-2\xi\ddot{\xi}\right),\nonumber \\
 & =\phi\partial_{i}^{2}\dot{N}_{L}-\xi\partial_{i}^{2}\dot{\eta}-\frac{1}{2}\phi\partial_{i}^{2}N-\frac{1}{2}\phi\ddot{\chi}-\frac{1}{2}\chi\ddot{\phi}+\xi\ddot{\xi},\label{h_jk_G_jk_flat1}
\end{align}
and with integration by parts on time differentiations (\ref{h_jk_G_jk_flat1})
yields 
\begin{equation}
h_{jk}\mathcal{G}_{jk}^{L}=-\dot{\phi}\partial_{i}^{2}N_{L}+\dot{\xi}\partial_{i}^{2}\eta+\xi\ddot{\xi}-\phi\ddot{\chi}-\frac{1}{2}\phi\partial_{i}^{2}N.\label{h_jk_G_jk_flat2}
\end{equation}

\subsubsection{Decomposed form of Einstein-Hilbert Action}

To get the final form of Einstein-Hilbert term, all the found results
(\ref{h_00_G_00}), (\ref{h_0j_G_0j}), (\ref{h_jk_G_jk}) are put
in (\ref{EH_comp}) to get 
\begin{align}
I_{E} & =-\frac{a}{2}\int d^{3}x\,\frac{t}{\ell}N\left[-\frac{t^{2}}{2\ell^{2}}\partial_{i}^{2}\phi-\frac{t}{2\ell^{2}}\left(\dot{\phi}+\dot{\chi}-2\partial_{i}^{2}N_{L}\right)-\frac{1}{\ell^{2}}\left(-N+\phi+\chi\right)\right]\nonumber \\
 & \phantom{=}-\frac{a}{2}\int d^{3}x\,\frac{t^{3}}{\ell^{3}}\left\{ \left(\dot{\xi}-\partial_{i}^{2}\eta\right)\partial_{i}^{2}\eta-\dot{\phi}\left(\partial_{j}^{2}N_{L}\right)+\frac{1}{t}\left[2\xi\partial_{i}^{2}\eta+\left(N-2\phi\right)\left(\partial_{j}^{2}N_{L}\right)\right]\right\} \nonumber \\
 & \phantom{=}-\frac{a}{2}\int d^{3}x\,\left(-\frac{t^{3}}{2\ell^{3}}\right)\left\{ \left(-2\phi\partial_{i}^{2}\dot{N}_{L}+2\xi\partial_{i}^{2}\dot{\eta}+\phi\partial_{i}^{2}N+\phi\ddot{\chi}+\chi\ddot{\phi}-2\xi\ddot{\xi}\right)\right.\label{I_E_comp1}\\
 & \phantom{\phantom{=-\frac{a}{2}\int d^{3}x\,\left(-\frac{t^{3}}{2\ell^{3}}\right)}}\left.+\frac{1}{t}\left(2\xi\partial_{i}^{2}\eta-6\xi\dot{\xi}-2\phi\partial_{i}^{2}N_{L}-\phi\dot{N}-\chi\dot{N}+3\phi\dot{\chi}+3\chi\dot{\phi}\right)\right\} .\nonumber 
\end{align}
To see the terms that may cancel or sum after integration by parts
operation, (\ref{I_E_comp1}) is written in the following form
\begin{align}
I_{E} & =-\frac{a}{2}\int d^{3}x\,\left(-\frac{t^{3}}{2\ell^{3}}\right)\left\{ \phi\ddot{\chi}+\chi\ddot{\phi}+2\phi\partial_{i}^{2}N+2\dot{\phi}\partial_{i}^{2}N_{L}-2\phi\partial_{i}^{2}\dot{N}_{L}\right.\nonumber \\
 & \phantom{=-\frac{a}{2}\int d^{3}x\,\left(-\frac{t^{3}}{2\ell^{3}}\right)}\left.+\left[2\left(\partial_{i}^{2}\eta\right)^{2}-2\dot{\xi}\partial_{i}^{2}\eta+2\xi\partial_{i}^{2}\dot{\eta}-2\xi\ddot{\xi}\right]\right.\nonumber \\
 & \phantom{=-\frac{a}{2}\int d^{3}x\,\left(-\frac{t^{3}}{2\ell^{3}}\right)}\left.+\frac{1}{t}\left(3\phi\dot{\chi}+3\chi\dot{\phi}+2\phi\partial_{j}^{2}N_{L}-2\xi\partial_{i}^{2}\eta-6\xi\dot{\xi}\right)\right.\label{I_E_comp2}\\
 & \phantom{=-\frac{a}{2}\int d^{3}x\,\left(-\frac{t^{3}}{2\ell^{3}}\right)}\left.+\frac{1}{t}\left[N\left(\dot{\phi}+\dot{\chi}\right)-\dot{N}\left(\phi+\chi\right)-4N\partial_{i}^{2}N_{L}\right]\right.\nonumber \\
 & \phantom{\phantom{=-\frac{a}{2}\int d^{3}x\,\left(-\frac{t^{3}}{2\ell^{3}}\right)}}\left.+\frac{2}{t^{2}}N\left(-N+\phi+\chi\right)\right\} .\nonumber 
\end{align}
After doing integration by parts in the terms $\chi\ddot{\phi}$ and
$\dot{N}\left(\phi+\chi\right)$, (\ref{I_E_comp2}) becomes 
\begin{align}
I_{E} & =-\frac{a}{2}\int d^{3}x\,\left(-\frac{t^{3}}{2\ell^{3}}\right)\left\{ \phi\ddot{\chi}+\phi\left(\ddot{\chi}+\frac{6}{t}\dot{\chi}+\frac{6}{t^{2}}\chi\right)+2\phi\partial_{i}^{2}N\right.\nonumber \\
 & \phantom{=-\frac{a}{2}\int d^{3}x\,\left(-\frac{t^{3}}{2\ell^{3}}\right)}-2\phi\partial_{i}^{2}\dot{N}_{L}-\frac{6}{t}\phi\partial_{i}^{2}N_{L}-2\phi\partial_{i}^{2}\dot{N}_{L}\nonumber \\
 & \phantom{=-\frac{a}{2}\int d^{3}x\,\left(-\frac{t^{3}}{2\ell^{3}}\right)}\left.+\left[2\left(\partial_{i}^{2}\eta\right)^{2}-2\dot{\xi}\partial_{i}^{2}\eta-2\dot{\xi}\partial_{i}^{2}\eta-\frac{6}{t}\xi\partial_{i}^{2}\eta+2\dot{\xi}^{2}+\frac{6}{t}\xi\dot{\xi}\right]\right.\nonumber \\
 & \phantom{=-\frac{a}{2}\int d^{3}x\,\left(-\frac{t^{3}}{2\ell^{3}}\right)}\left.+\frac{1}{t}\left(3\phi\dot{\chi}-3\dot{\chi}\phi-\frac{6}{t}\chi\phi+2\phi\partial_{j}^{2}N_{L}-2\xi\partial_{i}^{2}\eta-6\xi\dot{\xi}\right)\right.\label{I_E_comp3}\\
 & \phantom{=-\frac{a}{2}\int d^{3}x\,\left(-\frac{t^{3}}{2\ell^{3}}\right)}\left.+\frac{1}{t}\left[N\left(\dot{\phi}+\dot{\chi}\right)+N\left(\dot{\phi}+\dot{\chi}\right)+\frac{2}{t}N\left(\phi+\chi\right)-4N\partial_{i}^{2}N_{L}\right]\right.\nonumber \\
 & \phantom{\phantom{=-\frac{a}{2}\int d^{3}x\,\left(-\frac{t^{3}}{2\ell^{3}}\right)}}\left.+\frac{2}{t^{2}}N\left(-N+\phi+\chi\right)\right\} ,\nonumber 
\end{align}
and the final form of (\ref{I_E_comp3}) in terms of metric perturbation
functions is
\begin{align}
I_{E} & =-\frac{a}{2}\int d^{3}x\,\left(-\frac{t^{3}}{\ell^{3}}\right)\left\{ \phi\ddot{\chi}+\phi\partial_{i}^{2}\left(N-2\dot{N}_{L}\right)+\left(-\partial_{i}^{2}\eta+\dot{\xi}\right)^{2}\right.\nonumber \\
 & \phantom{=-\frac{a}{2}\int d^{3}x\,\left(-\frac{t^{3}}{\ell^{3}}\right)}\left.+\frac{1}{t}\left(3\phi\dot{\chi}+N\left(\dot{\phi}+\dot{\chi}\right)-2N\partial_{i}^{2}N_{L}-2\phi\partial_{i}^{2}N_{L}-4\xi\partial_{i}^{2}\eta\right)\right.\nonumber \\
 & \phantom{=-\frac{a}{2}\int d^{3}x\,\left(-\frac{t^{3}}{\ell^{3}}\right)}\left.+\frac{1}{t^{2}}\left(-N^{2}+2N\left(\phi+\chi\right)\right)\right\} ,\label{I_e_non_gauge}
\end{align}
which is not gauge invariant yet. In flat spacetime limit (\ref{I_e_non_gauge})
yields
\begin{equation}
I_{E}=\frac{a}{2}\int d^{3}x\,\left[\phi\ddot{\chi}+\phi\partial_{i}^{2}\left(N-2\dot{N}_{L}\right)+\left(-\partial_{i}^{2}\eta+\dot{\xi}\right)^{2}\right].\label{I_E_comp_flat}
\end{equation}

\subsubsection{The Gauge-Invariant form of the Einstein-Hilbert term:}

From (\ref{I_e_non_gauge}) the gauge-invariant form of the Einstein-Hilbert
action can be written. For this aim, first the functions are redefined
as $\phi\rightarrow\frac{\ell^{2}}{t^{2}}\phi$. With this redefinition
(\ref{I_e_non_gauge}) becomes
\begin{align}
I_{E} & =-\frac{a}{2}\int d^{3}x\,\left(-\frac{\ell}{t}\right)\left\{ \phi\ddot{\chi}+\phi\partial_{i}^{2}\left(N-2\dot{N}_{L}\right)+\left(-\partial_{i}^{2}\eta+\dot{\xi}\right)^{2}\right.\nonumber \\
 & \phantom{=-\frac{a}{2}\int d^{3}x\,\left(-\frac{\ell}{t}\right)}\left.+\frac{1}{t}\left(-\phi\dot{\chi}+N\left(\dot{\phi}+\dot{\chi}\right)-2N\partial_{i}^{2}N_{L}+2\phi\partial_{i}^{2}N_{L}\right)\right.\nonumber \\
 & \phantom{=-\frac{a}{2}\int d^{3}x\,\left(-\frac{\ell}{t}\right)}\left.-\frac{1}{t^{2}}N^{2}+\frac{4}{t^{2}}\xi^{2}-\frac{4}{t}\xi\dot{\xi}\right\} ,\label{I_E_comp4}
\end{align}
and doing integration by parts in the last term 
\begin{equation}
-\frac{4}{t^{2}}\xi\dot{\xi}=\partial_{0}\left(-\frac{4}{t^{2}}\xi\xi\right)-\frac{8}{t^{3}}\xi\xi+\frac{4}{t^{2}}\xi\dot{\xi},\Rightarrow\xi\dot{\xi}=\frac{1}{t}\xi\xi,\label{xi_ibp}
\end{equation}
where we have dropped the boundary term in the last equality. Then
the Einstein-Hilbert action becomes 
\begin{align}
I_{E} & =\left(-\frac{\ell}{t}\right)\left[\phi\ddot{\chi}+\phi\partial_{i}^{2}\left(N-2\dot{N}_{L}\right)+\left(-\partial_{i}^{2}\eta+\dot{\xi}\right)^{2}-\frac{1}{t^{2}}N^{2}\right]\nonumber \\
 & \phantom{=}-\frac{\ell}{t^{2}}\left[-\phi\dot{\chi}+N\left(\dot{\phi}+\dot{\chi}\right)-2N\partial_{i}^{2}N_{L}+2\phi\partial_{i}^{2}N_{L}\right],\label{I_E_comp5}
\end{align}
where the integral sign and the overall coefficients are suppressed.
Using (\ref{gauge_p_sigma}) and adding and subtracting the terms
$\frac{2}{t}\dot{\phi}N,\;\dot{\phi}^{2},\;\frac{1}{t^{2}}N^{2}$
yields
\begin{align}
I_{E} & =\left(-\frac{\ell}{t}\right)\left[\frac{t^{2}}{\ell^{2}}\left(\sigma^{2}+p^{2}\right)+\phi\ddot{\chi}+\phi\partial_{i}^{2}\left(N-2\dot{N}_{L}\right)-\frac{2}{t^{2}}N^{2}\right]\nonumber \\
 & \phantom{=}-\frac{\ell}{t^{2}}\left[-\phi\dot{\chi}+N\left(3\dot{\phi}+\dot{\chi}\right)-2N\partial_{i}^{2}N_{L}+2\phi\partial_{i}^{2}N_{L}-\dot{\phi}^{2}\right].\label{I_E_comp6}
\end{align}
Again doing integration by parts with respect to time in the following
terms terms $\frac{\ell}{t}\ddot{\chi}\phi,$ $\frac{2\ell}{t}\phi\nabla^{2}\dot{N_{L}}$
and adding the term $t\phi\nabla\dot{\phi}$ freely, since it is a
boundary term, to the action yields 
\begin{align}
I_{E} & =\left(-\frac{\ell}{t}\right)\left[\frac{t^{2}}{\ell^{2}}\left(\sigma^{2}+p^{2}\right)-\dot{\phi}\dot{\chi}+N\partial_{i}^{2}\phi+2\dot{\phi}\partial_{i}^{2}N_{L}-\frac{2}{t^{2}}N^{2}-t\phi\nabla\dot{\phi}\right]\nonumber \\
 & \phantom{=}-\frac{\ell}{t^{2}}\left[2N\dot{\phi}+N\dot{\phi}+N\dot{\chi}-2N\partial_{i}^{2}N_{L}-\dot{\phi}^{2}\right].\label{I_E_comp7}
\end{align}
With reordering the terms (\ref{I_E_comp7}) can be written as follows
\begin{align}
I_{E} & =\left(-\frac{\ell}{t}\right)\left[\frac{t^{2}}{\ell^{2}}\left(\sigma^{2}+p^{2}\right)-t\dot{\phi}\left(\frac{1}{t}\dot{\chi}-\frac{2}{t}\nabla^{2}N_{L}-\frac{2}{t^{2}}N+\frac{1}{t}\dot{\phi}+\nabla^{2}\phi\right)\right]\nonumber \\
 & \phantom{=}-\frac{\ell}{t}\left[N\left(\frac{1}{t}\dot{\chi}-\frac{2}{t}\nabla^{2}N_{L}-\frac{2}{t^{2}}N+\frac{1}{t}\dot{\phi}+\nabla^{2}\phi\right)\right],\label{I_E_comp8}
\end{align}
and putting (\ref{gauge_f}) in (\ref{I_E_comp8}) we obtain
\begin{equation}
I_{E}=\left(-\frac{\ell}{t}\right)\left[\frac{t^{2}}{\ell^{2}}\left(\sigma^{2}+p^{2}\right)-\frac{t^{2}}{\ell}\dot{\phi}\nabla^{2}f+\frac{t}{\ell}N\nabla^{2}f\right].\label{I_E_comp}
\end{equation}
Finally, using (\ref{gauge_p_sigma}) we get 
\begin{equation}
I_{E}=-\left[\frac{t}{\ell}\left(\sigma^{2}+p^{2}\right)-\frac{t^{2}}{\ell}\, p\nabla^{2}f\right].\label{I_E_comp9}
\end{equation}
We can also do some manipulations to write (\ref{I_E_comp9}) in terms
of the Ricci Scalar. When we look at the gauge-invariant Bianchi identity,
we see that the term $t\, f\,\nabla^{2}\dot{f}+f\,\nabla^{2}f$ must
be added to the action (\ref{I_E_comp9}). Since this term is boundary
term it can be added freely. Therefore, the action becomes 
\begin{equation}
I_{E}=\left(-\frac{t}{\ell}\right)\left[\left(\sigma^{2}+p^{2}\right)-t\, f\nabla^{2}p+t\, f\,\nabla^{2}\dot{f}+f\,\nabla^{2}f\right],\label{I_E_comp10}
\end{equation}
and using Bianchi identity (\ref{I_E_comp10}) becomes 
\begin{equation}
I_{E}=\left(-\frac{t}{\ell}\right)\left[\left(\sigma^{2}+p^{2}\right)+f\left(q+\dot{p}\right)\right].\label{I_E_comp11}
\end{equation}
Then, we add and subtract $f\,\nabla^{2}f$ in (\ref{I_E_comp11})
and using (\ref{gauge_inv_R_scalar}) we get
\begin{equation}
I_{E}=-\left[\frac{t}{\ell}\left(\sigma^{2}+p^{2}+f\,\nabla^{2}f\right)+\frac{\ell^{2}}{t^{2}}f\, R_{L}\right],\label{I_E_comp12}
\end{equation}
and in the formal form we have 
\begin{equation}
I_{E}=\frac{a}{2}\int\left[\frac{t}{\ell}\left(\sigma^{2}+p^{2}+f\,\nabla^{2}f\right)+\frac{\ell^{2}}{t^{2}}f\, R_{L}\right].\label{I_e_gauge}
\end{equation}
The flat spacetime limit of this action is 
\begin{equation}
I_{E}=\frac{1}{2\kappa}\int\left[\sigma^{2}+\dot{\phi}^{2}+\phi\,\nabla^{2}\phi+\phi\,\left(q-\square\phi\right)\right],\label{int_I_E_comp}
\end{equation}
where the flat space version of the gauge-invariant form of the Ricci
scalar (\ref{gauge_inv_R_flat2}) and $a=\frac{1}{\kappa}$ are used.
Then 
\begin{align}
I_{E} & =\frac{1}{2\kappa}\int\left[\sigma^{2}-\phi\ddot{\phi}+\phi\,\nabla^{2}\phi+\phi q-\phi\square\phi\right]\nonumber \\
 & =\frac{1}{2\kappa}\int\left[\sigma^{2}+\phi\left(-\partial_{0}^{2}+\nabla^{2}\right)\phi+\phi q-\phi\square\phi\right]\nonumber \\
 & =\frac{1}{2\kappa}\int\left(\sigma^{2}+\phi q\right),\label{I_e_gauge_flat}
\end{align}
where in the first line integration by parts is done and in the second
line the definition of the D'Alembertian $\square=-\partial_{0}^{2}+\nabla^{2}$
is used. Also, at the flat spacetime limit the coefficient $a$ becomes
$\frac{1}{\kappa}$, after taking $\Lambda=0$ in (\ref{coeff_a}).

\subsection{The $2\alpha+\beta$ Part of The Higher Derivative Action}

The $2\alpha+\beta$ of the action is 

\begin{equation}
I_{2\alpha+\beta}=-\frac{\left(2\alpha+\beta\right)}{2}\int d^{3}x\,\sqrt{\bar{g}}h_{\mu\nu}\left(\bar{g}^{\mu\nu}\Box-\nabla^{\mu}\nabla^{\nu}+2\Lambda\bar{g}^{\mu\nu}\right)R_{L},\label{2alpha_beta_int1}
\end{equation}
and doing integration by parts and dropping the boundary terms the
action becomes
\begin{equation}
I_{2\alpha+\beta}=-\frac{\left(2\alpha+\beta\right)}{2}\int d^{3}x\,\sqrt{\bar{g}}R_{L}\left(\Box h-\nabla^{\mu}\nabla^{\nu}h_{\mu\nu}+2\Lambda h\right),\label{2alpha_beta_int2}
\end{equation}
where the term in the parenthesis is $-R_{L}$. Therefore, the action
becomes, 

\begin{equation}
I_{2\alpha+\beta}=\frac{\left(2\alpha+\beta\right)}{2}\int d^{3}x\,\sqrt{\bar{g}}R_{L}^{2},\label{I_2a_+_b_gauge}
\end{equation}
since $R_{L}$ is already gauge-invariant this part of the action
is also gauge-invariant. The flat spacetime version of this action
is 
\begin{equation}
I_{2\alpha+\beta}=\frac{\left(2\alpha+\beta\right)}{2}\int d^{3}x\, R_{L}^{2},\label{I_2a+b_gauge_flat}
\end{equation}
where $\sqrt{-\bar{g}}=\frac{t^{3}}{\ell^{3}}=1$ is taken.

\subsection{The $\beta$ Part of the Higher Derivative Action}

Now, let us move to the $\beta$ term of (\ref{Linearized_action});
\begin{align}
I_{\beta} & =-\frac{\beta}{2}\int d^{3}x\,\sqrt{\bar{g}}h_{\mu\nu}\left(\Box\mathcal{G}_{L}^{\mu\nu}-\Lambda\bar{g}^{\mu\nu}R_{L}\right),\nonumber \\
 & =-\frac{\beta}{2}\int d^{3}x\,\sqrt{\bar{g}}\left[\left(\Box h_{\mu\nu}\right)\mathcal{G}_{L}^{\mu\nu}-\Lambda hR_{L}\right],\label{I_beta_action}
\end{align}
where in the second line the D'Alembertian is moved in front of the
metric perturbation and the boundary terms were dropped. Observe that,
the definition of $R_{\mu\nu}^{L}$ , (\ref{Linear_Ricci_ten}), involves
$\square h_{\mu\nu}$, therefore it can be written in terms of $\mathcal{G}_{\mu\nu}^{L}$.
From (\ref{Linear_Ricci_ten})
\begin{equation}
\Box h_{\mu\nu}=-2R_{\mu\nu}^{L}+\nabla^{\sigma}\nabla_{\mu}h_{\nu\sigma}+\nabla^{\sigma}\nabla_{\nu}h_{\mu\sigma}-\nabla_{\mu}\nabla_{\nu}h,\label{box_h_mu_nu1}
\end{equation}
and changing the order of the covariant derivatives yield
\begin{align}
\Box h_{\mu\nu} & =-2R_{\mu\nu}^{L}+\nabla^{\sigma}\nabla_{\mu}h_{\nu\sigma}+\nabla^{\sigma}\nabla_{\nu}h_{\mu\sigma}-\nabla_{\mu}\nabla_{\nu}h\nonumber \\
 & \phantom{=}+6\Lambda h_{\mu\nu}-2\Lambda g_{\mu\nu}h,\label{box_h_mu_nu2}
\end{align}
where we have used 
\begin{align}
\nabla^{\sigma}\nabla_{\mu}h_{\nu\sigma} & =\left[\nabla^{\sigma},\nabla_{\mu}\right]h_{\nu\sigma}+\nabla_{\mu}\nabla^{\sigma}h_{\nu\sigma}\nonumber \\
 & =\bar{R}_{\phantom{\sigma}\mu\nu}^{\sigma\phantom{\mu\nu}\lambda}h_{\lambda\sigma}+\bar{R}_{\phantom{\sigma}\mu\sigma}^{\sigma\phantom{\mu\sigma}\lambda}h_{\nu\lambda}+\nabla_{\mu}\nabla^{\sigma}h_{\nu\sigma}\nonumber \\
 & =3\Lambda h_{\mu\nu}-g_{\mu\nu}\Lambda h+\nabla_{\mu}\nabla^{\sigma}h_{\nu\sigma},\label{del_mu_del_nu_com}
\end{align}
and in the second line we have put the following identities
\begin{equation}
\bar{R}_{\mu\rho\nu\sigma}=\Lambda\left(\bar{g}_{\mu\nu}\bar{g}_{\rho\sigma}-\bar{g}_{\mu\sigma}\bar{g}_{\nu\rho}\right),\quad\bar{R}_{\mu\nu}=2\Lambda\bar{g}_{\mu\nu},\quad\bar{R}=6\Lambda.\label{back_curv_idens}
\end{equation}
Then,
\begin{align}
\Box h_{\mu\nu} & =-2R_{\mu\nu}^{L}+6\Lambda h_{\mu\nu}-2\Lambda g_{\mu\nu}h+\nabla_{\mu}\nabla^{\sigma}h_{\nu\sigma}+\nabla_{\nu}\nabla^{\sigma}h_{\mu\sigma}-\nabla_{\mu}\nabla_{\nu}h\nonumber \\
 & =-2\mathcal{G}_{\mu\nu}^{L}-\bar{g}_{\mu\nu}R_{L}+2\Lambda h_{\mu\nu}-2g_{\mu\nu}\Lambda h\nonumber \\
 & \phantom{=}+\nabla_{\mu}\nabla^{\sigma}h_{\nu\sigma}+\nabla_{\nu}\nabla^{\sigma}h_{\mu\sigma}-\nabla_{\mu}\nabla_{\nu}h,\label{box_h_Mu_nu3}
\end{align}
where in the second line we have used (\ref{Lin_Einstein_ten}). Then
the first term in (\ref{I_beta_action}) becomes,
\begin{align}
\left(\Box h_{\mu\nu}\right)\mathcal{G}_{L}^{\mu\nu} & =-2\mathcal{G}_{\mu\nu}^{L}\mathcal{G}_{L}^{\mu\nu}-\bar{g}_{\mu\nu}\mathcal{G}_{L}^{\mu\nu}R_{L}+2\Lambda h_{\mu\nu}\mathcal{G}_{L}^{\mu\nu}-2\Lambda\bar{g}_{\mu\nu}\mathcal{G}_{L}^{\mu\nu}h\nonumber \\
 & \phantom{=}+\mathcal{G}_{L}^{\mu\nu}\nabla_{\nu}\nabla^{\sigma}h_{\mu\sigma}+\mathcal{G}_{L}^{\mu\nu}\nabla_{\mu}\nabla^{\sigma}h_{\nu\sigma}-\mathcal{G}_{L}^{\mu\nu}\nabla_{\mu}\nabla_{\nu}h.\label{box_h_mu_nu_G_mu_nu1}
\end{align}
The last three terms in (\ref{box_h_mu_nu_G_mu_nu1}) becomes boundary
term and drops out by the use of Bianchi identity, $\nabla_{\mu}\mathcal{G}_{L}^{\mu\nu}=0$.
Also, note that 
\begin{align}
\bar{g}_{\mu\nu}\mathcal{G}_{L}^{\mu\nu} & =\bar{g}_{\mu\nu}\left(R_{L}^{\mu\nu}-\frac{1}{2}\bar{g}^{\mu\nu}R_{L}-2\Lambda h^{\mu\nu}\right)\nonumber \\
 & =\bar{g}_{\mu\nu}R_{L}^{\mu\nu}-\frac{3}{2}R_{L}-2\Lambda h,\label{bar_g_G1}
\end{align}
and using 
\begin{align}
R_{L}\equiv\left(g^{\mu\nu}R_{\mu\nu}\right)_{L} & =\bar{g}^{\mu\nu}R_{\mu\nu}^{L}-h^{\mu\nu}\bar{R}_{\mu\nu}\nonumber \\
 & =\bar{g}^{\mu\nu}R_{\mu\nu}^{L}-h^{\mu\nu}\left(2\Lambda\bar{g}_{\mu\nu}\right)\nonumber \\
 & =\bar{g}^{\mu\nu}R_{\mu\nu}^{L}-2\Lambda h,\label{R_L}\\
\bar{g}^{\mu\nu}R_{\mu\nu}^{L} & =R_{L}+2\Lambda h,\label{bar_g_R_L}
\end{align}
(\ref{bar_g_G1}) becomes
\begin{align}
\bar{g}_{\mu\nu}\mathcal{G}_{L}^{\mu\nu} & =-\frac{1}{2}R_{L}.\label{bar_g_G2}
\end{align}
Putting (\ref{bar_g_G2}) in (\ref{box_h_mu_nu_G_mu_nu1}) yields
\begin{align}
\left(\Box h_{\mu\nu}\right)\mathcal{G}_{L}^{\mu\nu} & =-2\mathcal{G}_{\mu\nu}^{L}\mathcal{G}_{L}^{\mu\nu}+\frac{1}{2}R_{L}^{2}+2\Lambda h_{\mu\nu}\mathcal{G}_{L}^{\mu\nu}+\Lambda hR_{L}.\label{box_h_mu_nu_G_mu_nu2}
\end{align}
The total action for the $\beta$ part of (\ref{Linearized_action})
becomes
\begin{align}
I_{\beta} & =-\frac{\beta}{2}\int d^{3}x\,\sqrt{\bar{g}}\left[-2\mathcal{G}_{\mu\nu}^{L}\mathcal{G}_{L}^{\mu\nu}+\frac{1}{2}R_{L}^{2}+\frac{2}{\ell^{2}}h_{\mu\nu}\mathcal{G}_{L}^{\mu\nu}\right],\label{I_b_non_gauge}
\end{align}
where $\Lambda\equiv\frac{1}{\ell^{2}}$ is used. The gauge-invariant
form of the last term is known since it is the same as Einstein-Hilbert
part. The middle term is already gauge-invariant. To have a total
gauge-invariant action the first term must be written in that form.

\subsubsection{The Gauge-Invariant form of $\mathcal{G}_{\mu\nu}^{L}\mathcal{G}_{L}^{\mu\nu}$:}

Decomposing $\beta\int d^{3}x\,\sqrt{\bar{g}}\left(\mathcal{G}_{\mu\nu}^{L}\mathcal{G}_{L}^{\mu\nu}\right)$
gives us
\begin{equation}
\beta\int d^{3}x\,\sqrt{\bar{g}}\mathcal{G}_{\mu\nu}^{L}\mathcal{G}_{L}^{\mu\nu}=\beta\int d^{3}x\,\frac{t}{\ell}\left[\left(\mathcal{G}_{00}^{L}\right)^{2}-2\left(\mathcal{G}_{0i}^{L}\right)^{2}+\left(\mathcal{G}_{ij}^{L}\right)^{2}\right].\label{G_square1}
\end{equation}
From (\ref{G_00_gauge}, \ref{G_0j_gauge}, \ref{G_jk_gauge}), we
have
\begin{equation}
\left(\mathcal{G}_{00}^{L}\right)^{2}=\frac{t^{2}}{4\ell^{2}}\left(\nabla^{2}f\right)^{2},\quad\left(\mathcal{G}_{0j}^{L}\right)^{2}=-\frac{t^{2}}{4\ell^{2}}\left(p\nabla^{2}p+\sigma\nabla^{2}\sigma\right),\quad\left(\mathcal{G}_{jk}^{L}\right)^{2}=\frac{t^{2}}{4\ell^{2}}\left[2\dot{\sigma}^{2}+q^{2}+\dot{p}^{2}\right],\label{G_square2}
\end{equation}
and using (\ref{G_square2}) in the action (\ref{G_square1}) we get,
\begin{align}
\sqrt{-\bar{g}}\mathcal{G}_{\mu\nu}^{L}\mathcal{G}_{L}^{\mu\nu} & =\frac{t^{3}}{4\ell^{3}}\left(\nabla^{2}f\right)^{2}+\frac{t^{3}}{2\ell^{3}}\left(p\nabla^{2}p+\sigma\nabla^{2}\sigma\right)\nonumber \\
 & \phantom{=}+\frac{t^{3}}{4\ell^{3}}\left[2\dot{\sigma}^{2}+q^{2}+\dot{p}^{2}\right],\label{G_mu_nu_square}
\end{align}
where $\sqrt{-\bar{g}}=\frac{\ell^{3}}{t^{3}}$ and note that we lowered
the upper indices by using the background metric.

\subsubsection{The Gauge-Invariant Form of the $I_{\beta}$ Action:}

Adding (\ref{G_mu_nu_square}), square of (\ref{gauge_inv_R_scalar})
and (\ref{I_e_gauge}) the action (\ref{I_b_non_gauge}) becomes
\begin{align}
I_{\beta} & =\frac{\beta}{2}\int d^{3}x\,\frac{t^{3}}{\ell^{3}}\left[\frac{1}{2}\left(\nabla^{2}f\right)^{2}+p\nabla^{2}p+\sigma\nabla^{2}\sigma+\dot{\sigma}^{2}+\frac{1}{2}q^{2}+\frac{1}{2}\dot{p}^{2}-\frac{1}{2}\left(q-\nabla^{2}f+\dot{p}\right)^{2}\right]\nonumber \\
 & \phantom{=}+\frac{\beta}{2}\int d^{3}x\,\frac{2}{\ell^{2}}\left(\frac{t}{\ell}\left(\sigma^{2}+p^{2}+f\,\nabla^{2}f\right)+\frac{\ell^{2}}{t^{2}}f\, R_{L}\right)\nonumber \\
 & =\frac{\beta}{2}\int d^{3}x\,\frac{t^{3}}{\ell^{3}}\left[p\nabla^{2}p+\sigma\nabla^{2}\sigma+\dot{\sigma}^{2}-q\dot{p}+q\nabla^{2}f+\dot{p}\nabla^{2}f\right]\nonumber \\
 & \phantom{=}+\frac{\beta}{2}\int d^{3}x\,\frac{2}{\ell^{2}}\left(\frac{t}{\ell}\left(\sigma^{2}+p^{2}+f\,\nabla^{2}f\right)+\frac{\ell^{2}}{t^{2}}f\, R_{L}\right).\label{I_b_gauge1}
\end{align}
In this action there is five gauge-invariant functions. By using the
gauge-invariant form of $R_{L}$ the number of functions can be reduced
to four. Here, it is preferred to eliminate the $q$ variable by using
$R_{L}=\frac{t^{3}}{\ell^{3}}\left(q-\nabla^{2}f+\dot{p}\right)\Rightarrow q=\frac{\ell^{3}}{t^{3}}R_{L}+\nabla^{2}f-\dot{p}$,
\begin{align}
I_{\beta} & =\frac{\beta}{2}\int d^{3}x\,\frac{t^{3}}{\ell^{3}}\left[p\nabla^{2}p+\sigma\nabla^{2}\sigma+\dot{\sigma}^{2}-\left(\frac{\ell^{3}}{t^{3}}R_{L}+\nabla^{2}f-\dot{p}\right)\dot{p}\right.\nonumber \\
 & \phantom{=\frac{\beta}{2}\int d^{3}x\,\frac{t^{3}}{\ell^{3}}}\left.+\left(\frac{\ell^{3}}{t^{3}}R_{L}+\nabla^{2}f-\dot{p}\right)\nabla^{2}f+\dot{p}\nabla^{2}f\right]\nonumber \\
 & \phantom{=}+\frac{\beta}{2}\int d^{3}x\,\frac{2}{\ell^{2}}\left(\frac{t}{\ell}\left(\sigma^{2}+p^{2}+f\,\nabla^{2}f\right)+\frac{\ell^{2}}{t^{2}}f\, R_{L}\right).\nonumber \\
 & =\frac{\beta}{2}\int d^{3}x\,\frac{t^{3}}{\ell^{3}}\left[p\nabla^{2}p+\sigma\nabla^{2}\sigma+\dot{\sigma}^{2}-\frac{\ell^{3}}{t^{3}}\dot{p}R_{L}-\dot{p}\nabla^{2}f+\dot{p}^{2}+\frac{\ell^{3}}{t^{3}}R_{L}\nabla^{2}f+\left(\nabla^{2}f\right)^{2}\right]\nonumber \\
 & \phantom{=}+\frac{\beta}{2}\int d^{3}x\,\frac{2}{\ell^{2}}\left(\frac{t}{\ell}\left(\sigma^{2}+p^{2}+f\,\nabla^{2}f\right)+\frac{\ell^{2}}{t^{2}}f\, R_{L}\right).\label{I_b_gauge2}
\end{align}
This action is the final result for the $\beta$ part of the total
action. Now everything is ready to write (\ref{Linearized_action})
in terms of the gauge-invariant functions.

Before going on, the flat spacetime limit of (\ref{I_b_gauge2}) can
be written as 
\begin{align}
I_{\beta} & =\frac{\beta}{2}\int d^{3}x\,\left[\dot{\phi}\nabla^{2}\dot{\phi}+\sigma\nabla^{2}\sigma+\dot{\sigma}^{2}-\ddot{\phi}R_{L}-\ddot{\phi}\nabla^{2}\phi+\ddot{\phi}^{2}+R_{L}\nabla^{2}\phi+\left(\nabla^{2}\phi\right)^{2}\right]\nonumber \\
 & =\frac{\beta}{2}\int d^{3}x\,\left[-\ddot{\phi}\nabla^{2}\phi+\sigma\nabla^{2}\sigma-\sigma\ddot{\sigma}-\ddot{\phi}\left(q-\square\phi\right)-\ddot{\phi}\nabla^{2}\phi+\ddot{\phi}^{2}+\left(q-\square\phi\right)\nabla^{2}\phi+\left(\nabla^{2}\phi\right)^{2}\right]\nonumber \\
 & =\frac{\beta}{2}\int d^{3}x\,\left[-\ddot{\phi}\left(\nabla^{2}-\partial_{0}^{2}\right)\phi+\sigma\left(\nabla^{2}-\partial_{0}^{2}\right)\sigma+q\left(-\partial_{0}^{2}+\nabla^{2}\right)\phi\right.\nonumber \\
 & \phantom{=\frac{\beta}{2}\int d^{3}x\,}\left.-\square\phi\left(-\partial_{0}^{2}+\nabla^{2}\right)\phi+\nabla^{2}\phi\left(-\partial_{0}^{2}+\nabla^{2}\right)\phi\right]\nonumber \\
 & =\frac{\beta}{2}\int d^{3}x\,\left[\square\phi\left(-\partial_{0}^{2}+\nabla^{2}\right)\phi+\sigma\square\sigma+q\square\phi-\left(\square\phi\right)^{2}\right],\label{I_b_gauge_flat1}
\end{align}
where in the second line we have done integration by parts and used
(\ref{gauge_inv_R_flat2}), in the third line some suitable combinations
have done and in the last line the definition of the D'Alembertian
operator ,$\square=-\partial_{0}^{2}+\nabla^{2}$, have been used.
Finally, for the flat spacetime limit (\ref{I_b_gauge2}) becomes
\begin{equation}
I_{\beta}=\frac{\beta}{2}\int d^{3}x\,\left[\sigma\square\sigma+q\square\phi\right].\label{I_b_gauge_flat2}
\end{equation}

\subsection{The Total Gauge-Invariant Action }

Summing (\ref{I_e_gauge}), (\ref{I_2a_+_b_gauge}) and (\ref{I_b_gauge2})
the total action can be written as 
\begin{align}
I & =\frac{1}{2}\int d^{3}x\,\left(a+\frac{2\beta}{\ell^{2}}\right)\left[\frac{t}{\ell}\left(\sigma^{2}+p^{2}+f\,\nabla^{2}f\right)+\frac{\ell^{2}}{t^{2}}f\, R_{L}\right]+\left(2\alpha+\beta\right)\frac{\ell^{3}}{t^{3}}R_{L}^{2}\nonumber \\
 & \phantom{=}+\frac{1}{2}\int d^{3}x\,\frac{t^{3}}{\ell^{3}}\beta\left[\dot{\sigma}^{2}+\sigma\nabla^{2}\sigma+\dot{p}^{2}+p\nabla^{2}p+\left(\nabla^{2}f\right)^{2}-\frac{\ell^{3}}{t^{3}}\dot{p}R_{L}+\frac{\ell^{3}}{t^{3}}R_{L}\nabla^{2}f-\dot{p}\nabla^{2}f\right].\label{total_action_gauge1}
\end{align}
This result is the gauge-invariant action for the general quadratic
curvature theory in three dimensions. This action can also be simplified
by defining $\varphi\equiv\nabla^{2}f\,$ and using (\ref{gauge_inv_R_scalar}).
For the first term 
\begin{align}
\frac{t}{\ell}\left(\sigma^{2}+p^{2}+f\,\nabla^{2}f\right)+\frac{\ell^{2}}{t^{2}}f\, R_{L} & =\frac{t}{\ell}\left[\sigma^{2}+p^{2}+f\,\nabla^{2}f+f\left(q-\nabla^{2}f+\dot{p}\right)\right]\nonumber \\
 & =\frac{t}{\ell}\left[\sigma^{2}+p^{2}+f\left(q+\dot{p}\right)\right],\label{total_first_term1}
\end{align}
using the Bianchi identity (\ref{Bianchi_identity_gauge1})
\begin{align}
\frac{t}{\ell}\left(\sigma^{2}+p^{2}+f\,\nabla^{2}f\right)+\frac{\ell^{2}}{t^{2}}f\, R_{L} & =\frac{t}{\ell}\left[\sigma^{2}+p^{2}+t\, f\,\nabla^{2}\dot{f}-t\, f\,\nabla^{2}p+f\,\nabla^{2}f\right],\label{total_first_term2}
\end{align}
and observe that $\frac{t^{2}}{\ell}\, f\,\nabla^{2}\dot{f}=\partial_{0}\left(\frac{t^{2}}{\ell}\, f\,\nabla^{2}f\right)-\frac{2t}{\ell}\, f\,\nabla^{2}f-\frac{t^{2}}{\ell}\,\dot{f}\,\nabla^{2}f=-\frac{2t}{\ell}\, f\,\nabla^{2}f-\frac{t^{2}}{\ell}\, f\,\nabla^{2}\dot{f}\Rightarrow\frac{t^{2}}{\ell}\, f\,\nabla^{2}\dot{f}=-\frac{t}{\ell}\, f\,\nabla^{2}f$
after dropping the boundary term. Then (\ref{total_first_term2})
becomes 
\begin{align}
\frac{t}{\ell}\left(\sigma^{2}+p^{2}+f\,\nabla^{2}f\right)+\frac{\ell^{2}}{t^{2}}f\, R_{L} & =\frac{t}{\ell}\left[\sigma^{2}+p^{2}-t\,\varphi\, p\right].\label{totla_first_term3}
\end{align}
The second term of (\ref{total_action_gauge1}) can be written as
\begin{align}
\frac{\ell^{3}}{t^{3}}R_{L}^{2} & =\frac{t^{5}}{\ell^{3}}\left(\dot{\varphi}-\nabla^{2}p\right)^{2},\label{total_second_term1}
\end{align}
where it has been used $R_{L}=\frac{t^{4}}{\ell^{3}}\nabla^{2}\left(\dot{f}-p\right)$
which comes from using Bianchi identity (\ref{Bianchi_identity_gauge1})
in the gauge-invariant form of Ricci scalar. The last term of (\ref{total_action_gauge1})
can be written without taking care of the $\sigma$ field by using
(\ref{Ricci_gauge_Bianchi}),
\begin{align}
\frac{t^{3}}{\ell^{3}}\left[\dot{p}^{2}+p\nabla^{2}p+\left(\nabla^{2}f\right)^{2}-\frac{\ell^{3}}{t^{3}}\dot{p}R_{L}+\frac{\ell^{3}}{t^{3}}R_{L}\nabla^{2}f-\dot{p}\nabla^{2}f\right] & =\frac{t^{3}}{\ell^{3}}\left[p\nabla^{2}p-\dot{p}q+\left(\dot{p}+q\right)\nabla^{2}f\right]\nonumber \\
=\frac{t^{3}}{\ell^{3}}\left[p\nabla^{2}p-\dot{p}q+t\nabla^{2}\left(\dot{f}-p\right)\nabla^{2}f+\left(\nabla^{2}f\right)^{2}\right]\nonumber \\
=\frac{t^{3}}{\ell^{3}}\left[p\nabla^{2}p-\dot{p}q-t\nabla^{2}p\nabla^{2}f-\left(\nabla^{2}f\right)^{2}\right],\label{total_third_term1}
\end{align}
where in the last line the integration by parts is used for the term
$\frac{t^{4}}{\ell^{3}}\nabla^{2}\dot{f}\nabla^{2}f=-\frac{2t^{3}}{\ell^{3}}\left(\nabla^{2}f\right)^{2}$.
Again using the Bianchi identity (\ref{Bianchi_identity_gauge1})
$q\dot{p}=t\nabla^{2}\left(\dot{f}-p\right)\dot{p}+\dot{p}\left(\nabla^{2}f\right)-\dot{p}^{2}$.
Then, (\ref{total_third_term1}) can be written as
\begin{align}
\frac{t^{3}}{\ell^{3}}\left[p\nabla^{2}p-t\dot{p}\nabla^{2}\dot{f}-t\dot{p}\nabla^{2}p-\dot{p}\left(\nabla^{2}f\right)+\dot{p}^{2}-t\nabla^{2}p\nabla^{2}f-\left(\nabla^{2}f\right)^{2}\right] & =\nonumber \\
\frac{t^{3}}{\ell^{3}}\left[p\nabla^{2}p-t\dot{p}\dot{\varphi}-t\dot{p}\nabla^{2}p-\dot{p}\varphi+\dot{p}^{2}-t\varphi\nabla^{2}p-\varphi^{2}\right] & =\nonumber \\
\frac{t^{3}}{\ell^{3}}\left[-p\nabla^{2}p-t\dot{p}\dot{\varphi}-\dot{p}\varphi+\dot{p}^{2}-t\varphi\nabla^{2}p-\varphi^{2}\right] & ,\label{total_third_term2}
\end{align}
where in the last line the integration by parts is used for $\frac{t^{4}}{\ell^{3}}\dot{p}\nabla^{2}p=-\frac{2t^{3}}{\ell^{3}}p\nabla^{2}p$.
The final form of (\ref{total_action_gauge1}) becomes with (\ref{total_first_term2}),
(\ref{total_second_term1}) and (\ref{total_third_term2})
\begin{align}
I & =\frac{1}{2}\int d^{3}x\,\left(a+\frac{2\beta}{\ell^{2}}\right)\left[\frac{t}{\ell}\left(p^{2}-tp\varphi\right)\right]+\frac{1}{2}\int d^{3}x\,\left(2\alpha+\beta\right)\frac{t^{5}}{\ell^{3}}\left(\dot{\varphi}-\nabla^{2}p\right)^{2}\nonumber \\
 & \phantom{=}+\frac{1}{2}\int d^{3}x\,\frac{t^{3}}{\ell^{3}}\beta\left[-p\nabla^{2}p-t\dot{p}\dot{\varphi}-\dot{p}\varphi+\dot{p}^{2}-t\varphi\nabla^{2}p-\varphi^{2}\right]+I_{\sigma},\label{total_action_gauge}
\end{align}
where we have defined 
\begin{equation}
I_{\sigma}=\frac{1}{2}\int d^{3}x\,\left[\beta\frac{t^{3}}{\ell^{3}}\left(\dot{\sigma}^{2}+\sigma\nabla^{2}\sigma\right)+\left(a+\frac{2\beta}{\ell^{2}}\right)\frac{t}{\ell}\sigma^{2}\right].\label{I_sigma_action}
\end{equation}
In the flat spacetime limit, (\ref{total_action_gauge}) becomes by
summing (\ref{I_e_gauge_flat}), (\ref{I_2a+b_gauge_flat}) and (\ref{I_b_gauge_flat2})
\begin{align}
I & =\frac{1}{2}\int d^{3}x\left[\frac{1}{\kappa}\phi q+\left(2\alpha+\beta\right)\left(q-\square\phi\right)^{2}+\beta q\square\phi\right]\nonumber \\
 & \phantom{=}+\frac{\beta}{2}\int d^{3}x\,\left[\sigma\square\sigma+\frac{1}{\kappa\beta}\sigma^{2}\right].\label{total_action_gauge_flat}
\end{align}

\subsection{Gauge Invariance of The Field Equations}

The field equations of the action can also be written in terms gauge-invariant
functions. In the field equation only the $\square\mathcal{G}_{\mu\nu}^{L}$
is unknown in terms of the gauge-invariant functions. Therefore in
this part only this term is written in terms of gauge invariant functions.
First the covariant derivatives are extracted

\begin{align}
\Box\mathcal{G}_{\mu\nu}^{L} & =\bar{g}^{\sigma\rho}\nabla_{\rho}\nabla_{\sigma}\mathcal{G}_{\mu\nu}^{L},\nonumber \\
 & =\bar{g}^{\sigma\rho}\nabla_{\rho}\left(\partial_{\sigma}\mathcal{G}_{\mu\nu}^{L}-\Gamma_{\sigma\mu}^{\lambda}\mathcal{G}_{\lambda\nu}^{L}-\Gamma_{\sigma\nu}^{\lambda}\mathcal{G}_{\mu\lambda}^{L}\right),\nonumber \\
 & =\frac{t^{2}}{\ell^{2}}\eta^{\sigma\rho}\left\{ \partial_{\rho}\left(\partial_{\sigma}\mathcal{G}_{\mu\nu}^{L}-\Gamma_{\sigma\mu}^{\lambda}\mathcal{G}_{\lambda\nu}^{L}-\Gamma_{\sigma\nu}^{\lambda}\mathcal{G}_{\mu\lambda}^{L}\right)-\Gamma_{\rho\sigma}^{\alpha}\left(\partial_{\alpha}\mathcal{G}_{\mu\nu}^{L}-\Gamma_{\alpha\mu}^{\lambda}\mathcal{G}_{\lambda\nu}^{L}-\Gamma_{\alpha\nu}^{\lambda}\mathcal{G}_{\mu\lambda}^{L}\right)\right.\nonumber \\
 & \phantom{=}\left.-\Gamma_{\rho\mu}^{\alpha}\left(\partial_{\sigma}\mathcal{G}_{\alpha\nu}^{L}-\Gamma_{\sigma\alpha}^{\lambda}\mathcal{G}_{\lambda\nu}^{L}-\Gamma_{\sigma\nu}^{\lambda}\mathcal{G}_{\alpha\lambda}^{L}\right)-\Gamma_{\rho\nu}^{\alpha}\left(\partial_{\sigma}\mathcal{G}_{\mu\alpha}^{L}-\Gamma_{\sigma\mu}^{\lambda}\mathcal{G}_{\lambda\alpha}^{L}-\Gamma_{\sigma\alpha}^{\lambda}\mathcal{G}_{\mu\lambda}^{L}\right)\right\} ,\nonumber \\
 & =\frac{t^{2}}{\ell^{2}}\eta^{\sigma\rho}\left\{ \left(\partial_{\rho}\partial_{\sigma}\mathcal{G}_{\mu\nu}^{L}-\mathcal{G}_{\lambda\nu}^{L}\partial_{\rho}\Gamma_{\sigma\mu}^{\lambda}-\Gamma_{\sigma\mu}^{\lambda}\partial_{\rho}\mathcal{G}_{\lambda\nu}^{L}-\mathcal{G}_{\mu\lambda}^{L}\partial_{\rho}\Gamma_{\sigma\nu}^{\lambda}-\Gamma_{\sigma\nu}^{\lambda}\partial_{\rho}\mathcal{G}_{\mu\lambda}^{L}\right)\right.\nonumber \\
 & \phantom{=}-\left(\Gamma_{\rho\sigma}^{\alpha}\partial_{\alpha}\mathcal{G}_{\mu\nu}^{L}-\Gamma_{\rho\sigma}^{\alpha}\Gamma_{\alpha\mu}^{\lambda}\mathcal{G}_{\lambda\nu}^{L}-\Gamma_{\rho\sigma}^{\alpha}\Gamma_{\alpha\nu}^{\lambda}\mathcal{G}_{\mu\lambda}^{L}\right)\nonumber \\
 & \phantom{=}-\left(\Gamma_{\rho\mu}^{\alpha}\partial_{\sigma}\mathcal{G}_{\alpha\nu}^{L}-\Gamma_{\rho\mu}^{\alpha}\Gamma_{\sigma\alpha}^{\lambda}\mathcal{G}_{\lambda\nu}^{L}-\Gamma_{\rho\mu}^{\alpha}\Gamma_{\sigma\nu}^{\lambda}\mathcal{G}_{\alpha\lambda}^{L}\right)\nonumber \\
 & \phantom{=}\left.-\left(\Gamma_{\rho\nu}^{\alpha}\partial_{\sigma}\mathcal{G}_{\mu\alpha}^{L}-\Gamma_{\rho\nu}^{\alpha}\Gamma_{\sigma\mu}^{\lambda}\mathcal{G}_{\lambda\alpha}^{L}-\Gamma_{\rho\nu}^{\alpha}\Gamma_{\sigma\alpha}^{\lambda}\mathcal{G}_{\mu\lambda}^{L}\right)\right\} ,\label{box_G_mu_nu1}
\end{align}
where we have used $\bar{g}_{\sigma\rho}=\frac{\ell^{2}}{t^{2}}\eta_{\sigma\rho}$
in the third line. Doing the summations in the repeated indices (\ref{box_G_mu_nu1})
becomes
\begin{align}
\Box\mathcal{G}_{\mu\nu}^{L} & =\frac{t^{2}}{\ell^{2}}\eta^{\sigma\rho}\left\{ \partial_{\rho}\partial_{\sigma}\mathcal{G}_{\mu\nu}^{L}-\left(\mathcal{G}_{\lambda\nu}^{L}\partial_{\rho}\Gamma_{\sigma\mu}^{\lambda}+\mathcal{G}_{\mu\lambda}^{L}\partial_{\rho}\Gamma_{\sigma\nu}^{\lambda}\right)\right.\nonumber \\
 & \phantom{=}\left.-\left(\Gamma_{\sigma\mu}^{\lambda}\partial_{\rho}\mathcal{G}_{\lambda\nu}^{L}+\Gamma_{\sigma\nu}^{\lambda}\partial_{\rho}\mathcal{G}_{\mu\lambda}^{L}+\Gamma_{\rho\sigma}^{\alpha}\partial_{\alpha}\mathcal{G}_{\mu\nu}^{L}+\Gamma_{\rho\mu}^{\alpha}\partial_{\sigma}\mathcal{G}_{\alpha\nu}^{L}+\Gamma_{\rho\nu}^{\alpha}\partial_{\sigma}\mathcal{G}_{\mu\alpha}^{L}\right)\right.\nonumber \\
 & \phantom{=}\left.+\Gamma_{\rho\sigma}^{\alpha}\left(\Gamma_{\alpha\mu}^{\lambda}\mathcal{G}_{\lambda\nu}^{L}+\Gamma_{\alpha\nu}^{\lambda}\mathcal{G}_{\mu\lambda}^{L}\right)+\left(\Gamma_{\rho\mu}^{\alpha}\Gamma_{\sigma\alpha}^{\lambda}\mathcal{G}_{\lambda\nu}^{L}+\Gamma_{\rho\mu}^{\alpha}\Gamma_{\sigma\nu}^{\lambda}\mathcal{G}_{\alpha\lambda}^{L}\right)\right.\nonumber \\
 & \phantom{=}\left.+\left(\Gamma_{\rho\nu}^{\alpha}\Gamma_{\sigma\mu}^{\lambda}\mathcal{G}_{\lambda\alpha}^{L}+\Gamma_{\rho\nu}^{\alpha}\Gamma_{\sigma\alpha}^{\lambda}\mathcal{G}_{\mu\lambda}^{L}\right)\right\} ,\nonumber \\
 & =\frac{t^{2}}{\ell^{2}}\left\{ \left(-\partial_{0}^{2}+\partial_{i}^{2}\right)\mathcal{G}_{\mu\nu}^{L}-\left(-\mathcal{G}_{\lambda\nu}^{L}\partial_{0}\Gamma_{0\mu}^{\lambda}+\mathcal{G}_{\lambda\nu}^{L}\partial_{i}\Gamma_{i\mu}^{\lambda}-\mathcal{G}_{\mu\lambda}^{L}\partial_{0}\Gamma_{0\nu}^{\lambda}+\mathcal{G}_{\mu\lambda}^{L}\partial_{i}\Gamma_{i\nu}^{\lambda}\right)\right.\nonumber \\
 & \phantom{=\frac{t^{2}}{\ell^{2}}}\left.-\left(-\Gamma_{0\mu}^{\lambda}\partial_{0}\mathcal{G}_{\lambda\nu}^{L}+\Gamma_{i\mu}^{\lambda}\partial_{i}\mathcal{G}_{\lambda\nu}^{L}-\Gamma_{0\nu}^{\lambda}\partial_{0}\mathcal{G}_{\mu\lambda}^{L}+\Gamma_{i\nu}^{\lambda}\partial_{i}\mathcal{G}_{\mu\lambda}^{L}-\Gamma_{00}^{\alpha}\partial_{\alpha}\mathcal{G}_{\mu\nu}^{L}+\Gamma_{ii}^{\alpha}\partial_{\alpha}\mathcal{G}_{\mu\nu}^{L}\right)\right.\nonumber \\
 & \phantom{=\frac{t^{2}}{\ell^{2}}}-\left(-\Gamma_{0\mu}^{\alpha}\partial_{0}\mathcal{G}_{\alpha\nu}^{L}+\Gamma_{i\mu}^{\alpha}\partial_{i}\mathcal{G}_{\alpha\nu}^{L}-\Gamma_{0\nu}^{\alpha}\partial_{0}\mathcal{G}_{\mu\alpha}^{L}+\Gamma_{i\nu}^{\alpha}\partial_{i}\mathcal{G}_{\mu\alpha}^{L}\right)\nonumber \\
 & \phantom{=\frac{t^{2}}{\ell^{2}}}+\left(-\Gamma_{00}^{\alpha}+\Gamma_{ii}^{\alpha}\right)\left(\Gamma_{\alpha\mu}^{\lambda}\mathcal{G}_{\lambda\nu}^{L}+\Gamma_{\alpha\nu}^{\lambda}\mathcal{G}_{\mu\lambda}^{L}\right)\label{box_G_mu_nu2}\\
 & \phantom{=\frac{t^{2}}{\ell^{2}}}\left.+\left(-\Gamma_{0\mu}^{\alpha}\Gamma_{0\alpha}^{\lambda}\mathcal{G}_{\lambda\nu}^{L}+\Gamma_{i\mu}^{\alpha}\Gamma_{i\alpha}^{\lambda}\mathcal{G}_{\lambda\nu}^{L}-\Gamma_{0\mu}^{\alpha}\Gamma_{0\nu}^{\lambda}\mathcal{G}_{\alpha\lambda}^{L}+\Gamma_{i\mu}^{\alpha}\Gamma_{i\nu}^{\lambda}\mathcal{G}_{\alpha\lambda}^{L}\right)\right.\nonumber \\
 & \phantom{=\frac{t^{2}}{a^{2}}}\left.+\left(-\Gamma_{0\nu}^{\alpha}\Gamma_{0\mu}^{\lambda}\mathcal{G}_{\lambda\alpha}^{L}+\Gamma_{i\nu}^{\alpha}\Gamma_{i\mu}^{\lambda}\mathcal{G}_{\lambda\alpha}^{L}-\Gamma_{0\nu}^{\alpha}\Gamma_{0\alpha}^{\lambda}\mathcal{G}_{\mu\lambda}^{L}+\Gamma_{i\nu}^{\alpha}\Gamma_{i\alpha}^{\lambda}\mathcal{G}_{\mu\lambda}^{L}\right)\right\} ,\nonumber 
\end{align}
and using Christoffel connections (\ref{comp_conec}), (\ref{box_G_mu_nu2})
yields 
\begin{align}
\Box\mathcal{G}_{\mu\nu}^{L} & =\frac{t^{2}}{\ell^{2}}\left\{ \left(-\partial_{0}^{2}+\partial_{i}^{2}\right)\mathcal{G}_{\mu\nu}^{L}-\left[2\mathcal{G}_{\mu\nu}^{L}\partial_{0}\left(\frac{1}{t}\right)+\mathcal{G}_{\lambda\nu}^{L}\partial_{i}\Gamma_{i\mu}^{\lambda}+\mathcal{G}_{\mu\lambda}^{L}\partial_{i}\Gamma_{i\nu}^{\lambda}\right]\right.\nonumber \\
 & \phantom{=\frac{t^{2}}{\ell^{2}}}\left.-\left(\frac{3}{t}\partial_{0}\mathcal{G}_{\mu\nu}^{L}+2\Gamma_{i\mu}^{\lambda}\partial_{i}\mathcal{G}_{\lambda\nu}^{L}+2\Gamma_{i\nu}^{\lambda}\partial_{i}\mathcal{G}_{\mu\lambda}^{L}-\frac{1}{t}\eta_{ii}\partial_{0}\mathcal{G}_{\mu\nu}^{L}\right)-\frac{2}{t}\partial_{0}\mathcal{G}_{\mu\nu}^{L}\right.\nonumber \\
 & \phantom{=\frac{t^{2}}{\ell^{2}}}\left.+\frac{2}{t^{2}}\mathcal{G}_{\mu\nu}^{L}+\left(-\frac{4}{t^{2}}\mathcal{G}_{\mu\nu}^{L}+\Gamma_{i\mu}^{\alpha}\Gamma_{i\alpha}^{\lambda}\mathcal{G}_{\lambda\nu}^{L}+\Gamma_{i\nu}^{\alpha}\Gamma_{i\alpha}^{\lambda}\mathcal{G}_{\mu\lambda}^{L}+2\Gamma_{i\mu}^{\alpha}\Gamma_{i\nu}^{\lambda}\mathcal{G}_{\alpha\lambda}^{L}\right)\right\} .\label{box_G_mu_nu3}
\end{align}
Doing the essential cancellations (\ref{box_G_mu_nu3}) becomes
\begin{align}
\Box\mathcal{G}_{\mu\nu}^{L} & =\frac{t^{2}}{\ell^{2}}\left\{ \left(-\partial_{0}^{2}-\frac{3}{t}\partial_{0}+\partial_{i}^{2}\right)\mathcal{G}_{\mu\nu}^{L}-\left(\mathcal{G}_{\lambda\nu}^{L}\partial_{i}\Gamma_{i\mu}^{\lambda}+\mathcal{G}_{\mu\lambda}^{L}\partial_{i}\Gamma_{i\nu}^{\lambda}\right)\right.\nonumber \\
 & \phantom{=\frac{t^{2}}{\ell^{2}}}\left.-\left(2\Gamma_{i\mu}^{\lambda}\partial_{i}\mathcal{G}_{\lambda\nu}^{L}+2\Gamma_{i\nu}^{\lambda}\partial_{i}\mathcal{G}_{\mu\lambda}^{L}\right)\right.\nonumber \\
 & \phantom{=\frac{t^{2}}{\ell^{2}}}\left.+\Gamma_{i\mu}^{\alpha}\Gamma_{i\alpha}^{\lambda}\mathcal{G}_{\lambda\nu}^{L}+\Gamma_{i\nu}^{\alpha}\Gamma_{i\alpha}^{\lambda}\mathcal{G}_{\mu\lambda}^{L}+2\Gamma_{i\mu}^{\alpha}\Gamma_{i\nu}^{\lambda}\mathcal{G}_{\alpha\lambda}^{L}\right\} .\label{box_G_mu_nu4}
\end{align}
Again doing summations in the repeated indices and dropping the derivatives
of the Christoffel connections, $\partial_{i}\Gamma_{\nu\sigma}^{\mu}=0$,
(\ref{box_G_mu_nu4}) yields

\begin{align}
\Box\mathcal{G}_{\mu\nu}^{L} & =\frac{t^{2}}{\ell^{2}}\left\{ \left(-\partial_{0}^{2}-\frac{3}{t}\partial_{0}+\partial_{i}^{2}\right)\mathcal{G}_{\mu\nu}^{L}-2\left(\Gamma_{i\mu}^{0}\partial_{i}\mathcal{G}_{0\nu}^{L}+\Gamma_{i\mu}^{j}\partial_{i}\mathcal{G}_{j\nu}^{L}+\Gamma_{i\nu}^{0}\partial_{i}\mathcal{G}_{\mu0}^{L}+\Gamma_{i\nu}^{j}\partial_{i}\mathcal{G}_{\mu j}^{L}\right)\right.\nonumber \\
 & \phantom{=\frac{t^{2}}{\ell^{2}}}\left.-\frac{1}{t}\Gamma_{i\mu}^{0}\mathcal{G}_{i\nu}^{L}-\frac{1}{t}\Gamma_{i\mu}^{i}\mathcal{G}_{0\nu}^{L}-\frac{1}{t}\Gamma_{i\nu}^{0}\mathcal{G}_{\mu i}^{L}-\frac{1}{t}\Gamma_{i\nu}^{i}\mathcal{G}_{\mu0}^{L}\right.\nonumber \\
 & \phantom{=\frac{t^{2}}{\ell^{2}}}\left.+2\Gamma_{i\mu}^{0}\Gamma_{i\nu}^{0}\mathcal{G}_{00}^{L}+2\Gamma_{i\mu}^{0}\Gamma_{i\nu}^{j}\mathcal{G}_{0j}^{L}+2\Gamma_{i\mu}^{j}\Gamma_{i\nu}^{0}\mathcal{G}_{j0}^{L}+2\Gamma_{i\mu}^{j}\Gamma_{i\nu}^{k}\mathcal{G}_{jk}^{L}\right\} .\label{box_G_mu_nu_dec}
\end{align}
By using (\ref{box_G_mu_nu_dec}) we can calculate the components
term by term.

\subsubsection{The $\square\mathcal{G}_{00}^{L}$ Term}

Setting $\mu=\nu=0$ in (\ref{box_G_mu_nu_dec}) and using (\ref{comp_conec})
we have
\begin{align}
\Box\mathcal{G}_{00}^{L} & =\frac{t^{2}}{\ell^{2}}\left\{ \left(-\partial_{0}^{2}-\frac{3}{t}\partial_{0}+\partial_{i}^{2}\right)\mathcal{G}_{00}^{L}-2\left(\Gamma_{i0}^{0}\partial_{i}\mathcal{G}_{00}^{L}+\Gamma_{i0}^{j}\partial_{i}\mathcal{G}_{j0}^{L}+\Gamma_{i0}^{0}\partial_{i}\mathcal{G}_{00}^{L}+\Gamma_{i0}^{j}\partial_{i}\mathcal{G}_{0j}^{L}\right)\right.\nonumber \\
 & \phantom{=\frac{t^{2}}{\ell^{2}}}\left.-\frac{1}{t}\Gamma_{i0}^{0}\mathcal{G}_{i0}^{L}-\frac{1}{t}\Gamma_{i0}^{i}\mathcal{G}_{00}^{L}-\frac{1}{t}\Gamma_{i0}^{0}\mathcal{G}_{0i}^{L}-\frac{1}{t}\Gamma_{i0}^{i}\mathcal{G}_{00}^{L}\right.\label{box_G_00_05}\\
 & \phantom{=\frac{t^{2}}{\ell^{2}}}\left.+2\Gamma_{i0}^{0}\Gamma_{i0}^{0}\mathcal{G}_{00}^{L}+2\Gamma_{i0}^{0}\Gamma_{i0}^{j}\mathcal{G}_{0j}^{L}+2\Gamma_{i0}^{j}\Gamma_{i0}^{0}\mathcal{G}_{j0}^{L}+2\Gamma_{i0}^{j}\Gamma_{i0}^{k}\mathcal{G}_{jk}^{L}\right\} ,\nonumber 
\end{align}
and after simple manipulations (\ref{box_G_00_05}) becomes
\begin{align}
\Box\mathcal{G}_{00}^{L} & =\frac{t^{2}}{\ell^{2}}\left\{ \left(-\partial_{0}^{2}-\frac{3}{t}\partial_{0}+\partial_{i}^{2}\right)\mathcal{G}_{00}^{L}-2\left(\Gamma_{i0}^{j}\partial_{i}\mathcal{G}_{j0}^{L}+\Gamma_{i0}^{j}\partial_{i}\mathcal{G}_{0j}^{L}\right)\right.\nonumber \\
 & \phantom{=\frac{t^{2}}{\ell^{2}}}\left.-\frac{2}{t}\Gamma_{i0}^{i}\mathcal{G}_{00}^{L}+2\Gamma_{i0}^{j}\Gamma_{i0}^{k}\mathcal{G}_{jk}^{L}\right\} ,\nonumber \\
 & =\frac{t^{2}}{\ell^{2}}\left\{ \left(-\partial_{0}^{2}-\frac{3}{t}\partial_{0}+\partial_{i}^{2}\right)\mathcal{G}_{00}^{L}-2\left(-\frac{2}{t}\delta_{i}^{j}\partial_{i}\mathcal{G}_{j0}^{L}\right)\right.\nonumber \\
 & \phantom{=\frac{t^{2}}{\ell^{2}}}\left.+\frac{2}{t^{2}}\delta_{i}^{i}\mathcal{G}_{00}^{L}+\frac{2}{t^{2}}\delta_{i}^{j}\delta_{i}^{k}\mathcal{G}_{jk}^{L}\right\} ,\nonumber \\
 & =\frac{t^{2}}{\ell^{2}}\left\{ \left(-\partial_{0}^{2}-\frac{3}{t}\partial_{0}+\partial_{i}^{2}\right)\mathcal{G}_{00}^{L}+\frac{4}{t}\partial_{i}\mathcal{G}_{i0}^{L}+\frac{4}{t^{2}}\mathcal{G}_{00}^{L}+\frac{2}{t^{2}}\mathcal{G}_{ii}^{L}\right\} .\label{box_G_00_1}
\end{align}
Putting (\ref{G_00_gauge}) and its derivatives, the derivative of
(\ref{G_0j_gauge}) and the trace of (\ref{G_jk_gauge}), which are
\begin{align}
\partial_{0}\mathcal{G}_{00}^{L} & =-\frac{1}{2\ell}\nabla^{2}f-\frac{t}{2\ell}\nabla^{2}\dot{f},\quad\partial_{0}^{2}\mathcal{G}_{00}^{L}=-\frac{1}{\ell}\nabla^{2}\dot{f}-\frac{t}{2\ell}\nabla^{2}\ddot{f},\nonumber \\
\partial_{i}\mathcal{G}_{i0}^{L} & =-\frac{t}{2\ell}\nabla^{2}p,\quad\mathcal{G}_{ii}^{L}=-\frac{t}{2\ell}\left(\dot{p}+q\right),\label{der_trace_G_00}
\end{align}
into (\ref{box_G_00_1}) yields
\begin{align}
\Box\mathcal{G}_{00}^{L} & =\frac{t^{3}}{2\ell^{3}}\left(\frac{2}{t}\nabla^{2}\dot{f}+\nabla^{2}\ddot{f}+\frac{3}{t^{2}}\nabla^{2}f+\frac{3}{t}\nabla^{2}\dot{f}-\nabla^{2}\nabla^{2}f\right)\nonumber \\
 & \phantom{=}+\frac{t^{3}}{2\ell^{3}}\left(-\frac{4}{t}\nabla^{2}p-\frac{4}{t^{2}}\nabla^{2}f-\frac{2}{t^{2}}\left(\dot{p}+q\right)\right),\nonumber \\
 & =\frac{t^{3}}{2\ell^{3}}\left(\frac{5}{t}\nabla^{2}\dot{f}+\nabla^{2}\ddot{f}-\frac{3}{t^{2}}\nabla^{2}f-\nabla^{2}\nabla^{2}f\right)\nonumber \\
 & \phantom{=}+\frac{t^{3}}{2\ell^{3}}\left(-\frac{4}{t}\nabla^{2}p+\frac{2}{t^{2}}\nabla^{2}f-\frac{2}{t^{2}}\left(\dot{p}+q\right)\right),\nonumber \\
 & =\frac{t^{3}}{2\ell^{3}}\left(\frac{5}{t}\nabla^{2}\dot{f}+\nabla^{2}\ddot{f}-\frac{3}{t^{2}}\nabla^{2}f-\nabla^{2}\nabla^{2}f\right)\nonumber \\
 & \phantom{=}+\frac{t^{3}}{2\ell^{3}}\left(-\frac{4}{t}\nabla^{2}p-\frac{2}{t^{2}}\left[\left(\dot{p}+q\right)-\nabla^{2}f\right]\right),\label{box_G_00_2}
\end{align}
and from Bianchi identity (\ref{Bianchi_identity_gauge1}) the last
term is $\frac{\ell^{3}}{t^{3}}R_{L}$ and the final result becomes
\begin{align}
\Box\mathcal{G}_{00}^{L} & =\frac{t^{3}}{2\ell^{3}}\left(\nabla^{2}\ddot{f}+\frac{5}{t}\nabla^{2}\dot{f}-\nabla^{2}\nabla^{2}f\right)\nonumber \\
 & \phantom{=}-\frac{t^{3}}{2\ell^{3}}\left(\frac{4}{t}\nabla^{2}p+\frac{3}{t^{2}}\nabla^{2}f+\frac{2\ell^{3}}{t^{5}}R_{L}\right).\label{box_G_00}
\end{align}
For the flat spacetime case (\ref{box_G_00}) yields
\begin{equation}
\Box\mathcal{G}_{00}^{L}=\frac{1}{2}\left(\nabla^{2}\ddot{f}-\nabla^{2}\nabla^{2}f\right)\label{box_G_00_flat}
\end{equation}

\subsubsection{The $\Box\mathcal{G}_{k0}^{L}$ Term}

Let us continue with $\Box\mathcal{G}_{k0}^{L}$ component. Inserting
$\mu=k$ and $\nu=0$, the equation (\ref{box_G_mu_nu_dec}) becomes,
\begin{align}
\Box\mathcal{G}_{k0}^{L} & =\frac{t^{2}}{\ell^{2}}\left\{ \left(-\partial_{0}^{2}-\frac{3}{t}\partial_{0}+\partial_{i}^{2}\right)\mathcal{G}_{k0}^{L}-2\left(\Gamma_{ik}^{0}\partial_{i}\mathcal{G}_{00}^{L}+\Gamma_{i0}^{j}\partial_{i}\mathcal{G}_{kj}^{L}\right)\right.\nonumber \\
 & \phantom{=\frac{t^{2}}{\ell^{2}}}\left.-\frac{1}{t}\Gamma_{ik}^{0}\mathcal{G}_{i0}^{L}-\frac{1}{t}\Gamma_{i0}^{i}\mathcal{G}_{k0}^{L}+2\Gamma_{ik}^{0}\Gamma_{i0}^{j}\mathcal{G}_{0j}^{L}\right\} ,\nonumber \\
 & =\frac{t^{2}}{\ell^{2}}\left\{ \left(-\partial_{0}^{2}-\frac{3}{t}\partial_{0}+\partial_{i}^{2}\right)\mathcal{G}_{k0}^{L}+\frac{2}{t}\left(\partial_{k}\mathcal{G}_{00}^{L}+\partial_{i}\mathcal{G}_{ki}^{L}\right)+\frac{5}{t^{2}}\mathcal{G}_{0k}^{L}\right\} .\label{box_G_k0_1}
\end{align}
Using (\ref{G_0j_gauge}) and its derivatives, which are
\begin{align}
\partial_{0}\mathcal{G}_{k0}^{L} & =-\frac{1}{2\ell}\left(\partial_{k}p+\epsilon_{kn}\partial_{n}\sigma\right)-\frac{t}{2\ell}\left(\partial_{k}\dot{p}+\epsilon_{kn}\partial_{n}\dot{\sigma}\right),\nonumber \\
\partial_{0}^{2}\mathcal{G}_{k0}^{L} & =-\frac{1}{\ell}\left(\partial_{k}\dot{p}+\epsilon_{kn}\partial_{n}\dot{\sigma}\right)-\frac{t}{2\ell}\left(\partial_{k}\ddot{p}+\epsilon_{kn}\partial_{n}\ddot{\sigma}\right),\label{der_G_k0}
\end{align}
the derivative of (\ref{G_jk_gauge}), that is 
\begin{equation}
\partial_{i}\mathcal{G}_{ik}^{L}=-\frac{t}{2\ell}\left(\partial_{k}\dot{p}+\epsilon_{kn}\partial_{n}\dot{\sigma}\right),\label{der_G_jk}
\end{equation}
and (\ref{comp_conec}) into (\ref{box_G_k0_1}) yields
\begin{align}
\frac{\ell^{2}}{t^{2}}\Box\mathcal{G}_{k0}^{L} & =\frac{1}{\ell}\left(\partial_{k}\dot{p}+\epsilon_{kn}\partial_{n}\dot{\sigma}\right)+\frac{t}{2\ell}\left(\partial_{k}\ddot{p}+\epsilon_{kn}\partial_{n}\ddot{\sigma}\right)\nonumber \\
 & \phantom{=}+\frac{3}{t}\frac{1}{2\ell}\left(\partial_{k}p+\epsilon_{kn}\partial_{n}\sigma\right)+\frac{3}{t}\frac{t}{2\ell}\left(\partial_{k}\dot{p}+\epsilon_{kn}\partial_{n}\dot{\sigma}\right)\nonumber \\
 & \phantom{=}-\frac{t}{2\ell}\left(\partial_{k}\partial_{i}^{2}p+\epsilon_{kn}\partial_{n}\partial_{i}^{2}\sigma\right)\label{box_G_k0_2}\\
 & \phantom{=}+\frac{2}{t}\left(-\frac{t}{2\ell}\partial_{k}\nabla^{2}f-\frac{t}{2\ell}\left(\partial_{k}\dot{p}+\epsilon_{kn}\partial_{n}\dot{\sigma}\right)\right)\nonumber \\
 & \phantom{=}-\frac{5}{t^{2}}\frac{t}{2\ell}\left(\partial_{k}p+\epsilon_{kn}\partial_{n}\sigma\right),\nonumber 
\end{align}
and 
\begin{align}
\Box\mathcal{G}_{k0}^{L} & =\frac{t^{3}}{2\ell^{3}}\partial_{k}\left(\ddot{p}+\frac{3}{t}\dot{p}-\nabla^{2}p-\frac{2}{t^{2}}p-\frac{2}{t}\nabla^{2}f\right)\nonumber \\
 & \phantom{=}+\frac{t^{3}}{2\ell^{3}}\epsilon_{kn}\partial_{n}\left(\ddot{\sigma}+\frac{3}{t}\dot{\sigma}-\nabla^{2}\sigma-\frac{2}{t^{2}}\sigma\right).\label{box_G_k0}
\end{align}
The flat spacetime version of (\ref{box_G_k0}) is 
\begin{equation}
\Box\mathcal{G}_{k0}^{L}=\frac{1}{2}\partial_{k}\left(\ddot{p}-\nabla^{2}p\right)+\frac{1}{2}\epsilon_{kn}\partial_{n}\left(\ddot{\sigma}-\nabla^{2}\sigma\right).\label{box_G_k0_flat}
\end{equation}

\subsubsection{The $\Box\mathcal{G}_{mn}^{L}$ Term}

The last component can be calculated in the same way. The indices
are renamed in (\ref{box_G_mu_nu_dec}) as $\mu=m$ and $\nu=n$.
After using (\ref{comp_conec})
\begin{align}
\frac{\ell^{2}}{t^{2}}\Box\mathcal{G}_{mn}^{L} & =\left(-\partial_{0}^{2}-\frac{3}{t}\partial_{0}+\partial_{i}^{2}\right)\mathcal{G}_{mn}^{L}+\frac{2}{t}\left(\partial_{m}\mathcal{G}_{0n}^{L}+\partial_{n}\mathcal{G}_{m0}^{L}\right)\nonumber \\
 & \phantom{=}+\frac{1}{t^{2}}\mathcal{G}_{mn}^{L}+\frac{1}{t^{2}}\mathcal{G}_{mn}^{L}-\frac{2}{t}\delta_{mn}\mathcal{G}_{00}^{L}.\label{box_G_mn_1}
\end{align}
After calculating the derivatives and doing suitable cancellations
the final answer comes out as 
\begin{align}
\frac{2\ell^{3}}{t^{3}}\Box\mathcal{G}_{mn}^{L} & =\left(\delta_{mn}+\hat{\partial}_{m}\hat{\partial}_{n}\right)\left(\ddot{q}+\frac{5}{t}\dot{q}+\frac{1}{t^{2}}q-\nabla^{2}q-\frac{2}{t^{2}}\nabla^{2}f\right)\nonumber \\
 & \phantom{=}-\hat{\partial}_{m}\hat{\partial}_{n}\left(\dddot{p}+\frac{5}{t}\ddot{p}+\frac{1}{t^{2}}\dot{p}-\nabla^{2}\dot{p}-\frac{4}{t}\nabla^{2}p-\frac{2}{t^{2}}\nabla^{2}f\right)\nonumber \\
 & \phantom{=}-\left(\epsilon_{mk}\hat{\partial}_{k}\hat{\partial}_{n}+\epsilon_{nk}\hat{\partial}_{k}\hat{\partial}_{m}\right)\left(\dddot{\sigma}+\frac{5}{t}\ddot{\sigma}+\frac{1}{t^{2}}\dot{\sigma}-\nabla^{2}\dot{\sigma}-\frac{2}{t}\nabla^{2}\sigma\right).\label{box_G_mn}
\end{align}
In the flat spacetime limit (\ref{box_G_mn}) becomes
\begin{align}
2\Box\mathcal{G}_{mn}^{L} & =\left(\delta_{mn}+\hat{\partial}_{m}\hat{\partial}_{n}\right)\left(\ddot{q}-\nabla^{2}q\right)-\hat{\partial}_{m}\hat{\partial}_{n}\left(\dddot{p}-\nabla^{2}\dot{p}\right)\nonumber \\
 & \phantom{=}-\left(\epsilon_{mk}\hat{\partial}_{k}\hat{\partial}_{n}+\epsilon_{nk}\hat{\partial}_{k}\hat{\partial}_{m}\right)\left(\dddot{\sigma}-\nabla^{2}\dot{\sigma}\right)\label{box_G_mn_flat}
\end{align}
With (\ref{box_G_00}), (\ref{box_G_k0}), (\ref{box_G_mn}) and the
results that are computed for $\mathcal{G}_{\mu\nu}^{L}$ and $R_{L}$
are enough to write the gauge-invariant form of the equations of motion.


\begin{thebibliography}{References}
\bibitem{Einstein} A.~Einstein,~Annalen~Phys.~\textbf{\noun{82}},
769-822,~(1916).

\bibitem{Grosser} M.~Grosser, \emph{{}``The Discovery of Neptun'',
}Harvard Uni. Press, Cambridge,~(1962).

\bibitem{Goldhaber} A.~S.~Goldhaber and M.~M.~Nieto, Rev.~Mod.~Phys.~\textbf{\noun{82}},
939,~(2010).

\bibitem{Verrier} U.~J.~J.~le Verrier, Ann.~Observ.~Imp.~Paris~\textbf{V},~1,~(1859).

\bibitem{Weinberg3} S.~Weinberg,~{}``\emph{Gravitation and Cosmology:}~\emph{
Principles and Applications} of \emph{The General Theory of Relativity}'',~Wiley,~New
York,~(1972).

\bibitem{Inverno}R.~d'Inverno,~{}``\emph{Introducing Einstein's
Relativity}'',~Clarendon Press,~Oxford,~(1995).

\bibitem{Carroll} S.~ Carroll, \emph{{}``Spacetime and Geometry'',}~Addison-Wesley,~(2004).

\bibitem{Risse} A.~G.~Risse \emph{et al.},~Astron.~J.~\textbf{116}~1009,~(1998).

\bibitem{Perlmutter} S.~Perlmutter \emph{et al.},~Astrophys.~J.~\textbf{517}~565,~(1999).

\bibitem{Hinterbichler} K.~Hinterbichler, {}``\emph{Theoretical
Aspects of Massive Gravity}'',~arXiv:~1105.3735~{[}hep-th{]},~(2011).

\bibitem{Rubakov} V.~A.~Rubakov and P.~G.~Tinyakov, Phys.Usp.~\textbf{51}
759,~(2008).

\bibitem{Weinberg} S.~Weinberg,~Rev.~Mod.~Phys.~\textbf{61}
1,~(1989).

\bibitem{Weinberg1} S.~Weinberg,~{}``\emph{The Quantum Theory
of Fields. Vol. 1: Foundations}''.~Cambridge,~UK:~Univ.~Press,\\~(1995).

\bibitem{Gupta} S.~N.~Gupta,~Phys.~Rev.~\textbf{96}~1683,~(1954).

\bibitem{Kraichnan} R.~H.~Kraichnan,~Phys.~Rev.~\textbf{98}~1118,~(1955).

\bibitem{Weinberg2} S.~Weinberg,~Phys.~Rev.~\textbf{138}~B988,~(1965).

\bibitem{Deser1} S.~Deser,~Gen.~Rel.~Grav~\textbf{1}~9,~(1970).

\bibitem{Boulware} D.~G.~Boulware and S.~Deser,~Ann.~Phys.~\textbf{89}~193,~(1975).

\bibitem{Fang} J.~Fang and C.~Fronsdal,~J.~Math.~Phys.~\textbf{20}~2264,~(1979).

\bibitem{Wald} R.~M.~Wald,~Phys.~Rev.~\textbf{D33}~3613,~(1986).

\bibitem{Stelle1} K.S.~Stelle, Phys. Rev. \textbf{D16}, 953,~(1977).

\bibitem{Stelle2} K.S.~Stelle, Gen.~Rel.~Grav.~\textbf{9}~353,~(1978).

\bibitem{Pauli1} W.~Pauli and M.~Fierz,~Helv.~Phys.~Acta~\textbf{12}~297,~(1939).

\bibitem{Pauli2} W.~Pauli and M.~Fierz,~Proc.~Roy.~Soc.~London~\textbf{A}~\textbf{173}~211,~(1939).

\bibitem{vanDam} H.~van~Dam and M.~J.~G.~Veltman,~Nucl.~Phys.~\textbf{B22}~397,~(1970).

\bibitem{Zakharov} V.~I.~Zakharov,~JETP~Lett.~\textbf{12},~312,~(1970).

\bibitem{Gullu1} \.{I}.~Gullu and B.~Tekin,~Phys.~Rev.~\textbf{D80}~064033,~(2009). 

\bibitem{Gullu2} \.{I}.~Gullu,~T.~Ç.~\c{S}i\c{s}man and B.~Tekin,~Phys.~Rev.~\textbf{D81}~104017,~(2010). 

\bibitem{higuchi} A.~Higuchi, 
 Nucl.\ Phys.\ B \textbf{282}, 397,~(1987). 

\bibitem{Porrati1} M.~Porrati, 
 Phys.\ Lett.\ B \textbf{498}, 92,~(2001).

\bibitem{kogan} I.~I.~Kogan, S.~Mouslopoulos and A.~Papazoglou,
 Phys.\ Lett.\ B \textbf{503},~173,~ (2001).

\bibitem{vainshtein} A.~Vainshtein, 
 Surveys High Energ.\ Phys.\ \textbf{20},~5,~(2006).

\bibitem{Djt1} S.~Deser, R.~Jackiw and S.~Templeton, 
Annals Phys.\ \textbf{140},~372,~(1982).

\bibitem{Djt2} S.~Deser, R.~Jackiw and S.~Templeton 
Phys.\ Rev.\ Lett.\ \textbf{48},~975,~(1982).

\bibitem{Bht1} E.~A.~Bergshoeff, O.~Hohm and P.~K.~Townsend,
Phys.~Rev.~Lett.~\textbf{102},~201301,~(2009).

\bibitem{Bht2} E.~A.~Bergshoeff, O.~Hohm and P.~K.~Townsend,~Phys.\ Rev.\ D
\textbf{79},~124042,~(2009).

\bibitem{Clement1} G.~Clement, 
 Class.\ Quant.\ Grav.\ \textbf{26},~105015,~(2009).

\bibitem{Sun1} Y.~Liu and Y.~W.~Sun, Phys.\ Rev.\ D \textbf{79},~126001,~(2009).

\bibitem{Sun2} Y.~Liu and Y.~W.~Sun, JHEP \textbf{0905},~039,~(2009).

\bibitem{Giribet1} E.~Ayon-Beato, G.~Giribet and M.~Hassaine,
JHEP \textbf{0905},~029,~(2009).

\bibitem{Oliva} J.~Oliva, D.~Tempo and R.~Troncoso,~JHEP \textbf{0907},~011,~(2009).

\bibitem{gurses} M.~Gurses,~Class.\ Quant.\ Grav.\ \textbf{27},~205018,~(2010).

\bibitem{Aliev_Ahmedov1} H.~Ahmedov and A.~N.~Aliev,~Phys.~Lett.~B
\textbf{694},~143,~(2010).

\bibitem{Aliev_Ahmedov2} H.~Ahmedov and A.~N.~Aliev,~Phys.~Rev.~Lett.~\textbf{106},~021301,~(2011).

\bibitem{Aliev_Ahmedov3} H.~Ahmedov and A.~N.~Aliev,~Phys.~Lett.~D
\textbf{83},~084032,~(2011).

\bibitem{Gurses_Sisman} M.~Gurses,~T.~Ç.~\c{S}i\c{s}man and B.~Tekin,~{}``\emph{Some
Exact Solutions of All f(Ricci) Theories in Three Dimensions}'',~arXiv:1112.6346~{[}hep-th{]},~(2011). 

\bibitem{Suat} S.~ Dengiz,~B.~Tekin,~Phys.~Rev.~\textbf{D84}~024033,~(2011).

\bibitem{Suat_Tanhayi} M.~R.~Tanhayi,~S.~ Dengiz,~B.~Tekin,~{}``\emph{Unitarity
of Weyl-Invariant New Massive Gravity and Generation of Graviton Mass
via Symmetry Breaking}'',~arXiv:1112.2338~{[}hep-th{]},~(2011).

\bibitem{Lu} H.~Lu,~C.~N.~Pope,~Phys.~Rev.~Lett.~\textbf{106}~181302~(2011). 

\bibitem{Sisman}S.~Deser,~H.~Liu,~H.~Lu,~C.~N.~Pope,~T.~C.~Sisman,~B.~Tekin,~Phys.~Rev.~D~\textbf{83}~061502,~(2011). 

\bibitem{Porrati2} M.~Porrati,~M.~M.~Roberts,~Phys.~Rev.~\textbf{D84}~024013,~(2011). 

\bibitem{Abbott} L.~F.~Abbott and S.~Deser,~Nucl.~Phys.~\textbf{B195},~76,~(1982).

\bibitem{Witten} E.~Witten,~{}``\emph{Quantum Gravity in de Sitter
Space}'',~arXiv:0106109~{[}hep-th{]},~(2001).

\bibitem{Strominger}A.~Strominger,~J.~High Energy Phys.~\textbf{10},~034,~(2001).

\bibitem{Griffiths} D.~Griffiths,~{}``\emph{Introduction to Elementary
Particles}'',~WILEY-VCH,~(2008).

\bibitem{Peskin} M.~E.~Peskin and D.~V.~Schroeder,~{}``\emph{An
Introduction to Quantum Field Theory}'',~Addison-Wesley,~(1995).

\bibitem{Reinhardt} G.~Reinhardt,~{}``\emph{Field Quantization}'',~Springer,~(1996).

\bibitem{Pais} A.~Pais and G.~E.~Uhlenbeck, Phys.\ Rev.\ \textbf{79},~145,~(1950).

\bibitem{Sarioglu} O.~Sarioglu and B.~Tekin, Class.\ Quant.\ Grav.\ \textbf{23},~7541,~(2006).

\bibitem{Ostrogradski} M.~Ostrogradski,~Mem.~Act.~St.~Petersbourg~\textbf{VI}~\textbf{4},~385,~(1850).

\bibitem{Simon} J.~Z.~Simon,~Phys.~Rev.~\textbf{D41},~3720,~(1990). 

\bibitem{nakason} M.~Nakasone and I.~Oda,~Prog.~Theor.~Phys.~\textbf{121}~1389,~(2009).

\bibitem{oda} I.~Oda, JHEP \textbf{0905},~064,~(2009).

\bibitem{nakasonoda} M.~Nakasone and I.~Oda, Phys. Rev. \textbf{D
79},~104012,~(2009).

\bibitem{Deser2}S.~Deser, Phys.\ Rev.\ Lett.\ \textbf{103},~101302,~(2009).

\bibitem{vas1} M.~A.~Vasiliev, 
 Nucl.\ Phys.\ B \textbf{616}, 106 (2001) {[}Erratum-ibid.\ B \textbf{652},~407,~(2003){]}.

\bibitem{vas2}X.~Bekaert,~S.~Cnockaert,~C.~Iazeolla and M.~A.~Vasiliev,~{}``Nonlinear
higher spin theories in various dimensions,''~arXiv:~0503128~{[}hep-th{]},~(2005).

\bibitem{dest} S.~Deser and B.~Tekin, 
 Class.\ Quant.\ Grav.\ \textbf{19},~L97,~(2002).

\bibitem{deserwaldron2} S.~Deser and A.~Waldron, 
 Phys.\ Lett.\ B \textbf{501},~134,~(2001).

\bibitem{duff} F.~A.~Dilkes, M.~J.~Duff, J.~T.~Liu and H.~Sati,
 Phys.\ Rev.\ Lett.\ \textbf{87},~041301,~ (2001).

\bibitem{bouldeser} D.~G.~Boulware and S.~Deser, 
 Phys.\ Rev.\ D \textbf{6},~3368,~(1972).

\bibitem{dt2} S.~Deser and B.~Tekin, 
Phys.\ Rev.\ D \textbf{67},~084009,~(2003).

\bibitem{deserwaldron} S.~Deser and A.~Waldron, 
 Nucl.\ Phys.\ B \textbf{607},~577,~(2001).

\bibitem{Arfken}G.~B.~Arfken and H.~J.~Weber,~{}``\emph{Mathematical
Methods for Physicists}'',~Elsevier~Academic~Press,~(2005).

\bibitem{andringa}R.~Andringa,~E.~A.~Bergshoeff,~M.~de~Roo,~O.~Hohm,~E.~Sezgin~and~P.~K.~Townsend,~Class.~Quant.\\~Grav.~\textbf{27},~025010,~(2010).

\bibitem{woodard} R.~P.~Woodard,~Lect.\ Notes Phys.\ \textbf{720},~403,~(2007).

\bibitem{smilga} A.~V.~Smilga,~Phys.\ Lett.\ B \textbf{632},~433,~(2006).

\bibitem{dt3} S.~Deser and B.~Tekin,~Class.\ Quant.\ Grav.\ \textbf{20},~4877,~(2003).

\bibitem{dt1}S.~Deser and B.~Tekin, 
Phys.\ Rev.\ Lett.\ \textbf{89},~101101,~(2002).

\bibitem{hindawi}A.~Hindawi,~B.~A.~Ovrut and D.~Waldram,~Phys.\ Rev.\ D
\textbf{53},~5583,~(1996).\end{thebibliography}
\end{document}